\documentclass[11pt]{cernrepdbg}
\usepackage{graphicx}
\usepackage{epsfig}
\usepackage{color}
\usepackage{cite}
\usepackage{dcolumn}
\usepackage{bm}
\usepackage{axodraw,rotating}
\usepackage{feynarts}
\usepackage{latexsym}
\usepackage{amsmath}
\usepackage{amssymb}
\usepackage{amsfonts}
\usepackage{amsbsy}

\makeatletter
\renewcommand\thesubsubsection{\thesubsection.\@arabic\c@subsubsection}
\renewcommand*\l@section[2]{%
  \ifnum \c@tocdepth >\z@
    \addpenalty\@secpenalty
    \addvspace{0.9em \@plus\p@}%
    \setlength\@tempdima{1.5em}%
    \begingroup
      \parindent \z@ \rightskip \@pnumwidth
      \parfillskip -\@pnumwidth
      \leavevmode \bfseries
      \advance\leftskip\@tempdima
      \hskip -\leftskip
      #1\nobreak\hfil \nobreak\hb@xt@\@pnumwidth{\hss #2}\par
    \endgroup
  \fi}
\makeatother

\begin{document}
\bibliographystyle{lesHouches}

\title{\centering{THE NLO MULTILEG WORKING GROUP:\\
          \textbf{Summary Report}}}

\author{\underline{Convenors}:
Z.~Bern$^1$, 
S.~Dittmaier$^2$,  
L.~Dixon$^3$, 
G.~Heinrich$^4$, 
J.~Huston$^5$,  
B.~Kersevan$^{6,7}$,
Z.~Kunszt$^8$,
D.E.~Soper$^{9,10}$ 
\\
\underline{Contributing authors}:
Z.~Bern$^1$, 
C.~Bernicot$^{11}$, 
T.~Binoth$^{12}$,
F.~Boudjema$^{11}$,
R.~Britto$^{13}$,
J.~Campbell$^{14}$,
M.~Czakon$^{15}$,
A.~Denner$^{16}$,
G.~Dissertori$^8$,
S.~Dittmaier$^2$,  
L.~Dixon$^3$, 
G.~Duplan\v ci\' c$^{17}$,
R.K.~Ellis$^{18}$,
R.~Frederix$^{19}$,
T.~Gehrmann$^{20}$,
A.~Gehrmann--De Ridder$^8$,
W.T.~Giele$^{18}$,
E.W.N.~Glover$^4$,
J.P.~Guillet$^{11}$,
G.~Heinrich$^4$,
S.~Kallweit$^2$,
S.~Karg$^{21}$,
N.~Kauer$^{15}$,
D.A.~Kosower$^{22}$,
F.~Krauss$^4$,
Z.~Kunszt$^8$,
N.D.~Le$^{11}$, 
P.~Mastrolia$^{10,20}$,
A.~Mitov$^{23}$,
S.~Moch$^{24}$,
S.~Odaka$^{25}$,
G.~Ossola$^{26}$,
C.G.~Papadopoulos$^{26}$,
E.~Pilon$^{11}$,
R.~Pittau$^{27,28}$,
T.~Reiter$^{12}$,
G.~Sanguinetti$^{11}$,
S.~Schumann$^{12}$,
C.~Schwinn$^{21}$,
P.Z.~Skands$^{10}$,
D.E.~Soper$^{9,10}$,
H.~Stenzel$^{29}$,
P.~Uwer$^{30}$,
S.~Weinzierl$^{31}$,
G.~Zanderighi$^{32}$
\\
\mbox{} }
\institute{\centering{\small
$^1$ Department of Physics and Astronomy, 
UCLA, Los Angeles, CA 90095-1547, USA\\ 
$^2$ Max-Planck-Institut f\"ur Physik (Werner-Heisenberg-Institut),
  D-80805 M\"unchen, Germany\\
$^3$ Stanford Linear Accelerator Center, 
Stanford University, Stanford, CA 94309, USA\\ 
$^4$ Institute of Particle Physics Phenomenology,
        University of Durham, Durham, DH1 3LE, UK\\
$^5$ Michigan State University, East Lansing, Michigan 48824, USA\\
$^6$ Jozef Stefan Institute, Jamova 39, SI-1000 Ljubljana, Slovenia\\ 
$^7$ Faculty of Mathematicsand Physics, University of Ljubljana, 
Jadranska 19a, SI-1000 Ljubljana, Slovenia\\ 
$^8$ Institute for Theoretical Physics, ETH, CH-8093 Zurich, Switzerland\\
$^9$ University of Oregon, Eugene, Oregon 97403, USA\\
$^{10}$ CERN, CH-1211 Geneva 23, Switzerland\\
$^{11}$ LAPTH, CNRS and Universit{\'e} de Savoie, 
B.P.~110, Annecy-le-Vieux 74951, France\\
$^{12}$ The University of Edinburgh, J.C.M.B., The King's Buildings,
Edinburgh EH9 3JZ, Scotland\\
$^{13}$ Institute for Theoretical Physics, University of Amsterdam,
1018 XE Amsterdam, The Netherlands\\
$^{14}$ Department of Physics and Astronomy, University of Glasgow, Glasgow
G12 8QQ, UK\\
$^{15}$ Institut f\"ur Theoretische Physik und Astrophysik, 
Universit\"at W\"urzburg, D-97074 W\"urzburg, Germany\\
$^{16}$ Paul Scherrer Institut, W\"urenlingen und Villigen, 
CH-5232 Villigen PSI, Switzerland\\
$^{17}$ Theoretical Physics Division, Rudjer Boskovic Institute,
HR-10002 Zagreb, Croatia\\
$^{18}$ Theoretical Physics, Fermilab MS106, Batavia, IL 60510, USA\\
$^{19}$ CP3, Universit\'e catholique de Louvain, 
B-1348 Louvain-la-Neuve, Belgium\\
$^{20}$ Institut f\"ur Theoretische Physik, Universit\"at Z\"urich,
CH-8057 Z\"urich, Switzerland\\
$^{21}$ Institute for Theoretical Physics E, RWTH Aachen, D-52056 Aachen, 
Germany \\
$^{22}$ Institut de Physique Theorique, CEA-Saclay
F--91191 Gif-sur-Yvette cedex, France\\
$^{23}$ Department of Mathematical Sciences, University of
Liverpool, Liverpool L69 3BX, United Kingdom\\
$^{24}$ Deutsches Elektronensynchrotron DESY,
Platanenallee 6, D-15738 Zeuthen, Germany\\
$^{25}$ High Energy Accelerator Research Organization (KEK),
1-1 Oho, Tsukuba, Ibaraki 305-0801, Japan\\
$^{26}$ Institute of Nuclear Physics, NCSR Demokritos,
15310 Athens, Greece \\
$^{27}$ Departamento de F\'{i}sica Te\'orica y del Cosmos,
Centro Andaluz de F\'{i}sica de Part\'{i}culas Elementales (CAFPE),
Universidad de Granada, E-18071 Granada, Spain \\
$^{28}$ Dipartimento di Fisica Teorica, Universit\` a di Torino
and INFN sezione di Torino, 
Torino, Italy\\
$^{29}$ II.~Physikalisches Institut, Justus-Liebig Universit\"at Giessen,
D-35392 Giessen, Germany\\
$^{30}$ Institut f\"ur Theoretische Teilchenphysik,
Universit\"at Karlsruhe, D-76128 Karlsruhe, Germany\\
$^{31}$ Institut f{\"u}r Physik, Universit{\"a}t Mainz,
D-55099 Mainz, Germany\\
$^{32}$ The Rudolf Peierls Centre for Theoretical Physics, University of 
Oxford, UK\\
}}
 
\maketitle

 \begin{center}
   \textit{Report of the NLO Multileg Working Group 
     for the Workshop ``Physics at TeV
     Colliders'', Les Houches, France, 11--29 June, 2007.  }
\end{center}
\newpage
  
\setcounter{tocdepth}{1}
\tableofcontents
\setcounter{footnote}{0}

\section[Introduction]{INTRODUCTION}



%
The LHC will be a very complex environment with most of the interesting physics signals, and their
 backgrounds, consisting of multi-parton (and lepton/photon) final states. The ATLAS and CMS
 experiments will measure these final states with negligible statistical error, even in the early
 running, and in many cases with systematic errors smaller than those achieved by the experiments at
 the Tevatron (see the contribution in these proceedings from G. Dissertori). The luminosity
uncertainty and the uncertainty in the parton distribution functions (PDFs)
can be minimized by the normalization of the physics process of interest to certain Standard Model (SM) benchmark processes, such as $W$, $Z$, and $t\overline{t}$ production. Thus, it is important to have theoretical predictions at the same or better precision as the experimental measurements. In many cases, SM backgrounds to non-SM physics can be extrapolated from background-rich to signal-rich regions, but a definite determination of the background often requires an accurate knowledge of the background cross sections.  An accurate knowledge of a cross section requires its calculation to at least  next-to-leading order (NLO). 

There are many tools for constructing basically any complex final state at the LHC at leading order (LO). When interfaced to parton shower Monte Carlo programs, such predictions can provide a qualitative prediction of both inclusive and exclusive final states. There are several different interfaces between fixed order (both LO and NLO) matrix element and parton shower Monte Carlo programs, with a benchmark comparison reported in this workshop. 

A realistic theoretical description of complex final states, though,  exists only at NLO\footnote{Unless otherwise stated, the terms LO, NLO , NNLO refer to the order in perturbative QCD only.}, with the current limit of such calculations being $2\rightarrow 3$ and $2\rightarrow 4$ processes (see below).  At LO, calculations often have large scale dependence, a sensitivity to kinematic cuts, and a poor modeling of jet structure. These deficiencies are most often remedied at NLO.   NLO parton level calculations can serve as useful benchmarks by themselves, as well as providing an even more complete event description when interfaced with parton shower Monte Carlo programs, or when resummation effects are included. For the crucial benchmark processes mentioned above 
($W$,~$Z$ and $t\overline{t}$ production), it is useful to go beyond NLO to NNLO. This has been done for $W$ and $Z$ production, including the calculation of  differential rapidity distributions, and is expected for $t\overline{t}$, $Z/\gamma$+jet and $W$+jet production in the near future. Progress towards $t\overline{t}$ is reported in the contributions from M.~Czakon, A.~Mitov and S.~Moch. 

Even at NLO, the calculation of $2\rightarrow 3$ (and $2\rightarrow 4$)  processes is extremely time- and theorist-consuming, so clear priority needs to be established for those processes most needed for the LHC. In the 2005 Les Houches proceedings, such a realistic NLO wishlist was established (see Table~\ref{tab:wishlist}).  
It is gratifying that 3 of the 8 processes (and some which were not listed, 
for example the one-loop interference between gluon fusion and weak boson fusion
in Higgs plus dijet production~\cite{Bredenstein:2008tm,Andersen:2007mp}
), 
have been calculated in the intervening two years, but daunting to know that 5 remain and a new process has been added.  As noted in the table, three groups have calculated $WW$+jet since Les Houches 2005 and a detailed comparison of the results is presented in these Les Houches proceedings. 
 In addition to the  new NLO calculation, several processes beyond NLO also have been added to the list.
 
\noindent  
The new processes that have been added are:
\begin{itemize}





\item $pp \rightarrow b\bar{b}b\bar{b}$

There are several interesting physics signatures involving two b-pairs in the final state, such as $b\bar{b}\,H(\rightarrow b\bar{b})$ and hidden valley signatures where $Z$ bosons may decay to multiple $b$-quarks. Related to this calculation is the production of 4jets, which is less interesting experimentally, but a benchmark calculation from a theoretical point of view. 
\end{itemize}
\noindent
The calculations beyond NLO added to the 2007 version of the list are: 
\begin{itemize}
\item $gg \rightarrow W^*W^*$ ${\cal O}(\alpha^2\alpha_s^3)$

This subprocess is important for understanding the backgrounds for $H\rightarrow W^{(*)} W^{(*)}$. 
\item NNLO $pp\to t\bar{t}$ 

This process is important for the use of $t\bar{t}$ production at the LHC as a precision benchmark. 

\item NNLO to weak vector boson fusion (VBF) and $Z/\gamma$+jet 

VBF production of a Higgs boson is essential for measuring the coupling of the Higgs to bosons. $Z/\gamma$+jet is an essential experimental process that is used to understand the jet energy scale. It will also be useful for PDF determination. 

\item In addition, to further reduce the theoretical uncertainty for the benchmark $W/Z$ processes, a combined NNLO QCD and NLO electroweak (EW) calculation is needed. The cross sections are known separately to NNLO QCD and to NLO EW, but a combined calculation will improve the accuracy of the result. 
\end{itemize}

\begin{table}
  \begin{center}
     \begin{tabular}{|l|l|}
\hline\hline
Process & Comments\\
($V\in\{Z,W,\gamma\}$)& \\
\hline
Calculations completed since Les Houches 2005&\\
\hline
&\\
1. $pp\to VV$jet & $WW$jet completed by
Dittmaier/Kallweit/Uwer~\cite{Dittmaier:2007th};\\
 &
Campbell/Ellis/Zanderighi~\cite{Campbell:2007ev}\\ 
 &
and Binoth/Karg/Kauer/Sanguinetti (in progress)\\
2. $pp \to$ Higgs+2jets & NLO QCD to the $gg$ channel \\
& completed by Campbell/Ellis/Zanderighi~\cite{Campbell:2006xx};\\
& NLO QCD+EW to the VBF channel\\
& completed by Ciccolini/Denner/Dittmaier~\cite{Ciccolini:2007jr,Ciccolini:2007ec}\\
3. $pp\to V\,V\,V$ & $ZZZ$ completed by Lazopoulos/Melnikov/Petriello~\cite{Lazopoulos:2007ix} \\
 & and $WWZ$ by Hankele/Zeppenfeld~\cite{Hankele:2007sb}\\
&\\
 \hline 
Calculations remaining from Les Houches 2005&\\
\hline
&\\
4. $pp\to t\bar{t}\,b\bar{b}$ &  relevant for $t\bar{t}H$ \\
5. $pp\to t\bar{t}$+2jets & relevant for $t\bar{t}H$  \\ 
6. $pp\to VV\,b\bar{b}$,  & relevant for VBF $\rightarrow H\rightarrow VV$,~$t\bar{t}H$ \\
7. $pp\to VV$+2jets  & relevant for VBF $\rightarrow H\rightarrow VV$ \\
& VBF contributions calculated by \\
& (Bozzi/)J\"ager/Oleari/Zeppenfeld~\cite{Jager:2006zc,Jager:2006cp,Bozzi:2007ur}
\\
8. $pp\to V$+3jets & various new physics signatures\\
&\\
\hline
NLO calculations added to list in 2007&\\
\hline
&\\
9. $pp\to b\bar{b}b\bar{b}$ & Higgs and new physics signatures \\
&\\
\hline
Calculations beyond NLO added in 2007&\\
\hline
&\\
10. $gg\to W^*W^*$ ${\cal O}(\alpha^2\alpha_s^3)$& backgrounds to Higgs\\
11. NNLO $pp\to t\bar{t}$ & normalization of a benchmark process\\
12. NNLO to VBF and $Z/\gamma$+jet  & Higgs couplings and SM benchmark\\
&\\
\hline 
Calculations including electroweak effects&\\
\hline
&\\
13. NNLO QCD+NLO EW for $W/Z$ & precision calculation of a SM benchmark\\
&\\
\hline
\hline
\end{tabular}
\end{center}
\caption{The updated experimenter's wishlist for LHC processes 
\label{tab:wishlist}}
\end{table}
It is also daunting to realize that all of the three finished calculations from the 2005 list remain private code. To be truly useful, such calculations need to be available in programs accessible to experimenters. Most useful is if the event 4-vectors and event weight outputs can be stored in ROOT n-tuple format, so that experimental analysis cuts can be easily applied in a manner similar to what is used for the actual data, and so that results do not have to be re-generated if the analysis cuts change. In such a format, it is also easy to store not only the nominal event weight, generated with the central PDF of a NLO set, but also the weights for the set of error PDFs as well. In such a manner, the PDF uncertainty for any event configuration can be easily established, at the expense of a larger n-tuple size. Such a modification is being carried out for the MCFM program.

The calculation of complex multi-parton final states results in the generation of many 
subtraction terms for soft and/or collinear real radiation (e.g. Catani-Seymour dipole
or antenna subtraction terms), 
and each of these in turn requires a counter-event to be generated for the Monte Carlo evaluation of the matrix element. Thus, for example, in MCFM for 
$W$+2jets (and for Higgs+2jets as well), there are 24 counter-term events for each real event.  The net result is the requirement of a large amount of CPU time for computing such cross sections, and the need for many GB of disk space for storing the results in ntuples. These requirements will become even more extreme as the complexity of the calculations increases. 

Although most of the NLO calculations for multi-particle production so far are 
private code tailored to the particular process at hand, there is 
a clear effort towards more automatisation and making results available 
to the community. 
Several agreements have been made during the workshop to facilitate comparisons 
and to make at least certain building blocks entering NLO calculations
publicly available:
\begin{itemize}
\item Les Houches accord on master integrals: the aim is 
to have a library of one-loop integrals, finite as well as  divergent ones, 
which can be used by anybody using a method which requires scalar master integrals. 
It has been agreed that the format for the labelling of the integrals respectively 
their arguments should follow the LoopTools~\cite{vanOldenborgh:1990yc,Hahn:1998yk} 
conventions, as the infrared  finite integrals are already available in LoopTools. 
The infrared divergent ones recently have been classified and listed in~\cite{Ellis:2007qk} 
and can be found in analytic form at {\tt http://qcdloop.fnal.gov}.
The final aim is a webpage containing
\medskip
\begin{enumerate}
\item a collection of scalar one-loop integrals in analytic form,
\item benchmark points and comments which kinematic regions have been tested,
\item code to calculate the Laurent series of each integral at points 
specified by the user,
\item ideally also various codes for the reduction to master integrals.
\end{enumerate}
\medskip
This webpage is in Wiki format, such that contributions can be added easily.
The location of the webpage is\\
 {\tt http://www.ippp.dur.ac.uk/LoopForge/index.php/Main\_Page},
and input is eagerly awaited.
\item If an amplitude is published in an analytic form, numerical values at 
some benchmark points  should be given to facilitate cross-checks by other groups. 
\end{itemize}

All of the 2005 NLO wishlist processes that have been completed to date
relied on traditional Feynman diagrams for the loop amplitudes. On the
other hand, as the complexity of the final-states grows further, it may 
prove necessary to adopt as well new approaches and methods. At the 2007 
session of Les Houches, several such approaches were under discussion 
and development, primarily those based on the general analytic structure 
of amplitudes.  These methods include recursive techniques at both tree 
and loop level; the use of (generalized) unitarity in four dimensions,
and in $4-2\epsilon$ dimensions (the latter in the context of dimensional
regularization); and automated solutions for coefficients of one-loop 
integrals, which is also connected with generalized unitarity. Complex 
final states possess intricate kinematic regions in which either the 
amplitude itself becomes singular, or a particular representation of it 
becomes numerically unstable. The general identification of such regions,
and methods for dealing with potential instabilities, are also areas of
active interest, which are not unrelated to the use of analyticity to 
construct loop amplitudes.

Even with the rapid progress we have been seeing in the last few years, there are NLO cross sections of interest that will not be completed in a timely manner for the LHC. One question is whether we can provide any approximations/estimates of the uncalculated NLO matrix elements based on experiences with simpler calculations. Table~\ref{tab:kfac} shows the K-factors (NLO/LO) tabulated for some important processes at the Tevatron and LHC. Of course, K-factors are a simplified way of presenting the effects of NLO corrections (depending on both scale choice and PDF used for example), but the table provides some interesting insights. For example, it appears that processes that involve a large color annihilation (for example $gg \rightarrow$ Higgs) tend to have large K-factors for scales typically chosen to evaluate the matrix elements. The addition of extra legs in the final state tends to result in a smaller K-factor. For example,  the K-factor for Higgs+2jets is smaller than for Higgs+1jet, which in turn is smaller than that for inclusive Higgs production. The same is true for the K-factor for $W$+2jet  being less than that for  $W$+1jet and the K-factor for $t\bar{t}$+1jet being less than that for  $t\bar{t}$.  Can we generalize this to estimate that the NLO corrections for $W$+3jets and $t\bar{t}$+2jets will be smaller still? 
\begin{table}[h]
\begin{center}
\begin{tabular}{|l|l|l|c|c|c|c|c|c|}
\hline

 & \multicolumn{2}{|l|}{Typical scales} & 
 \multicolumn{3}{|c|}{Tevatron $K$-factor} & \multicolumn{3}{|c|}{LHC $K$-factor} \\ 
  & \multicolumn{2}{|l|}{\quad} & 
 \multicolumn{3}{|c|}{} & \multicolumn{3}{|c|}{} \\
Process & $\mu_0$ & $\mu_1$ &
 ${\cal K}(\mu_0)$ & ${\cal K}(\mu_1)$ & ${\cal K}^\prime(\mu_0)$ &
 ${\cal K}(\mu_0)$ & ${\cal K}(\mu_1)$ & ${\cal K}^\prime(\mu_0)$ \\
\hline
&&&&&&&&\\
$W$                & $m_W$ & $2m_W$		    & 1.33 & 1.31 & 1.21 & 1.15 & 1.05 & 1.15 \\
$W$+1jet          & $m_W$ & $ p_T^{\rm jet}$ & 1.42 & 1.20 & 1.43 & 1.21 & 1.32 & 1.42 \\
$W$+2jets & $m_W$ & $ p_T^{\rm jet}$	    & 1.16 & 0.91 & 1.29 & 0.89 & 0.88 & 1.10 \\
$WW$+jet                & $m_W$ & $2m_W$		    & 1.19 & 1.37 & 1.26 & 1.33 & 1.40 & 1.42 \\
$t{\bar t}$        & $m_t$ & $2m_t$		    & 1.08 & 1.31 & 1.24 & 1.40 & 1.59 & 1.48 \\
$t{\bar t}$+1jet        & $m_t$ & $2m_t$		    & 1.13 & 1.43 & 1.37 & 0.97 & 1.29 & 1.10 \\
$b{\bar b}$        & $m_b$ & $2m_b$		    & 1.20 & 1.21 & 2.10 & 0.98 & 0.84 & 2.51 \\
Higgs      & $m_H$ & $ p_T^{\rm jet}$ & 2.33 & -- & 2.33 & 1.72 & -- & 2.32 \\
Higgs via VBF      & $m_H$ & $ p_T^{\rm jet}$ & 1.07 & 0.97 & 1.07 & 1.23 & 1.34 & 1.09 \\
Higgs+1jet     & $m_H$ & $ p_T^{\rm jet}$ & 2.02 & -- & 2.13 & 1.47 & -- & 1.90 \\
Higgs+2jets     & $m_H$ & $ p_T^{\rm jet}$ & -- & -- & -- & 1.15 & -- & -- \\
&&&&&&&&\\
\hline
\end{tabular}
\caption{\label{tab:kfac}$K$-factors for various processes at the Tevatron
and the LHC calculated using a selection of input parameters. In all cases, the CTEQ6M
PDF set is used at NLO. ${\cal K}$ uses the CTEQ6L1 set at leading order, whilst
${\cal K}^\prime$ uses the same set, CTEQ6M, as at NLO.
For most of the processes listed, jets satisfy the requirements $p_T>15$~GeV/c and $|\eta|<2.5$ ($5.0$) at the
Tevatron (LHC). For Higgs+1,2jets, a jet cut of 40 GeV/c and $|\eta|<4.5$ has been applied. 
A cut of $p_{T}^{\mathrm{jet}}>20~GeV/c$ has been applied for the $t\bar{t}$+jet process, and
a cut of $p_{T}^{\mathrm{jet}}>50~GeV/c$ for $WW$+jet.
In the $W$(Higgs)+2jets process the jets are separated by $\Delta R>0.52$, whilst the
VBF calculations are performed for a Higgs boson of mass $120$~GeV. In each case the value of the $K$-factor is compared at two often-used scale choices, where the
scale indicated is used for both renormalization and factorization scales.}
\end{center}
\end{table}

The dream of experimentalists is for every NLO parton level  calculation to  come packaged with a complete parton shower for the  partons produced in the NLO hard scattering process. So far, this  exists for a few not-too-complicated processes, but it is not so easy  to arrange this for each given NLO parton level calculation. To make  this process easier, it will be useful to have a very systematic  shower with a simple structure that can be matched to the structure  of the NLO calculation. Two programs discussed at the workshop, and  represented by contributions later in this section, may help. One  
would naturally match to a NLO calculation with  antenna subtractions. The other 
would naturally match to a NLO calculation with  the widely used Catani-Seymour dipole subtractions.

For many physics processes, though, we will have to continue to rely upon LO parton shower Monte Carlo programs (interfaced with exact LO matrix element calculations).  In many instances, a large part of the difference between LO and NLO predictions is the use of LO PDFs for the former and NLO PDFs for the latter. Nominally, the choice indicated above is correct, but LO PDFs can differ from their NLO counterparts by a significant amount due to the influence of DIS data on the global fits. The LO PDFs often are changed in such a manner as to lead to significant deviations of LO predictions with LO PDFs from NLO predictions with NLO PDFs, in some kinematic regions. One solution that has been discussed is the use of NLO PDFs with LO Monte Carlos. This solves the problem mentioned above, but can lead to additional problems, for example with predictions for low mass objects at the LHC. The solution adopted by several groups, and presented at this workshop, is the development of $\it modified$ LO PDFs, including the best features (for use in LO Monte Carlos) of the LO and NLO PDFs. It will be useful/important to tabulate the K-factors using these modified LO PDFs. 

For the maximal exploitation of physics, there are also requirements on the experimental side. We suggest that cross sections at the LHC should be quoted at the hadron level, and where possible with the estimated parton-to-hadron corrections, so that any theoretical prediction (parton or hadron level) can easily be compared after the fact to the archived data~\cite{Campbell:2006wx}.  
Also, the experimental data needs to be quoted only for the range of measurement, rather than extrapolated to the full cross section; for example, measurements of $W\rightarrow e \nu$ should be quoted for the range of electron transverse momentum and rapidity and of missing transverse energy actually used in the triggering and analysis, rather than performing an extrapolation to the full $W$ cross sections. Such recommendations were the exception (CDF $W$+jets) rather than the rule at the Tevatron and a clear model needs to be set for the LHC.

The structure of this report is as follows. First a review on expected cross sections and 
uncertainties at the LHC from an experimental point of view is given to set the stage. 
Then various new approaches to the calculation of tree-level and one-loop 
multi-leg amplitudes are presented, followed by a section on 
``improvements on standard techniques", with particular emphasis on 
the analysis of singularities which can create numerical instabilities 
when integrating  multi-particle one-loop amplitudes.
Section III contains various results, first a tuned comparison of different 
NLO calculations for $pp\to WW$+jet, then results 
pointing towards the $t\bar{t}$ cross section at NNLO, and finally NNLO predictions 
for hadronic event shapes in $e^+e^-$ annihilation.
The latter is not of direct relevance for the LHC, but is a benchmark calculation 
in what concerns the construction of NNLO Monte Carlo programs in the presence 
of a complicated infrared singularity structure. 
The report is closed by a section on parton showers, addressing  
the matching of  parton showers with multi-leg LO matrix elements 
as well as the matching  with partonic NLO calculations, 
which is of primordial interest at present and future TeV colliders.

%


\section[Measurements of hard processes at the LHC]
{MEASUREMENTS OF HARD PROCESSES AT THE LHC%
\protect\footnote{Contributed by: G.~Dissertori}}
{\graphicspath{{dissertori/}}

%
%
%
%
%
%
%
%
%


\subsection{Introduction}
 \label{sec:dissertori_intro}

 We  are approaching the start-up of the world's most powerful particle accelerator
 ever built. It is expected that CERN's Large Hadron Collider (LHC) will
 start its operation in 2008. Thanks to the unprecedented energies and luminosities, it
 will give particle physicists the possibility to explore the TeV energy
 range for the first time and hopefully discover new phenomena, which go beyond the
 so successful Standard Model (SM). Among the most prominent new physics scenarios 
 are the appearance of one (or several) Higgs bosons, of supersymmetric particles
 and of signatures for the existence of extra spatial dimensions.
 
However, before entering the discovery regime, considerable efforts will be
invested in the measurements of SM processes. We are sure that these have to be seen 
and thus they can serve as a proof for a working detector 
(a necessary requirement before any claim of discovery is made).
Indeed, some of the SM processes are also excellent tools to calibrate
parts of the detector. However, such measurements are also interesting in their own right. 
We will be able to challenge the SM predictions at unprecedented energy and momentum transfer scales,
by measuring cross sections and event features for minimum-bias events, jet production, 
W and Z production with their leptonic decays, as well as top quark production.
This will allow to check the validity of the Monte Carlo generators, both at the highest energy scales 
and at small momentum transfers, such as in models for the omnipresent underlying event. 
The parton distribution functions (pdfs) can be further constrained or measured for the first time in
kinematic ranges not accessible at HERA. Important tools for pdf studies will be 
jet+photon production or Drell-Yan processes. Finally, SM processes such as W/Z+jets, multi-jet and top pair 
production will be important backgrounds to a large number of searches for new physics and therefore have to
be understood in detail. 

The very early goals to be pursued by the experiments, once the first data are on tape, are
three-fold~: (a) It will be of utmost importance to commission and calibrate the detectors
in situ, with physics processes as outlined below. The trigger performance has to be understood
in as unbiased a manner as possible, by analyzing the trigger rates of minimum-bias events,
jet events for various thresholds, single and di-lepton as well as single and di-photon
events. (b) It will be necessary to measure the main SM processes and 
(c)  prepare the road for possible discoveries. It is instructive to recall the 
event statistics collected for different types of processes. 
For an integrated luminosity of $1\,\mathrm{fb}^{-1}$ per experiment, we expect about $10^7$
$\mathrm{W}\rightarrow \mathrm{e}\nu$ events on tape, a factor of ten less $\mathrm{Z}\rightarrow \mathrm{e}^+ \mathrm{e}^-$ 
and some $10^5$ $\mathrm{t}\bar{\mathrm{t}}\rightarrow \mu + X$ events. If a trigger bandwidth of about 10\%
is assumed for  QCD jets with transverse momentum $p_{\mathrm{T}} > 150$ GeV,  
$\mathrm{b}\bar{\mathrm{b}}\rightarrow \mu + X$ and minimum-bias events, 
we will write about $10^6$ events to tape, for each of these channels. Also the existence
of supersymmetric particles,
for example gluinos with $m_{\tilde{\mathrm{g}}}\approx 1$ TeV,  or a
Higgs with $m_{\mathrm{H}} \approx 130$ GeV,
would result in sizeable event statistics ($10^2 - 10^3$).
This means that the statistical uncertainties
will be negligible after a few days, for most of the physics cases. The analysis results will be dominated
by systematic uncertainties, be it the detailed understanding of the detector response, theoretical
uncertainties or the uncertainty from the luminosity measurements.

Concerning the experimentally achievable precision, it is worth noting that the numerous quality checks during 
construction and beam tests of series detector modules let us conclude that the detectors as built should
give a good starting-point performance. Furthermore, cosmic ray muons, beam-gas interactions and 
beam halo muons are available as commissioning and calibration tools already before the first real proton-proton
collisions. Finally, with such first collisions in hand, the trigger and data acquisition systems will be timed-in,
the data coherence checked, sub-systems synchronized and reconstruction algorithms debugged
and calibrated. The electromagnetic and hadronic calorimeters will be calibrated with first
physics events. For example, the initial crystal inter-calibration precision of about 4\%
for the CMS ECAL will be improved to about 2\% by using the $\phi$-symmetry of the energy
deposition in minimum-bias and jet events.
Later the ultimate precision ($\approx 0.5\%$) and the absolute
calibration will be obtained using $\mathrm{Z}\rightarrow \mathrm{e}^+ \mathrm{e}^-$ decays and
the $E/p$ measurements for isolated electrons, such as in $\mathrm{W}\rightarrow \mathrm{e}\nu$ decays
\cite{Bayatian:2006zz}.
The latter requires a well understood tracking system. The uniformity of the 
hadronic calorimeters can be checked with single pions and jets. In order to obtain the
jet energy scale (JES) to a few per-cent precision or better, physics processes such as 
$\gamma + \mathrm{jet}$, 
$\mathrm{Z}(\rightarrow \ell\ell) + \mathrm{jet}$ or $\mathrm{W}\rightarrow$ 2 jets in top pair events
will be analyzed.
Finally, the tracker and muon system alignment will be carried out with generic tracks, isolated muons or
$\mathrm{Z}\rightarrow \mu^+ \mu^-$ decays. Regarding all these calibration
and alignment efforts, the ultimate statistical precision should be 
achieved very quickly in most cases. Then systematic effects
have to be faced, which, eg., implies that pushing the tracker $R\phi$ alignment from an initial
$100\,\mu$m to about $10\,\mu$m might involve at least one year of data taking. More detailed
reviews of the initial detectors and their performance can be found in Refs.\  \cite{Froidevaux:2006rg} and
\cite{Gianotti:2005fm}.

The anticipated detector performance leads to the following estimates for the reconstruction
precision of the most important physics objects~:
\begin{itemize}
	\item Isolated electrons and photons can be reconstructed with a relative energy resolution characterized
	         by a stochastic term (which is proportional to $1/\sqrt{E}$) of a few per-cent and an aimed-for
	         0.5\% constant term. Typically isolation requirements are defined by putting a cone 
	         around the electron/photon and counting the additional electromagnetic and hadronic energy
	         and/or track transverse momentum within this cone. The optimal cone size in $\eta-\phi$ 
	         space\footnote{Here $\eta$ denotes the pseudo-rapidity and $\phi$ the azimuthal angle around the beam pipe.}
	         depends on the particular analysis and event topology. For typical acceptance cuts, 
	         such as a transverse momentum above 10-20 GeV and $|\eta| < 2.5$, electrons and photons
	         can be expected to be reconstructed with excellent angular resolution, high efficiency ($\ge 90\%$)
	           and small backgrounds. Again, the precise values depend very much on the final state topology and the
	         corresponding tightness of the selection cuts. Most importantly, the systematic uncertainty 
	         on the reconstruction efficiency should be controllable at the 1-2\% level, using in-situ measurements
	         such $\mathrm{Z}\rightarrow \mathrm{e}^+ \mathrm{e}^-$ decays, 
	         with one of the electrons serving as tag lepton 
	         and the other one as probe object for which the efficiency is determined.          
	\item  Isolated muons, with similar acceptance cuts as mentioned above for electrons, should be 
	         reconstructed with a relative transverse momentum resolution of 1 - 5\% and excellent angular
	           resolution up to several hundreds of GeV. Again, a systematic uncertainty 
   	         on the reconstruction efficiency of 1-2\% appears to be achievable.
	\item Hadronic jets will be reconstructed up to pseudo-rapidities of 4.5 - 5, with good angular resolution.
	         The energy resolution depends rather strongly on the specific calorimeter performance. 
	            For example, in the
	           case of ATLAS (CMS) a stochastic term of the order of 
	            50 - 60\% (100 - 150\%) is to be expected when energy deposits in
	           projective calorimeter towers
	            are used for the jet clustering procedure. Important improvements on the CMS jet energy 
	            resolution are expected
	           from new approaches such as particle flow algorithms. Well above the trigger thresholds jets
	            will be reconstructed with very high efficiency; the challenge is the understanding
	            of the efficiency turn-on curves. In contrast to leptons, for jets the experimental 
	             systematic uncertainties are much more sizeable and difficult to control. A more detailed discussion 
	             will follow below. 
	             
	             A further important question is the lowest $p_T$ threshold above which jets
	              can be reconstructed reliably. Contrary to the naive expectation that only high-$p_T$ objects
	              (around 100 GeV and higher) are relevant, it turns out that many physics channels require jets
	               to be reconstructed with rather low transverse momentum of $\sim 20 - 30$ GeV. One reason for this is
	               the importance of jet veto requirements in searches for new physics, such as in the
	               $\mathrm{H}\rightarrow\mathrm{W}\mathrm{W}^*\rightarrow2\ell\,2\nu$ channel, where a jet
	               veto is necessary to reduce the top background. The experimental difficulties related to the 
	               understanding of the low-$p_T$ jet response\footnote{The jet response is defined as the
	                ratio of the reconstructed  and the ``true" jet momentum.}, the thresholds due to 
	                noise suppression, the impact of the underlying event and additional pile-up events and
	                ultimately the knowledge of the JES lead to the conclusion that it will
	                 be extremely challenging, if not impossible, to reliably reconstruct jets below a $p_T$ of 30 GeV.
	                In addition, also the theoretical predictions are challenged by very low-$p_T$ effects, as 
	                 for example induced by jet veto requirements. Here fixed-order calculations may have to be
	                 supplemented by resummations of large logarithms. 
	\item  Finally, the missing transverse energy will be a very important "indirect"
	                 observable, which is constructed from measurements of other quantities, such as
	                  all calorimeter energy deposits. Many searches for new physics, such as Supersymmetry,
	                   rely very much on this observable. However, it turns out that 
	                    it is also an extremely difficult quantity to measure, since it is sensitive to almost
	              every detail of the detector performance. Here it is even more difficult to give estimates
	                of the expected systematic uncertainties. Also, the reconstruction performance depends
	                very much on the details of the particular final state, such as the number of  jets and/or leptons
	                 in the event, the existence of\ ``true" missing energy, e.g.\ from neutrinos, the amount
	                  of pile-up events and in general the overall transverse energy deposited in the detector.
	                  The very first data will be of paramount importance for a timely understanding of this quantity.
\end{itemize}
More detailed discussions of the expected detector and reconstruction performance can be found in recent
reviews (\cite{Froidevaux:2006rg}, \cite{Gianotti:2005fm}), for ATLAS in Ref.\ \cite{AtlasPTDR}  and for CMS in
its Physics Technical Design Reports (PTDR), Vol.\ 1 \cite{Bayatian:2006zz} and Vol.\ 2 \cite{Ball:2007zza}. 

In the following I will concentrate on the early physics reach of the LHC experiments,
i.e.\ on measurements to be performed on the first few hundred pb$^{-1}$ up to 1 fb$^{-1}$ of integrated
luminosity. Many reviews exist on this topic, 
such as Refs.\ \cite{Gianotti:2005fm,Dissertori:2005he,DittmarLaThuile:2007,Sphicas:2003sb} to mention only a few.
Most of the results presented here are taken from the CMS PTDR Vol.\ 2 \cite{Ball:2007zza}, because it
represents the most recent comprehensive overview compiled by one of the LHC experiments.


\subsection{Jet production}
 \label{sec:dissertori_jets}

Because of its extremely large cross section, the inclusive
dijet production (pp $\rightarrow 2$ jets + anything) 
completely dominates over all other expected LHC processes with large momentum transfer.
At lowest order in perturbative Quantum Chromodynamics (QCD),
it is described as a $2 \rightarrow 2$ scattering of partons (quarks and gluons),
with only partons in the initial, intermediate and final state. 
Depending on the exchanged transverse momentum (or generally the
energy scale of the scattering process), the final state will consist of
more or less energetic "jets" which arise from the fragmentation of the outgoing partons.
Indeed, soft scattering processes, which give the largest contribution to the
total inelastic proton-proton cross section, are most likely, leading to final states
with hundreds of soft (i.e.\ below a few GeV) charged and neutral hadrons, uniformly
distributed over most of the experimental acceptance in pseudo-rapidity. Since these
are the most likely processes to occur, they 
are triggered on with the least stringent requirements and thus called "minimum-bias" events.
For the same reason they also represent the typical pile-up events
which can occur simultaneously with other triggered proton-proton collisions. 
Therefore very early measurements of the production rates\footnote{Currently the extrapolations from
the TEVATRON up to the LHC energies suffer from large uncertainties. For example, various
Monte Carlo generators predict charged track multiplicities which differ by more than 30\%.} and the charged particle
distributions will be extremely important, in particular for the tuning of the widely used Monte Carlo
generators. Here I will not discuss further this class of measurements, but rather concentrate
on the parton scattering at large transverse momentum. Examples of envisaged studies of
minimum-bias events can be found in \cite{Ball:2007zza}.

For outgoing partons with transverse momentum well above the QCD fragmentation scale 
($\Lambda \sim 1$ GeV) the picture of jet production arises, namely well collimated bundles of
particles, leading to isolated clusters of deposited energy in the calorimeters. Several algorithms
exist for the clustering of the final state objects (simulated particles, calorimeter towers, charged
tracks) into jets with a well defined four-momentum, which in the optimal case closely matches 
the four-momentum of the original scattered parton. Examples of commonly used prescriptions
are the Iterative Cone, Midpoint Cone, SISCone and $k_T$ algorithms. In particular, the latter two
algorithms recently receive a lot of theoretical and experimental attention, mainly because of their
property of being infrared and collinear safe to all orders of perturbation theory. A detailed discussion
of those jet algorithms is given elsewhere in these proceedings, as well as in \cite{Salam:2007xv,
Salam:2008qq,Ellis:2007ib} and references therein.

For the measurement of the inclusive jet cross section we simply count the 
number of jets inside a fixed pseudo-rapidity  region as a function of jet $p_T$. 
For a second typical measurement, the dijet cross section, events are selected in which the two highest 
$p_T$ jets, the leading jets, are both inside a specified pseudo-rapidity region and 
counted as a function of the dijet (invariant) mass.  Both cases are inclusive processes
dominated by the $2\rightarrow 2$ QCD scattering of partons. The distinction between 
inclusive jets and dijets is only in a different way of measuring the same process. For 
a common choice of the $\eta$ region, events selected by the dijet analysis 
are a subset of the events selected by the inclusive jet analysis, but the number of events in the two analyses coming from QCD is expected to be close at  high $p_T$. The steeply falling cross sections 
are shown in Fig.\ \ref{fig:dissertori_jetxsec}. For the inclusive jet case, the spectrum roughly follows a 
power law, however, with increasing power for increasing $p_T$, ie.,
the power increases from about 6 at $p_T=150$ GeV  to about 13 at $p_T = 3$ TeV
and keeps on increasing with jet $p_T$.

\begin{figure}[htb]
\begin{center}
\begin{tabular}{ll}
\includegraphics[width=0.47\textwidth]{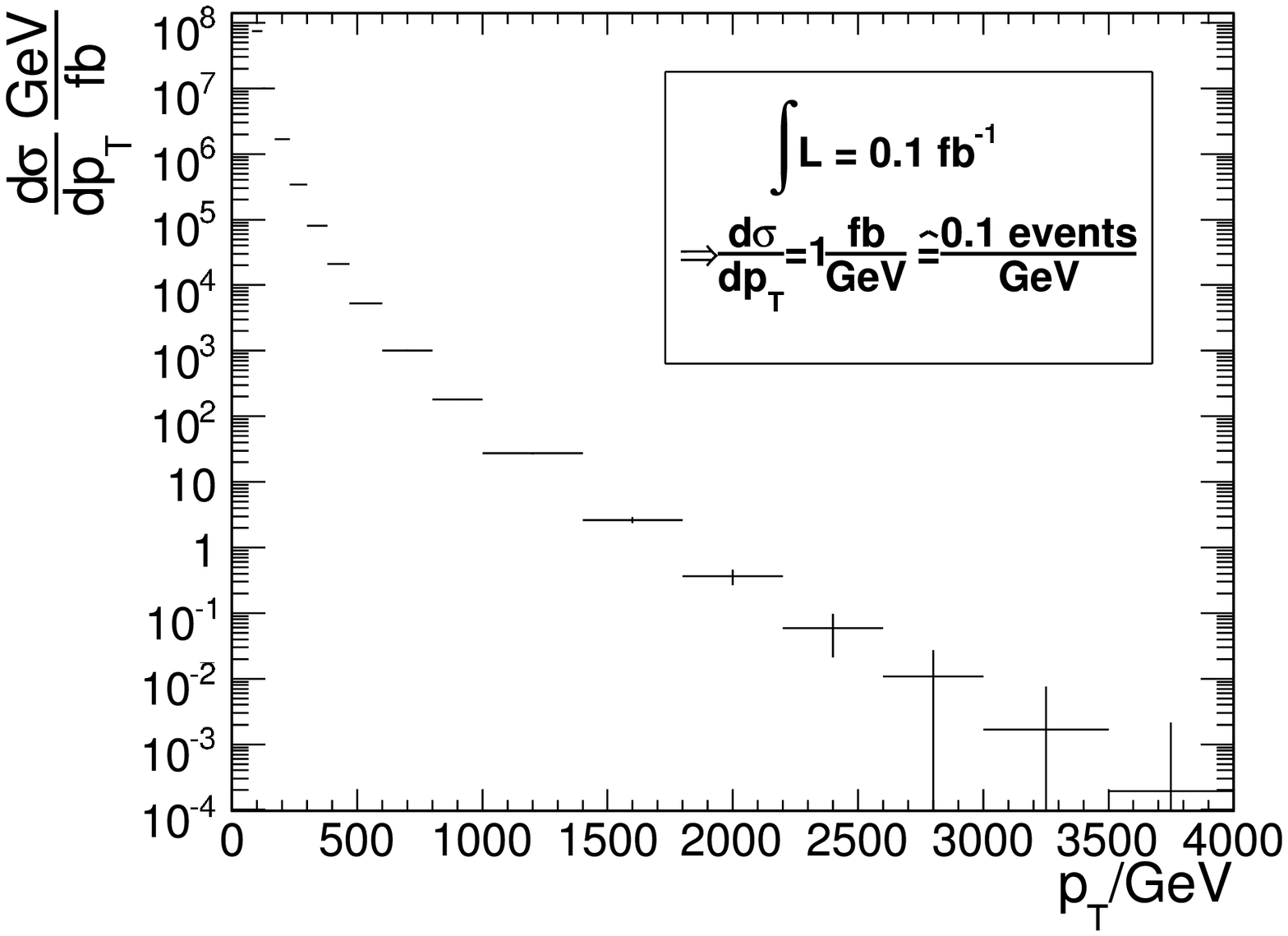} &
\includegraphics[width=0.53\textwidth]{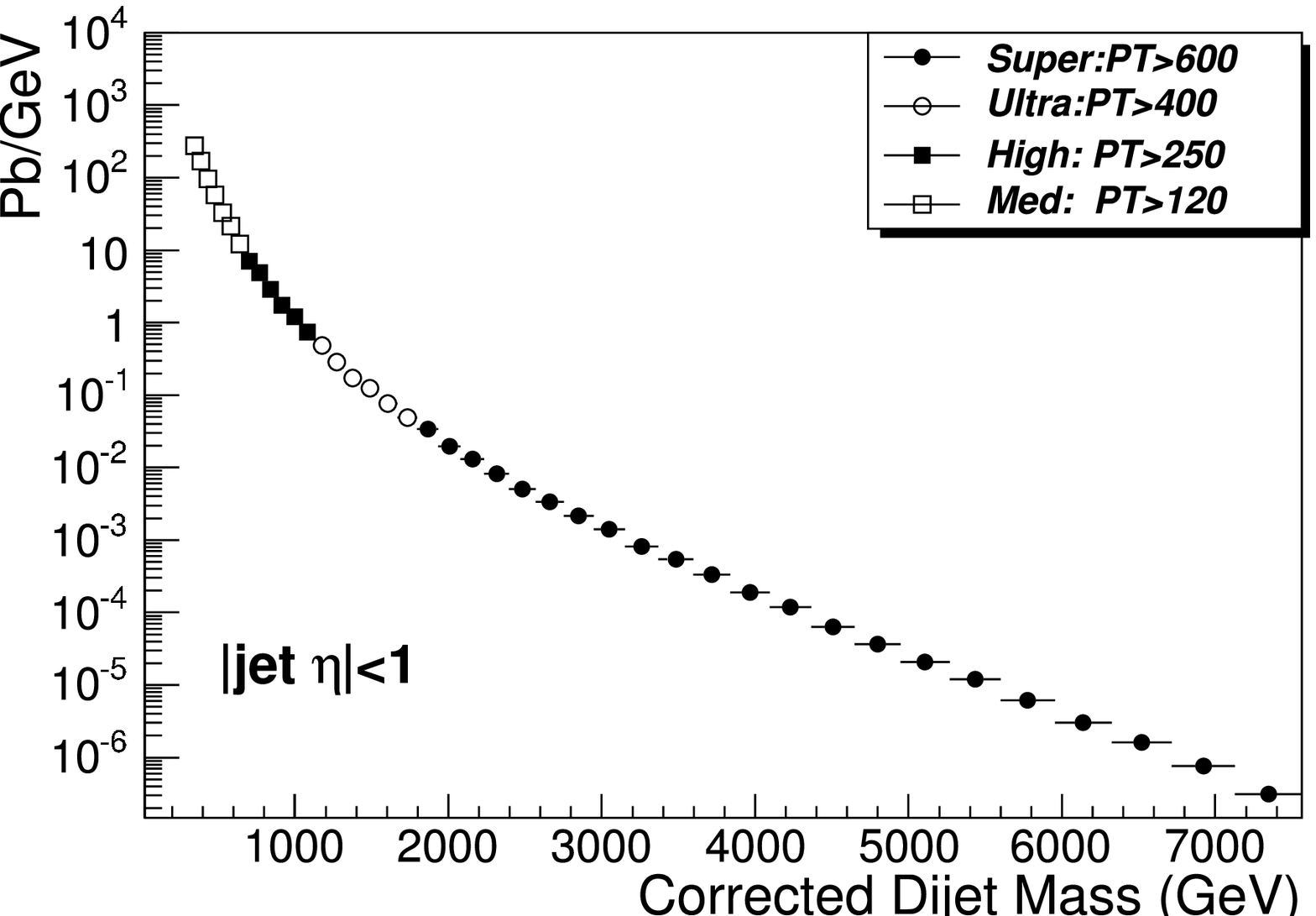} 
\end{tabular}
 \caption{Inclusive jet (left) and dijet (right) cross section measurements as foreseen by CMS \cite{Ball:2007zza}.
  The central cross section values are taken from a leading-order calculation in dependence of the transverse 
  momenta of the hard interaction. The insert on the right plot indicates various trigger paths.
}
\label{fig:dissertori_jetxsec}
\end{center}
\end{figure}

It can be seen that even for very small integrated luminosities the statistical uncertainties will be
negligible, up to very high jet momenta. Thus the TEVATRON reach in terms of highest
momenta and therefore sensitivity to new physics, such as contact interactions or heavy resonances,
will be quickly surpassed. For 1 fb$^{-1}$, the inclusive cross section
for central jet production (ie.\ jet pseudo-rapidities below $\sim 1$) will be known statistically to
better than 1\% up to a $p_T$ of 1 TeV, and the statistical errors on the dijet cross section will
be below 5\% up to dijet masses of 3 TeV.

The real challenge for these measurements will be the determination and control of the jet energy
scale. As mentioned above, the cross sections are steeply falling as a function of jet $p_T$. 
Therefore any relative uncertainty on the jet $p_T$ will translate
into a $n$-times larger relative uncertainty on the cross section, where $n$ indicates the power 
of the spectrum in a specified $p_T$ region, ie. $d\sigma/dp_T \propto p_T^{-n}$. For example, 
a $5\%$ uncertainty on the energy scale for jets around 100-200 GeV of transverse momentum 
induces a $30\%$ uncertainty on the inclusive jet cross section. This is also shown in 
Fig.\ \ref{fig:dissertori_jetxsecerr} (left), here for the case of a 3\% JES uncertainty. 
As a comparison, in Fig.\ \ref{fig:dissertori_jetxsecerr} (right) we see the expected theoretical uncertainties
on the inclusive jet cross section from the propagation of pdf uncertainties. These are below the
10\% level up to a jet $p_T$ of 1 TeV, thus much smaller than the experimental systematics from 
the JES. Therefore it is obvious that a measurement of the inclusive jet cross section
will not allow to constrain the pdfs, unless the JES is known to 2\% or better. This is definitely
beyond reach for the early phase of the LHC, and might remain a huge challenge even later.
Furthermore, because of these large experimental uncertainties, it might turn out that
 the currently known next-to-leading order (NLO) perturbative QCD calculation
of the hard scattering process is precise enough for a comparison to data. However, with 
better experimental control at a later stage and/or other definitions of observables (see below) the need for
going to next-to-next-to-leading order (NNLO) might arise.

\begin{figure}[htb]
\begin{center}
\begin{tabular}{ll}
\includegraphics[width=0.45\textwidth]{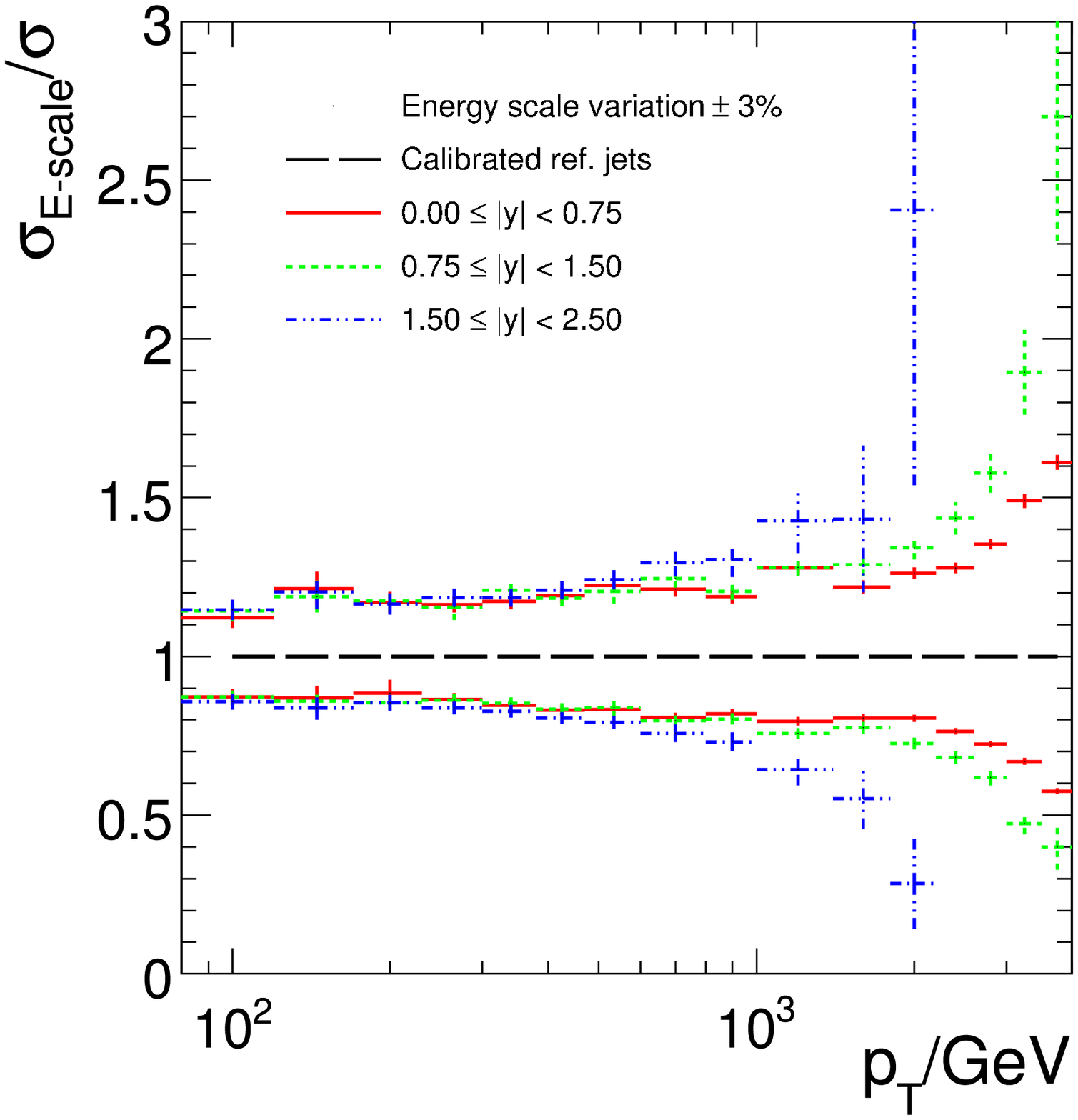} &
\includegraphics[width=0.45\textwidth]{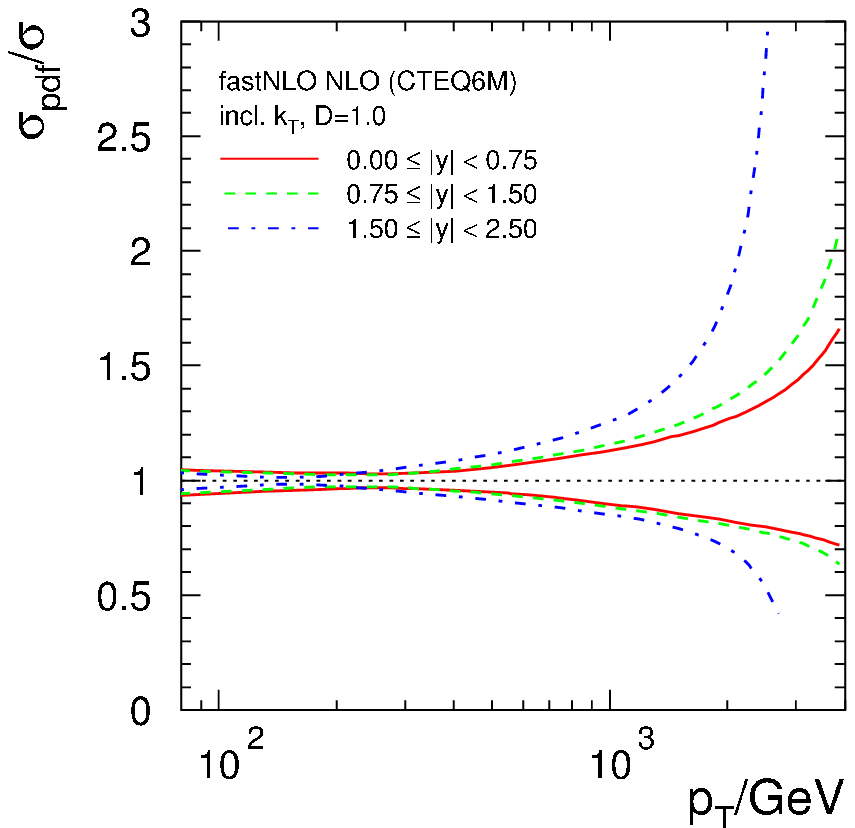} 
\end{tabular}
 \caption{Left: Relative systematic uncertainties of the inclusive jet cross sections for the $k_T$ algorithm 
 versus jet $p_T$ due to a change in the JES of $\pm 3\%$ for three bins in rapidity, $y$. 
 The error  bars indicate the statistical uncertainty. Right: relative uncertainties propagated from the error sets of the CTEQ6M \cite{Pumplin:2002vw} pdfs, for the same regions in rapidity. Plots taken from 
   \cite{Ball:2007zza}. 
}
\label{fig:dissertori_jetxsecerr}
\end{center}
\end{figure}

Obviously, the knowledge of the JES also has a strong impact on the achievable precision of the
dijet cross section measurement, as shown in Fig.\ \ref{fig:dissertori_dijetmasserr} (left). However, the
problem can be avoided by performing relative instead of absolute cross section measurements. A well suited
observable is the dijet ratio $N(|\eta| < |\eta_{\mathrm{in}}|) / N(|\eta_{\mathrm{in}}| < |\eta| < |\eta_{\mathrm{out}}|)$,
ie., the ratio of the number of dijet events within an inner region $|\eta| < |\eta_{\mathrm{in}}|$ to
the number of dijet events within an outer region
$|\eta_{\mathrm{in}}| < |\eta| < |\eta_{\mathrm{out}}|$. Both leading jets of the dijet event must satisfy the $|\eta|$ cuts.
In Ref.\ \cite{Ball:2007zza} the values chosen were $\eta_{\mathrm{in}} = 0.5$ and $\eta_{\mathrm{out}} = 1$,
whereas in a recent update \cite{CMS_AN_07_039} of the CMS studies 
on inclusive and dijet production they have been
increased to 0.7 and 1.3, respectively. The dijet ratio
has two interesting features. First, it is very sensitive to new physics, such as contact interactions or the
production of a heavy resonance, because those lead to jets at more central rapidities than in genuine QCD 
dijet events. Second, in the ratio we can expect many systematic uncertainties to cancel. For example, the
luminosity uncertainty completely disappears in the ratio. More importantly, also the JES 
uncertainty is strongly reduced, since the dijet ratio is sensitive only to the relative knowledge of the
scale as a function of rapidity, but not to the absolute scale any more. This is well illustrated in
Fig.\ \ref{fig:dissertori_dijetmasserr} (right), where the JES uncertainty is shown to be reduced to 
about 3\%. In this figure also the sensitivity to new contact interactions at various
scales is indicated.  Hence we have a nice example of a ratio measurement where systematic uncertainties
are reduced. Having an observable in hand with experimental systematic uncertainties at the level of 5\%
or less, it might become relevant to obtain a NNLO prediction for jet production. 


\begin{figure}[htb]
\begin{center}
\begin{tabular}{ll}
\includegraphics[width=0.45\textwidth]{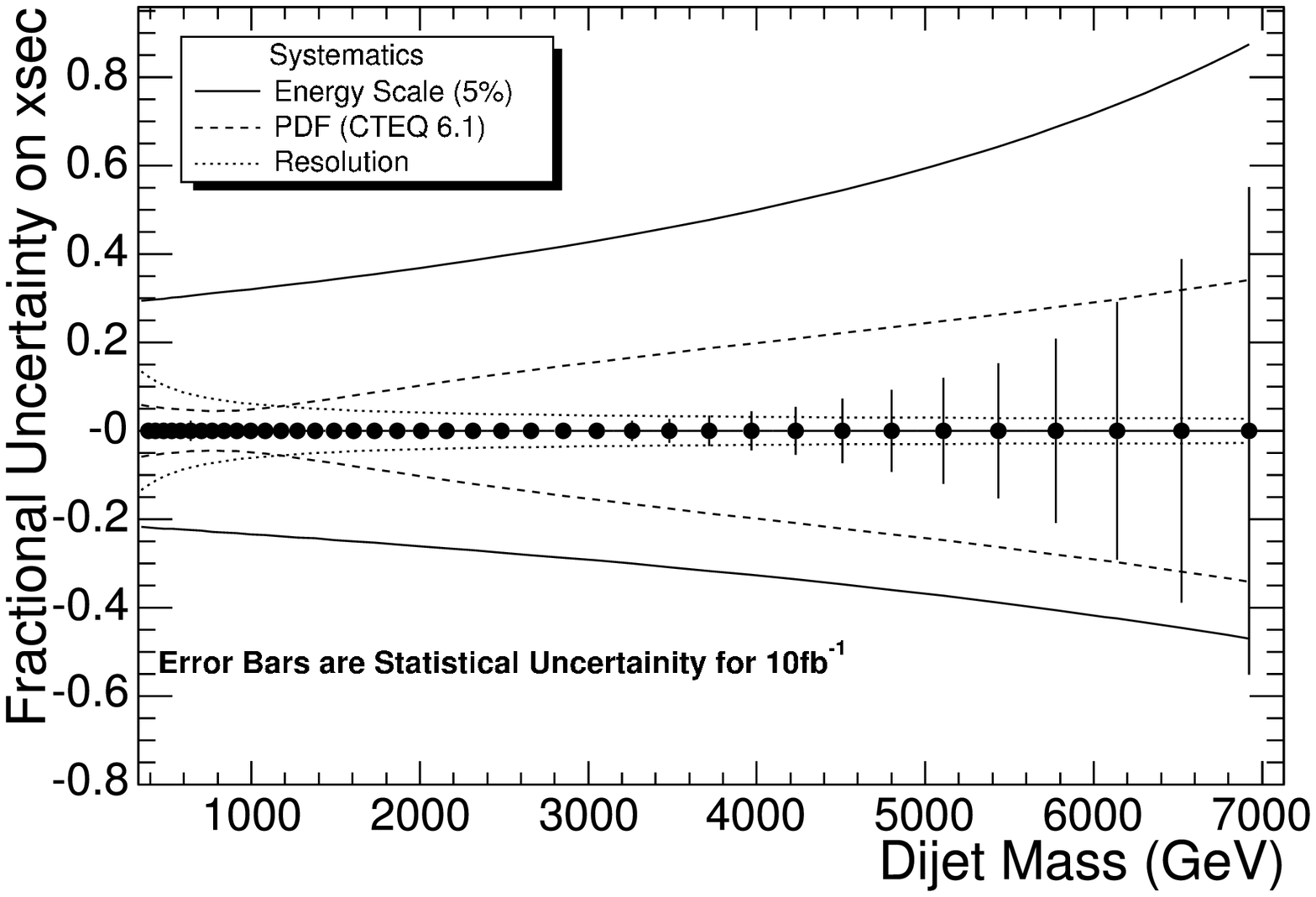} &
\includegraphics[width=0.45\textwidth]{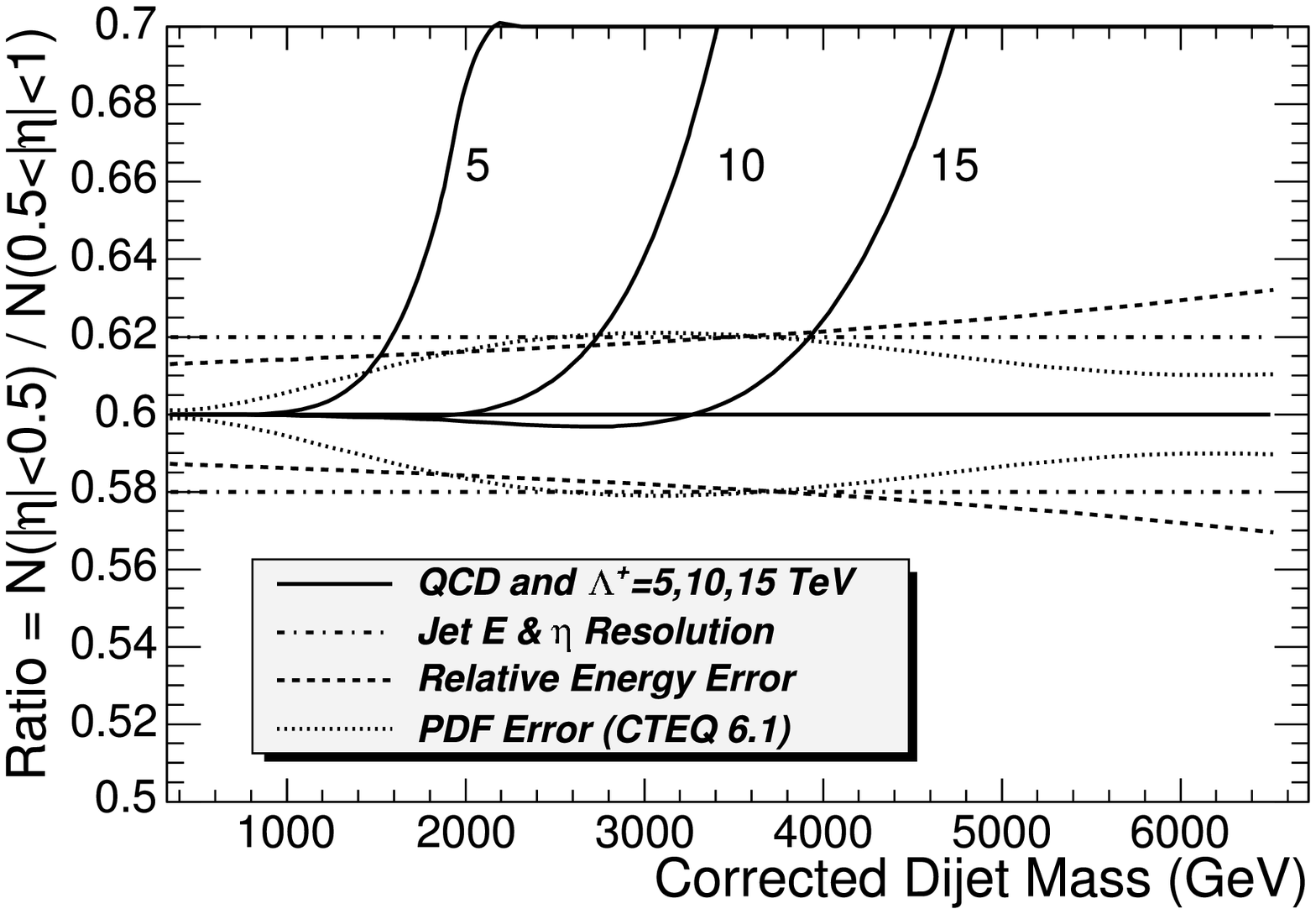} 
\end{tabular}
 \caption{Left~: Systematic uncertainty on the dijet cross section due to JES 
(solid curve), parton distributions (dashed curve) and calorimeter energy and $\eta$ resolution 
(dotted curve), compared to the statistical uncertainties for 10 fb$^{-1}$ (errorbars). 
Right~: Systematic bounds on the dijet ratio from uncertainties in the relative JES 
(dashed curve), parton distributions (dotted curve)  and calorimeter energy and $\eta$ resolution
(dot dashed curve), compared to the expectations of QCD and three contact interaction scales 
(solid line and curves). Plots taken from  \cite{Ball:2007zza}. 
}
\label{fig:dissertori_dijetmasserr}
\end{center}
\end{figure}

As we have seen above, the JES is the dominant source of uncertainty in jet cross section measurements.
Obviously, it is also important for many other analyses and searches which involve jet final states and
possibly invariant mass reconstructions with jets. Therefore major efforts are devoted by the experimental
collaborations to prepare the tools for obtaining JES corrections, both from the Monte Carlo simulations
and, more importantly, from the data themselves. Currently approaches are followed which are inspired
by the TEVATRON experience \cite{Bhatti:2005ai, Abbott:1998xw}. The correction procedure is split into several
steps, such as offset corrections (noise, thresholds, pile-up), relative corrections as a function of $\eta$,
absolute corrections within a restricted $\eta$-region, corrections to the parton level, flavour-specific corrections
etc. At the LHC startup we will have to rely on Monte Carlo corrections only, but with the first data coming in
it will be possible to switch to data-driven corrections. At a later stage, after a lot of effort will have
gone into the careful tuning of the Monte Carlo simulations, it might be feasible to use Monte Carlo 
corrections again. A rough estimate for the early JES uncertainty evolution in CMS is 10\% at start-up,
7\% after 100 pb$^{-1}$ and 5\% after  1 fb$^{-1}$ \cite{HarrisPrivate}.  Certainly it will be difficult and
require time to obtain a detailed understanding of the non-Gaussian tails in the jet energy resolution. 

Concerning data-driven JES corrections, one of the best channels is $\gamma +$jet production. At leading
order, the photon and the jet are produced back-to-back, thus the precisely measured photon energy
can be used to balance the jet energy. Real life is more difficult, mainly because of additional 
QCD radiation and the large background from jets faking a photon. These can be suppressed
very strongly  with
tight selection and isolation cuts (eg., no additional third jet with a transverse energy beyond a certain
threshold and tight requirements on additional charged and neutral energy in a cone around the photon). 
The need to understand well the photon-faking jet background 
and the photon fragmentation is avoided by using the channel Z$(\rightarrow \ell\ell) + $jet, with 
electrons or muons, however, at the price of a lower cross section. 

Besides being a tool for obtaining JES
corrections, both $\gamma +$jet and Z + jet processes will also be important handles for constraining
the gluon pdf. It appears feasible to probe the gluon pdf at Bjorken-$x$ values between about 0.0005 and 0.2
with a few per-cent statistical errors after only 1 fb$^{-1}$ of integrated luminosity \cite{DittmarJerusalem:1997}. The
$x$ value is well determined using the lepton or photon kinematics only, thus it does not suffer from the
less precise measurement of the jet momentum. Of course, in order to consistently constrain NNLO pdf 
sets (which should become more and more relevant with time), 
a NNLO calculation of the hard scattering part of the process
is needed. Whereas this appears beyond reach for the $\gamma +$jet case, the Z+jet process might
be tractable within the not-too-far future. As discussed below, Z+jet (as well as W+jet) production is a very
important background to many searches, therefore having a NNLO prediction should be very
valuable, also as a benchmark for Monte Carlo generators which combine leading order (LO) and/or
NLO matrix elements with parton shower models.


\subsection{Vector boson production}
 \label{sec:dissertori_vbosons}

The production of vector bosons (W and Z), triggered on with their subsequent leptonic
decays, will be among the most important and most precise tests of the SM at the LHC.
The leptonic channels, mainly electrons and muons, can be reconstructed very cleanly, 
at high statistics, with excellent resolution and efficiency 
and very small backgrounds. At the same time, the theoretical
predictions are known to high accuracy, as discussed in more detail below. 
This precision will be useful for constraining pdfs, by measuring the rapidity dependence of the
Z production cross section, in particular when going to large rapidities and thus probing low $x$ values.
As proposed in \cite{Dittmar:1997md}, this process will serve as a standard candle for 
determining to high precision (at the few per-cent level) the proton-proton
luminosity or alternatively the parton-parton luminosity. Finally, it will be attempted to improve
 on the current precision of the W mass. Besides that, W and Z production will be an important experimenter's tool. As mentioned already earlier, Z and W decays to leptons will be used to understand and calibrate various sub-detectors, measure the lepton reconstruction efficiencies and control even the missing transverse energy measurement. 

Below I will first discuss the inclusive case, concentrating on 
resonant production. Then I will highlight some issues for the W and Z production in 
association with jets. Although being highly interesting processes, di-boson production 
will not be discussed here, since for integrated luminosities up to 1 fb$^{-1}$ the statistical
precision will be the limiting factor for these measurements and only allow first
proofs of existence and rough validations of the model expectations.


\subsubsection{Inclusive W and Z production}
 \label{sec:dissertori_inclWZ}

Inclusive W and Z production currently is and probably will remain the
theoretically best known process at the LHC. Predictions are available
at NNLO in perturbative QCD, fully differential in the vector boson and
even the lepton momenta \cite{Melnikov:2006kv}. Figure
\ref{fig:dissertori_zrap} (left) shows the Z rapidity distribution at various
orders in perturbation theory. We see that the shape stabilizes when going
to higher orders and that the NNLO prediction nicely
falls within the uncertainty band of the NLO expansion, giving confidence
in the good convergence of the perturbation series. More importantly, 
the renormalization scale uncertainty is strongly reduced at NNLO, to 
a level of about 1\% for Z rapidities below 3. A renormalization scale uncertainty
even below 1\% can be obtained for ratio observables such as 
$\sigma($W$^+)$/$\sigma($W$^-)$ and $\sigma($W)/$\sigma($Z), possibly
as a function of rapidity. Again, ratio measurements are interesting also from the
experimental point of few, since many systematic uncertainties cancel completely
or to a large extend. The prospect of a precise measurement and 
knowing the hard scattering part of the process so well
means that we have a tool for precisely constraining pdfs (or couplings and masses, in
a more general sense). Indeed, when 
taking the full theoretical prediction for the W and Z production cross section, 
ie., the convolution of pdfs and hard scattering part,
its uncertainty is dominated by the limited knowledge of the pdfs, currently estimated
to be around 5-7\% \cite{Dittmar:2005ed,Tung:2007bm}. This will then also limit the proton-proton luminosity to
a precision of this size, unless the pdfs are further constrained, mainly by the
rapidity dependence of the cross section, as for example shown in Ref.\ \cite{Dittmar:2005ed}.
It is worth noting that at this level of precision also electro-weak corrections
have to be considered \cite{Dittmaier:2001ay,Baur:1998kt,Kuhn:2007qc}.

An important point to make in this context is the importance of having differential
cross section predictions. If we take resonant W and Z production at central
vector boson rapidity, we probe $x$ values of around 0.006, a region rather well
constrained by the current pdf fits. However, for larger rapidities
we probe more and more the small $x$ region, which is less well known, 
eg., at leading order and for a Z rapidity 
of 3 we need (anti-)quark pdfs at $x=0.12$ and $x=0.0003$. Experimentally,
because of the detector acceptance, we can only access a limited sub-region of the full 
phase space. This means that when measuring a total cross section, we have
to extrapolate the measurement to the full acceptance (eg., full rapidity), which introduces a 
model dependence, especially on the poorly known low-$x$ region. On the other hand,
having differential predictions, we can compute exactly the same quantity as we
measure, thus eliminating any extrapolation uncertainty. Similarly, for constraining
NLO (NNLO) pdfs, exactly the same acceptance cuts (on the leptons) as in the data can now be applied on
the available NLO (NNLO) predictions. Of course, with more and more differential higher-order 
predictions becoming available, this kind of argument applies to any
cross section measurement (and/or deduced determination of physics quantities such
as couplings, masses, pdfs), namely that we should compare the measurements
and predictions for the experimentally accessible acceptance and avoid un-necessary extrapolations,
which will not teach us anything new and only introduce additional uncertainties.

\begin{figure}[htb]
\begin{center}
\begin{tabular}{lr}
\includegraphics[width=0.50\textwidth]{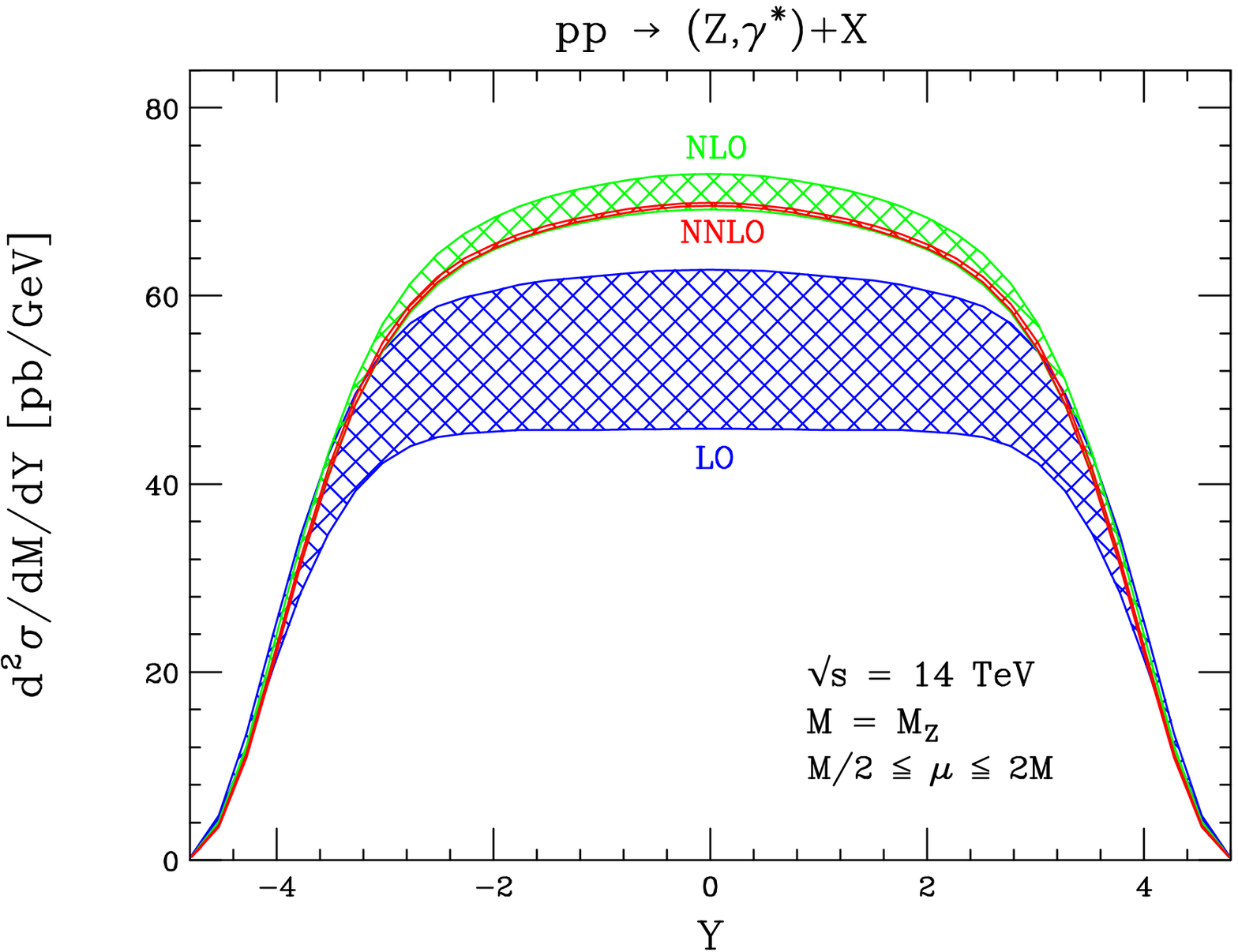} &
\includegraphics[width=0.40\textwidth]{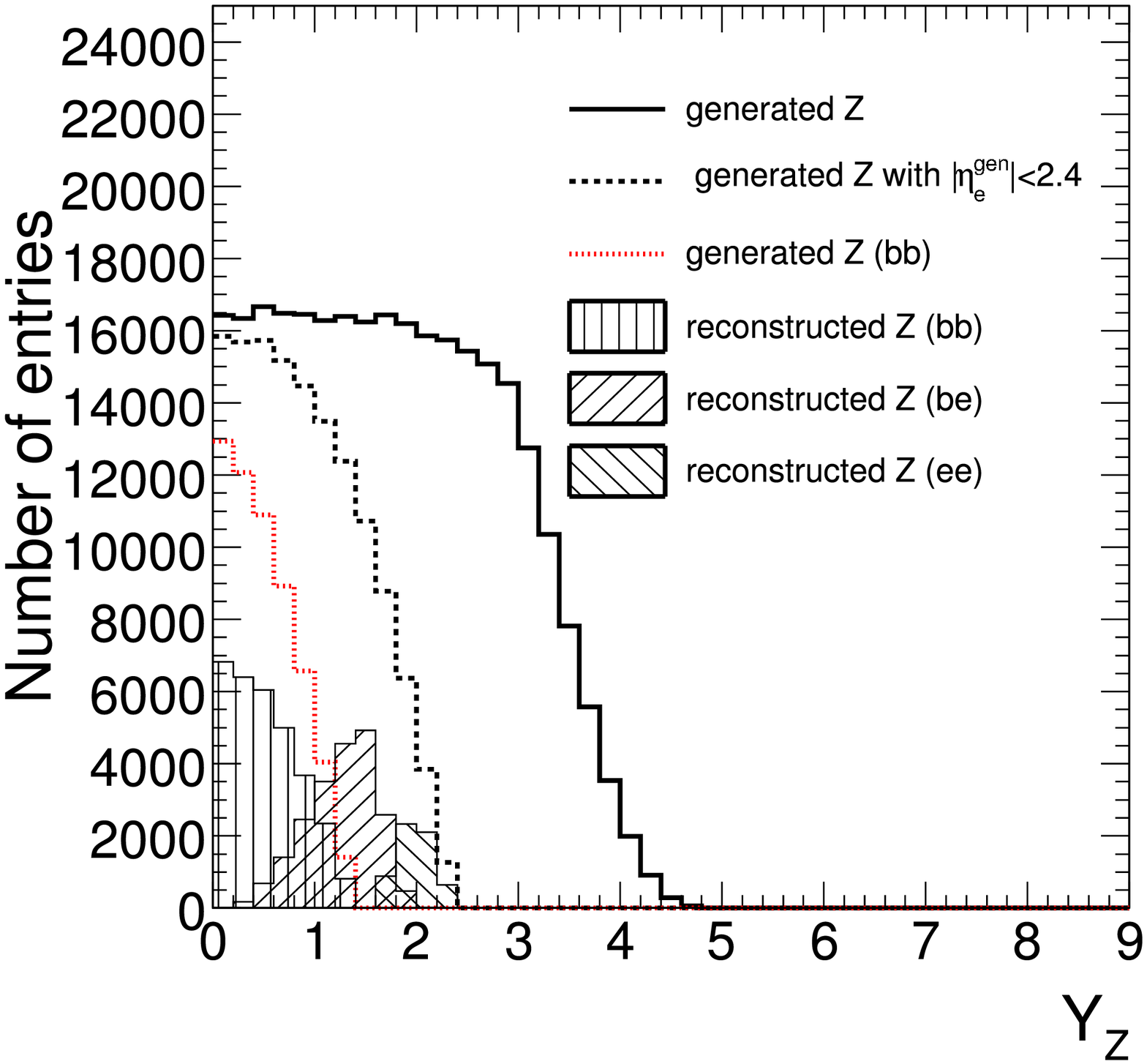} 
\end{tabular}
 \caption{Left~: QCD predictions at various orders of perturbation theory for the Z rapidity
 distribution at the LHC. The shaded bands indicate the renormalization scale uncertainty.
 Plot taken from \cite{Anastasiou:2003ds}.
Right~: Generated rapidity distribution for all Z candidates and for those where both electrons were 
generated within the geometrical acceptance of the CMS electromagnetic calorimeter (b=barrel, e=endcap).
Also shown is the rapidity distribution of the finally accepted Z events. Plot taken from  \cite{Ball:2007zza}. 
}
\label{fig:dissertori_zrap}
\end{center}
\end{figure}

As mentioned above, the experimental reconstruction of W and Z production is rather straight forward.
Leptons are required to have a minimum $p_T$ of about 20 GeV, within a pseudo-rapidity of 2.5 (cf.~Fig. \ref{fig:dissertori_zrap}, right).
In the Z case the mass peak allows for further event selections and background estimations. However,
the neutrino in the W decay leads to missing energy, which obviously is reconstructed less precisely.
Instead of an invariant mass peak only the transverse W mass can be reconstructed, with larger
backgrounds than for the Z. 
Here it is interesting to mention that a jet veto can help to control better
the QCD backgrounds and to improve the resolution of the missing transverse energy reconstruction.
However, a jet veto introduces sensitivity to low-$p_T$ QCD radiation, thus comparing the 
measurement to a calculation for the same acceptance cuts will only be meaningful if soft-$p_T$
resummation effects are taken into account in the predictions. Fortunately, with the Z+jet process
we have an experimental handle to study these issues rather precisely (see also below), since the 
radiation pattern in W+jet and Z+jet events is very similar.
In Ref.\ \cite{Ball:2007zza} it has been shown that 
reconstruction efficiencies and ultimately cross section measurements with systematic uncertainties
around 2\% (or better) should be possible, excluding the luminosity uncertainty. 


\subsubsection{W/Z+jets production}
 \label{sec:dissertori_WZjets}

Vector bosons produced in association with jets lead to final states with
high-$p_T$ leptons, jets and possibly missing transverse energy. Such a topology
is also expected for many searches, in particular for squark and gluino
production and subsequent cascade decays. Obviously it will be important to
understand these SM processes as quickly as possible and validate the
available Monte Carlo generators, which typically combine LO matrix elements with parton showers.
A standard observable will be the W/Z cross section as a function of the associated
leading jet transverse momentum or the number of additional jets. Obviously, such measurements
will suffer from the same JES uncertainties as the QCD measurements discussed above, and thus
constitute only limited calibration tools during the early data taking.
The problem can be reduced by defining clever ratios of cross sections, involving different
vector bosons and/or number of additional jets, or by normalizing the predictions to the data
in limited regions of the phase space (eg.\ for small jet multiplicity and extrapolating to larger multiplicities).
A completely different approach is to take a more inclusive look at this process, in the sense
that the Z transverse momentum is measured from the lepton kinematics, which is possible
at high statistical and, more importantly, high experimental accuracy (cf.\ Fig.\ \ref{fig:dissertori_zpt}). 
This distribution can be understood as the convolution
of the Z+0/1/2/$\ldots$jets distributions, therefore any model intended to describe Z+jets production
has necessarily to reproduce the Z $p_T$ distribution over its full range. As mentioned above,
in this context it would be highly desirable to have a NNLO prediction, possibly matched with a 
resummation calculation, for a comparison to the precise data and as benchmark for other approximations,
implemented in Monte Carlo simulations.

\begin{figure}[htb]
\begin{center}
\begin{tabular}{ll}
\includegraphics[width=0.45\textwidth]{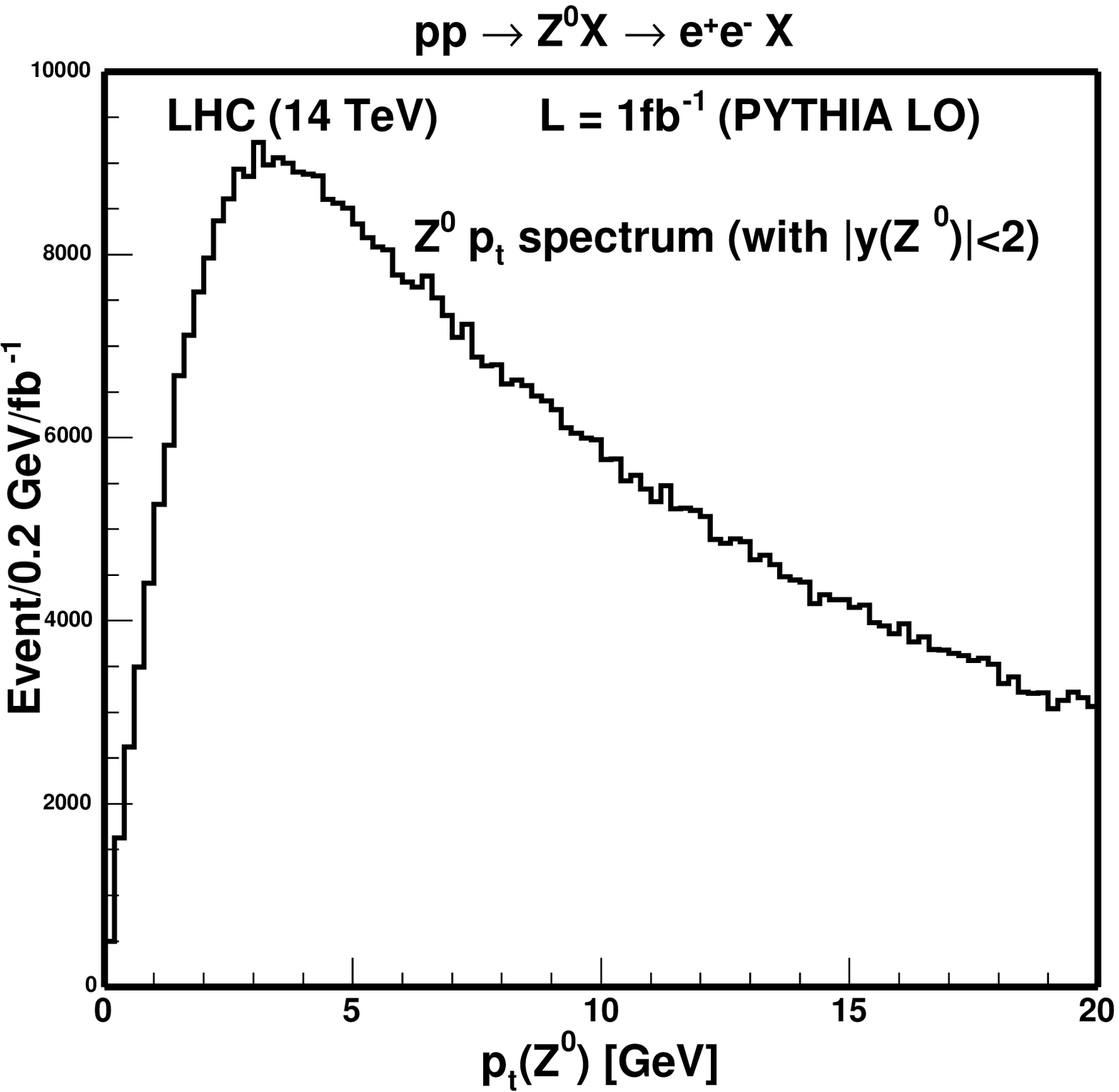} &
\includegraphics[width=0.45\textwidth]{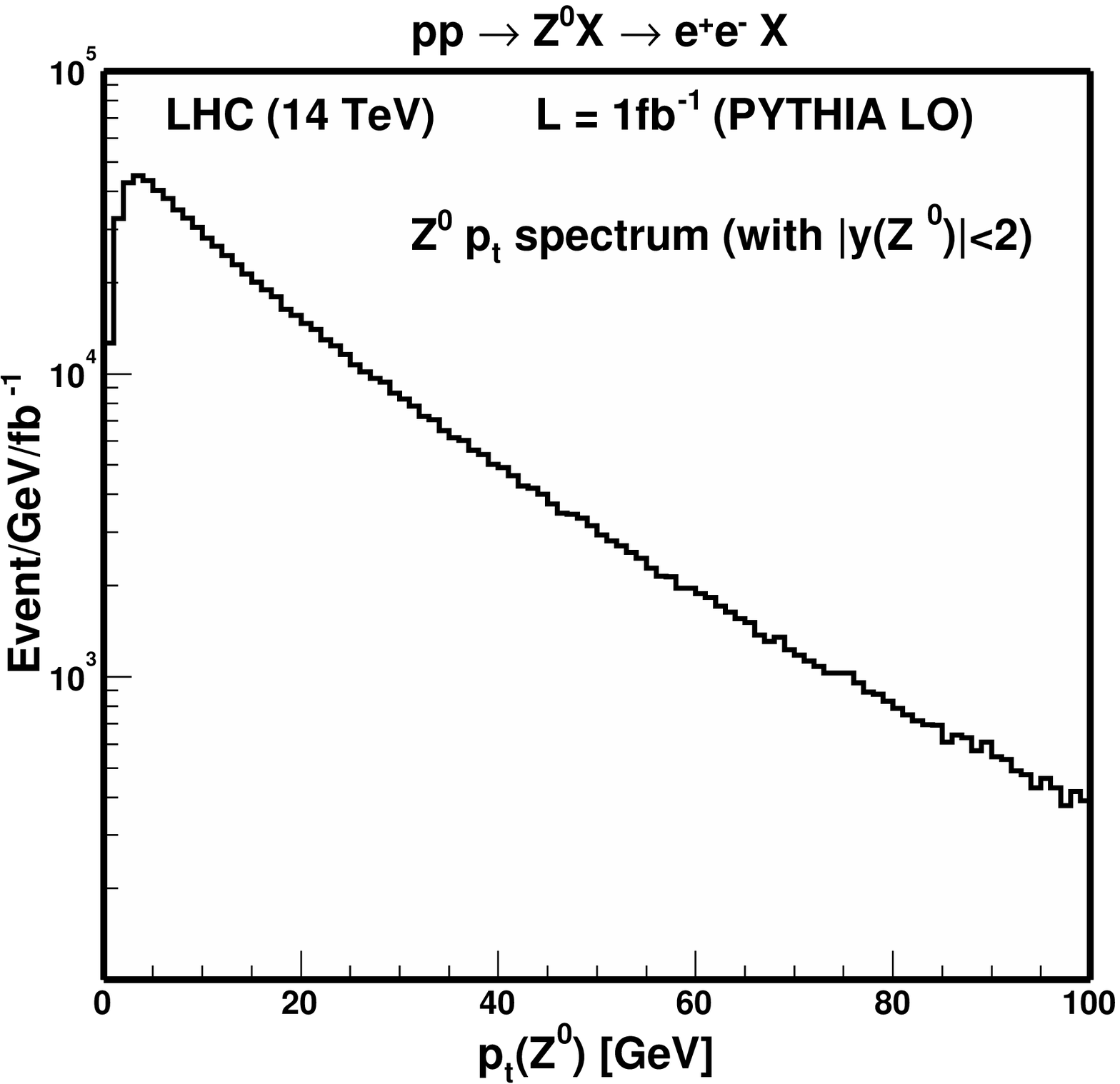} 
\end{tabular}
 \caption{Expected transverse momentum distribution for Z production at the LHC, for a $p_T$ range
 up to 20 GeV (left) and 100 GeV (right). The $p_T$ of the Z is
 reconstructed from the lepton kinematics. The fluctuations in the spectra  indicate the statistical precision achievable with 1 fb$^{-1}$ of integrated luminosity.
 Plots taken from \cite{Dittmar:2005ed}.
}
\label{fig:dissertori_zpt}
\end{center}
\end{figure}


\subsection{Top pair production}
 \label{sec:dissertori_top}
 
The top quark is produced very abundantly at the LHC. With 1 fb$^{-1}$ of integrated
luminosity, we should already have a couple of thousand clean signal events on tape
in the di-lepton channel, and a factor of 10 more in the single lepton channel (lepton+jets
channel) \cite{Ball:2007zza}. The physics case for the study of top production is very 
rich and can not be discussed in detail here. For example, a recent review can be found in Ref.\ \cite{D'hondt:2007aj}. 
Combining many different channels, a top mass measurement with a precision of
1 GeV might be achieved, which together with a precise W mass measurement constitutes an important indirect constraint of SM predictions and its extensions. 
The production cross section (for single and top-pair production) will be an important measurement,
again for testing the SM predictions and because top production is a copious background to a large number
of new physics searches. In the single muon+X channel, the top-pair production cross section
will soon (ie.\ with about 1 fb$^{-1}$) be measured with a statistical precision of 1\%. The total 
uncertainty of 10-15\% (excluding the luminosity uncertainty) will be dominated by systematics,
most notably due to the knowledge of the b-tagging efficiency. At the moment it seems difficult
to reduce this uncertainty to below 10\% \cite{Ball:2007zza}, even for much larger integrated luminosities.
Therefore this should be seen as a benchmark value to be challenged by the theoretical
predictions. Efforts are under way to compute the NNLO corrections to top-pair production and
it will be interesting to compare the ultimately achievable theoretical precision to the
experimental accuracy. Precise higher order predictions (possibly including resummation), both
for inclusive top and top+jets production, should also be
very valuable for obtaining precise background estimates, such as in Higgs searches. Although
it will be tried to calibrate the backgrounds with the data themselves, by using background-enriched
samples for the normalization \cite{Kauer:2004fg, Davatz:2006wr}, the theoretical predictions are still
needed for the extrapolation
from the background-rich to the signal-enriched regions of phase space. A good theoretical 
precision will lead to reduced systematics on the background, which will be most relevant for
searches with small signal-to-noise ratios.
It is worth mentioning that for the measurement of the b-jet cross section similar observations
hold as for the top, ie., the statistical error will soon be negligible, whereas the systematic
uncertainty is expected to be around 15-20\%, dominated by the JES.

Finally, top production will
become an extremely valuable calibration tool. The mass peak can already be
reconstructed with much less than 1 fb$^{-1}$, even without b-tagging requirements.
With a clean sample in hand, it can be exploited for controlling the b-tagging efficiency
and serve as a closure test for the JES corrections determined from other processes.
Concerning the JES, the mass of the hadronically decaying W serves as a calibration handle.
CMS expects that for intermediate jet $p_T$ values this sample could lead to JES uncertainties
around or below 3\% \cite{Ball:2007zza}.


\subsection{Conclusions}

I have summarized the experimental and theoretical prospects for some
of the most important measurements of SM processes at the LHC, namely
jet, vector boson and top production. The early benchmark measurements
will include the inclusive jet cross section, the dijet cross section and the
dijet ratio, photon/Z plus jet production, the Z rapidity distribution, ratios
of W and Z cross sections, the Z transverse momentum distribution and top
pair production. I have indicated the expected uncertainties of the
measurements and shown how these processes serve as tools for
the understanding of the detector, for the control of backgrounds and for
the validation and tuning of Monte Carlo generators. Particularly interesting
are ratio measurements, because otherwise important systematic uncertainties
cancel out in this case. With differential predictions at higher order in perturbation
theory in hand, I have highlighted the importance of comparing theory and
experiment for the same acceptance cuts, thus avoiding extrapolation errors.  
It is important to have (differential)  NLO predictions, possibly combined
with resummation calculations such as implemented in the Monte Carlo generator 
MC@NLO \cite{Frixione:2002ik,Frixione:2003ei}, for as many processes as possible. For the
cases where this appears to be difficult to achieve, LO plus parton shower
approaches might still be very valuable tools. However, higher order
predictions, up to NNLO, should be aimed for as benchmarks, at least in a few
cases. I have identified dijet, Z+jet and top production as most interesting
cases for investing the efforts towards NNLO calculations.

\subsection*{Acknowledgements}
I would like to thank the organizers for inviting me to this most stimulating
workshop, my colleagues within the CMS and ATLAS collaborations
for many interesting discussions on the topic and Ch.\ Anastasiou and M.\ Dittmar
 for useful comments on the manuscript.


%
}

\part[NEW APPROACHES]{NEW APPROACHES}

\section[On-shell recursion relations]{ON-SHELL RECURSION RELATIONS%
\protect\footnote{Contributed by: C.~Schwinn, S.~Weinzierl}}
{
%
%
%
%
%

\subsection{Introduction}

The efficient calculation of scattering amplitudes with many external legs
is a challenging task and needed for phenomenological studies at TeV colliders.
In the past years, various new methods for efficient calculations in QCD have
been introduced, originally motivated by the relation of QCD amplitudes to twistor string theory~\cite{Witten:2003nn}.  
These methods include the diagrammatic rules of
Cachazo, Svr\v{c}ek and Witten (CSW)~\cite{Cachazo:2004kj}, where tree level QCD
amplitudes are constructed from vertices that are off-shell continuations of
maximal helicity violating (MHV) amplitudes~\cite{Parke:1986gb}, and the
recursion relations of Britto, Cachazo, Feng and Witten~(BCFW)~\cite{Britto:2004ap,Britto:2005fq} that construct
scattering amplitudes from on-shell amplitudes with external momenta shifted into the complex plane.  
These developments have triggered significant research and numerous applications towards Born amplitudes in 
QCD~\cite{Luo:2005rx,Luo:2005my,Britto:2005dg,Badger:2005zh,Badger:2005jv,Forde:2005ue,Quigley:2005cu,Risager:2005vk,Draggiotis:2005wq,Vaman:2005dt,Ozeren:2006ft,Schwinn:2005pi,Schwinn:2006ca,Dinsdale:2006sq,Duhr:2006iq,Draggiotis:2006er,deFlorian:2006ek,deFlorian:2006vu,Rodrigo:2005eu,Ferrario:2006np,Schwinn:2007ee,Hall:2007mz}.
In addition, when combined with the unitarity method~\cite{Bern:1994zx,Bern:1995cg}
the recursion relations have proven very useful for one-loop calculations in 
QCD~\cite{Bidder:2004tx,Bidder:2004vx,Bidder:2005ri,Bedford:2004nh,Britto:2005ha,Bern:2005hs,Bern:2005ji,Bern:2005cq,Bern:2005hh,Forde:2005hh,Berger:2006ci,Berger:2006vq,Berger:2006sh,Britto:2006sj,Xiao:2006vr,Su:2006vs,Xiao:2006vt,Binoth:2006hk,Binoth:2007ca,Ossola:2006us,Ossola:2007bb,Anastasiou:2006jv,Anastasiou:2006gt,Mastrolia:2006ki,Britto:2006fc,Badger:2007si,Forde:2007mi,Ossola:2007ax,Britto:2007tt,Kilgore:2007qr}.
Here, we would like to review the basics of the on-shell recursion relations for Born QCD amplitudes and the proof of its validity.


\subsection{Helicity amplitudes and colour decomposition}

It is a well-known fact that the complexity of a calculation based on Feynman diagrams
grows factorially with the number of external particles.
In order to keep the size of intermediate expressions under control, a divide-and-conquer strategy
has been proven useful: One divides the quantity to be calculated into smaller pieces and calculates the
small pieces separately.

One first observes that it is not necessary to square the amplitude and sum over the spins and helicities
analytically. It is sufficient to do this numerically. This avoids obtaining $O(N^2)$ terms from
an expression with $O(N)$ terms.
The individual amplitudes have to be calculated in a helicity or spin basis.
This is straightforward for massless fermions. The two-component Weyl spinors
provide a convenient basis:
\begin{eqnarray}
 | p \pm \rangle & = & \frac{1}{2} \left( 1 \pm \gamma_5 \right) u(p).
\end{eqnarray}
In the literature there are different notations for Weyl spinors.
Apart from the bra-ket-notation there is the notation with dotted and un-dotted indices:
The relation between the two notations is the following:
\begin{eqnarray}
|p+\rangle = p_B,          \;\;\;\;\;\; \langle p+| = p_{\dot{A}}, \;\;\;\;\;\;
|p-\rangle = p^{\dot{B}},  \;\;\;\;\;\; \langle p-| = p^A. 
\end{eqnarray}
Spinor products are defined as
\begin{eqnarray}
\langle p q \rangle = \langle p - | q + \rangle, 
 & &
\left[ p q \right] = \langle p + | q - \rangle,
\end{eqnarray}
and take value in the complex numbers. 
It was a major break-through, when it was realised that also gluon polarisation vectors
can be expressed in terms of 
two-component Weyl spinors\cite{Berends:1981rb,DeCausmaecker:1982bg,Gunion:1985vc,Kleiss:1985yh,Kleiss:1986qc,Xu:1987xb,Gastmans:1990xh}.
The polarisation vectors of external gluons can be chosen as
\begin{eqnarray}
\varepsilon_{\mu}^{+}(k,q) = \frac{\langle q-|\gamma_{\mu}|k-\rangle}{\sqrt{2} \langle q- | k + \rangle},
 & &
\varepsilon_{\mu}^{-}(k,q) = \frac{\langle q+|\gamma_{\mu}|k+\rangle}{\sqrt{2} \langle k + | q - \rangle},
\end{eqnarray}
where $k$ is the momentum of the gluon and $q$ is an arbitrary light-like reference momentum.
The dependence on the arbitrary reference momentum $q$ will drop out in gauge
invariant quantities.

The second observation is related to the fact, that individual helicity amplitudes can be decomposed
into group-theoretical factors (carrying the colour structures)
multiplied by kinematic functions called partial amplitudes
\cite{Cvitanovic:1980bu,Berends:1987cv,Mangano:1987xk,Kosower:1987ic,Bern:1990ux}. 
These partial amplitudes do not contain any colour information and are gauge-invariant objects. 
In the pure gluonic case tree level amplitudes with $n$ external gluons may be written in the form
\begin{eqnarray}
{\cal A}_{n}(1,...,n) & = & g^{n-2} \sum\limits_{\sigma \in S_{n}/Z_{n}} 2 \; \mbox{Tr} \left(
 T^{a_{\sigma(1)}} ... T^{a_{\sigma(n)}} \right)
 A_{n}\left( \sigma(1), ..., \sigma(n) \right), 
\end{eqnarray}
where the sum is over all non-cyclic permutations of the external gluon legs
and the normalisation of the colour matrices is $\mbox{Tr}\;T^a T^b = \delta^{a b}/2$.
The quantities $A_n$ on the r.h.s. are the partial amplitudes and contain the kinematic information.
They are colour-ordered, e.g. only diagrams with a particular cyclic ordering of the gluons contribute.
In general, the colour factors are combinations of open strings 
$\left( T^{a_1} ... T^{a_n} \right)_{i_q j_{\bar{q}}}$
and closed strings
$\mbox{Tr} \left( T^{b_1} ... T^{b_m} \right)$
of colour matrices. These building blocks form a basis in colour space.
The choice of the basis for the colour structures is not unique, and several proposals
for bases can be found in the literature \cite{DelDuca:1999rs,Maltoni:2002mq,Weinzierl:2005dd}.


\subsection{Spinor space versus momentum space}

It will be useful to discuss the relationship between spinor space and
complexified momentum space.
Let us first fix our conventions. The metric tensor is $g_{\mu \nu} = \mbox{diag}(+1,-1,-1,-1)$.
A null-vector satisfies
\begin{eqnarray}
\left( p_0 \right)^2 - \left( p_1 \right)^2 - \left( p_2 \right)^2 - \left( p_3 \right)^2 & = & 0.
\end{eqnarray}
This relation holds also for complex $p_\mu$.
In complexified momentum space it is possible to choose a basis consisting only of 
null-vectors:
\begin{eqnarray}
e_1 = (1,0,0,1),
\;\;\;
e_2 = (0,1,i,0),
\;\;\;
e_3 = (0,1,-i,0),
\;\;\;
e_4 = (1,0,0,-1),
\end{eqnarray}
is an example of such a basis.
Light-cone coordinates are defined as follows:
\begin{eqnarray}
p_+ = p_0 + p_3, \;\;\; p_- = p_0 - p_3, \;\;\; p_{\bot} = p_1 + i p_2,
                                         \;\;\; p_{\bot^\ast} = p_1 - i p_2.
\end{eqnarray}
Note that $p_{\bot^\ast}$ does not involve a complex conjugation of
$p_1$ or $p_2$.
A convenient representation for the Dirac matrices is the 
Weyl representation:
\begin{eqnarray}
 \gamma^{\mu} = \left(\begin{array}{cc}
 0 & \sigma^{\mu} \\
 \bar{\sigma}^{\mu} & 0 \\
\end{array} \right),
 \;\;\;
\gamma_{5} = i \gamma^0 \gamma^1 \gamma^2 \gamma^3 
           = \left(\begin{array}{cc}
 1 & 0 \\
 0 & -1 \\
\end{array} \right),
 \;\;\;
\sigma_{A \dot{B}}^{\mu} = \left( 1 , - \vec{\sigma} \right),
 \;\;\;
\bar{\sigma}^{\mu \dot{A} B} = \left( 1 ,  \vec{\sigma} \right),
\end{eqnarray}
with $\vec{\sigma}=(\sigma_x,\sigma_y,\sigma_z)$ being the Pauli matrices.
A Weyl spinor $p_A$ is an element of a complex two-dimensional vector space $S$,
and similar a spinor $p_{\dot{B}}$ is an element of (another) complex two-dimensional vector space $S'$.
We will think of $p_A$ and $p_{\dot{B}}$ as independent quantities.
The dual space to $S$ will be denoted by $\bar{S}$, its elements by $p^A$.
Similarly, we denote the dual space to $S'$ by $\bar{S}'$ and its elements by $p^{\dot{B}}$.
The two-dimensional antisymmetric tensor provides an isomorphism between $S$ and $\bar{S}$ as well as between
$S'$ and $\bar{S}'$:
\begin{eqnarray}
\label{weinzierl_iso}
 p^A = \varepsilon^{AB} p_B,
 \;\;\;
 p_B = p^A \varepsilon_{AB},
 \;\;\;
 p^{\dot{A}} = \varepsilon^{\dot{A}\dot{B}} p_{\dot{B}},
 \;\;\;
 p_{\dot{B}} = p^{\dot{A}} \varepsilon_{\dot{A}\dot{B}}.
\end{eqnarray}
We take the two-dimensional antisymmetric tensor as
\begin{eqnarray}
\varepsilon^{AB} = \varepsilon_{AB} = \varepsilon^{\dot{A}\dot{B}} = \varepsilon_{\dot{A}\dot{B}}
 =
\left(\begin{array}{cc}
 0 & 1\\
 -1 & 0 \\
\end{array} \right).
\end{eqnarray}
Spinors are solutions of the Dirac equation, therefore we have for massless Weyl spinors
\begin{eqnarray}
\label{weinzierl_dirac}
p_\mu \bar{\sigma}^\mu \left| p+ \right\rangle = 0,
 \;\;\;\;\;\;
p_\mu \sigma^\mu \left| p- \right\rangle = 0,
 \;\;\;\;\;\;
\left\langle p+ \right| p_\mu \bar{\sigma}^\mu  = 0,
 \;\;\;\;\;\;
\left\langle p- \right| p_\mu \sigma^\mu  = 0. 
\end{eqnarray}
As normalisation we take for massless spinors
\begin{eqnarray}
\label{weinzierl_spinor_norm}
\langle p - | \sigma_{\mu} | p - \rangle = 2 p_{\mu},
 & &
\langle p + | \bar{\sigma}_{\mu} | p + \rangle = 2 p_{\mu}.
\end{eqnarray}
The solutions to eqs.~(\ref{weinzierl_dirac}), (\ref{weinzierl_spinor_norm}) and (\ref{weinzierl_iso}) are
\begin{eqnarray}
\label{weinzierl_spinors}
\left| p+ \right\rangle = \frac{e^{i \left(\alpha - \frac{1}{2} \phi\right)}}{\sqrt{\left| p_+ \right|}} \left( \begin{array}{c}
  -p_{\bot^\ast} \\ p_+ \end{array} \right),
 & &
\left| p- \right\rangle = \frac{e^{-i \left( \alpha + \frac{1}{2} \phi\right)}}{\sqrt{\left| p_+ \right|}} \left( \begin{array}{c}
  p_+ \\ p_\bot \end{array} \right),
 \nonumber \\
\left\langle p+ \right| = \frac{e^{-i\left(\alpha+ \frac{1}{2} \phi\right)}}{\sqrt{\left| p_+ \right|}} 
 \left( -p_\bot, p_+ \right),
 & &
\left\langle p- \right| = \frac{e^{i\left(\alpha - \frac{1}{2} \phi\right)}}{\sqrt{\left| p_+ \right|}}
 \left( p_+, p_{\bot^\ast} \right).
\end{eqnarray}
Here $\alpha$ is an arbitrary phase and $\phi$ is the phase of $p_+ = \left| p_+ \right| e^{i\phi}$.
The spinors corresponding to a four-vector $p_\mu$ are only determined up to a phase.
With these spinors we have
\begin{eqnarray}
\left\langle p q \right\rangle \left[ q p \right] & = & 2 p \cdot q.
\end{eqnarray}
It is worth to note that the relation $\bar{u}(p) = u(p)^\dagger \gamma^0$,
or equivalently
\begin{eqnarray}
\left| p+ \right\rangle^\dagger = \left\langle p+ \right|,
 & &
\left| p- \right\rangle^\dagger = \left\langle p- \right|,
\end{eqnarray}
holds only for real $p_\mu$ and positive $p_+$ (e.g. $\phi=0$), since
\begin{eqnarray}
\left| p+ \right\rangle^\dagger = \frac{e^{-i \left(\alpha - \frac{1}{2} \phi\right)}}{\sqrt{\left|p_+\right|}} 
                                 \left( \left(-p_{\bot^\ast}\right)^\ast, \left.p_+\right.^\ast \right),
& &
\left| p- \right\rangle^\dagger = \frac{e^{i \left( \alpha + \frac{1}{2} \phi\right)}}{\sqrt{\left|p_+\right|}}  
                                 \left( \left.p_+\right.^\ast, \left.p_\bot\right.^\ast \right).
\end{eqnarray}
Here the upper asterisk denotes the usual complex conjugation.
A pair of spinors $(p_{\dot{A}},p_B)$ determines a (unique) null-vector through
\begin{eqnarray}
\label{weinzierl_spinor_map}
 p_\mu & = & \frac{1}{2} p_{\dot{A}} \bar{\sigma}_\mu^{\dot{A}B} p_B 
 = \frac{1}{2} \langle p + | \bar{\sigma}_{\mu} | p + \rangle.
\end{eqnarray}
This is just eq.~(\ref{weinzierl_spinor_norm}) written reversely.
For arbitrary $p_{\dot{A}}$ and $p_B$ the four-vector $p_\mu$ will be in general complex.
While eq.~(\ref{weinzierl_spinors}) defines a map from complexified momentum space to 
the spinor space $S$ and $S'$, which is unique up to a phase,
eq.~(\ref{weinzierl_spinor_map}) goes in the reverse direction:
It defines a map from the space $S' \times S$ to complexified momentum space.
In this context it is worth to observe that if we change $p_{\dot{A}}$ or $p_B$ (but not both)
by a linear transformation as
\begin{eqnarray}
p_{\dot{A}} \rightarrow p_{\dot{A}} + z q_{\dot{A}}
\;\;\;
\mbox{or}
\;\;\;
p_B \rightarrow p_B - z q_B,
\end{eqnarray}
the resulting four-vector $p_\mu(z)$ will be a linear function of $z$.
Note however that a linear change in $p_\mu$ as in $p_\mu \rightarrow p_\mu + z q_\mu$
with a subsequent application of eq.~(\ref{weinzierl_spinors})
will not result in a linear change in $p_{\dot{A}}$ nor $p_B$.


\subsection{On-shell recursion relations}

In the previous section we have seen that we can associate to any
null-vector $p_\mu$ a pair of spinors $(p_{\dot{A}}, p_B)$.
From this  pair we can reconstruct the original four-vector through eq.~(\ref{weinzierl_spinor_map}).
To state the on-shell recursion relations it is best not to view the partial amplitude $A_n$ as a function
of the four-momenta, but to replace each four-vector by a pair of two-component Weyl spinors.
Therefore the partial amplitude $A_n$, being originally a function of the momenta $k_j$ and helicities
$\lambda_j$,
can equally be viewed as a function of the Weyl spinors $k_A^j$, $k_{\dot{B}}^j$ and the helicities
$\lambda_j$:
\begin{eqnarray}
 A_n(k_1,\lambda_1,...,k_n,\lambda_n) & = & 
 A_n( k_A^1, k_{\dot{B}}^1, \lambda_1, ..., k_A^n, k_{\dot{B}}^n, \lambda_n).
\end{eqnarray}
Let us now consider the $n$-gluon amplitude.
For the recursion relation we single out two particles $i$ and $j$.
If $(\lambda_i,\lambda_j) \neq (-,+)$ we have the following recurrence relation:
\begin{eqnarray}
\label{weinzierl_on_shell_recursion}
\lefteqn{
A_n\left( k_A^1, k_{\dot{B}}^1, \lambda_1, ..., k_A^n, k_{\dot{B}}^n, \lambda_n\right)
 = 
 } & & \\
 & & 
 \sum\limits_{partitions} \sum\limits_{\lambda=\pm}
  A_{L}\left( ..., \hat{k}_A^i, k_{\dot{B}}^i, \lambda_i, 
              ..., 
              i \hat{K}_A, i \hat{K}_{\dot{B}}, -\lambda
              \right)
  \frac{i}{K^2} 
  A_{R}\left(
                  \hat{K}_A, \hat{K}_{\dot{B}}, \lambda,
                  ...,
                  k_A^j, \hat{k}_{\dot{B}}^j, \lambda_j, 
                  ... \right). 
 \nonumber 
\end{eqnarray}
where the sum is over all partitions such that particle $i$ is on the left and particle $j$ is on the right.
The momentum $K$ is given as the sum over all unshifted momenta of the original 
external particles, which are part of $A_L$.
In eq.~(\ref{weinzierl_on_shell_recursion}) the shifted spinors
$\hat{k}_A^i$, $\hat{k}_{\dot{B}}^j$, $\hat{K}_A$ and $\hat{K}_{\dot{B}}$ are given by
\begin{eqnarray} 
\label{weinzierl_holomorphic_1}
 \hat{k}_A^i = k_A^i - z k_A^j, 
 \;\;\;
 \hat{k}_{\dot{B}}^j = k_{\dot{B}}^j + z k_{\dot{B}}^i,
 \;\;\;
 \hat{K}_A = \frac{K_{A\dot{B}} k_i^{\dot{B}}}{\sqrt{\left\langle i+ \left| K \right| j+ \right\rangle}},
 \;\;\;
 \hat{K}_{\dot{B}} = \frac{k_j^A K_{A\dot{B}}}{\sqrt{\left\langle i+ \left| K \right| j+ \right\rangle}},
\end{eqnarray}
and
\begin{eqnarray}
 z = \frac{K^2}{\left\langle i+ \left| K \right| j+ \right\rangle}.
\end{eqnarray}
Here we shifted $k_A^i$ and $k_{\dot{B}}^j$, while $k_{\dot{A}}^i$ and $k_B^j$ were left untouched.
We could equally well have used the other choice:
Shifting $k_{\dot{A}}^i$ and $k_B^j$, while leaving $k_A^i$ and $k_{\dot{B}}^j$ unmodified.
In this case one obtains a recursion relation valid for the helicity combinations
$(\lambda_i,\lambda_j) \neq (+,-)$.
Therefore for all helicity combinations of $(\lambda_i,\lambda_j)$ there is at least one valid
recursion relation. 
Applying this recursion relation to the six-gluon amplitude $A_6(1^-,2^-,3^-,4^+,5^+,6^+)$
with three positive and three negative helicities, 
we choose $(i,j)=(6,1)$. In this case only two diagrams need to be calculated
and
we obtain the compact result
\begin{eqnarray}
\lefteqn{
A_6(1^-,2^-,3^-,4^+,5^+,6^+) = } & & 
 \nonumber \\
 & &
 4 i \left[
           \frac{\langle 6+ | 1+2 | 3+ \rangle^3}{[ 61 ] [ 12 ] \langle 34 \rangle \langle 45 \rangle s_{126} \langle 2+ | 1+6 | 5+ \rangle}
         + \frac{\langle 4+ | 5+6 | 1+ \rangle^3}{[ 23 ] [ 34 ] \langle 56 \rangle \langle 61 \rangle s_{156} \langle 2+ | 1+6 | 5+ \rangle}
     \right].
\end{eqnarray}


\subsection{Quarks, massive or massless}

QCD does not consist solely of gluons, but contains the quarks as well.
Let us now discuss the general case of the inclusion of massive quarks.
All formulae will have a smooth limit $m\rightarrow 0$, therefore the case
of massless quarks will need no further discussion.
For massive fermions we have to consider Dirac spinors.
We can take them as
\begin{eqnarray}
 u(\pm) = \frac{1}{\langle p^\flat \mp | q \pm \rangle} \left( p\!\!\!/ + m \right) | q \pm \rangle,
 & &
\bar{u}(\pm) = \frac{1}{\langle q \mp | p^\flat \pm \rangle} \langle q \mp | \left( p\!\!\!/ + m \right),
 \nonumber \\
 v(\pm) = \frac{1}{\langle p^\flat \mp | q \pm \rangle} \left( p\!\!\!/ - m \right) | q \pm \rangle,
&&
\bar{v}(\pm) = \frac{1}{\langle q \mp | p^\flat \pm \rangle} \langle q \mp | \left( p\!\!\!/ - m \right).
\end{eqnarray}
Here, $p$ is the momentum of the fermion and $|q+\rangle$ and $\langle q+|$ are two independent Weyl spinors
used as reference spinors. These two spinors define
a light-like four-vector $q^\mu = \frac{1}{2} \langle q+ | \gamma^\mu | q+ \rangle$, which in turn is used
to associate to any not necessarily light-like four-vector $p$ a light-like four-vector $p^\flat$:
\begin{eqnarray}
 p^\flat & = & p - \frac{p^2}{2 p \cdot q} q.
\end{eqnarray}
The reference spinors are related to the quantisation axis of the spin for 
the fermion, and the individual amplitudes with label $+$ or $-$ will therefore refer to this spin axis.
From the Dirac spinors we can reconstruct the four-vector $p^\mu$ as follows:
\begin{eqnarray}
 p^\mu & = & \frac{1}{4} \sum\limits_{\lambda} \bar{u}(\lambda) \gamma^\mu u(-\lambda).
\end{eqnarray}
For the recursion relation, we again single out two particles $i$ and $j$, which need not be massless, with
four-momenta $p_i$ and $p_j$.
To these two four-momenta we associate two light-like four-momenta $l_i$ and $l_j$ as follows \cite{delAguila:2004nf,vanHameren:2005ed}:
\begin{eqnarray}
l_i = \frac{1}{1-\alpha_i \alpha_j} \left( p_i - \alpha_j p_j \right), 
\;\;\;
l_j = \frac{1}{1-\alpha_i \alpha_j} \left( -\alpha_i p_i + p_j \right), 
\;\;\;
\alpha_k = \frac{2p_ip_j-\mbox{sign}(2p_ip_j)\sqrt{\Delta}}{2p_k^2}.
\end{eqnarray}
with $\Delta = \left( 2p_ip_j \right)^2 - 4p_i^2 p_j^2$.
These light-like four-vectors define massless spinors
$|l_i+ \rangle$, $\langle l_i+ |$, $|l_j+ \rangle$ and $\langle l_j+ |$.
If particle $i$ is a massive quark or anti-quark, we use 
$|l_j+ \rangle$ and $\langle l_j+ |$ as reference spinors for particle $i$.
If particle $j$ is a massive quark or anti-quark, we use
$|l_i+ \rangle$ and  $\langle l_i+ |$ as reference
spinors for particle $j$.
We have the recursion relation
\begin{eqnarray}
\lefteqn{
A_n\left( u_1(-), \bar{u}_1(+), \lambda_1, ..., u_n(-), \bar{u}_n(+), \lambda_n\right)
 = 
 } & & \\
 & & 
\hspace*{20mm}
 \sum\limits_{partitions} \sum\limits_{\lambda=\pm}
  A_{L}\left( ..., u_i'(-), \bar{u}_i(+), \lambda_i, ..., 
              i v_K'(-), i \bar{v}_K'(+), -\lambda
              \right)
 \nonumber \\
 & &
\hspace*{20mm}
  \times
  \frac{i}{K^2-m_k^2} 
  A_{R}\left( u_K'(-), \bar{u}_K'(+), \lambda,
              ..., u_j(-), \bar{u}_j'(+), \lambda_j, ...
              \right).
 \nonumber 
\end{eqnarray}
Here we denote by $k$ the intermediate particle where we factorise the amplitude, and by
$K$ the off-shell four-momentum flowing through this propagator in the
unshifted amplitude.
We shift the Dirac spinors as follows:
\begin{eqnarray}
\label{weinzierl_holomorphic_2}
 {u_i}'(-) = u_i(-) - z | l_j + \rangle,
 \;\;\;
 {\bar{u}_j}'(+) = \bar{u}_j(+) + z \langle l_i + |,
 \;\;\;
 z & = & \frac{K^2-m_k^2}{\langle l_i+| K | l_j+ \rangle}.
\end{eqnarray}
For the intermediate particle $k$ we define 
the polarisations with respect to the reference spinors $|l_j + \rangle$ and $\langle l_i + |$:
\begin{eqnarray}
 {u_K}'(-) = \frac{1}{\langle K^\flat + | l_i - \rangle} \left( K\!\!\!/' + m_k \right) \left| l_i - \right\rangle,
 & &
 {\bar{u}_K}'(+) = \frac{1}{\langle l_j- | K^\flat + \rangle} \left\langle l_j- \right| \left( K\!\!\!/' + m_k \right),
\end{eqnarray}
where
\begin{eqnarray}
 {K'}^\mu = K^\mu - \frac{z}{2} \langle l_i+ | \gamma^\mu | l_j + \rangle,
 & &
 {K^\flat}^\mu = K^\mu - \frac{1}{2} \frac{K^2}{\langle l_i+| K | l_j+ \rangle} \langle l_i+ | \gamma^\mu | l_j + \rangle.
\end{eqnarray}
The recursion relation is valid for $(\lambda_i,\lambda_j) \neq (-,+)$
with the following exceptions:
\begin{itemize}
\item Particles $i$ and $j$ cannot belong to the same fermion line.
\item The combinations $(q_i^+,g_j^+)$, $(\bar{q}_i^+,g_j^+)$, $(g_i^-,q_j^-)$ 
and $(g_i^-,\bar{q}_j^-)$ are excluded.
\item If $i$ is massive, the combinations
$(q_i^+,q_j'{}^+)$, $(q_i^+,\bar{q}_j'{}^+)$,  
$(\bar{q}_i^+,q_j'{}^+)$ and $(\bar{q}_i^+,\bar{q}_j'{}^+)$ are excluded.  
\item If $j$ is massive, the combinations
$(q_i^-,q_j'{}^-)$, $(q_i^-,\bar{q}_j'{}^-)$,  
$(\bar{q}_i^-,q_j'{}^-)$ and $(\bar{q}_i^-,\bar{q}_j'{}^-)$ are excluded.  
\end{itemize}
Instead of shifting ${u_i}(-)$ and ${\bar{u}_j}(+)$, we can alternatively shift
$\bar{u}_i(+)$ and $u_j(-)$:
\begin{eqnarray}
\label{weinzierl_anti_holomorphic_2}
 {\bar{u}_i}'(+) = \bar{u}_i(+) - z \langle l_j+ |,
 \;\;\;
 {u_j}'(-) = u_j(-) + z | l_i + \rangle,
 \;\;\;
 z = \frac{K^2-m_k^2}{\langle l_j+| K | l_i+ \rangle}.
\end{eqnarray}
For the intermediate particle $k$ we define 
the polarisations now with respect to the reference spinors $|l_i + \rangle$ and $\langle l_j + |$:
\begin{eqnarray}
 {u_K}'(-) = \frac{1}{\langle K^\flat + | l_j - \rangle} \left( K\!\!\!/' + m_k \right) \left| l_j - \right\rangle,
 & &
 {\bar{u}_K}'(+) = \frac{1}{\langle l_i- | K^\flat + \rangle} \left\langle l_i- \right| \left( K\!\!\!/' + m_k \right),
\end{eqnarray}
where
\begin{eqnarray}
 {K'}^\mu = K^\mu - \frac{z}{2} \langle l_j+ | \gamma^\mu | l_i + \rangle,
 & &
 {K^\flat}^\mu = K^\mu - \frac{1}{2} \frac{K^2}{\langle l_j+| K | l_i+ \rangle} \langle l_j+ | \gamma^\mu | l_i + \rangle.
\end{eqnarray}
Doing so, we obtain a recursion relation valid for
$(\lambda_i,\lambda_j) \neq (+,-)$
with the following exceptions:
\begin{itemize}
\item Particles $i$ and $j$ cannot belong to the same fermion line.
\item The combinations $(g_i^+,q_j^+)$, $(g_i^+,\bar{q}_j^+)$, $(q_i^-,g_j^-)$ 
and $(\bar{q}_i^-,g_j^-)$ are excluded.
\item If $j$ is massive, the combinations
$(q_i^+,q_j'{}^+)$, $(q_i^+,\bar{q}_j'{}^+)$,  
$(\bar{q}_i^+,q_j'{}^+)$ and $(\bar{q}_i^+,\bar{q}_j'{}^+)$ are excluded.  
\item If $i$ is massive, the combinations
$(q_i^-,q_j'{}^-)$, $(q_i^-,\bar{q}_j'{}^-)$,  
$(\bar{q}_i^-,q_j'{}^-)$ and $(\bar{q}_i^-,\bar{q}_j'{}^-)$ are excluded.  
\end{itemize}
As we are free to choose the particles $i$ and $j$, we can compute all Born helicity amplitudes 
in QCD with two-particle shifts via recursion relations, except the ones which involve only
massive quarks or anti-quarks.
Amplitudes consisting solely of massive quarks and anti-quarks and with more than six particles 
may be calculated recursively if one allows more general shifts, where more than
two particles are shifted.


\subsection{Proof of the on-shell recursion relations}

For the proof \cite{Britto:2005fq,Badger:2005zh,Risager:2005vk,Draggiotis:2005wq,Vaman:2005dt,Schwinn:2007ee}
of the on-shell recursion relation 
we discuss as an example the case of the holomorphic shift as in eq.~(\ref{weinzierl_holomorphic_1}) 
or eq.~(\ref{weinzierl_holomorphic_2}).
One considers the function
\begin{eqnarray}
A(z) & = &
A_n\left( ..., {u_i}'(-), \bar{u}_i(+), \lambda_i, ..., u_j(-), {\bar{u}_j}'(+), \lambda_j, ... \right)
\end{eqnarray}
of one variable $z$,
where the $z$-dependence enters through 
\begin{eqnarray}
 {u_i}'(-) = u_i(-) - z | l_j + \rangle,
 & &
 {\bar{u}_j}'(+) = \bar{u}_j(+) + z \langle l_i + |.
\end{eqnarray}
The function $A(z)$ is a rational function of $z$, which has only simple poles in $z$.
This follows from the Feynman rules and the factorisation properties of amplitudes.
Therefore, if $A(z)$ vanishes for $z \rightarrow \infty$, $A(z)$ is given 
by Cauchy's theorem as the sum over its residues. This is just the right hand side of the recursion relation.
The essential ingredient for the proof is the vanishing of $A(z)$ at $z \rightarrow \infty$.
If $(\lambda_i,\lambda_j)=(+,-)$ it can be shown that each individual Feynman diagram 
vanishes for $z \rightarrow \infty$. 
For the helicity combinations $(+,+)$ and $(-,-)$ one first constructs a supplementary recursion relation
based on three-particle shifts and deduces from this representation the large $z$-behaviour of $A(z)$.
This establishes the recursion relation for these helicity combinations with the exceptions indicated above.
The proof for the anti-holomorphic shift as in eq.~(\ref{weinzierl_anti_holomorphic_2}) proceeds analogously.

%
}


\section[On-shell recursion to determine rational terms]
{ON-SHELL RECURSION TO DETERMINE RATIONAL TERMS%
\protect\footnote{Contributed by: Z.~Bern, L.J.~Dixon}}
{\graphicspath{{bern/}}
%
%
%
%
%
On-shell methods offer an auspicious approach for dealing with the
rapid growth in complexity of loop amplitudes as the number of
particles in the process increases.  These methods rely on the
unitarity of the theory~\cite{Landau:1959fi,Cutkosky:1960sp} which
requires that the poles and branch cuts of amplitudes correspond to
physical propagation of particles.  On-shell methods are presently
undergoing intense development for use at loop level (see, for
example, refs.~\cite{Berger:2006ci, Britto:2006sj, Berger:2006vq,
Badger:2007si, Ossola:2006us, Anastasiou:2006jv, Mastrolia:2006ki, 
Britto:2006fc, Anastasiou:2006gt, Brandhuber:2007vm, Forde:2007mi, 
Ellis:2007br, BjerrumBohr:2007vu, Ossola:2007ax, Britto:2007tt}).  
Their advantage lies in the relatively mild growth in complexity
as the number of external particles increases, 
{\it effectively reducing loop calculations to tree-like calculations}.

On-shell methods fall into two basic categories: the unitarity
method~\cite{Bern:1994zx, Bern:1994cg} which constructs amplitudes
based on their branch cuts, and on-shell recursion~\cite{Britto:2004ap,
Britto:2005fq} which constructs amplitudes from their poles. In this section we
discuss using on-shell recursion as a means for computing rational
terms of one-loop amplitudes.  The loop-level construction is based
directly on the construction of tree-level recursion relations by
Britto, Cachazo, Feng and Witten (BCFW), though a number of new
features are present.  Further discussion of the unitarity method
approach, as well as other new methods exploiting on-shell
conditions on intermediate states~\cite{Britto:2004nc, Britto:2005ha,
Britto:2006sj,Anastasiou:2006gt, Ossola:2006us, Forde:2007mi,
Britto:2007tt} may be found in other sections of this report.
Introductions to on-shell methods may be found in various
reviews~\cite{Bern:1996je, Cachazo:2005ga, Bern:2007dw}.  Earlier
reviews of spinor methods, which are profitably used in
conjunction with on-shell methods, may be found in 
refs.~\cite{Mangano:1990by,Dixon:1996wi}.

In the context of the unitarity method, it is convenient to divide the
amplitudes into pieces that contain branch cuts, plus rational 
(non-cut-containing) pieces.
When using dimensional regularization, the branch-cut containing 
pieces may be computed by ignoring the distinction
between $D=4-2\epsilon$ dimensions and four-dimensions in the numerators
of the loop-momentum integrands~\cite{Bern:1994zx, Bern:1994cg}.  
This observation allows
powerful four-dimensional spinor techniques to be used to greatly
simplify the on-shell tree amplitudes appearing in the unitarity cuts.
However, if one wants to obtain also the rational terms directly from
the cuts~\cite{Bern:1995db,Bern:1996je}, then the $(-2\epsilon)$
dimensional contributions are crucial: dropping these pieces leaves
undetermined additive rational terms.  (The branch cuts can determine
rational terms at ${\cal O}(\epsilon^0)$ because they develop branch
cuts at ${\cal O}(\epsilon)$.)  
By using amplitudes valid in $D=4-2\epsilon$ dimensions in the 
unitarity cuts, all rational terms are
kept\footnote{In the language of dispersion
  relations~\cite{Mandelstam:1958xc, Mandelstam:1959bc}, this
  reconstruction is possible because the dispersion integrals converge
  with dimensional regularization~\cite{vanNeerven:1985xr}.}  but at the
cost of more complicated expressions. It has been
pointed out~\cite{Xiao:2006vr, Su:2006vs,
  Xiao:2006vt, Binoth:2006hk} that the rational terms are relatively
easy to obtain from Feynman diagrams because they do not require the
full set of tensor integrals.  In addition, Brandhuber {\em et.~al.} have
argued that the rational terms can be obtained from a set of
counterterms~\cite{Brandhuber:2007vm}.  Britto and Feng have
recently given a complete set of formul\ae\ for constructing loop
amplitudes, including their rational terms~\cite{Britto:2007tt},
following earlier work~\cite{Britto:2004nc, Britto:2005ha,
  Anastasiou:2006jv, Britto:2006sj, Mastrolia:2006ki, Britto:2006fc,
  Anastasiou:2006gt}.

An early version of on-shell methods
was used to compute the one-loop matrix elements needed for 
the NLO QCD corrections to 
$e^+ e^- \rightarrow \gamma^*,Z \rightarrow$ 4 jets and $pp \rightarrow W,Z$ + 2
jets~\cite{Bern:1997sc}.  They have also been used to obtain analytic
expressions for the complete one-loop six-gluon amplitude~\cite{Bern:1994zx,
  Bern:1994cg, Bidder:2004tx, Bidder:2005ri, Britto:2005ha,
  Britto:2006sj, Berger:2006ci, Berger:2006vq, Xiao:2006vr,
  Su:2006vs,Xiao:2006vt} as well as a variety of helicity configurations
for $n$-gluon amplitudes~\cite{Bern:2005ji, Forde:2005hh, Berger:2006ci,
  Berger:2006vq,Badger:2007si}.  The results confirm the mild growth in
complexity of these methods as the number of external particles grows.

A crucial next step for applying these methods to LHC physics is the
construction of automated programs to compute the large number of
phenomenologically interesting high-multiplicity processes.  As
discussed in other sections of this report, such automated programs
are in the midst of being
constructed~\cite{Ellis:2007br,Ossola:2007ax}, using the integration
machinery of Ossola, Papadopoulos, and Pittau~\cite{Ossola:2006us}.
The recent numerical implementation by Ellis, Giele and 
Kunszt~\cite{Ellis:2007br} of the
unitarity method presently makes use of $D=4$ simplifications and
hence does not contain rational terms.   The
program of Ossola, Papadopoulos, and Pittau~\cite{Ossola:2007ax} 
can be used to obtain the rational terms, but currently requires
one-loop Feynman diagrams to capture these terms, instead of more efficient 
on-shell tree amplitudes.


On-shell recursion offers an efficient alternative for constructing
one-loop rational terms directly from their known factorization
properties, in much the same way as the BCFW recursion relations can
be used to obtain tree-level amplitudes.  However, a number of new
issues arise at loop level that must be dealt with first to have a
practical method.  These issues include the appearance of branch cuts,
spurious singularities, the behavior of loop amplitudes under large
complex deformations and in some cases, `unreal poles', which are
present with complex but not real momenta.  More practical issues are
automation and numerical stability.  Here we
briefly summarize the construction of rational terms via on-shell
recursion~\cite{Bern:2005hs, Bern:2005ji, Bern:2005cq, Berger:2006ci,
Berger:2006vq, Badger:2007si}, 
describing in particular a simple modification making 
it straightforward to automate.

\begin{figure}
\begin{center}
\includegraphics[width=0.3\textwidth]{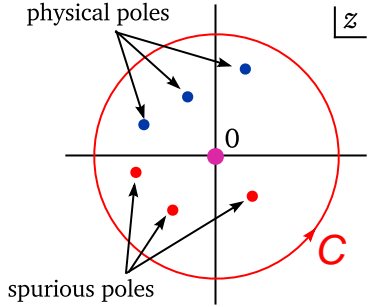}
 \caption{Using Cauchy's theorem, rational terms in loop
amplitudes can be reconstructed from residues at poles in the 
complex plane.  The poles are of two types: physical and spurious.
All pole locations are known {\it a priori}.
Residues at physical poles follow from factorization onto
lower-point amplitudes.  Residues at spurious poles
cancel against corresponding contributions from the cut parts,
and so they can be inferred from four-dimensional cuts.}
\label{LoopCauchyFigure}
\end{center}
\end{figure}

In general, any one-loop amplitude can be divided into two pieces, 
\begin{equation}
A_n^{(1)} = c_\Gamma \Bigl[C_n + R_n \Bigr] \,,
\label{PureCutDecomp}
\end{equation}
where $C_n$ are the `cut-containing terms' possessing logarithms,
polylogarithms, and associated $\pi$'s. 
The rational terms, denoted by $R_n$, are
defined by setting these (poly)logarithmic terms to zero,
\begin{equation}
R_n \equiv {1\over c_\Gamma} A_n
 \Bigr|_{\rm rat} \equiv {1\over c_\Gamma} A_n
  \biggr|_{\ln, {\rm Li}_2, \pi \rightarrow 0} \,.
\label{RatDef}
\end{equation}
Let us assume that the cut-containing terms $C_n$ of the particular
amplitude under consideration have already been computed using
four-dimensional unitarity. This leaves the problem of computing the
rational terms $R_n$.

On-shell recursion relations can be derived by considering complex
on-shell deformations of amplitudes $A(z)$, which are
characterized by a single complex parameter $z$~\cite{Britto:2005fq}.
The $z$-dependence allows us to use standard complex
variable theory to construct amplitudes via Cauchy's Theorem.  To
set up an on-shell recursion relation for $R_n$ consider 
the effect of shifting some set of external momenta $k_i
\rightarrow k_i(z)$, such that the on-shell conditions $[k_i(z)]^2 =
m_i^2$ and the original momentum conservation are satisfied.  In the
massless case, it is particularly convenient to shift the momenta of
two external legs, say, $j$ and $l$,
\begin{eqnarray}
&k_j^\mu &\rightarrow k_j^\mu(z) = k_j^\mu - 
      {z\over2}{\langle{j^-}|{\gamma^\mu}|{l^-}\rangle}, \nonumber\\
&k_l^\mu &\rightarrow k_l^\mu(z) = k_l^\mu + 
      {z\over2}{\langle{j^-}|{\gamma^\mu}|{l^-} \rangle} \,,
\label{MomShift}
\end{eqnarray}
where $z$ is a complex parameter and $|{i^+}\rangle$ and
$|{i^-}\rangle$ are Weyl spinors of positive and negative chirality,
following the notation of ref.~\cite{Mangano:1990by}.  In terms
of these spinors, the shift is 
\begin{equation}
|{j^-}\rangle \rightarrow  |{j^-}\rangle - z \,|{l^-}\rangle\,, \hskip 1 cm 
|{l^+}\rangle \rightarrow  |{l^+}\rangle + z \,|{j^+}\rangle\,.
\label{SpinorShift}
\end{equation}
We denote the shift in eqs.~(\ref{MomShift}) and~(\ref{SpinorShift})
as a $[j, l \rangle$ shift.

The on-shell recursion relations follow from evaluating the contour
integral,
\begin{equation}
{1\over 2\pi i} \oint_C dz \,{R_n(z)\over z} \,,
\end{equation}
where the contour is taken around the circle at infinity, as depicted
in fig.~\ref{LoopCauchyFigure}, and $R_n(z)$ is $R_n$ evaluated at the
shifted momenta~(\ref{MomShift}).  If the rational terms under
consideration vanish as $z\rightarrow \infty$, the contour integral
vanishes and we obtain a relationship between the desired rational
contributions at $z=0$, and a sum over residues of the poles of
$R_n(z)$, located at $z_{\alpha}$,
\begin{equation}
R_n(0) = -\sum_{{\rm poles}\ \alpha} 
 {\rm Res}_{z=z_\alpha}  {R_n(z)\over z} \,.
\label{TreeResidueSum}
\end{equation}
If $R_n(z)$ does not vanish as $z\rightarrow\infty$, 
then there are additional contributions.  A systematic
strategy for computing such large $z$ contributions using 
auxiliary recursion relations was presented in 
ref.~\cite{Berger:2006ci}, to which we refer the reader.

\begin{figure}
\begin{center}
\includegraphics[width=.8\textwidth]{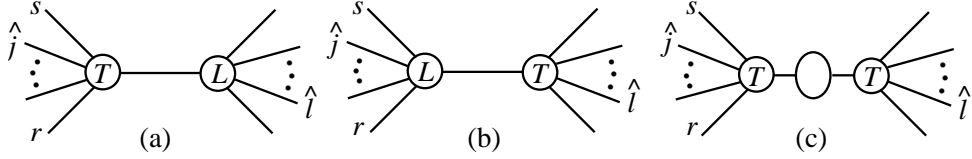}
 \caption{Diagrammatic contributions to on-shell recursion at one-loop
for a $[j, l \rangle$ shift.  The labels `T' and `L' refer to tree and
one-loop vertices corresponding to the rational parts of lower-point
on-shell amplitudes.}
\label{LoopGenericFigure}
\end{center}
\end{figure}

As illustrated in fig.~\ref{LoopCauchyFigure}, the poles in one-loop
rational terms fall into two categories: the physical poles, which are
present in the full amplitudes; and the spurious poles, which cancel
against poles in the cut-containing terms.

Residues at physical poles are dictated by factorization onto lower-point
amplitudes.  They may be computed using the recursive 
diagrams\footnote{`Unreal' poles, which do not correspond to
factorizations with real momenta, may be avoided by choosing 
appropriate shifts~\cite{Berger:2006ci}.}
in fig.~\ref{LoopGenericFigure},
\def\indentA{\hskip 7mm}
\begin{eqnarray}
R^D_n &\!\! \equiv\!\! & \ 
 -\sum_{{\rm phys.\ poles}\ \{r,s\}} {\rm Res}_{z=z_{rs}} {R_n(z)\over z}
\nonumber\\
&\!\!=\!\!& 
\sum_{r,s,h} \Biggl\{
 A^{\rm tree}_L(z= z_{rs})
\, {i\over K_{r\cdots s}^2} \,
R_R(z= z_{rs})
+ R_L(z = z_{rs}) \,
{i\over K_{r\cdots s}^2} \, 
A^{\rm tree}_R(z= z_{rs}) 
\nonumber\\
&& \null \hskip 1. cm
+ A_L^{\rm tree}(z= z_{rs})  
\, {i R_{\cal F} \over K_{r\cdots s}^2 } \,
A^{\rm tree}_R(z= z_{rs}) \Biggr\} 
\,.  \label{RationalRecursion}
\end{eqnarray}
The `vertices' $R_L$ and $R_R$ in this recursion relation are the pure
rational parts --- using the definition~(\ref{RatDef}) --- of the
lower-point, on-shell one-loop amplitudes.
The `vertices' $A^{\rm tree}_L$ and $A^{\rm tree}_R$ are on-shell 
tree amplitudes.  The subscripts $L$ and $R$ on the vertices indicate
their location to the left or right of the central propagator in 
fig.~\ref{LoopGenericFigure}.  In the vertices the shift variable $z$
is frozen to the values
\begin{equation}
z_{rs} = {K_{r\cdots s}^2 \over \langle{j^-}| \; {\slash \hskip -.3 cm 
     K_{r\cdots s}} \, |{l^-}\rangle } \,,
\end{equation}
corresponding to the location of the poles in $z$, coming from shifted
propagators.  The rational part $R_{\cal F}$ of the factorization 
function ${\cal F}$~\cite{Bern:1995ix}
only contributes in multi-particle channels, and only if the tree
amplitude contains a pole in that channel.  Generically we have a double
sum, labeled by $r,s$, over recursive diagrams, with legs $j$ and $l$
always appearing on opposite sides of the pole.  There is also a sum
over the helicity $h$ of the intermediate state.  The superscript $D$ on
$R^D_n$ indicates that this set of recursive diagrammatic
contributions is not the whole rational part, as discussed below.

It is interesting to note the similarity of the one-loop recursion 
relation~(\ref{RationalRecursion}), to the corresponding tree-level 
recursion relation~\cite{Britto:2005fq},
\begin{equation}
A_n^{\rm tree} = 
\sum_{r,s,h} A^{\rm tree}_L(z = z_{rs}) \,
{i\over K_{r\cdots s}^2} \, 
A^{\rm tree}_R(z= z_{rs}) \,.
\end{equation}
Thus loop-level recursive diagrams echo the simplicity
of tree-level recursion.

\begin{figure}
\begin{center}
\includegraphics[width=.3\textwidth]{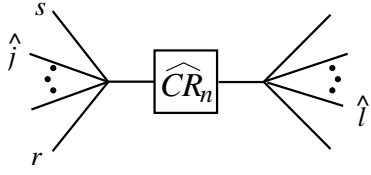}
 \caption{Diagrammatic representation of one-loop overlap terms
for a $[j, l \rangle$ shift. The channels correspond to physical 
poles and remove double counts induced by cut completion.}
\label{OverlapGenericFigure}
\end{center}
\end{figure}

One way to deal with the spurious poles is to start by finding a 
`cut completion' 
$\hat{C}_n$~\cite{Bern:2005cq, Berger:2006ci, Berger:2006vq, 
Bern:2007dw, Badger:2007si}.
One adds certain rational terms $\widehat{CR}_n$ to $C_n$, such that the 
spurious poles in $\hat{C}_n(z)$ cancel entirely.  Because physical
amplitudes cannot have spurious singularities, the remaining rational 
terms, $\hat{R}_n(z)$, must also be free of these spurious singularities.
This cut completion makes it unnecessary to compute residues at
spurious poles. It is rather helpful when deriving compact analytic
expressions for the amplitudes.  It does introduce additional 
`overlap diagrams', as depicted in fig.~\ref{OverlapGenericFigure}.
These diagrams correct for the contributions of $\widehat{CR}_n$ in
physical factorization limits.  They are simple to compute from the 
residue of $\widehat{CR}_n$ at each physical pole $z_{rs}$.

Following the cut-completion procedure, a variety
of rational terms with an arbitrary number of external legs have been
constructed~\cite{Forde:2005hh, Berger:2006ci, Berger:2006vq, Badger:2007si}, 
giving complete amplitudes when combined with the previously-computed
cut-containing parts~\cite{Bern:1994zx, Bern:1994cg, Bedford:2004nh,
Bidder:2005ri, Bern:2005hh, Badger:2007si}.  
More generally, it should be possible to
form a set of cut completions using integral functions of the type
given in ref.~\cite{Campbell:1996zw} to absorb the spurious
singularities.

For the purposes of automation in a numerical program, another approach
is preferable~\cite{BlackHat}.  It is simpler to obtain the residues
at the spurious poles directly from the cut parts, calculated
from the four-dimensional unitarity method.  Because a complete
amplitude is free of spurious poles, any spurious pole found in
the rational parts must cancel a spurious pole in the cut
parts.  To get the full rational part,
\begin{equation}
R_n = R^D_n + R^S_n \, , 
\label{totrec}
\end{equation}
we add to the recursive diagrams $R^D_n$
some `spurious' contributions $R^S_n$, 
evaluated by means of the cut terms $C_n(z)$,
\begin{equation}
R^S_n =  
 -\sum_{{\rm spur.\ poles}\ \beta} {\rm Res}_{z=z_\beta} {R_n(z)\over z}
= \sum_{{\rm spur.\ poles}\ \beta} {\rm Res}_{z=z_\beta} {C_n(z)\over z} \,.
\label{spureqn}
\end{equation}
The spurious poles $\beta$ can be classified systematically in terms
of the vanishing loci, $\Delta(z)=0$, of shifted Gram determinants $\Delta$
associated with box, triangle and bubble functions.
(In the massless case, the bubble Gram determinant does not generate
a spurious pole.)


To illustrate this modified procedure, consider the
five-gluon amplitude $A_5^{(1),s}(1^-,2^-,3^+,4^+,5^+)$,
with a scalar in the loop.  The
construction of the rational terms in this amplitude, using on-shell
recursion with cut completion, has already been discussed in some detail
elsewhere~\cite{Bern:2005cq,Bern:2007dw}.  Here we describe the new
approach for obtaining these terms.

The cut part of the amplitude~\cite{Bern:1993mq} is 
\begin{eqnarray}
C_5 &\!\!= \!\!& 
 -{i\over6} {\langle 1\,2 \rangle^3 \over \langle2\,3\rangle
   \langle3\,4\rangle \langle4\,5\rangle \langle5\,1\rangle}  
   \biggl[ \ln\biggl( {-s_{23} \over \mu^2} \biggr) 
         + \ln\biggl( {-s_{51} \over \mu^2} \biggr) \biggr] \nonumber \\
& & \null
  -{i\over 3} {[3\,4] \langle4\,1\rangle \langle2\,4\rangle [4\,5]
     (\langle2\,3\rangle [3\,4] \langle4\,1\rangle 
            +\langle2\,4\rangle [4\,5] \langle5\,1\rangle ) 
                 \over\langle3\,4\rangle \langle4\,5\rangle }
     {\ln \biggl({-s_{23}\over -s_{51}}\biggr) \over (s_{51} -s_{23})^3} 
+ \cdots 
\label{Cutfive}
\end{eqnarray}
where $\mu^2$ is a scale and `$\cdots$' signifies that we are dropping
terms not pertinent for our discussion. The spinor inner
products and kinematic invariants are defined as,
\begin{equation}
\langle a \, b \rangle \equiv \langle a^-\,| \, b^+\rangle \,, \hskip 2 cm 
[ a \, b ] \equiv \langle a^+\,| \, b^-\rangle \,, \hskip  2 cm 
s_{ab} \equiv (k_a + k_b)^2 \,.
\end{equation}
%

\begin{figure}
\begin{center}
\includegraphics[width=.5\textwidth]{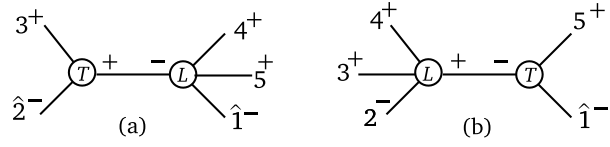}
 \caption{
Non-vanishing recursive diagrams for the rational terms of 
$A_5^{(1),s}(1^-,2^-,3^+,4^+,5^+)$, using a $[1,2\rangle$ shift.
}
\label{RecursmmpppFigure}
\end{center}
\end{figure}

The rational terms are determined by evaluating the recursive
diagrams, plus the rational residues of the cut terms
at the spurious poles.  Here we use the $[1,2\rangle$ shift.
(As discussed in ref.~\cite{Bern:2005cq, Berger:2006ci}, for this
shift there are no additional contributions from either large $z$ 
behavior or unreal poles.)  With the $[1,2\rangle$ shift, 
the non-vanishing recursive diagrams are depicted in
fig.~\ref{RecursmmpppFigure}.  A simple computation of these
diagrams (see section 5.1 of ref.~\cite{Bern:2007dw}) gives,
\begin{equation}
D_5^{\rm (a)} = 
i \biggl( {1\over 3 \epsilon} + {8 \over 9} \biggl) 
{\langle1\,2\rangle^3 \over \langle2\,3\rangle \langle3\,4\rangle
      \langle4\,5\rangle \langle5\,1\rangle}
 \,, \hskip 1.5 cm 
D_5^{\rm (b)}  = - {i \over 3} 
           {[2\,4] [3\,5]^3 \over \langle3\,4\rangle 
         [1\,2] [1\,5] [2\,3]^2} \,,
\label{D5ab}
\end{equation}
as the recursive contributions.  

We still need to account for the residues at the spurious poles.
In the present example with a $[1,2\rangle$ shift, 
the only such pole comes from solving $s_{51}(z) - s_{23}(z) = 0$ 
(corresponding to a shifted two-mass triangle Gram determinant).
The solution is,
\begin{equation}
z_s \equiv {s_{51}  - s_{23}
   \over  \langle1\,5\rangle [ 5\, 2 ] + \langle1\,3\rangle [3\, 2]}
    =  {s_{51}  - s_{23} \over \langle 1\, 4 \rangle [2\, 4] } \,.
\label{PoleLocation}
\end{equation}
To obtain the residue, we start from the logarithmic terms of
eq.~(\ref{Cutfive}), and perform the $[1,2\rangle$ shift
eq.~(\ref{SpinorShift}), yielding,
\begin{eqnarray}
{C_5(z) \over z}
 &\!\! = \!\!& -{i\over 3}
     {[3\,4] \langle4\,1\rangle (\langle2\,4\rangle + z \langle 1\,4\rangle)
      [4\,5]  ((\langle2\,3\rangle + z \langle1\,3\rangle)
              [3\,4] \langle4\,1\rangle 
          + (\langle2\,4\rangle + z \langle1\,4\rangle)
                [4\,5] \langle5\,1\rangle ) 
                     \over\langle3\,4\rangle \langle4\,5\rangle}\nonumber
  \\
 && \hskip .5 cm \times
   {\ln \biggl({(\langle 2\, 3 \rangle + z \langle 1\, 3 \rangle) [3 \, 2]
         \over \langle5 \,1 \rangle ([1 \,5] - z [2\, 5])}\biggr) 
         \over z \, ( s_{51} - s_{23}  
                   - z \langle1 \,4 \rangle [2\, 4] )  )^3} 
\, + \, \cdots \,, 
\label{PureCutFiveZ}
\end{eqnarray}
where we have kept only the term contributing to the spurious 
residue at $z_s$. 

The residue needed for eq.~(\ref{spureqn}) 
can be extracted straightforwardly, by series expanding both the
logarithm and its coefficient in eq.~(\ref{PureCutFiveZ})
around $z=z_s$.  Cleaning up the result of this residue evaluation, 
we find,
\begin{eqnarray}
S_5^{\rm (a)} &\!\! =\!\!& {\rm Res}_{z=z_s} {C_5(z)\over z}
\nonumber\\
&\!\! =\!\!&
- {i\over6} {\langle1\,2\rangle^2\langle1\,4\rangle [3\,4]
           \over \langle1\,5\rangle \langle2\,3\rangle \langle3\,4\rangle
             \langle4\,5\rangle [2\,3] }
+
{i\over6} {\langle1\,4\rangle [3\,4] [3\,5] \bigl(\langle1\,4\rangle
        [3\,4] -\langle1\,5\rangle [3\,5] \bigr)
           \over\langle1\,5\rangle \langle3\,4\rangle \langle4\,5\rangle
            [1\,5] [2\,3]^2}  \nonumber\\
&&\hskip0cm \null 
 -{i\over 6} {s_{51} + s_{23} \over s_{23} s_{51} (s_{51} -  s_{23})^2}
  \, {[3\, 4] \langle4\,1\rangle \langle2\,4\rangle [4\, 5]
 \Bigl(\langle2\,3\rangle [3\,4] \langle4\,1\rangle
     + \langle2\,4\rangle [4\,5] \langle5\,1\rangle \Bigr)
             \over\langle3\,4\rangle \langle4\,5\rangle} \,. \hskip .5 cm 
\label{SpuriousResidue}
\end{eqnarray}
The total rational part,
\begin{equation}
 R_5 = R_5^D + R_5^S
     = D_5^{\rm (a)} + D_5^{\rm (b)} + S_5^{\rm (a)} \,,
\label{totrec5}
\end{equation}
matches the result obtained in
refs.~\cite{Bern:2005cq,Bern:2007dw} using a cut completion.  
The complete amplitude is obtained by summing the cut (\ref{Cutfive})
and rational (\ref{totrec5}) contributions.

The modified construction described here is amenable to automation.
In a numerical program, instead of obtaining the residues at spurious poles
by series expansion, we may compute them by numerically evaluating
the cut terms at several points around each pole.  
The automation and numerical implementation of on-shell recursion to 
amplitudes of interest for LHC phenomenology will be described
elsewhere~\cite{BlackHat}.

\subsection*{Acknowledgments}
We are grateful to Carola Berger, Fernando Febres Cordero, 
Darren Forde, Harald Ita, David Kosower and Daniel Ma\^itre
for collaboration on the topics discussed here.

}


\section[Four- and D-dimensional unitarity cuts]
{FOUR- AND D-DIMENSIONAL UNITARITY CUTS%
\protect\footnote{Contributed by: R. Britto, P. Mastrolia}}
{

\subsection{Four-dimensional unitarity cuts}

The application of unitarity as an on-shell method of calculation,
as introduced in~\cite{Bern:1994zx}, is based on 
the principles that  products of on-shell tree-level amplitudes
produce functions with the correct branch cuts in
all channels~\cite{Landau,Mandelstam,Cutkosky,Eden}, and that
any one-loop amplitude can be expressed as a linear combination of
of scalar (i.e. trivial numerator) master integrals
\cite{'tHooft:1978xw,Bern:1992em,Bern:1993kr,Tarasov:1996br,Binoth:1999sp,
Duplancic:2003tv}.
Given the independent knowledge of the master integrals,  
to compute any amplitude it is sufficient to evaluate  the 
coefficients of such a decomposition.

For one-loop amplitudes, systematic techniques have been developed to
extract the coefficients algebraically, preserving gauge invariance at
every intermediate stage of the computation.
The use of {\it four-dimensional}
states and momenta in the cuts enables the construction of the 
polylogarithmic terms in the amplitudes, 
which are fixed by their branch cuts,
but generically drops rational
terms, which have to be recovered independently.

Some recent developments of unitarity-based methods
apply generalized unitarity cuts to amplitudes and master integrals.
The coefficients are then extracted by matching the generalized cuts. 
Generalized unitarity corresponds
to requiring more than two internal particles to be on-shell, and
the fulfillment of these constraints can only be realized through 
complex kinematics.
Complex kinematics are the key for the exploration 
of singularities of amplitudes and the use of
factorization information to reconstruct amplitudes recursively,
since the singularities of a scattering amplitude are determined 
by lower-point amplitudes in the case of poles and by lower-loop
ones in the case of cuts~\cite{Parke:1986gb,Mangano:1990by,Dixon:1996wi,Bern:1996je}.

A notable application 
of complex momenta within generalized unitarity
is the
quadruple cut, which allows for an immediate and purely algebraic determination of
the coefficients of box functions~\cite{Britto:2004nc}. 
Every box coefficient is simply determined
by the product of the four tree-level amplitudes sitting at each corner,
evaluated at the two particular values of the loop momentum which fulfill the four equations imposed
by the vanishing of the cut denominators.
Double and triple unitarity cuts 
have led to direct 
techniques for extracting triangle and bubble integral 
coefficients analytically~\cite{Britto:2005ha,Britto:2006sj,Mastrolia:2006ki}.
In cases where fewer than four denominators are cut, the loop momentum
is not frozen, so some explicit integration over the phase space is
still required.
In ~\cite{Britto:2005ha,Britto:2006sj,Mastrolia:2006ki}, double or
triple cut phase-space integration has been reduced to  
extraction of residues in spinor variables, and, in the case of a
triple cut, residues in a Feynman parameter.
This approach has been used to compute analytically the final
contributions to the cut-constructible part of the
the six-gluon amplitude~\cite{Britto:2005ha,Britto:2006sj},
and the complete six-photon amplitudes~\cite{Binoth:2006hk,Binoth:2007ca}.

In general, one can compute
$n$-point ($n \ge 4$) coefficients from quadruple cuts,
three-point coefficients from triple-cuts, 
and two-point coefficients from double-cuts, by avoiding 
the conventional tensor reduction.
As it turns out,
given the decomposition of any amplitude in terms of master integrals, 
the coefficient of any $n$-point master integral
can be recovered from the $n$-particle cut.
Obviously, any $n$-particle cut may also detect higher-point master
integrals, 
which appear with different analytic structures 
for they come from the Landau poles specific
to each of the master integrals. This is indeed the case for the
usual (double) unitarity cut, which can be used exclusively
to derive box, triangle, and bubble coefficients.  In cases with
massive particles, it is useful to apply a generalized cut to find the
coefficient of the 1-point (tadpole) master integral.

The algorithm of \cite{Britto:2005ha,Britto:2006sj} for evaluating any finite unitarity cut involves a change of coordinates that brings the loop momentum variable into the spinor formalism.  The idea is that the final integrals always localize to some poles in the region of
integration.   Phase space integration is thus reduced to a sequence of algebraic manipulations, up to an integration over a single Feynman parameter, which is responsible for logarithms.  Ultimately, even this integration does not need to be carried out, since it is possible to match integrands at an early stage of the calculation.
The procedure naturally leads to a clean separation of the master integrals, allowing for an individual calculation of the corresponding coefficients.

 By now, explicit analytic formulas for the results of 
  unitarity-based methods are available
 \cite{Britto:2004nc,Britto:2006fc,Forde:2007mi,Britto:2007tt,Kilgore:2007qr}.  Coefficients of the master integrals are listed directly in terms of tree-level input data.  
All integration and reduction can now be avoided.
 Although it may not be a significant distinction in terms of
 the final results, we note that the derivations of \cite{Britto:2004nc,Forde:2007mi,Kilgore:2007qr} used generalized cuts, while those of \cite{Britto:2006fc,Britto:2007tt} used ordinary double cuts.

\subsection{D-dimensional unitarity}

Full one-loop amplitudes can be
reconstructed from unitarity cuts in  $D=4-2\epsilon$ dimensions
\cite{vanNeerven:1985xr,Bern:1995db}.  In the $D$-dimensional
unitarity method, there is no need to distinguish ``rational'' and
``cut-constructible'' parts of the amplitude.  Contributions that
might be called ``rational'' 
(after expanding around $\epsilon = 0$)
appear here as $\epsilon$-dependent terms
in the coefficients of the master integrals
(before expanding around $\epsilon = 0$).  

A systematic $D$-dimensional unitarity double-cut
method was proposed in  \cite{Anastasiou:2006jv,Anastasiou:2006gt}, reducing  one-loop  amplitudes to master integrals
for arbitrary values of the dimension parameter.
Coefficients of the master integrals can be extracted without fully
carrying out the $D$-dimensional phase space integrals.
Only a four dimensional (massive) integration is explicitly required.  That can
be performed by four-dimensional unitarity techniques or any other
available alternative.
The remaining integral, which gives rise to the $\epsilon$-dependence
of the cut-amplitude, is mapped to phase-space integrals in
$4+2n-2\epsilon$ dimensions, where $n$ is a positive integer. With
recursive dimensional shift identities, similar to the ones in loop
integration, the cut-amplitude is reduced in terms of bubble, triangle, box
and pentagon cut master integrals in $4-2\epsilon$ dimensions.
The reduction is valid for an arbitrary number of dimensions.
Expanding  in $\epsilon$ gives both the (poly)logarithmic and
rational part of the amplitude at ${\cal O}(\epsilon^0)$ and higher; 
these contributions are required in cross-sections beyond the 
next-to-leading order in the relevant coupling strength.

 Generalized unitarity cuts
are possible and useful in $D$ dimensions as well
\cite{Brandhuber:2005jw, Mastrolia:2006ki}.
The benefits of the double-cut integration of  
\cite{Britto:2005ha,Britto:2006sj,Anastasiou:2006jv,Anastasiou:2006gt}
have been extended to the evaluation 
of triple cuts \cite{Mastrolia:2006ki},
for the direct extraction of
triangle and higher-point-function coefficients from
any one-loop amplitude in arbitrary dimensions.
Accordingly,
the triple cut is treated as a difference of two double cuts
with the same particle content, and
the same propagator carrying respectively causal 
and anti-causal prescription in each of the two cuts.
The triple cut phase space for a massless particle in $D$ dimensions
is written as a convolution of 
a four-dimensional triple cut of a massive particle, and an
integration over the corresponding mass parameter, which 
plays the role of a $(-2\epsilon)$-dimensional scale.
Just as in the case of the double-cut \cite{Anastasiou:2006gt,Anastasiou:2006jv},
to perform the four-dimensional integration, one combines
the method of spinor integration 
of massive phase-space integrals, and 
an integration over the Feynman parameter.
But, in the case of the triple-cut,
after Feynman parametrization, by combining back the two double-cuts,
the parametric integration is reduced to the extraction of residues
to the branch points in correspondence of the zeroes
of a  standard quadratic function  in the Feynman parameter.
It is that standard quadratic function (or rather, its roots) that carry the analytic information 
characterizing each master integral, therefore determining 
its own generalized cuts.
The final integration over the dimensional scale parameter
is mapped directly to the triple cut of 
master integrals, possibly with shifted dimensions.

\subsection{Mathematica package for spinor formalism}

Recently, the package S@M (Spinors@Mathematica) was released
\cite{Maitre:2007jq}.  It implements the spinor-helicity formalism in Mathematica. 
The package allows the use of complex-spinor algebra 
along with the multi-purpose features of Mathematica, and it is suitable
for the algebraic manipulation and integration 
of products of tree amplitudes with
complex spinors sewn in generalized unitarity cuts.

%
}


\section[Comments on unitarity based one-loop algorithms]
{COMMENTS ON UNITARITY BASED ONE-LOOP ALGORITHMS%
\protect\footnote{Contributed by: R.K.~Ellis, W.T.~Giele, Z.~Kunszt}}
{
\def\PL #1 #2 #3 {{\it Phys. Lett.} {\bf#1} (#3) #2}
\def\NP #1 #2 #3 {{\it Nucl. Phys.} {\bf#1} (#3) #2}
\def\ZP #1 #2 #3 {{\it Z. Phys.} {\bf#1} (#3) #2}
\def\PRL #1 #2 #3 {{\it Phys. Rev. Lett.} {\bf #1} (#3) #2}
\def\PR #1 #2 #3 {{\it Phys. Rev.} {\bf#1} (#3) #2}
\def\MPL #1 #2 #3 {{\it Mod. Phys. Lett.} {\bf#1} (#3) #2}
\def\RMP #1 #2 #3 {{\it Rev.~Mod. Phys.} {\bf#1} (#3) #2}

\newcommand{ \WG}[1]{{\bf WG:} {\em #1 }}
\newcommand{ \ZK}[1]{{\bf ZK:} {\em #1 }}
\newcommand{ \RKE}[1]{{\bf RKE:} {\em #1 }}

\newcommand{\e}{\epsilon}

\newcommand{\beqn}{\begin{eqnarray}}
\newcommand{\eeqn}{\end{eqnarray}}
\newcommand{\beqns}{\begin{eqnarray*}}
\newcommand{\eeqns}{\end{eqnarray*}}
\newcommand{\nnnr}{\nonumber}
\def\vspaceinarray{\nonumber ~&~&~\\}

\newcommand{\beq}{\begin{equation}}
\newcommand{\eeq}{\end{equation}}
\newcommand{\beqa}{\begin{eqnarray}}
\newcommand{\eeqa}{\end{eqnarray}}
\newcommand{\s}{\slash\hspace{-5pt}}
\newcommand{\nn}{\nonumber \\}
\newcommand{\ve}{\varepsilon}
\bibliographystyle{lesHouches}

%
%
%
%
%

{\large\bf }  
\subsection{Introduction}
At the LHC deviations from the Standard Model will likely show up in  
observables of complex multi-particle final states. It is important
to understand the Standard Model predictions and uncertainties
for these complicated final states. Leading-order Monte Carlo (LO-MC) 
programs give a first estimate. However, to understand the
uncertainties we need at least a next-to-leading order Monte Carlo (NLO-MC).

The basic calculational framework for both tree-level amplitudes
(needed for the LO-MC) and one-loop amplitudes (needed for NLO-MC)
is the perturbative expansion in Feynman diagrams. This immediately
gives us a straightforward algorithm suitable for
numerical implementation. However, such implementations are not satisfactory from
a numerical standpoint. The number of Feynman
diagrams grows faster than factorial with the number of external
particles involved in the scattering process. As a consequence the
number of multiplications, and therefore the computer time needed to
evaluate a phase space point, will grow at least as fast.  
 
In computer science, algorithms with factorial growth 
are called exponential or factorial
algorithms or simply E-algorithms \cite{Sipser}.    
Such algorithms are not considered optimal, i.e. the number
of external particles we can calculate becomes quickly limited
by computer resources. 
In contrast, the other class of algorithms with polynomial growth
in the number of external legs are called P-algorithms.
Such algorithms are highly desirable as the added computational
effort needed to go from $N$ to $(N+1)$ external particles is
$\left(\frac{N+1}{N}\right)^\alpha$. This means the limiting
factor for these types of algorithms in scattering amplitude
calculations is often human resources instead of computer resources.
In the subsequent sections we will argue that for numerical solutions,
especially in the era of LHC physics,
the complexity of the algorithms are an important consideration.

\subsection{Tree-level algorithms of polynomial complexity}

The number of Feynman graphs grows very fast with the number
of external legs. For a tree-level $N$-gluon scattering
the number of individual Feynman graphs is approximately
$N^{(N-3)}$ (within 5\% accuracy up to 16 gluons) \cite{Kleiss:1988ne}. 
This means
that to extend the LO-MC from $2\ \mbox{gluon}\rightarrow 5\ \mbox{gluon}$ to 
$2\ \mbox{gluon}\rightarrow 6\ \mbox{gluon}$, the number
of multiplications increases by at least a factor of 13.  
Several LO-MC are available for the
numerical evaluation of arbitrary tree-level processes
in the Standard Model and some of its extensions. 
Most of these packages are based on simple
Feynman diagram evaluations. We call these 
Numerically Implemented Exponential (NIE) algorithms. 
A prominent representative in this class of algorithms is MadGraph \cite{Stelzer:1994ta}.

By using currents instead of amplitudes in Feynman diagram calculations
one can construct recursion relations \cite{Berends:1987me}. 
This method re-uses recurring groups of off-shell Feynman graphs in an
optimal manner. Because this leads to a more factorized way of calculating
the scattering amplitude one can immediately extend the analytic
calculations to more complex processes such as vector boson production
with up to 6 partons \cite{Berends:1988yn,Berends:1990ax} and 7 parton
processes \cite{Berends:1989hf}. 

Another consequence of the recursion relations is the formulation of
an algorithm of polynomial
complexity. For a tree-level $N$-gluon process the number of 
multiplications grows as $N^4$ \cite{Kleiss:1988ne}.
This means
that to extend the LO-MC from $2\ \mbox{gluon}\rightarrow 5\ \mbox{gluon}$ to 
$2\ \mbox{gluon}\rightarrow 6\ \mbox{gluon}$ the increase in the
number of multiplications is only 1.7 (compared to 13 for standard
Feynman graph calculations). We will denote the LO-MC programs
based on recursive type of evaluation Numerically Implemented 
Polynomial (or NIP) algorithms. A prominent representative
is the ALPGEN program \cite{Mangano:2002ea}.

As is clear from the discussion we have reached a point for LO-MC
where the problem
of numerically calculating the scattering amplitudes can be
considered solved.

\subsection{Toward one-loop  algorithms of polynomial complexity}

The LO-MC prediction at LHC type of energies for QCD and/or Electro-Weak
processes are rather qualitative. One estimates the magnitude of the 
cross section and predicts the shape for an observables. The
NLO-MC will give us a first real estimate of the expected normalization
and will give an order $\alpha_S$ correction to the shape. Within the
perturbative context this allows us to estimate the uncertainties on
the predictions with some confidence. 

The one-loop amplitude of the basic  $2\ \mbox{gluon}\rightarrow 2\ \mbox{gluon}$
was already calculated analytically in 1986 \cite{Ellis:1985er}
using the standard Feynman diagram calculation.  One can extend this method
brute force with modern day computers. Using a combination of 
e.g. QGRAF \cite{Nogueira:1991ex} and FORM \cite{Vermaseren:2000nd}
one can generate and manipulate the Feynman graphs giving tensor coefficients
times tensor integrals. The tensor integrals can be determined using 
Pasasarino-Veltman reduction \cite{Passarino:1978jh} or other techniques.
This then can be straightforwardly implemented in a numerical code for e.g.
$2\ \mbox{gluon}\rightarrow 4\ \mbox{gluon}$ \cite{Ellis:2006ss}. The evaluation
of a single phase space point for this process is of the order 9 second (10,000
times slower as the $2\ \mbox{gluon}\rightarrow 2\ \mbox{gluon}$ one-loop
amplitude generated using the same procedure). It is clear that such a direct
approach using Feynman diagrams is severely affected by the factorial growth
in complexity. One needs badly a polynomial complexity calculational approach.

It can be shown that any dimensional regulated
multi-loop amplitude is fully reconstructible using unitarity cuts \cite{Neerven:1985xr}.
Because the unitarity cuts factorizes the one-loop amplitudes into a product of two tree-level amplitudes
this proves the existence of a polynomial complexity algorithm for one-loop calculations.
This was exploited in the analytic calculation of the 
$e^+e^-\rightarrow$ 4 partons one-loop amplitude \cite{Bern:1997sc}
\footnote{The 5 gluon one-loop was calculated using string inspired methods \cite{Bern:1993mq}.}. 
The method applies four-dimensional unitarity
cuts, thereby it only partly reconstructs the one-loop amplitude through unitarity, 
the so-called
cut-constructible part. The missing part is referred to as the rational part and is determined by other methods.
The applied 4-dimensional unitarity method has no direct numerical equivalent, but it is  explicitly
demonstrated that such methods of polynomial complexity work very well within the context of analytic multi-leg
one-loop calculations.

The first numerically implementable method came from the so-called quadruple cut method \cite{Britto:2004nc}.
While presented as an analytic method to calculate coefficients of the 4-point scalar master
integrals for multi-gluon processes, it has a direct numerical implementation. The numerical
procedure can be used to calculate the box coefficients for any multi-particle scattering process.
By applying the quadruple cut the one-loop graph breaks down into four tree-level amplitudes. This
is therefore instantly a NIP algorithm for calculating the coefficients of the 4-point master integrals.
From the unitarity constraints, i.e. the four cut propagators have to be numerically zero, one gets only
two complex solutions for the loop-momentum. By evaluating the product of the four tree-level graphs 
using the two complex loop momenta solutions, one gets the coefficient by simply averaging over the two solutions.

The numerical implementation of the method is extremely fast and simple, showing the potential
of a full numerical implementable unitarity method. To achieve this one also has to calculate
the coefficients of the other 3 master integrals (the 1-, 2- and 3-point scalar integrals).
A direct generalization of the quadruple cut method becomes complicated because of overlapping 
contributions. By applying a triple cut to determine the 3-point coefficient one has to take 
into account that part of this contribution is also in the quadruple cut. Disentangling these
overlapping contributions proves to be not that straightforward.

For a one-loop amplitude one can construct a general parametric form of the integrand 
and determine its coefficients by demanding different combinations of sets of propagators to be zero 
(i.e. cutting the lines) for both the parametric form of the amplitude
and the expression obtained using Feynman graphs \cite{Ossola:2006us}. 
This method is purely algebraic as it works on the integrand level.
When setting four propagators to zero this method is identical to the
quadruple cut method. However, we now get in addition the full loop 
dependence of the integrand of the 4-point master functions through
its parametric form. This allows one to simply determine the triple
cut contribution of the parametric 4-point integrand and hence we
know the subtraction term.

Using this method to construct the subtraction terms it is now straightforward
to formulate a numerical implementable algorithm of polynomial complexity for
the cut constructible part \cite{Ellis:2007br}. Because we determine the coefficients
of the 2-, 3- and 4-point parametric form of integrands by the equivalent of unitarity
cuts, the actual one-loop amplitude factorizes in a product of two, three or four
tree-level amplitudes. That is, we can determine the full parametric form of the 
integrand from tree-level amplitudes. The final
loop integration over the parametric form is straightforward and gives us the
three scalar master integrals and their respective coefficients. This method now
extends the polynomial complexity algorithm of the quadruple cut method to include
also the triple and double cut contributions. As a demonstration we used this method to numerically
evaluate multi-gluon scattering amplitudes. We found using a single standard processor
the following results: 
the $2\ \mbox{gluon}\rightarrow 2\ \mbox{gluon}$ at 9 seconds/10,000 events,
the $2\ \mbox{gluon}\rightarrow 3\ \mbox{gluon}$ at 35 seconds/10,000 events
and the $2\ \mbox{gluon}\rightarrow 4\ \mbox{gluon}$ at 107 seconds/10,000 events.
This can be approximated by $N^6/450$ seconds/10,000 events, which by extrapolation would give around 
260 seconds/10,000 events for $2\ \mbox{gluon}\rightarrow 5\ \mbox{gluon}$.
These evaluation times are more than sufficient for use in NLO-MC generators, even
on a modest single processor system.

\subsection{Conclusions: the rational part}

The final step is a numerical suitable algorithm for the rational part 
of the one-loop amplitude. This is the final hurdle in achieving a full
solution of polynomial complexity for numerical one-loop amplitude evaluations.
Three methods exist in the literature. The first method determines the rational
part of the tensor integrals. These rational parts can then be contracted in with
the tensor coefficients to give the full one-loop rational part \cite{Xiao:2006vr,Binoth:2006hk}.
This method goes back to the Feynman diagram expansion and leads to an
algorithm of factorial complexity. This negates all progress made with the
determination of the cut constructible part using numerical unitarity techniques.

The other two methods are more analytic in concept, but should in principle be suitable
for a numerical implementation. The  so-called bootstrap method sets
up a recursive procedure for the rational part \cite{Berger:2006ci} similar to the 
tree-level unitarity based recursion relations \cite{Britto:2005fq}. This makes the
method of polynomial complexity. However, in its current formulation it is not
suitable for numerical implementation. The reason is that both the rational and
cut constructible part of the one-loop amplitude contain so-called spurious
poles. When adding the two parts together these spurious poles cancel. This means
that for the construction of an unitarity based recursion relation in the rational
part these spurious poles have to be removed. This procedure is called cut-completion,
i.e. make both cut-constructible and rational part free of spurious poles. Then the
rational part contains only physical poles and a unitarity based tree-level like recurrence
relation for the rational part is constructible. Unfortunately the cut-completion
procedure requires analytic knowledge of the spurious terms, which up to now have
only be determined by explicit analytic calculation of the cut-constructible part.

One can in principle retrieve the full one-loop amplitude by applying
$D$-dimensional unitarity cuts \cite{Bern:1995db,Anastasiou:2006gt}. 
Such an implementation is  per construction of polynomial complexity.
It requires the calculation of the $D$-dimensional tree-level amplitudes. 
This can be implemented by restricting oneself to massive scalar internal particles 
where the mass in generated by the extra-dimensional length of the loop-momentum. 
In this manner the extra-dimensional part of the loop-momentum can be integrated out. 
After that one can read off the appropriate master integral coefficients and rational part.
The required scalar internal particles  restrict this method at the moment
to purely gluonic scattering amplitudes.
In its current implementation this method is restricted to analytic applications for purely gluonic one-loop scattering amplitudes. 

It is clear from the discussions that a numerical algorithm of polynomial complexity
is the only issue left in fully solving one-loop calculations in a similar way
tree-level calculations have been solved. Achieving this final step would open the
way to a multitude of NLO-MC generators for processes such as for example
$PP\rightarrow t\bar{t}$ +2 jets, $PP\rightarrow t\bar{t}\, +\, b\bar{b}$ and
$PP\rightarrow$ Vector-Boson + 3, 4 jets. 

A NIP implementation for the rational part has to exist. Its construction 
in the near future is of great importance to make the first step towards 
more complicated NLO-MC programs relevant for the LHC phenomenology. 
%
%
}


\section[Physical applications of the OPP method to compute
one-loop amplitudes]
{PHYSICAL APPLICATIONS OF THE OPP METHOD TO COMPUTE
ONE-LOOP AMPLITUDES%
\protect\footnote{Contributed by: G.~Ossola, C.G.~Papadopoulos, R.~Pittau}}
{\graphicspath{{papadopoulos/}}
%
\newcommand{\nc}{\newcommand}
\nc{\eqn}[1]{Eq.~\ref{eq:#1}}
\nc{\bqa}{\begin{eqnarray}}
\nc{\eqa}{\end{eqnarray}}
\newcommand{\nl}{\nonumber \\}
\def\db#1{\bar D_{#1}}
\def\zb#1{\bar Z_{#1}}
\def\d#1{D_{#1}}
\def\tld#1{\tilde {#1}}
\def\slh#1{\rlap / {#1}}
\def\eqn#1{Eq.~(\ref{#1})}
\def\eqns#1#2{Eqs.~(\ref{#1}) and~(\ref{#2})}
\def\eqnss#1#2{Eqs.~(\ref{#1})-(\ref{#2})}
\def\fig#1{Fig.~{\ref{#1}}}
\def\figs#1#2{Figs.~\ref{#1} and~\ref{#2}}
\def\sec#1{Section~{\ref{#1}}}
\def\app#1{Appendix~\ref{#1}}
\def\tab#1{Table~\ref{#1}}
%
%
%
%
%
%
\def\beq{\begin{equation}}
\def\eeq{\end{equation}}
\def\bea{\begin{eqnarray}}
\def\eea{\end{eqnarray}}

\def\dofourfigs#1#2#3#4#5{\centerline{
\epsfxsize=#1\epsfig{file=#2, width=7.5cm,height=7.0cm, angle=0}
\hspace{0cm}
\hfil
\epsfxsize=#1\epsfig{file=#3,  width=7.5cm, height=7.0cm, angle=0}}

\vspace{0.5cm}
\centerline{
\epsfxsize=#1\epsfig{file=#4, width=7.5cm,height=7.0cm, angle=0}
\hspace{0cm}
\hfil
\epsfxsize=#1\epsfig{file=#5,  width=7.5cm, height=7.0cm, angle=0}}
}

\def\dosixfigs#1#2#3#4#5#6#7{\centerline{
\epsfxsize=#1\epsfig{file=#2, width=6.5cm,height=5.5cm, angle=0}
\hspace{0cm}
\hfil
\epsfxsize=#1\epsfig{file=#3,  width=6.5cm, height=5.5cm, angle=0}}

\vspace{0.5cm}
\centerline{
\epsfxsize=#1\epsfig{file=#4, width=6.5cm,height=5.5cm, angle=0}
\hspace{0cm}
\hfil
\epsfxsize=#1\epsfig{file=#5,  width=6.5cm, height=5.5cm, angle=0}}

\vspace{0.5cm}
\centerline{
\epsfxsize=#1\epsfig{file=#6, width=6.5cm,height=6.cm, angle=0}
\hspace{0cm}
\hfil
\epsfxsize=#1\epsfig{file=#7, width=6.5cm, height=6.cm, angle=0}}
}

\def\dotwofigsa#1#2#3{\centerline{
\epsfxsize=#1\epsfig{file=#2, width=3cm,height=3.8cm, angle=0}
\hspace{0cm}
\hfil
\epsfxsize=#1\epsfig{file=#3,  width=8cm, height=7.5cm, angle=0}}
}

\def\dotwofigs#1#2#3{\centerline{
\epsfxsize=#1\epsfig{file=#2, width=6cm,height=8cm, angle=0}
\hspace{0cm}
\hfil
\epsfxsize=#1\epsfig{file=#3,  width=7cm, height=8cm, angle=0}}
}

\def\dofig#1#2{\centerline{
\epsfxsize=#1\epsfig{file=#2, width=12cm,height=9cm, angle=0}
\hspace{0cm}
}}

\def\dofigc#1#2{\centerline{
\epsfxsize=#1\epsfig{file=#2, width=15cm,height=8cm, angle=0}
\hspace{0cm}
}}

\def\dofigb#1#2{\centerline{
\epsfxsize=#1\epsfig{file=#2, width=15cm,height=8cm, angle=0}
\hspace{0cm}
}}

\def\dofiga#1#2{\centerline{
\epsfxsize=#1\epsfig{file=#2, width=4cm,height=4cm, angle=0}
\hspace{0cm}
}}

\newcommand{\dedouble}{ \stackrel{ \leftrightarrow }{ \partial } }
\newcommand{\deR}{ \stackrel{ \rightarrow }{ \partial } }
\newcommand{\deL}{ \stackrel{ \leftarrow }{ \partial } }
\newcommand{\ci}{{\cal I}}
\newcommand{\ca}{{\cal A}}
\newcommand{\Wp}{W^{\prime}}
\newcommand{\vep}{\varepsilon}
\newcommand{\kk}{{\bf k}}
\newcommand{\pp}{{\bf p}}
\newcommand{\hs}{{\hat s}}
\newcommand{\proj}{\frac{1}{2}\;(\eta_{\mu\alpha}\eta_{\nu\beta}
+  \eta_{\mu\beta}\eta_{\nu\alpha} - \eta_{\mu\nu}\eta_{\alpha\beta})}
\newcommand{\projm}{\frac{1}{2}\;(\eta_{\mu\alpha}\eta_{\nu\beta}
+  \eta_{\mu\beta}\eta_{\nu\alpha})
- \frac{1}{3}\;\eta_{\mu\nu}\eta_{\alpha\beta}}

\def\lsim{\raise0.3ex\hbox{$\;<$\kern-0.75em\raise-1.1ex\hbox{$\sim\;$}}}

\def\gsim{\raise0.3ex\hbox{$\;>$\kern-0.75em\raise-1.1ex\hbox{$\sim\;$}}}

\def\Frac#1#2{\frac{\displaystyle{#1}}{\displaystyle{#2}}}
\def\no{\nonumber\\}
\renewcommand{\thefootnote}{\fnsymbol{footnote}}
%
%
%
%
%
%
%
%
%
\subsection{Introduction \label{sect1}}
%
%
In two recent papers~\cite{Ossola:2006us,Ossola:2007bb},
we proposed a reduction
technique (OPP) for arbitrary one-loop sub-amplitudes at {\it the
integrand level}~\cite{delAguila:2004nf}
by exploiting numerically the set of kinematical
equations for the integration momentum, that extend the quadruple,
triple and double cuts used in the unitarity-cut method~\cite{Bern:1994cg,Britto:2004nc,Ellis:2007br,Kilgore:2007qr}.
The method requires a minimal information about the form of the one-loop
(sub-)amplitude and therefore it is well suited for a numerical
implementation. The method works for any set of internal and/or
external masses, so that one is able to study the full electroweak
model, without being limited to massless theories.

In Section~\ref{sect2} we outline the basics features of the method.
In Section~\ref{sect3} we describe a numerically stable implementation of
the OPP algorithm, in a form of a {\tt FORTRAN90} code,
{\tt CutTools}~\cite{Ossola:2007ax}.
In the last section, we compute, as an application,
the one-loop QCD corrections to the process $p p \to ZZZ$ at the LHC,
also showing distributions for physically interesting quantities.
\subsection{The OPP method\label{sect2}}
%
%
The starting point of the OPP reduction method is the general expression for the
{\it integrand} of a generic $m$-point
one-loop (sub-)amplitude
\bqa
\label{eq:1}
A(\bar q)= \frac{N(q)}{\db{0}\db{1}\cdots \db{m-1}}\,,~~~
\db{i} = ({\bar q} + p_i)^2-m_i^2\,,~~~ p_0 \ne 0\,.
\eqa
In the previous equation, we use a bar to denote objects living
in $n=~4+\epsilon$  dimensions, and $\bar q^2= q^2+ \tld{q}^2$, where
$\tld{q}^2$ is $\epsilon$-dimensional and $(\tld{q} \cdot q) = 0$.
$N(q)$ is the $4$-dimensional part of the
numerator function of the amplitude. If needed,
the $\epsilon$-dimensional part of the numerator should be treated
separately, as explained later.
$N(q)$ depends on the $4$-dimensional denominators
$\d{i} = ({q} + p_i)^2-m_i^2$ as follows
\bqa
\label{eq:2}
N(q) &=&
\sum_{i_0 < i_1 < i_2 < i_3}^{m-1}
\left[
          d( i_0 i_1 i_2 i_3 ) +
     \tld{d}(q;i_0 i_1 i_2 i_3)
\right]
\prod_{i \ne i_0, i_1, i_2, i_3}^{m-1} \d{i} \nl
     &+&
\sum_{i_0 < i_1 < i_2 }^{m-1}
\left[
          c( i_0 i_1 i_2) +
     \tld{c}(q;i_0 i_1 i_2)
\right]
\prod_{i \ne i_0, i_1, i_2}^{m-1} \d{i} \nl
     &+&
\sum_{i_0 < i_1 }^{m-1}
\left[
          b(i_0 i_1) +
     \tld{b}(q;i_0 i_1)
\right]
\prod_{i \ne i_0, i_1}^{m-1} \d{i} \nl
     &+&
\sum_{i_0}^{m-1}
\left[
          a(i_0) +
     \tld{a}(q;i_0)
\right]
\prod_{i \ne i_0}^{m-1} \d{i} \nl
     &+& \tld{P}(q)
\prod_{i}^{m-1} \d{i}\,. \eqa
Inserted back in \eqn{eq:1}, this expression
simply states the multi-pole nature of any $m$-point one-loop amplitude,
that, clearly, contains a pole for any
propagator in the loop, thus one has terms ranging from 1 to $m$ poles.
 Notice that the term with no poles, namely that one proportional to
$\tld{P}(q)$ is polynomial and vanishes upon integration
in dimensional regularization; therefore does not contribute to the amplitude,
as it should be.
  The coefficients of the poles can be further split in two pieces.
A piece that still depend on $q$ (the terms
$\tld{d},\tld{c},\tld{b},\tld{a}$), that vanishes upon integration,
and a piece that do not depend on q (the terms $d,c,b,a$).
 Such a separation is always possible and the
latter set of coefficients is immediately
interpretable as the ensemble of the
coefficients of all possible 4, 3, 2, 1-point
one-loop functions contributing to the amplitude.

 Once \eqn{eq:2} is established, the task of computing the one-loop amplitude
is then reduced to the algebraical problem of fitting
the coefficients $d,c,b,a$ by evaluating the function $N(q)$
a sufficient number of times, at different values of $q$,
and then inverting the system.
That can be achieved quite efficiently by singling out
particular choices of $q$ such that, systematically,
4, 3, 2 or 1 among all possible denominators $\d{i}$ vanishes.
 Then the system of equations is solved iteratively.
First one determines all possible 4-point functions,
then the 3-point functions and so on.
 For example, calling $q_0^\pm$ the 2 (in general complex) solutions for which
\bqa \d{0}= \d{1}= \d{2}=\d{3} = 0\,, \eqa (there are 2 solutions because
of the quadratic nature of the propagators) and since the functional
form of $\tld{d}(q;0123)$ is known, one directly finds the coefficient
of the box diagram containing the above 4 denominators through
the two simple equations
\bqa
N(q_0^\pm) &=& [d(0123) + \tld{d}(q_0^\pm;0123)] \prod_{i\ne 0,1,2,3}
\d{i} (q_0^\pm)
\,.
\eqa
This algorithm also works in the case of
complex denominators, namely with complex masses.
 Notice that the described procedure can be performed
{\em at the amplitude level}. One does not need to
repeat the work for all Feynman diagrams, provided their sum is known:
we just suppose to be able to compute $N(q)$ numerically.

 The described procedure works in 4 dimensions.
However, even when starting from a perfectly finite tensor integral,
the tensor reduction may eventually lead to integrals
that need to be regularized (we use dimensional regularization).
 Such tensors are finite, but tensor reduction iteratively leads to
rank $m$ $m$-point tensors with $ 1 \le m \le 5 $, that are
ultraviolet divergent when $m \le 4$.
For this reason, we introduced, in \eqn{eq:1}, the $d$-dimensional
denominators $\db{i}$, that differs by an amount $\tld{q}^2$ from
their $4$-dimensional counterparts
\bqa
\db{i}= \d{i} + \tld{q}^2\,.
\eqa
The result of this is a mismatch in the cancellation
of the $d$-dimensional denominators of \eqn{eq:1} with the $4$-dimensional
ones of \eqn{eq:2}. The rational part of the amplitude, called
$R_1$~\cite{Ossola:2008xq}, comes from
such a lack of cancellation. A different source of Rational Terms,
called $R_2$, can also be generated from the $\epsilon$-dimensional part
of $N(q)$ (that is missing in \eqn{eq:1}).
For the time being, it should be added by hand by looking at the
analytical structure of the Feynman Diagrams of via
a dedicated set of Feynman Rules.
Examples on how to compute $R_2$ are reported in~\cite{Ossola:2008xq}
and~\cite{Pittau:1997mv,Pittau:1996ez}.
The Rational Terms $R_1$ are generated by the following
extra integrals, introduced in~\cite{Ossola:2006us,Ossola:2007bb}
\bqa \label{eq:ratexp}
\int d^n \bar{q}
\frac{\tld{q}^2}{\db{i}\db{j}}             &=& - \frac{i \pi^2}{2}
\left[m_i^2+m_j^2-\frac{(p_i-p_j)^2}{3} \right]   +
{\cal O}(\epsilon)\,, \nl
\int d^n \bar{q}
\frac{\tld{q}^2}{\db{i}\db{j}\db{k}}       &=& - \frac{i \pi^2}{2} +
{\cal O}(\epsilon)\,,~~~
\int d^n \bar{q}
\frac{\tld{q}^4}{\db{i}\db{j}\db{k} \db{l}} ~=~ - \frac{i \pi^2}{6} +
\cal{O}(\epsilon)\,.
\eqa
The coefficients of the above integrals can  be computed
by looking at the implicit mass
dependence (namely reconstructing the $\tld{q}^2$ dependence) in
the coefficients $d,c,b$ of the one-loop functions,
once $\tld{q}^2$ is reintroduced through the
mass shift $m_i^2 \to m_i^2 -\tld{q}^2$.
One gets
\bqa
b(ij;\tld{q}^2) &=&   b(ij)
                     +\tld{q}^2 b^{(2)}(ij) \,,~~~
c(ijk;\tld{q}^2) ~=~   c(ijk)
                     +\tld{q}^2 c^{(2)}(ijk)\,.
\eqa
Furthermore, by defining
\bqa
{\cal D}^{(m)}(q,\tld{q}^2) \equiv \sum_{i_0 < i_1 < i_2 < i_3}^{m-1}
\left[
          d( i_0 i_1 i_2 i_3;\tld{q}^2 ) +
     \tld{d}(q;i_0 i_1 i_2 i_3;\tld{q}^2)
\right]
\prod_{i \ne i_0, i_1, i_2, i_3}^{m-1} \db{i} \,,
\eqa
the following expansion holds
\bqa
\label{bigd}
{\cal D}^{(m)}(q,\tld{q}^2)= \sum_{j= 2}^{m} \tld{q}^{(2j-4)} d^{(2j-4)}(q)\,,
\eqa
where the last coefficient is independent on $q$
\bqa
d^{(2m-4)}(q) = d^{(2m-4)}\,.
\eqa
In practice, once the $4$-dimensional
coefficients have been determined, one can redo
the fits for different values of $\tld{q}^2$, in order to determine
$b^{(2)}(ij)$, $c^{(2)}(ijk)$ and  $d^{(2m-4)}$.
Such three quantities are the coefficients of the three
extra scalar integrals listed in \eqn{eq:ratexp}, respectively.
Therefore, the OPP method allows an easy and purely numerical
computation of the Rational Terms of type $R_1$.
\subsection{{\tt CutTools} and the problem of the Numerical Inaccuracies
\label{sect3}}
%
%
A {\tt FORTRAN90} program  ({\tt CutTools})
implementing the OPP method can be found
in~\cite{Ossola:2007ax}, to which we refer for more details.
We just mention that the only information needed by the code is
the number and type of contributing propagators and the
numerator function $N(q)$ (and its maximum rank). A particularly interesting
feature of the OPP technique, also implemented in {\tt CutTools},
is that it allows a natural numerical check of
the accuracy of the whole procedure. Given the paramount importance
of this issue in practical calculations, we describe it here
in some detail.

During the fitting procedure to determine the coefficients,
numerical inaccuracies may occur due to
\begin{itemize}
\item[1)] appearance of Gram determinants in the solutions for which
          4, 3, 2 or 1 denominators vanish;
\item[2)] vanishing of some of the remaining denominators,
          when computed at a given solution;
\item[3)] instabilities occurring when solving
          systems of linear equations;
\end{itemize}
In principle, each of these three sources of instabilities
can be cured by performing a proper expansion
around the problematic (i.e.  {\em exceptional}) Phase-Space point.
However, this often results in a huge amount of work that, in addition,
spoils the generality of the algorithm. Furthermore, one is anyway
left with the problem of choosing a separation criterion
to identify the region where applying the proper expansion rather than
the general algorithm.

 The solution implemented in {\tt CutTools} is, instead, of
a purely numerical nature and
relies on a unique feature of the OPP method: the fact that
the reduction is performed at the {\rm integral} level.
 In detail, the OPP reduction is obtained when, as in \eqn{eq:2},
the numerator function $N(q)$ is rewritten in terms of denominators.
Therefore $N(q)$ computed for some arbitrary value of $q$ by
using the l. h. s. of \eqn{eq:2} should always be {\em numerically} equal to
the result obtained by using the expansion in the r. h. s.
 This is a very stringent test that is applied in {\tt CutTools}
for any Phase-Space point.
 When, in an {\em exceptional} Phase-Space point, these two numbers differ more
than a user defined quantity,
the coefficients of the loop functions {\em for that particular point}
are recomputed by
using multi-precision routines (with up to 2000 digits)
contained in {\tt CutTools}~\cite{multilib1,multilib2}.
 The only price to be payed by the user is writing, beside the
normal ones (namely written in double-precision), a multi-precision version
of the routines computing $N(q)$.
The described procedure ensures that the coefficients of the scalar loop
functions are computed with a precision defined by the user.
Finally, one should mention that, usually, only very few points
are potentially dangerous, namely {\em exceptional}, so that a limited fraction of additional {\tt CPU} time
is used to cure the numerical instabilities, therefore compensating the fact
that the multi-precision routines are by far much slower than the normal ones.
This procedure has been shown to work rather well in practice,
as we shall see in the next section.
\subsection{$pp \to ZZZ$ at one-loop\label{sect4}}

The calculation is composed of two parts: the evaluation of virtual corrections, namely one-loop contributions obtained by adding a virtual particle to the tree-order diagrams, and corrections from the real emission of one additional massless particle from initial and final states, which is necessary in order to control and cancel infrared singularities.
The virtual corrections are computed using the {\tt OPP} reduction\cite{Ossola:2006us,Ossola:2007bb}. In particular, we make use of {\tt CutTools} \cite{Ossola:2007ax}.
Concerning the contributions coming from real emission we used the dipole
subtraction method~\cite{Catani:1996vz} to isolate the soft and collinear
divergences and checked the results using the phase space slicing
method~\cite{Giele:1993dj} with soft and collinear cutoffs, as outlined in~\cite{Harris:2001sx}.

These results have also been recently presented, following a very different approach, by Lazopoulos \emph{et al} in Ref.~\cite{Lazopoulos:2007ix}. A more complete study, that will also include the case of $W^{+}W^{-}Z$, $W^{\pm} Z Z$, and $W^{+}W^{-}W^{\pm}$ production, will be presented in a forthcoming publication \cite{vvv}.

Let us begin with the evaluation of the virtual QCD corrections to the process $q {\bar q} \to Z Z Z$.
We consider the process
\begin{equation}
q (p_1) + {\bar q}(p_2) \longrightarrow Z (p_3) + Z (p_4)+ Z (p_5)
\end{equation}
All momenta are chosen to be incoming, such that $\sum_i p_i =0$.

\begin{figure}[bh] \
     \begin{center}
      \begin{picture}(60,60)(0,0)
        \ArrowLine(0,60)(30,60)
        \ArrowLine(30,0)(0,0)
        \ArrowLine(30,30)(30,0)
    \ArrowLine(30,60)(30,30)
        \DashLine(30,0)(60,0){3}
        \DashLine(30,30)(60,30){3}
        \DashLine(30,60)(60,60){3}
    \Text(-10,0)[]{$p_2$}
    \Text(-10,60)[]{$p_1$}
    \Text(70,0)[]{$p_3$}
    \Text(70,30)[]{$p_4$}
    \Text(70,60)[]{$p_5$}
    \end{picture}
     \end{center} \caption{Tree-level structure contributing to  $q {\bar q} \to Z Z Z$.} \label{ztree}
\end{figure}
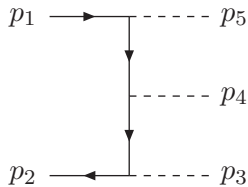

At the tree-level, there are six contributions to this process, obtained by the diagram
illustrated in Fig.~\ref{ztree} by permuting the final legs in all possible ways.
One-loop corrections are obtained by adding a virtual gluon to the tree-level structures,
as depicted in Fig.~\ref{fig1}. Each of the eight diagram of Fig.~\ref{fig1}
should be evaluated for six permutations of the final particles: overall this calculation
involves the reduction of 48 diagrams.

\begin{figure}[t]
\begin{center} \begin{picture}(300,50)(0,0)
        \ArrowLine(0,50)(25,50)
        \ArrowLine(25,0)(0,0)
        \Line(25,0)(25,50)
        \DashLine(25,0)(50,0){3}
        \DashLine(25,25)(50,25){3}
        \DashLine(25,50)(50,50){3}
        \GlueArc(25,37.5)(6,90,270){2}{3}

        \ArrowLine(75,50)(100,50)
        \ArrowLine(100,0)(75,0)
        \Line(100,0)(100,50)
        \DashLine(100,50)(125,50){3}
        \DashLine(100,0)(125,0){3}
        \DashLine(100,25)(125,25){3}
    \GlueArc(100,12.5)(6,90,270){2}{3}

    \ArrowLine(150,50)(175,50)
        \ArrowLine(175,0)(150,0)
        \Line(175,0)(175,50)
        \DashLine(200,50)(175,50){3}
        \DashLine(200,0)(175,0){3}
        \DashLine(200,25)(175,25){3}
    \GlueArc(175,50)(17.5,180,270){3}{4}

        \ArrowLine(220,50)(250,50)
        \ArrowLine(250,0)(220,0)
        \Line(250,0)(250,50)
        \DashLine(250,50)(275,50){3}
        \DashLine(250,0)(275,0){3}
        \DashLine(250,25)(275,25){3}
    \GlueArc(250,0)(17.5,90,180){3}{4}
 \end{picture}
\begin{picture}(300,5)(0,0)
 \end{picture}
\begin{picture}(300,50)(0,0)
        \ArrowLine(0,50)(25,50)
        \ArrowLine(25,0)(0,0)
        \Line(25,0)(25,50)
        \DashLine(25,0)(50,0){3}
        \DashLine(25,25)(50,25){3}
        \DashLine(25,50)(50,50){3}
    \GlueArc(25,25)(12.5,90,270){3}{5}

        \ArrowLine(75,50)(100,50)
        \ArrowLine(100,0)(75,0)
        \Line(100,0)(100,50)
        \DashLine(100,50)(125,50){3}
        \DashLine(100,0)(125,0){3}
        \DashLine(100,25)(125,25){3}
        \Gluon(100,37.5)(80,0){3}{6}

    \ArrowLine(150,50)(175,50)
        \ArrowLine(175,0)(150,0)
        \Line(175,0)(175,50)
        \DashLine(200,50)(175,50){3}
        \DashLine(200,0)(175,0){3}
        \DashLine(200,25)(175,25){3}
        \Gluon(175,12.5)(155,50){3}{6}

        \ArrowLine(220,50)(250,50)
        \ArrowLine(250,0)(220,0)
        \Line(250,0)(250,50)
        \DashLine(250,50)(275,50){3}
        \DashLine(250,0)(275,0){3}
        \DashLine(250,25)(275,25){3}
        \Gluon(230,0)(230,50){3}{7}

      \end{picture}
     \end{center} \caption{Diagrams contributing to virtual QCD corrections
to $q {\bar q} \to Z Z Z$} \label{fig1}
\end{figure}
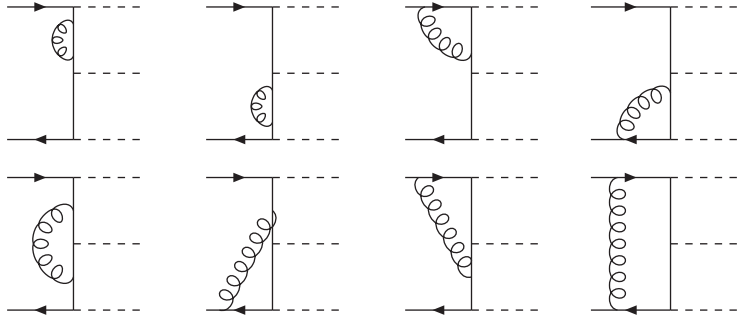

We perform a reduction to scalar integrals using the {\tt OPP} reduction method~\cite{Ossola:2006us,Ossola:2007bb}.
As described in Section~\ref{sect2}, we need to provide the numerical value of the numerator
of the integrand in the loop integrals.
The numerator function $N(q)$ can be expressed in terms of 4-dimensional denominators $\d{i}$
according to the decomposition of Eq.~(\ref{eq:2}).
For the particular case of five denominators, that is the relevant case for the process studied in this paper, we have $m=5$ and the indices range from $0$ to $4$.
Next, simply by evaluating the numerator function $N(q)$ for a given set of values of $q$, we can extract all the coefficients in Eq.~(\ref{eq:2}).

The coefficients determined in this manner should be multiplied by the corresponding scalar
integrals. Since, in the process that we are studying, no $q$-dependent massive propagator appears, we will only need massless scalar integrals. They are computed using the package {\tt OneLOop} written by A.~van~Hameren~\cite{vanHameren:2005ed}.

As an example, let us consider the pentagon diagram (the last diagram of Fig.~\ref{fig1}).
In our notation, the integrand will read
\begin{equation} \label{a5}
 A_5(q) = \frac{N_5(q)}{[q^2][(q+p_1)^2][(q+p_1+p_5)^2][(q-p_2-p_3)^2][(q-p_2)^2]}
\end{equation}
with
{\small
\begin{equation} \label{n5}
 N_5(q) = \left\{ {\bar u}(p_2)\,\gamma^\alpha\,P_{(q-p_2)} \, V^Z_3\, P_{(q-p_2-p_3)}\, V^Z_4\, P_{(q+p_1+p_5)}\, V^Z_5\, P_{(q+p_1)}\, \gamma^\alpha\, u(p_1) \right\}
\end{equation}
}

The function $P(q)$ is the numerator of the quark propagator
$$P_{(q)} = \displaystyle{\not} q + m\, , $$
while $V_i^Z = V^Z \cdot \epsilon_i$ ,
namely the contraction between he polarization vector of the \emph{i-th} $Z$ boson $\epsilon_i$
and the $\gamma$-matrix in the vertex $Zq{\bar q}$
\begin{equation}
 V^Z_\mu = ie\gamma_{\mu}( g_{f}^{-}  \omega_{-} + g_{f}^{+} \omega_{+})
\end{equation}
where
\begin{equation}
 g_{f}^{+} = -\frac{s}{c} Q_{f} \,\,\,\, , \,\,\,\,  g_{f}^{-} = \frac{I_{W,f}^{3}-s^{2}Q_{f}}{sc} \,\,\,\,
, \,\,\,\,  \omega_{\pm} = (1 \pm \gamma^5)/2\, .
\end{equation}
For any fixed value $q_0$ of integration momentum, and for a given phase-space point,
$N_5(q_0)$ is simply the trace of a string of known matrices. After choosing a representation for Dirac matrices and spinors, we evaluate $N(q)$ by performing a naive matrix multiplication. By providing this input to the reduction algorithm, we can compute all the coefficients of the scalar integrals (in other words, the ``cut-constructible'' part of the calculation).

The last step is the calculation of Rational Terms. As explained in Section~\ref{sect2}, part of this
contribution, that we call $R_1$, is automatically included by the to the reduction algorithm.
The second term $R_2$, coming from the $\epsilon$-dimensional part of $N_5(q)$, has been added by hand by looking at the Feynman Diagram and turns out to be proportional to the tree-order amplitude.

In the same fashion, we can repeat the calculation for the other seven diagrams. However, our method allows for
a further simplification: for each fixed permutation of the final legs, only the q-dependent denominators of Eq.~(\ref{a5}) will appear in the remaining diagrams. Therefore, we can combine all diagrams in a single numerator function and perform the reduction directly for the sum of such diagrams, allowing for a one-shot evaluation of the resulting scalar coefficients.

We checked that our results, both for poles and finite parts, agree with the results obtained by the authors of Ref.~\cite{Lazopoulos:2007ix}.

In what concerns the real emission, we only have to
deal with initial state singularities, where we distinguish
$q\bar{q}$ and $qg$ initial states. For the $qg$ initial state, no soft
singularity is present because the corresponding tree-level contribution
vanishes.
We recall that the structure of the NLO partonic cross sections
is as follows:
\bea
\sigma_{q\bar{q}}^{NLO} &=& \int\limits_{VVV}\Bigl[ d\sigma_{q\bar{q}}^B + d\sigma_{q\bar{q}}^V
+ d\sigma_{q\bar{q}}^C + \int\limits_{g} d\sigma_{q\bar{q}}^A \Bigr]  +
 \int\limits_{VVVg} \Bigl[ d\sigma_{q\bar{q}}^R - d\sigma_{q\bar{q}}^A\Bigr]\nonumber\\
\sigma_{gq}^{NLO} &=&  \int\limits_{VVV}
\Bigl[ + d\sigma_{gq}^C \int\limits_{g} d\sigma_{gq}^A  \Bigr]
+ \int\limits_{VVVg} \Bigl[ d\sigma_{gq}^R - d\sigma_{gq}^A \Bigr]\;,
\eea
where $d\sigma^B,d\sigma^V,d\sigma^C,d\sigma^R,d\sigma^A$ are respectively the
Born cross section, the virtual, virtual counterterm, real and real-subtraction
cross sections.
For the $q\bar{q}$ initial state two dipoles are needed as subtraction terms.
If $p_6$ is the  momentum which can become soft or collinear,
the dipole term for gluon emission off the quark  is  given by
\bea
\mathcal{D}^{q_1g_6,\bar{q}_2} &=&
  \frac{8\pi \alpha_s C_F}{2 \tilde{x}\,p_1\cdot p_6}
  \left( \frac{1+\tilde{x}^2}{1-\tilde{x}} \right)
   |\mathcal{M}_{q\bar{q}}^B(\{\tilde{p}\})|^2\label{dipoleqq}\\
\tilde{x} &=& \frac{p_1\cdot p_2 - p_2\cdot p_6 - p_1\cdot p_6}{p_1\cdot p_2}\nonumber
\eea
where the $\{\tilde{p}\}$ are redefined momenta,
$\{\tilde{p}_j\}=\{ \tilde{p}_{16},\tilde{p}_2,\tilde{p}_3,\tilde{p}_4,\tilde{p}_5 \}$,
 which are again on-shell
and go to $\{p_1,\ldots,p_5 \}$ in the singular limit,
e.g. $\tilde{p}_{16}=\tilde{x}\,p_1$.
The regularised real emission part then reads
\bea
  d\sigma_{q\bar{q}}^R - d\sigma_{q\bar{q}}^A &=&
\frac{1}{6}\frac{1}{N} \frac{1}{2 s_{12}} \Bigl[ C_F \,
| \mathcal{M}_{q\bar{q}}^R(\{p_j\})|^2
 - \mathcal{D}^{q_1g_6,\bar{q}_2}
 - \mathcal{D}^{\bar{q}_2g_6,q_1}  \Bigr] d\Phi_{VVVg}\nonumber\;,
\eea
where the factor $1/6$ accounts for the three identical bosons in the final state.
More details can be found in~\cite{Catani:1996vz,vvv}.

The hadronic differential cross section with hadron momenta $P_1$ and $P_2$
is the sum over all partonic initial states
convoluted with the parton distribution functions
\bea
d\sigma(P_1,P_2) = \sum\limits_{ab} \int dz_1 dz_2 f_a(z_1,\mu_F) f_b(z_2,\mu_F) d\sigma_{ab}(z_1P_1,z_2P_2)\;,
\eea
where the sum runs over the partonic configurations
$q\bar{q}$, $\bar{q}q$, $gq$, $qg$, $g\bar{q}$, $\bar{q}g$.

\subsubsection{Numerical results} \label{numbers}

As an explicit example we present the numerical results for the case
$u\bar{u}\to ZZZ$ for $\sqrt{s}=14$ TeV and using
CTEQ6L1\cite{Pumplin:2002vw}. The tree-order cross section has been
evaluated using the {\tt HELAC} event
generator\cite{Kanaki:2000ey,Kanaki:2000ms,Cafarella:2007pc}. In the
following table the results in fb are presented for the tree-order
cross section $\sigma_0$, the ratio of the virtual to the tree-level
cross section, and the real contribution, combining $5-$ and
$6-$point contributions, as described above, for all channels, i.e.,
$u\bar{u}, ug, g\bar{u}$, for different values of the
factorization(renormalization) scale ($\mu=\mu_F=\mu_R$).

\begin{center}
\begin{tabular}{|c|c|c|c|c|}
  \hline

  scale & $\sigma_0$ & $\sigma_V / \sigma_0$ & $\sigma_R$ & $\sigma_{NLO}$
  \\ \hline
  $\mu=M_Z$  & 1.481(5) & 0.536(1) & 0.238(2) & 2.512(2) \\
  $\mu=2M_Z$ & 1.487(5) & 0.481(1) & 0.232(2) & 2.434(2) \\
  $\mu=3M_Z$ & 1.477(5) & 0.452(1) & 0.232(2) & 2.376(2) \\
  $\mu=4M_Z$ & 1.479(5) & 0.436(1) & 0.232(2) & 2.355(2) \\
  $\mu=5M_Z$ & 1.479(5) & 0.424(1) & 0.237(2) & 2.343(2) \\
  \hline
\end{tabular}
\end{center}

As it is evident from these results, the $K-$factor is quite sizable
$(1.58-1.69)$, whereas the dependence on the scale $\mu$ is for both
cases quite weak, due mainly to the electroweak character of the
process.

\begin{figure}[hb]
\begin{center}
\begin{rotate}{90}\hspace*{25mm}{\small
$\sigma$ [fb]}\end{rotate}
\psfig{figure=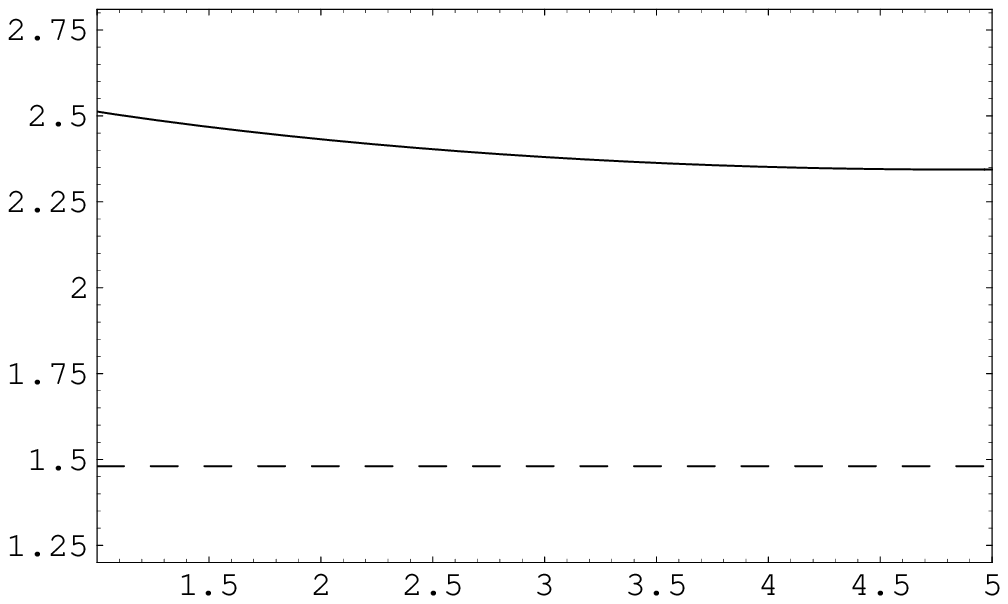,width=3.5truein}\\
\hspace*{5mm}
\small{$\mu / M_Z$}
\caption{Scale-dependence of the cross section $\sigma_{NLO}$ (solid line)
compared with  the tree-level cross section $\sigma_0$ (dashed line).
The scale is reported in the plot in units of $M_Z$, from $\mu=M_Z$
to  $\mu=5M_Z$. }
\end{center}
\end{figure}

\subsection{Conclusions} \label{conclusions}

We presented a new method for NLO processes (OPP), in which the
reduction to known integrals is performed at the integrand level.
The method has been successfully tested in a number of applications,
the latest being the production of three Z bosons at the LHC.

The efficiency of the method is quite good. It can be further
improved if the numerical evaluation of the integrand in the
one-loop amplitude, by means of recursion relations, without relying
on Feynman diagrams, is developed~\cite{Draggiotis:2006er}.

In general, the speed, the precision and the simplicity of
the OPP method, make it a very good candidate for the construction of
a universal NLO calculator/event-generator.

%

}

\part[IMPROVEMENTS ON STANDARD TECHNIQUES]{IMPROVEMENTS ON STANDARD TECHNIQUES}

\section[GOLEM: a semi-numerical approach to one-loop amplitudes]
{GOLEM: A SEMI-NUMERICAL APPROACH TO ONE-LOOP AMPLITUDES%
\protect\footnote{Contributed by: C.~Bernicot, T.~Binoth, J.-Ph.~Guillet, 
G.~Heinrich, E.~Pilon, T.~Reiter}}
{\graphicspath{{heinrich/}}
%
%
%
%
%

\subsection{Introduction}
The first collision data from the Large Hadron Collider (LHC) at CERN 
are expected in a couple of months, giving us the 
opportunity to explore unprecedented  energies and luminosities.
However, in order that a discovery of New Physics can be claimed, 
it is of crucial importance to have the Standard Model physics under 
control. This includes e.g. understanding of the detectors, 
the underlying event,
the luminosity determination, the jet energy scale~\cite{dissertori}. 
For most of these issues, an interplay between measurements and 
precise theory predictions is mandatory. 
In a hadron collider environment, multi-particle/jet final states 
will be produced in abundance. Therefore considerable effort 
needs to be spent to make predictions for multi-particle processes 
beyond the leading order. While the calculation of one-loop 
five-point amplitudes can be considered as the state of the art 
at the moment, the first complete cross section 
for  six-point processes at hadron colliders still awaits its completion. 
Many different approaches to multi-particle production 
have been developed in the last few years, 
most of them being described in these proceedings.
For other reviews and very 
recent developments, see e.g.~\cite{Bern:2007dw,Ossola:2007ax,Giele:2008ve}.

Here we will focus on a method implemented in the program 
{\tt GOLEM} (General One-Loop Evaluator of Matrix elements), 
which is based on a semi-numerical evaluation of building blocks 
stemming from the reduction of one-loop Feynman diagrams~\cite{Binoth:2005ff}. 
The main features of the formalism are the following:
\begin{itemize}
\item It is valid for massive and massless particles
\item For $N>5$ external legs, the reduction of rank $R$ $N$-point integrals 
is done algebraically, reducing the rank and the number of propagators 
at the same time in each reduction step.
For $N\leq 5$ we worked out form factor representations
which allow to avoid inverse Gram determinants in exceptional 
kinematic regions.
\item The infrared divergences are easily extracted analytically 
in terms of triangles.
\item The rational parts of the amplitudes are obtained 
as by-products and can be projected out.
\item 
The program has an analytic and a numerical branch:
it can perform a complete reduction to scalar integrals, 
represented in terms of analytic functions: such a complete reduction 
introduces inverse Gram determinants, but 
this branch can be chosen safely in phase space regions 
where the Gram determinants are sufficiently large (which is the bulk 
of the phase space). 
As the evaluation of analytic functions is fast, this speeds up the 
program considerably as compared to a purely numerical approach.
Near exceptional phase space points, the program allows to 
stop the reduction {\it before} dangerous denominators are produced. 
The building blocks to evaluate in this case are finite 
three- and four-point functions with Feynman parameters in the numerator.
As a brute-force numerical evaluation of the four-point 
functions is rather slow, 
we have worked out one-dimensional integral representations, 
whose numerical evaluation is extremely fast. 
Details will be given in the following section.
\end{itemize}
We  have implemented the reduction  
in algebraic manipulation programs and have 
obtained fully analytical results for several 
amplitudes using these 
methods~\cite{Andersen:2007mp,Binoth:2007ca,Binoth:2006mf,Binoth:2006ym,Binoth:2003xk,Binoth:2001vma}. 
Without having efficient and 
automated simplification methods to reduce the size of big 
analytic expressions, the fully analytic approach 
based on form factors suffers from 
factorial complexity and therefore does not seem to be 
appropriate for 6-point processes. The semi-numerical 
reduction is preferable in this case.
For the calculation of 
the rational terms alone the situation
is different, as the form factor representations simplify
considerably when restricted to terms which can generate 
rational parts~\cite{Binoth:2006hk}.

\subsection{Results}

Below we will describe applications of our method to 
one-loop six-point amplitudes and explain in detail 
certain features which guarantee a fast and numerically robust 
evaluation in all phase space regions.

\subsubsection{The GOLEM numerical library}
In the {\tt GOLEM} library, the strategy is to evaluate numerically higher dimensional
 three- and four-point functions in phase space regions where numerical instabilities 
 arise due to spurious singularities. 
 To be specific, these integrals are 
six- and eight-dimensional four-point functions $I_4^{D+2},I_4^{D+4}$, 
 and four- and six-dimensional 
three-point functions $I_3^{D},I_3^{D+2}$, 
with or without Feynman parameters in the numerator. 
While the triangles are two-dimensional integrals in Feynman parameter space, 
the boxes a priori involve integration over three Feynman parameters. 
As numerical integrations in multi-dimensional parameter space are 
rather slow, we worked out one-dimensional integral representations 
for these integrals, whose evaluation is both fast and precise.
In \cite{Binoth:2002xh,Binoth:2005ff} we have already presented 
other methods for the numerical evaluation of Feynman parameter integrals,
but the one-dimensional representations discussed here are preferable,
as they are much faster.  

As an example, let us consider the case where two massive particles scatter 
into two light particles via a fermion loop. The two ingoing particles have a 
small velocity. In this kinematic region, the Gram determinant is small. 
In this case, we have to evaluate  four-point functions with two adjacent 
massive  legs, and with Feynman parameters in the numerator.
In fig. \ref{plot_number1} we plot the six-dimensional  four-point 
function with two adjacent massive legs, $I_{4,\rm{adj}}^{6}$,  
against the absolute value of the coefficient $B$ which is 
proportional to the ratio $det(G)/det(S)$, for a trajectory of points 
with $10^{-15} \leq |B|\leq 10^{-3}$. 

In the {\tt GOLEM} library, there is a 
cut $c$ which allows to split the phase space regions where 
the four-point function is 
evaluated analytically from those where it is evaluated numerically. 
The larger the cut, the longer the evaluation takes, as more calls 
of the numerical integration routine are made.
On the other hand, if the cut is too small, the analytical evaluation causes 
a loss of precision of several digits.

As an illustration, we compute $I^{6}_{4}(z_{1} z_{2}^{2})$ and we 
plot the real and imaginary parts for different values of the cut $c$\,: 
$c = 10^{-1}$ (Fig. \ref{plot_number1}), $c = 10^{-3}$ (Fig. \ref{plot_number2}) 
and $c = 10^{-5}$ (Fig. \ref{plot_number3}).
\begin{figure}[htb!]
\begin{center}
\includegraphics[scale=0.55]{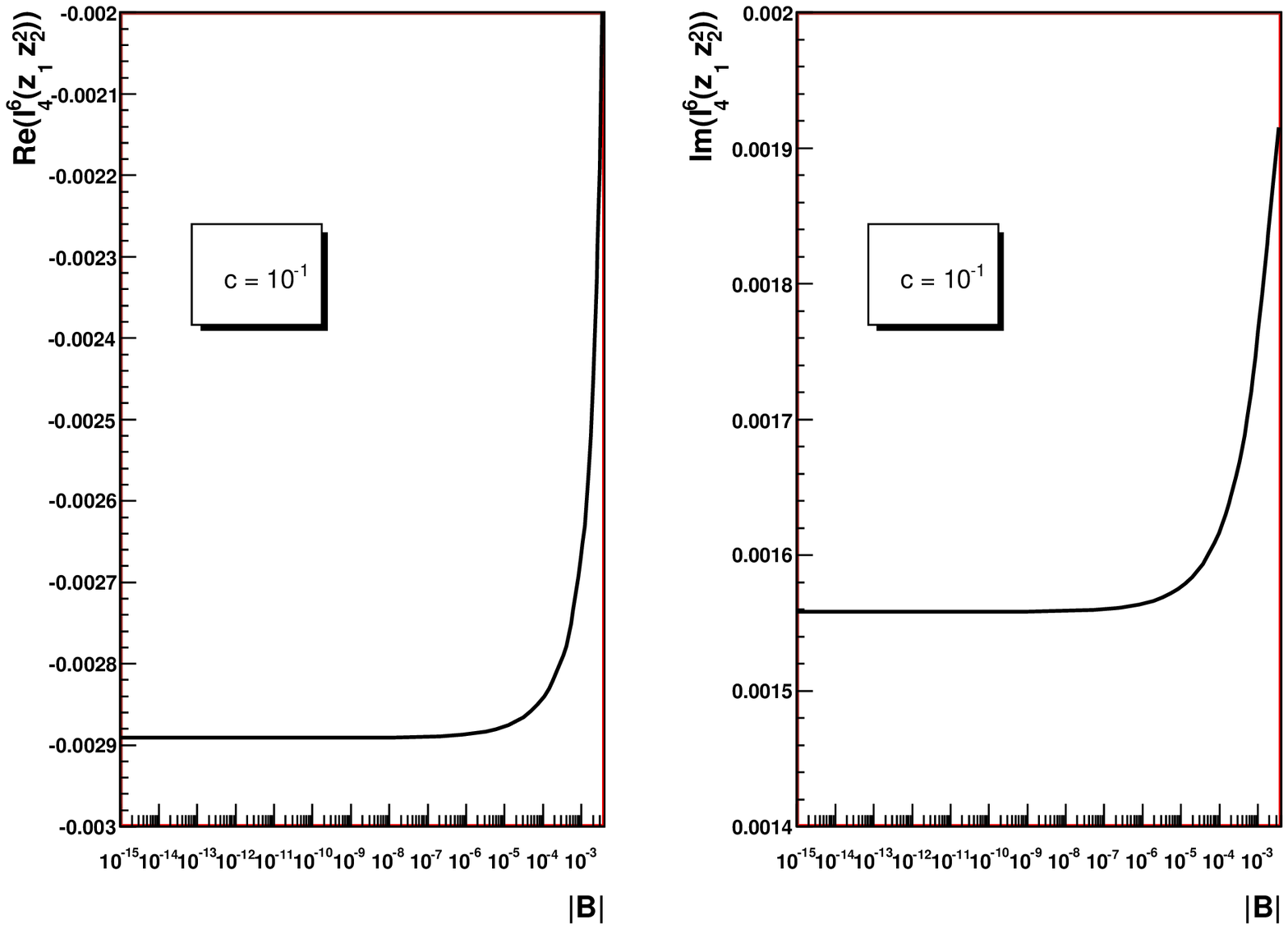}
\caption{\textit {The six-dimensional 
four-point function with three Feynman parameters in the numerator, 
$I^{6}_{4}(z_{1} z_{2}^{2})$, with two adjacent massive legs  and 
the cut $c=10^{-1}$}.} \label{plot_number1}
\end{center}
\end{figure}

\begin{figure}[httb!]
\begin{center}
\includegraphics[scale=0.55]{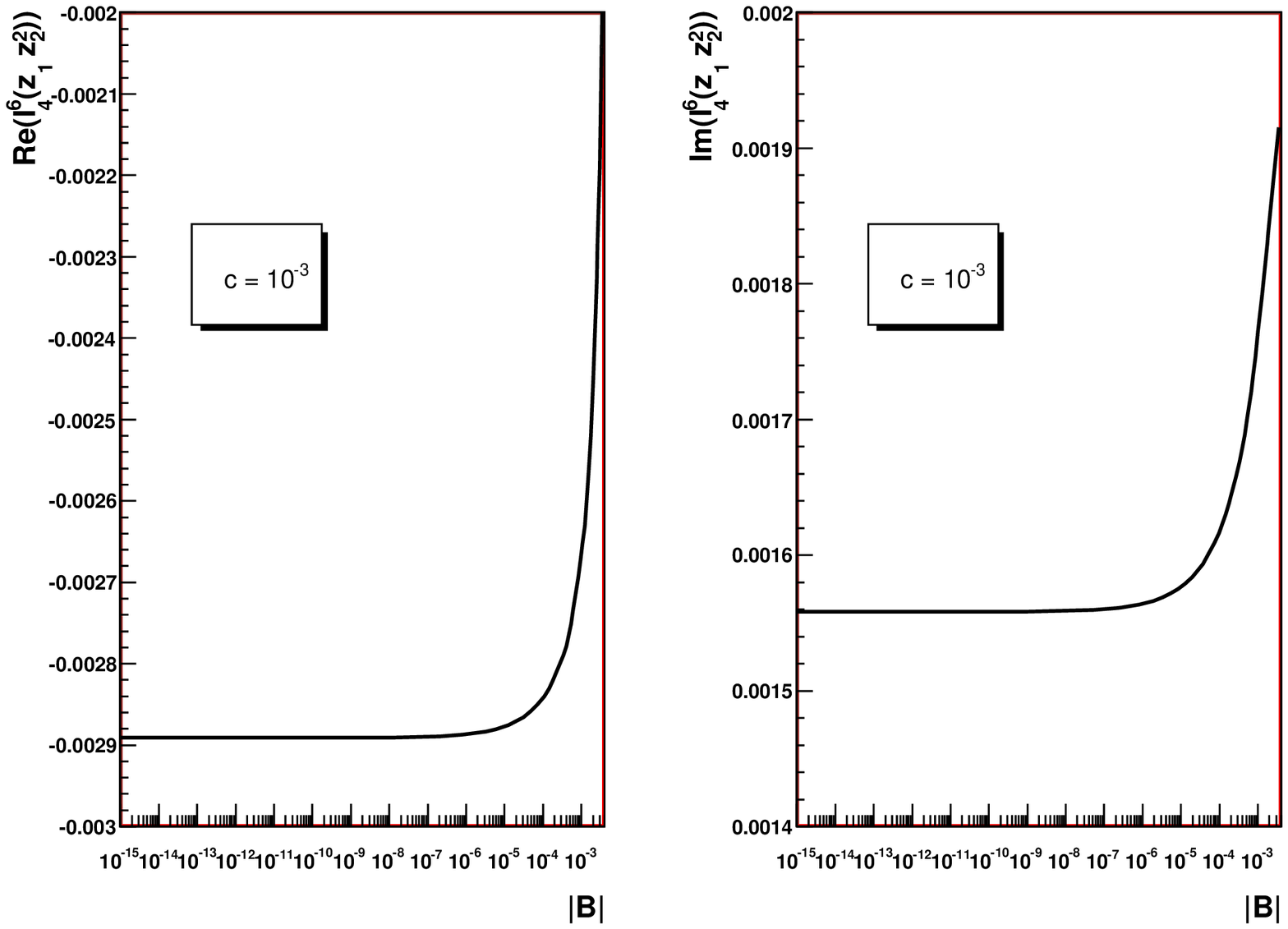}
\caption{\textit{The six-dimensional 
four-point function with three Feynman parameters in the numerator, 
$I^{6}_{4}(z_{1} z_{2}^{2})$, with two adjacent massive legs  and 
the cut $c=10^{-3}$.}} \label{plot_number2}
\end{center}
\end{figure}

\begin{figure}[httb!]
\begin{center}
\includegraphics[scale=0.55]{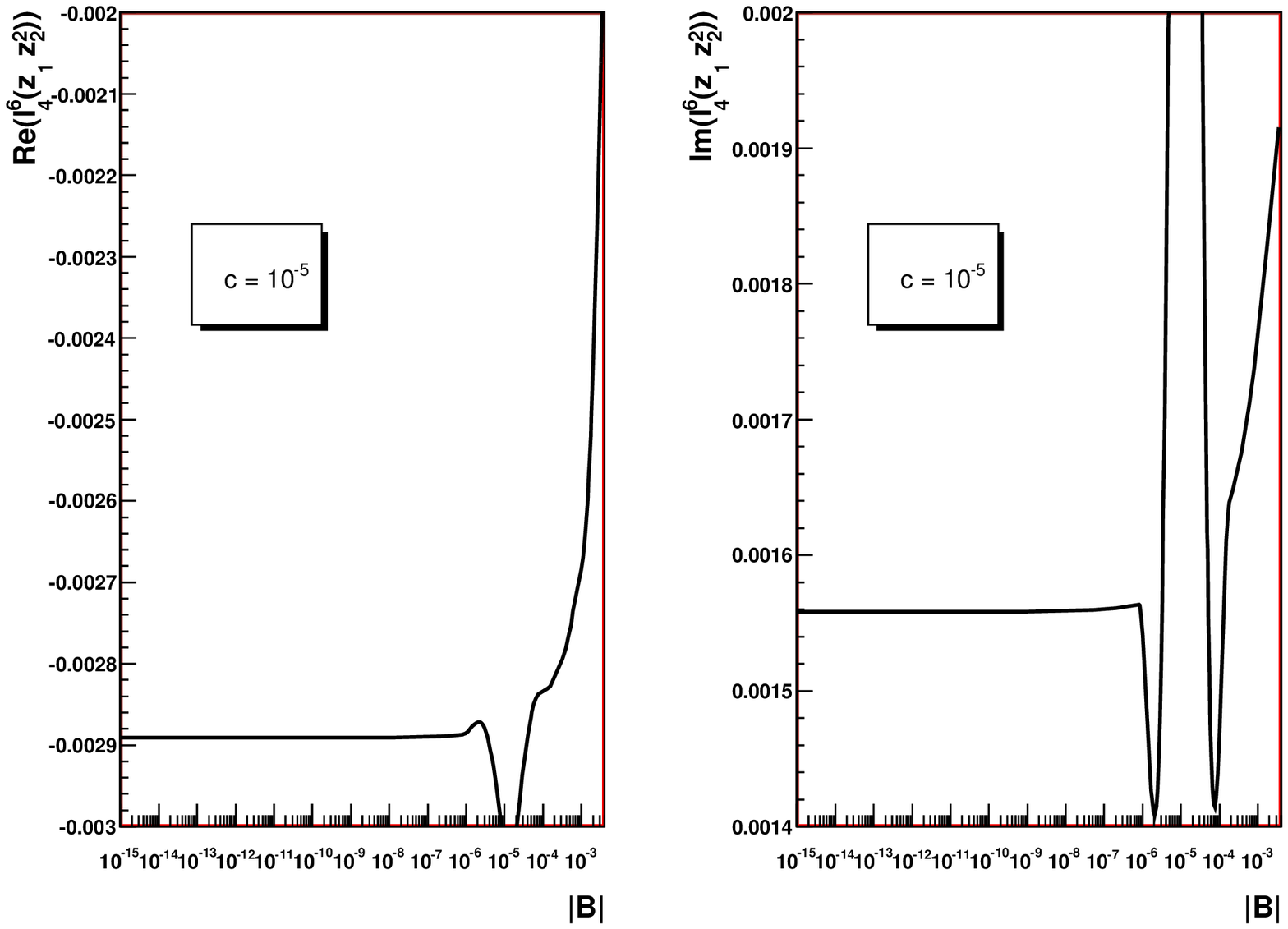}
\caption{\textit {The six-dimensional 
four-point function with three Feynman parameters in the numerator, 
$I^{6}_{4}(z_{1} z_{2}^{2})$, with two adjacent massive legs  and 
the cut $c=10^{-5}$.}} \label{plot_number3}
\end{center}
\end{figure}
In the case at hand, the CPU time does not vary very much with the cut, 
the evaluation time ranges from 0.14\,s 
(on an Intel Pentium M 1.3 GHz) for $c=10^{-1}$ to 0.10\,s for $c=10^{-5}$. 
However, this statement is hard to generalise 
to all possible situations occurring in a 
calculation of a complex multi-leg amplitude. 
In any case, the cut $c$ allows to adjust the trade-off between 
speed and precision. 

\subsubsection{The $u\bar{u}\rightarrow{}d\bar{d}s\bar{s}$ amplitude}
With our method we calculated the one-loop 
six-quark amplitude
\begin{equation}
{\mathcal A}(u(p_1, \lambda_1), \bar{u}(p_2, \lambda_2)
\rightarrow
d(p_3, \lambda_3), \bar{d}(p_4, \lambda_4),
 s(p_5, \lambda_5), \bar{s}(p_6, \lambda_6))
\end{equation}
in massless QCD.
The calculation has been carried out using spinor helicity
amplitudes in the 't~Hooft-Veltman scheme.
We have chosen a convenient colour basis,
which allows to split the amplitude as follows
\begin{equation}
\sum_{\lambda}
\sum_{i=1}^6
{\mathbf C}^iA^{\lambda}_i(p_1,\ldots,p_6),
\end{equation}
where $A_c$ are the helicity and colour subamplitudes. In
particular we chose the colour structures
\begin{equation}
({\mathbf C}^1,
{\mathbf C}^2,
{\mathbf C}^3,
{\mathbf C}^4,
{\mathbf C}^5,
{\mathbf C}^6) =
(\delta_{c_1}^{c_2}\delta_{c_4}^{c_3}\delta_{c_6}^{c_5},
\delta_{c_1}^{c_2}\delta_{c_4}^{c_5}\delta_{c_6}^{c_3},
\delta_{c_1}^{c_5}\delta_{c_4}^{c_2}\delta_{c_6}^{c_3},
\delta_{c_1}^{c_5}\delta_{c_4}^{c_3}\delta_{c_6}^{c_2},
\delta_{c_1}^{c_3}\delta_{c_4}^{c_5}\delta_{c_6}^{c_2},
\delta_{c_1}^{c_3}\delta_{c_4}^{c_2}\delta_{c_6}^{c_5}).
\end{equation}
In our notation $\lambda$ is the vector $(\lambda_1,\ldots,\lambda_6)$,
and $\lambda_j=\pm1$ is the helicity of the particle with momentum $p_j$
of which the colour index is $c_j$. In the six-quark amplitude one can
identify two independent helicites $\lambda^a=(+,+,+,+,+,+)$ and
$\lambda^b=(+,+,+,+,-,-)$; all other helicities are either identically
zero or related to $\lambda^a$ or $\lambda^b$ by parity invariance, which
is exploited in our calculation.

We generated the Feynman diagrams for this process with
\emph{QGraf}\,\cite{Nogueira:1991ex} and reduced the tensor integrals
using \emph{FORM}~\cite{Vermaseren:2006ag,Vermaseren:2000nd}
to form factors as defined in~\cite{Binoth:2005ff}. 
We deal with the spinor algebra by
completing spinor lines to traces, e.g. for an arbitrary product $\Gamma$
of Dirac matrices we use
\begin{equation}
\left\langle p_i^+\right\vert\Gamma\left\vert p_j^+\right\rangle=
\frac{1}{2[p_jq]\langle qp_i\rangle}
{\mathrm{tr}}\{(1+\gamma_5){\not{\!p}}_j\!\!\not{\!q}\!\not{\!p}_i\Gamma\}.
\end{equation}
With the help of \emph{FORM} and \emph{Java} code the expressions for the
diagrams are transformed into a \emph{Fortran90} program. The {\tt Golem90}
library is used for the numerical evaluation of the form factors. In this
approach we found it advantageous to treat the spinor traces numerically
as well,  in order to keep the expressions more compact.

The code returns the subamplitudes in the form
\begin{equation}
A^\lambda_i(p_1,\ldots,p_6)=\frac{g_s^6}{4\pi^2}\frac{1}{s}%
\left(\frac{A}{\varepsilon^2}+\frac{B}{\varepsilon}+C
+{\mathcal O}(\varepsilon)\right)
\end{equation}
for each of the six colour structures and for all non-zero helicities,
where $A$, $B$ and $C$ are complex coefficients. As an example 
we plot in Figure~\ref{fig:sixquarks:rotplot;golem} 
the quantity $s\vert A_c^\lambda\vert\alpha_s^{-3}$ for one colour structure
${\mathbf C}^1$ and the two helicity configurations  
$\lambda^a$ and $\lambda^b$. The initial state momenta are chosen to be
along the $z$-axis while the final state momenta have been
rotated about the $y$-axis by an angle $\theta$. For $\theta=0$ the momenta
are chosen as in Ref.~\cite{Nagy:2006xy}:
\begin{eqnarray}
\vec{p}_3&=&(33.5, 15.9, 25.0)\nonumber\\
\vec{p}_4&=&(-12.5, 15.3, 0.3)\nonumber\\
\vec{p}_5&=&(-10.0, -18.0, -3.3)\nonumber\\
\vec{p}_6&=&(-11.0, -13.2, -22.0)
\end{eqnarray}
In the chosen units the renormalisation scale is $\mu=1$.
The amplitude has been evaluated at 50 successive points between
$\theta=0$ and $\theta=2\pi$ ($\theta=0,0.126,0.252,\ldots$), which took
2.4~seconds per point and helicity on an Intel Pentium~4 CPU (3.2\,GHz).
Table~\ref{tbl:sixquarks:numvalues;golem} shows the numerical values of
all coefficients for the point $\theta=0$. 
\begin{figure}
\begin{center}
\includegraphics[width=0.7\textwidth]{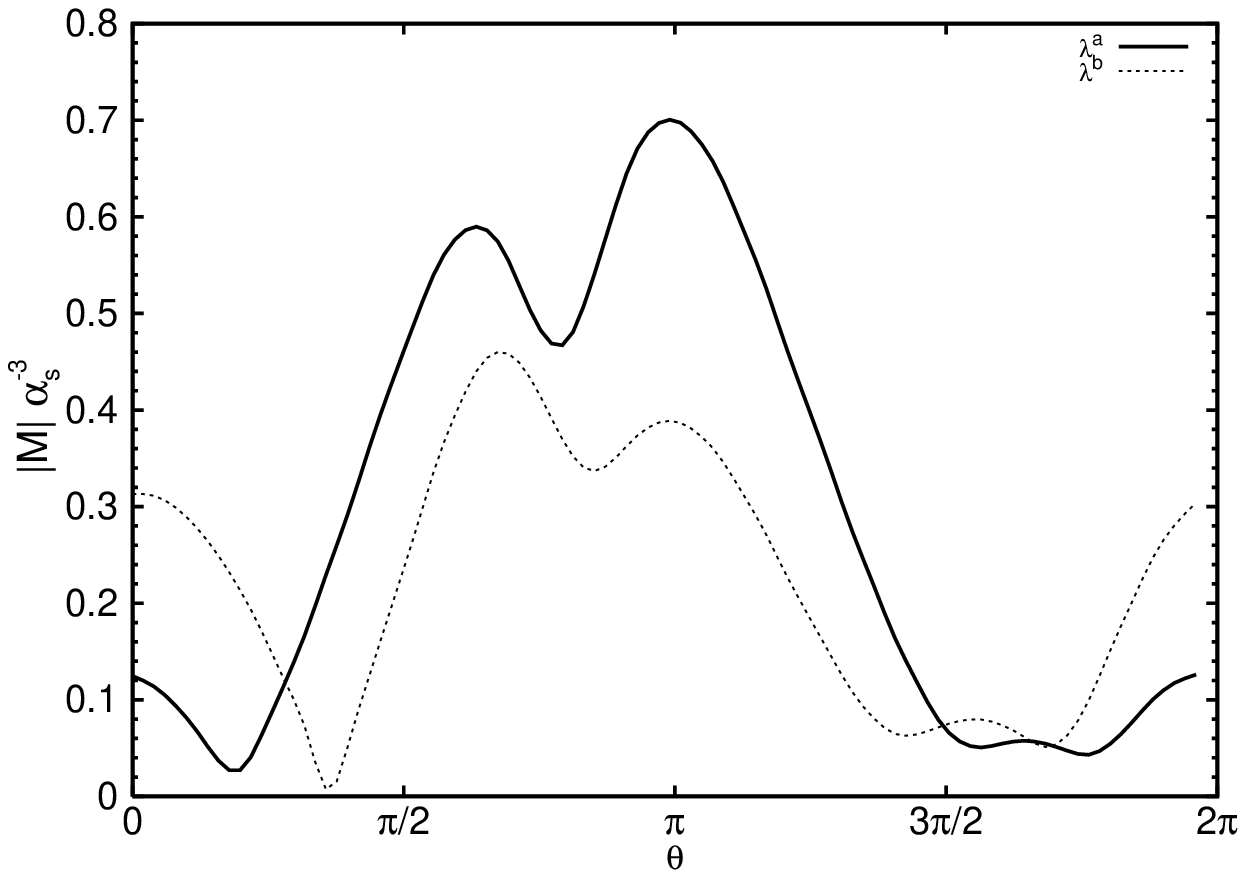}
\caption{The six-quark amplitude. The finite parts of the 
Laurent expansion in~$\varepsilon$ of
$s\vert A_1^{\lambda^a}\vert\alpha_s^{-3}$ (solid)
and $s\vert A_1^{\lambda^b}\vert\alpha_s^{-3}$ (dashed)
are plotted for a
kinematic point defined in the text, 
where the final state momenta have been rotated about
the $y$-axis by an angle $\theta$.}
\label{fig:sixquarks:rotplot;golem}
\end{center}
\end{figure}


\begin{table}
\begin{center}
\begin{tabular}{lrrr}
$c$ & \multicolumn{1}{c}{$A$} & \multicolumn{1}{c}{$B$} &
\multicolumn{1}{c}{$C$} \\
\hline
1&$-0.0029670-0.0036065i$&$ 0.0203701+ 0.0281510i$&$-0.0659100-0.1057940i$\\
2&$ 0.0042784+ 0.0049474i$&$-0.0191448-0.0420120i$&$ 0.0338141+ 0.1820798i$\\
3&$-0.0123663-0.0186981i$&$ 0.1171088+ 0.1401148i$&$-0.4902357-0.4754639i$\\
4&$ 0.0051836+ 0.0066459i$&$-0.0462621-0.0477458i$&$ 0.1803702+ 0.1706208i$\\
5&$-0.0143367-0.0137603i$&$ 0.1282264+ 0.1049820i$&$-0.5199953-0.3972433i$\\
6&$ 0.0083400+ 0.0100456i$&$-0.0745825-0.0730179i$&$ 0.2929410+ 0.2459317i$
\end{tabular}
\end{center}
\caption{Six-quark amplitude. Numerical values of the virtual part
$A_c^{\lambda^a}(\{p_j\}_{j=1\ldots 6})\alpha_s^{-3}$
for the kinematics given in the text and~$\theta=0$.}
\label{tbl:sixquarks:numvalues;golem}
\end{table}

\clearpage

\subsection*{Acknowledgements}
This research was supported  by the UK Science and Technology Facilities 
Council.
 
%
}


\section[Issues with the Landau singularities]
{ISSUES WITH THE LANDAU SINGULARITIES%
\protect\footnote{Contributed by: C.~Bernicot, F.~Boudjema, J.P.~Guillet, 
N.D.~Le, E.~Pilon}}
{\graphicspath{{boudjema/}}

\newcommand{\caln}{{\cal N}}
\newcommand{\dsp}{\displaystyle}
\def\mathswitchr#1{\relax\ifmmode{\mathrm{#1}}\else$\mathrm{#1}$\fi}
\def\mathswitch#1{\relax\ifmmode#1\else$#1$\fi}
\newcommand{\Pe}{\mathswitchr e}
\newcommand{\Pep}{\mathswitchr {e^+}}
\newcommand{\Pem}{\mathswitchr {e^-}}
\newcommand{\Pd}{\mathswitchr d}
\newcommand{\Pu}{\mathswitchr u}
\newcommand{\PW}{\mathswitchr W}
\newcommand{\PZ}{\mathswitchr Z}
\newcommand{\PH}{\mathswitchr H}
\newcommand{\rd}{\mathrm{d}}
\newcommand{\ri}{\mathrm{i}}
\newcommand{\ina}{i_1}
\newcommand{\inb}{i_2}
\newcommand{\inc}{i_3}
\newcommand{\ind}{i_4}
\newcommand{\ine}{i_5}
\newcommand{\ing}{i_6}
\newcommand{\Zadj}{\tilde Z}
\newcommand{\Zadjadj}{\,\smash{\tilde{\!\tilde Z}}\vphantom{\tilde Z}}
\newcommand{\Ymod}{X}
\newcommand{\Ymodadj}{\tilde{\Ymod}}
\newcommand{\Ymodadjadj}{\,\smash{\tilde{\!\Ymodadj}}\vphantom{\Ymodadj}}
\newcommand{\Gramdet}{|Z|}
\def\nl{\nonumber\\*}
\newcommand{\MZ}{\mathswitch {M_\PZ}}
\newcommand{\Mu}{\mathswitch {m_\Pu}}
\newcommand{\GeV}{\unskip\,\mathrm{GeV}}
\newcommand{\ra}{\rightarrow}

%
%
%
%

\newcommand{\A}{{\cal A}}
\newcommand{\h}{{\cal H}}
\newcommand{\s}{{\cal S}}
\newcommand{\W}{{\cal W}}
\newcommand{\BH}{\mathbf B(\cal H)}
\newcommand{\KH}{\cal  K(\cal H)}
\newcommand{\Real}{\mathbb R}
\newcommand{\Complex}{\mathbb C}
\newcommand{\Field}{\mathbb F}
\newcommand{\RPlus}{[0,\infty)}
\newcommand{\norm}[1]{\left\Vert#1\right\Vert}
\newcommand{\essnorm}[1]{\norm{#1}_{\text{\rm\normalshape ess}}}
\newcommand{\abs}[1]{\left\vert#1\right\vert}
\newcommand{\set}[1]{\left\{#1\right\}}
\newcommand{\seq}[1]{\left<#1\right>}
\newcommand{\eps}{\varepsilon}
\newcommand{\To}{\longrightarrow}
\newcommand{\RE}{\operatorname{Re}}
\newcommand{\IM}{\operatorname{Im}}
\newcommand{\Poly}{{\cal{P}}(E)}
\newcommand{\EssD}{{\cal{D}}}
\newcommand{\be}{\begin{equation}}
\newcommand{\ee}{\end{equation}}
\newcommand{\bea}{\begin{eqnarray}}
\newcommand{\eea}{\end{eqnarray}}
\newcommand{\ben}{\begin{enumerate}}
\newcommand{\een}{\end{enumerate}}
\newcommand{\nn}{\nonumber}
\newcommand{\crn}{\nonumber \\}
\newcommand{\non}{\nonumber}
\newcommand{\noi}{\noindent}
\newcommand{\al}{\alpha}
\newcommand{\la}{\lambda}
\newcommand{\bet}{\beta}
\newcommand{\ga}{\gamma}
\newcommand{\va}{\varphi}
\newcommand{\om}{\omega}
\newcommand{\pa}{\partial}
\newcommand{\fr}{\frac}
\newcommand{\bc}{\begin{center}}
\newcommand{\ec}{\end{center}}
\newcommand{\Ga}{\Gamma}
\newcommand{\de}{\delta}
\newcommand{\De}{\Delta}
\newcommand{\ep}{\epsilon}
\newcommand{\varep}{\varepsilon}
\newcommand{\ka}{\kappa}
\newcommand{\La}{\Lambda}
\newcommand{\si}{\sigma}
\newcommand{\Si}{\Sigma}
\newcommand{\ta}{\tau}
\newcommand{\up}{\upsilon}
\newcommand{\Up}{\Upsilon}
\newcommand{\ze}{\zeta}
\newcommand{\ps}{\psi}
\newcommand{\Ps}{\Psi}
\newcommand{\ph}{\phi}
\newcommand{\vph}{\varphi}
\newcommand{\Ph}{\Phi}
\newcommand{\Om}{\Omega}

\subsection{Introduction}
Cross sections involving a large number of external particles can
contain numerical instabilities which must be carefully located
and controlled. At tree-level one can mention integration over a
$t$-channel pole if the integration variables are not properly
chosen. The crossing of a  resonance might also be problematic.
Beside these physical situations there might be fake singularities
specific to the way one has set up the amplitude; one example is
the  singularity brought about  by an unlucky choice of a
reference vector at the helicity amplitude level. These problems
are exacerbated at the loop level since the loop integrals can
also develop singularities. A prominent example is the occurrence
of vanishing inverse Gram determinants: see for example the
contribution of Denner and Dittmaier. The latter is a fake
singularity that can be met for some special, and simple,
kinematical conditions on the phase space of the external
particles having to do with how one has chosen one's (independent)
basis for the loop integrals and how one has subsequently
expressed the other loop integrals in this basis. Loop integrals
can also have {\em true} singularities that have an underlying
physical origin. They depend on the dynamics of the problem.
Thresholds are one example, though harmless and trivial to locate.
These types of singularities belong to the general class of Landau
singularities. The physical singularity can be revealed by
studying the analytic properties of the  scalar integral. Here we
study the case of one-loop integrals. In particular we will review
how the conditions for having such singularities can be derived,
especially in a format that is conducive to an easy implementation
in a computer code. When such a singularity is present it is
important to inquire whether this singularity is integrable or
not. We rederive here the singular part. We then consider two
specific complementary examples taken from the recent literature.
The first one, the electroweak corrections to 
$pp \rightarrow b \bar b H$, reveals a Landau
singularity having to do with massive, indeed unstable, particles
in the loop. In this case the singularity is smoothed out by the
width of the unstable particles. The second is the $6$-photon
amplitude which involves massless states, both internally and
externally. In this case the Landau determinant is a quadratic
function whose square root is proportional to the Gram
determinant.

\subsection{Conditions for a Landau singularity and the nature of the singularity}
\begin{minipage}[l]{0.48\textwidth}
Consider the one-loop process $F_1(p_1)+F_2(p_2)+\ldots
+F_N(p_N)\to 0,$ where $F_i$ stands for either a scalar, fermion or
vector field with momentum $p_i$ as in the figure opposite. The
internal momentum for each propagator is  $q_i$ with $i=1,\ldots
N$. Each momentum $q_i$ is associated with one Feynman parameter
$x_i$ respectively. The scalar loop integral reads
\bea
T^{N}_{0}&\equiv & \int\fr{d^Dq}{(2\pi)^Di}\fr{1}{D_1D_2\cdots
D_{N}},
\nonumber \\
D_i&=&q_i^2-m_i^2+i\epsilon, \,\,\, q_i=q+r_i, \nonumber \\
r_i&=&\sum_{j=1}^{i}p_j,\,\,\, i=1,\ldots,N,
\eea
\end{minipage}
\begin{minipage}[r]{0.48\textwidth}
\hspace*{0.15\textwidth}\includegraphics[width=0.8\textwidth]{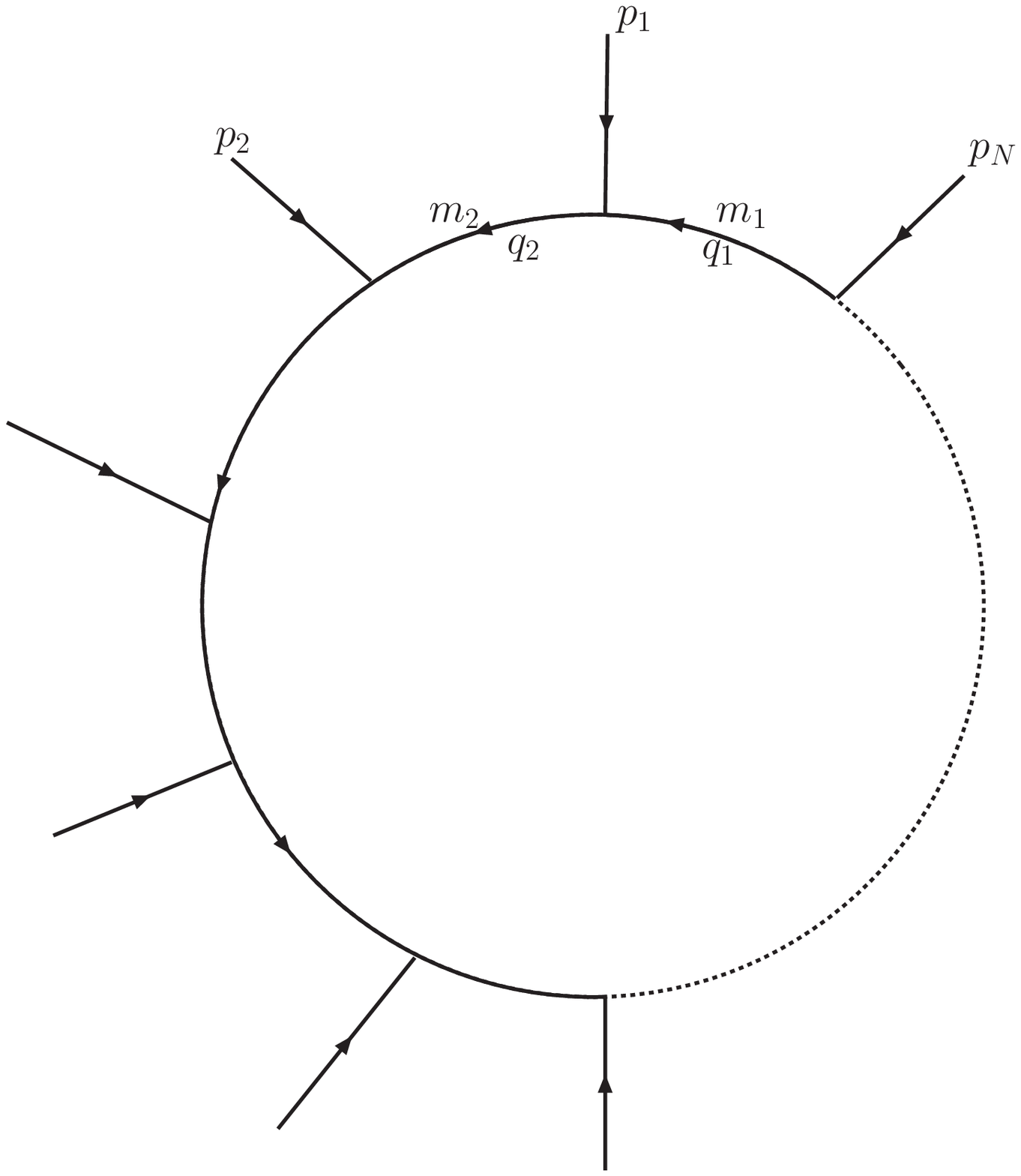}
\end{minipage}

The Feynman parameter representation reads
\bea
T^{N}_{0}=\Gamma(N)\int_0^\infty dx_1\cdots dx_N\delta(\sum_{i=1}^{N}x_i-1)\int\fr{d^Dq}{(2\pi)^Di}\fr{1}{(x_1D_1+x_2D_2+\cdots x_ND_{N})^N}.\label{eq_TN0}
\eea
Because of the Dirac delta function, the integration boundary in
the Feynman parameter space are $x_i=0$, $i=1,\ldots ,N$. Thus the
only important condition on $x_i$ is that they are \underline{not
negative.} The singularities are given by the Landau conditions
\cite{Landau:1959fi,Smatrix}
\bea
\left\{
\begin{array}{ll}
\forall i \,\,\, x_i(q_i^2-m_i^2)=0,\\
\sum_{i=1}^Nx_iq_i=0.\label{landau_eqN}
\end{array}
\right.
\eea
If eq. (\ref{landau_eqN}) has a solution $x_i>0$ for every
$i\in\{1,\ldots ,N\}$, {\it i.e. all particles in the loop are
simultaneously on-shell}, then  the integral $T^N_{0}$ has a
leading Landau singularity (LLS). If a solution exists but with
some $x_i=0$ while the other  $x_i$'s are positive, the Landau
condition corresponds to a lower-order Landau singularity
(LOLS).

By introducing the matrix $Q$, under the condition $q_i^2=m_i^2$,
\bea
Q_{ij}=2q_i.q_j=m_i^2+m_j^2-(q_i-q_j)^2=m_i^2+m_j^2-(r_i-r_j)^2;\,\,\,
i,j\in\{1,2,\ldots,M\},
\eea
%
the conditions to have a  Landau singularity in the physical
region are
\bea
\left\{
\begin{array}{ll}
\det(Q)=0,\\
x_i>0, \,\,\, i=1,\ldots,M.\label{landau_cond0}
\end{array}
\right.
\eea
For $M=N$ one has a leading singularity, otherwise if $M<N$ this
is a subleading singularity.  If some internal (external)
particles are massless, as in the case of six-photon scattering,
then some $Q_{ij}$ are zero, and the above conditions can be easily
checked. However, if the internal particles are massive then it is
difficult to check these conditions explicitly, especially if $M$
is large. In this case, we can rewrite the above conditions as
follows
\bea
\left\{
\begin{array}{ll}
\det(Q)=0,\\
x_j=\det(\hat{Q}_{jM})/\det(\hat{Q}_{MM})>0,\,\,\,
j=1,\ldots,M-1,\label{landau_cond1}
\end{array}
\right.
\eea
where $\hat{Q}_{ij}$ is obtained from $Q$ by discarding row $i$
and column $j$ from $Q$. Note that
$\det(\hat{Q}_{MM})=d[\det(Q)]/dQ_{MM}$. If $\det(\hat{Q}_{MM})=0$
then the second condition in (\ref{landau_cond1}) becomes
$\det(\hat{Q}_{jM})=0$ with $j=1,\ldots,M-1$.    There may be
cases, as we will encounter in section~\ref{sec:6photons}, where
the Landau determinant $\det(Q)$ has a quadratic form. These
special situations have to be handled with care.

\noindent The existence of a Landau singularity corresponds to an
eigenvector of $Q$ with zero eigenvalue.  In general, $Q$ has $N$
real eigenvalues  $\la_1$, \ldots, $\la_N$. Consider the case
where $Q$ has \underline{only one} (non-degenerate) very small
eigenvalue $\la_N\ll 1$. To leading order
\bea
\la_N=\fr{a_0}{a_1},\qquad a_{1}=\la_1\la_2\ldots \la_{N-1}\neq 0, \qquad
a_0=\det(Q).
\eea
With $V=\{x_1^0,x_2^0,\ldots,x_N^0\}$  the eigenvector
corresponding to $\la_N$, we define $\upsilon^2=V\cdot V$.  We will
assume that $\la_i>0$ for $i=1,\ldots,K$ and $\la_j<0$ for
$j=K+1,\ldots,N-1$ with $0\le K\le N-1$. It can then be shown that
in $D$ dimensions,
\bea
T^{N}_{0}&=&\fr{(-1)^Ne^{i\pi(N-K-1)/2}\upsilon}{2^{3D/2-N}\sqrt{(-1)^{N-K-1}a_1}}
\fr{\pi^{(N-D-1)/2}\Gamma((N-D+1)/2)}{(\la_N\upsilon^2-i\eps)^{(N-D+1)/2}}.
\eea
This result holds provided $a_1\neq 0 $ and $ N-D+1>0$.  For  the
box, $N=4$, $D=4$, $a_0\to 0$ and $a_1\neq 0$ we get
\bea
(T_0^4)_{div}=\fr{e^{i\pi(3-K)/2}}{4\sqrt{(-1)^{3-K}\det(Q_4)-i\eps}}.\label{eq_T04}
\eea
This shows that $(T_0^4)_{div}$ is integrable but its
\underline{square is not}.

\noindent In the case $N=3$ (the triangle), $D=4$, it is possible
to derive
\bea
T^{3}_{0}=\fr{-e^{i\pi(2-K)/2}\upsilon}{8\pi\sqrt{(-1)^{2-K}\la_1\la_2}}\ln(\la_3\upsilon^2-i\eps)
\eea
$T_0^3$ and its square are therefore integrable.

\subsection{$gg\to b\bar{b}H$}

 The first example we study is the electroweak corrections to $pp \ra b \bar
b H$\cite{Boudjema:2007uh} where the one-loop amplitude squared,
which is all that remains in the limit of vanishing bottom Yukawa
coupling, develops a Landau singularity which represents the
rescattering of the top pair and their decay into a $W$ pair that
produces the Higgs through $WW$ fusion. As we will see, in this
example, introducing the width of the internal top and $W$
particles smoothes the singularity. There is a leading Landau
singularity present in the box diagram shown in Fig.
\ref{fig-gg-to-bbH} that occurs for some specific values of the
kinematic variables.

\begin{figure}[htb]
\begin{center}
\includegraphics[width=0.4\textwidth]{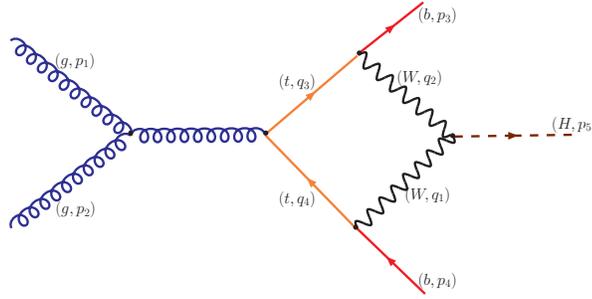}
\caption{\label{fig-gg-to-bbH}{\em A box diagram contributing to
$gg \ra b \bar b H$  that can develop a Landau singularity for
$M_H\ge 2M_W$ and $\sqrt{s}\ge 2m_t$, i.e. all the four particles
in the loop can be simultaneously on-shell.}}
\end{center}
\end{figure}

With $g(p_1)+g(p_2) \to b(p_3)+\bar{b}(p_4)+H(p_5),\;\;\; s=
(p_1+p_2)^2,\,\, s_1=(p_3+p_5)^2,\,\, s_2=(p_4+p_5)^2$, and the
on-shell conditions $p_1^2=p_2^2=0$, $p_3^2=p_4^2=m_b^2=0$,
$p_5^2=M_H^2$, fixing $s$ and $M_H$, the scalar box integral is a
function of two variables $s_{1,2}$
\bea
T_0^4(s_1,s_2)=D_0(M_H^2,0,s,0,s_1,s_2,M_W^2,M_W^2,m_t^2,m_t^2).
\eea
The kinematically allowed region is
\bea
M_H^2\le  s_{1} \le s,\;\; M_H^2\fr{s}{s_1}\le s_2  \le
M_H^2+s-s_1.\label{phys_region_ggbbH}
\eea
The reduced  matrix, $S^{(4)}$, which is equivalent in this case
to the $Q$ matrix for studying the Landau singularity, is given by
\bea S_{ij}^{(4)}=\left( \begin{array}{cccc}
1 & \fr{2M_W^2-M_H^2}{2M_W^2} & \fr{m_t^2+M_W^2-s_1}{2M_Wm_t} & \fr{M_W^2+m_t^2}{2M_Wm_t}\\
\fr{2M_W^2-M_H^2}{2M_W^2} & 1 & \fr{M_W^2+m_t^2}{2M_Wm_t} & \fr{m_t^2+M_W^2-s_2}{2M_Wm_t}\\
\fr{m_t^2+M_W^2-s_1}{2M_Wm_t} & \fr{M_W^2+m_t^2}{2M_Wm_t} & 1 & \fr{2m_t^2-s}{2m_t^2}\\
\fr{M_W^2+m_t^2}{2M_Wm_t} & \fr{m_t^2+M_W^2-s_2}{2M_Wm_t} & \fr{2m_t^2-s}{2m_t^2} & 1\\
\end{array}\right), \;\; S^{(4)}_{ij}=\frac{Q_{ij}}{2m_i m_j}.\eea
The singularity corresponds to $\det(S^{(4)})=0$. The determinant
is a quadratic function of $s_1,s_2$ when $s$ and all internal
masses are fixed. The Landau determinant, the real and imaginary
parts of $T_0^4$ are displayed in Fig.~\ref{box_diag_3D_plots} for
$\sqrt{s}=353\;$GeV, $M_H=165\;$GeV, $m_t=174\;$GeV,
$M_W=80.3766\;$GeV. We clearly see that the Landau determinant
vanishes inside the phase space and leads to regions of
instability exhibiting leading and lower-order Landau
singularities in the real and imaginary parts of the scalar
integral.
\begin{figure}[htb]
\begin{center}
\mbox{
\includegraphics[width=0.45\textwidth]{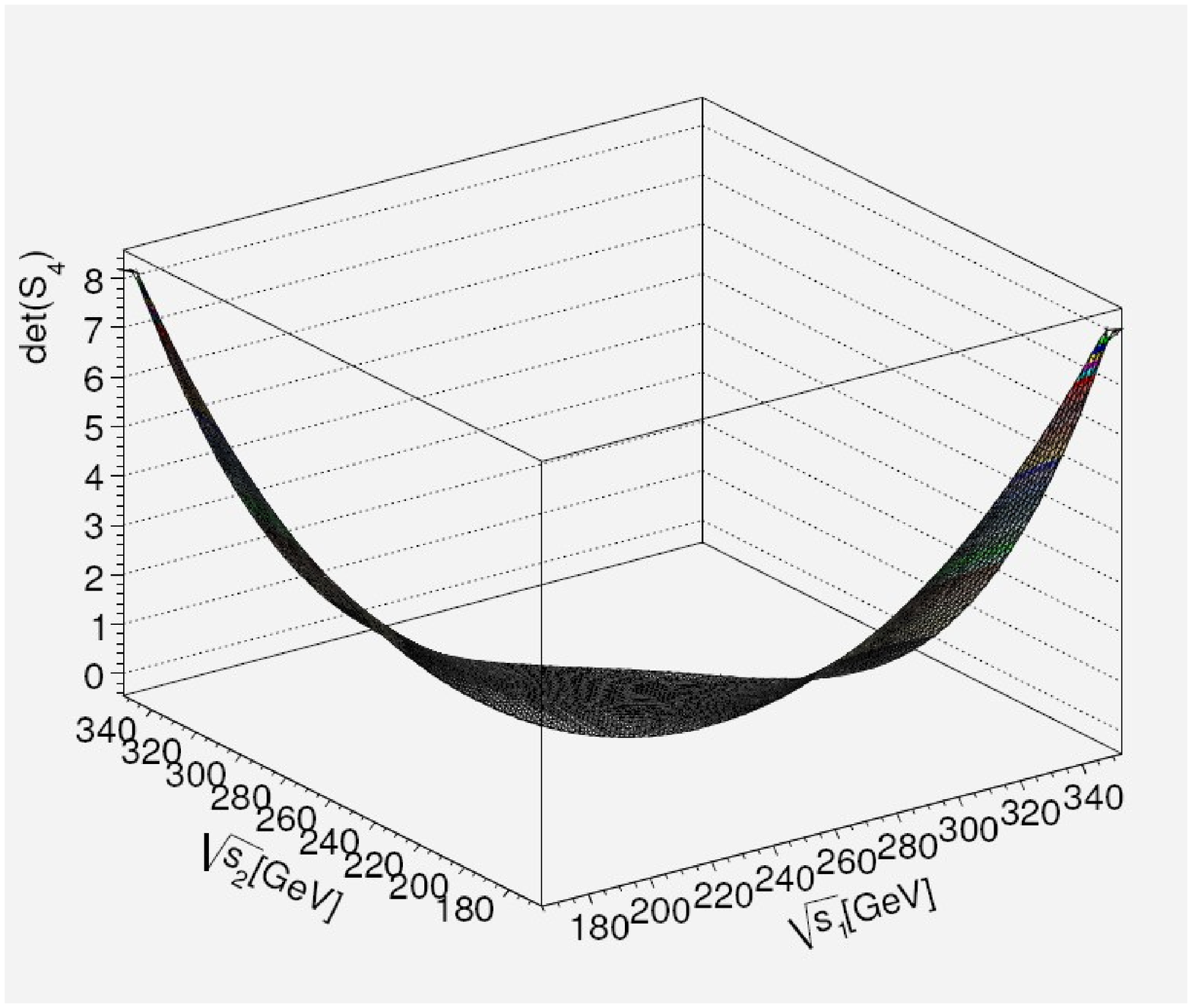}}
\mbox{\includegraphics[width=0.45\textwidth]{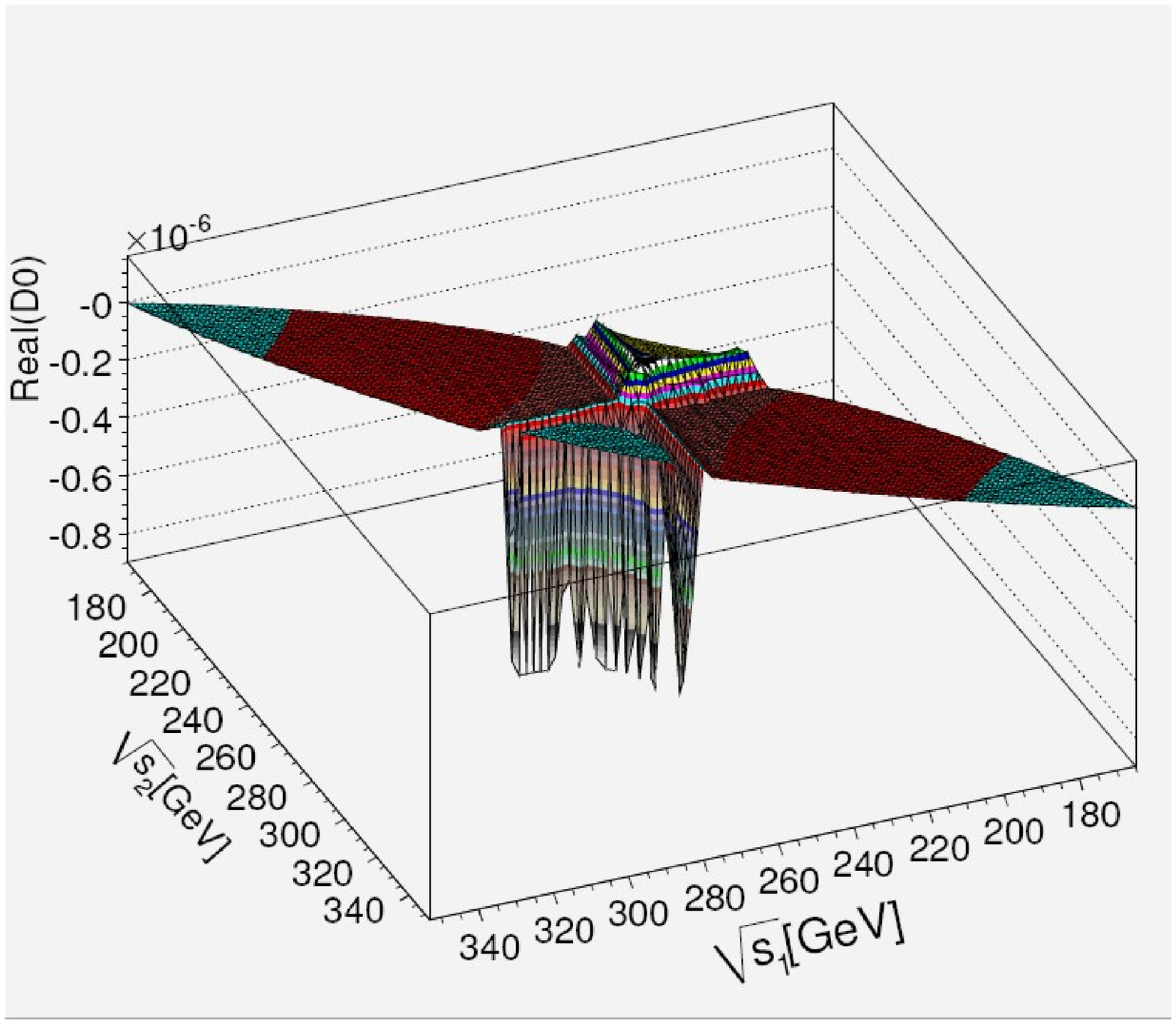}
\hspace*{0.075\textwidth}
\includegraphics[width=0.45\textwidth]{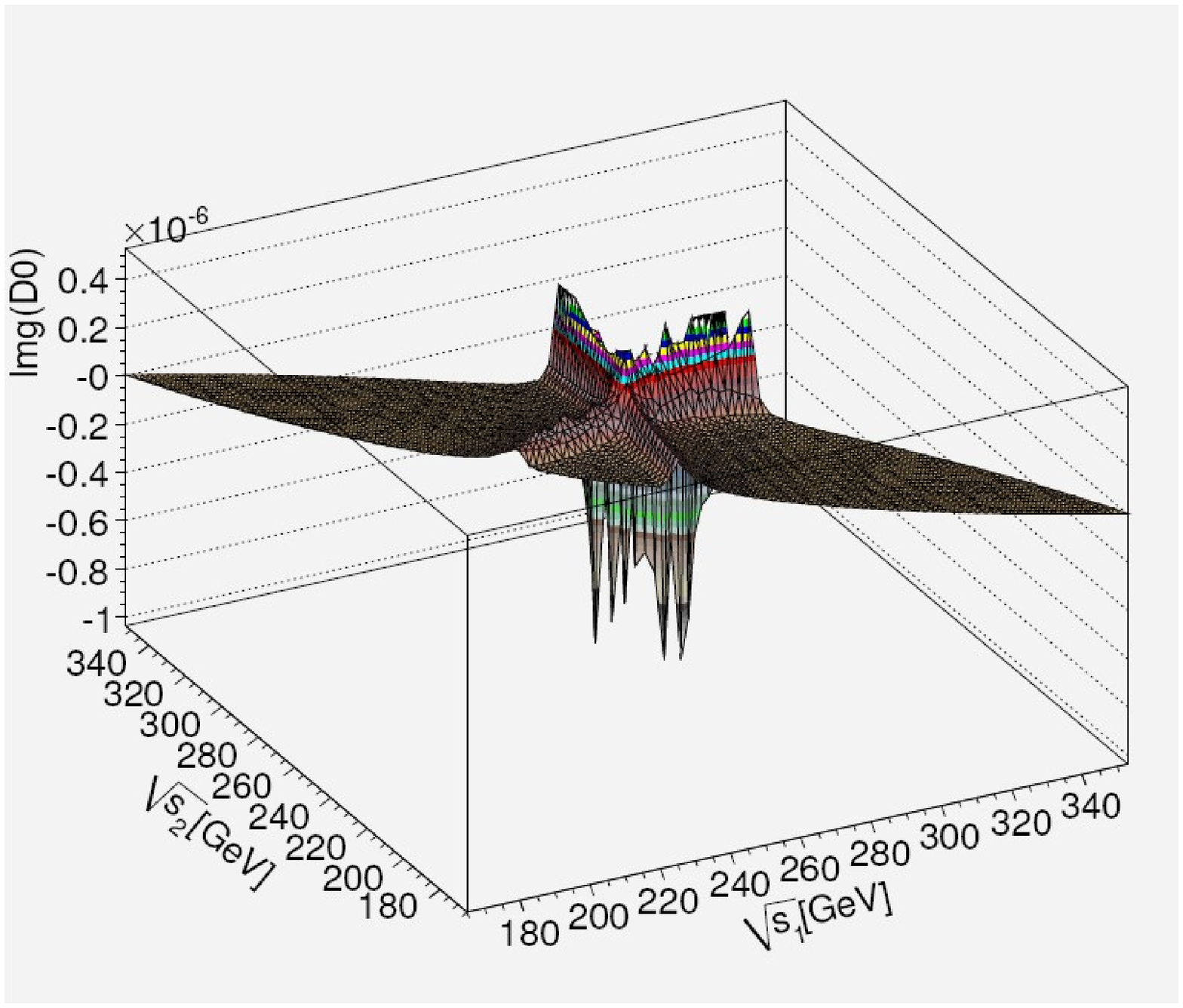}}
\caption{\label{box_diag_3D_plots}{\em The Landau determinant as a
function of $s_1$ and $s_2$ (upper figure). The real and imaginary
parts of $D_0$ as a function of $s_1$ and $s_2$.}}
\end{center}
\end{figure}
To  investigate the structure of the singularities in more detail
let us fix $\sqrt{s_1}=\sqrt{2(m_t^2+M_W^2)}\approx 271.06\;$GeV,
so that the properties are studied for the single variable $s_2$.
The results are shown in Fig.~\ref{gbbH_img_real_det0_2D}.

\begin{figure}[h]
\begin{center}
\includegraphics[width=0.48\textwidth]{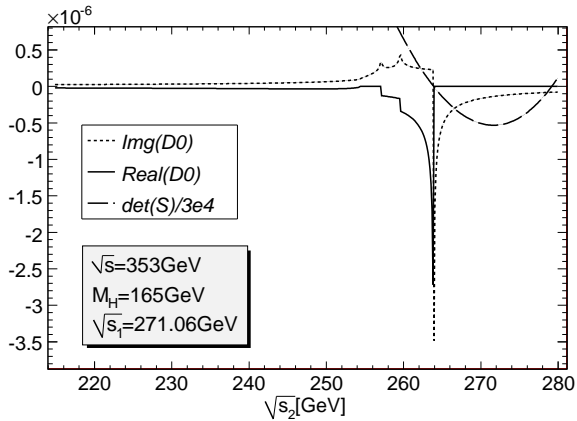}
\caption{\label{gbbH_img_real_det0_2D}{\em The imaginary, real
parts of $D_0$ and the Landau determinant as functions of $s_2$.}}
\end{center}
\end{figure}
From Fig.~\ref{gbbH_img_real_det0_2D} we see that there are four
discontinuities in the function representing the real part of the
scalar integral in the variable $\sqrt{s_2}$. As $s_2$ increases
we first encounter a discontinuity  at the normal threshold
$\sqrt{s_2}=m_t+M_W=254.38\;$GeV. This corresponds to the solution
(for the Feynman parameters) $x_{1,3}=0$ and $x_{2,4}>0$ of the
Landau equations. The second discontinuity occurs at the anomalous
threshold $\sqrt{s_2}=257.09\;$GeV of a reduced triangle diagram.
This corresponds to the solution $x_{3}=0$ and $x_{1,2,4}>0$ of
the Landau equations.  The condition of vanishing determinant
$\det(S_3)=0$ for this triangle has two solutions
\bea
s_2=\fr{1}{2M_W^2} \left( M_H^2(m_t^2+M_W^2)\mp
M_H\sqrt{M_H^2-4M_W^2}(m_t^2-M_W^2) \right)\label{landaupole3_x3}
\eea
which gives $\sqrt{s_2}=257.09\;$GeV(inside of phase space) and
$297.86\;$GeV(outside of phase space). We can also check that the
former value satisfies the sign condition in (\ref{landau_cond1})
while the latter does not. Note that one of the conditions for
this anomalous threshold to occur in the physical region is
$M_H\ge 2M_W$, see Eq.~(\ref{landaupole3_x3}). The same phenomenon
happens for the third discontinuity at $\sqrt{s_2}=259.58\;$GeV
which corresponds to the anomalous threshold of the reduced three
point function obtained from the box diagram by contracting to a
point the $x_1$ line. The last singular discontinuity is the
leading Landau singularity. The condition $\det(S_4)=0$ for the
box has two solutions which numerically correspond to
$\sqrt{s_2}=263.88\;$GeV or $\sqrt{s_2}=279.18\;$GeV. Both values
are inside the phase space, see Fig.~\ref{gbbH_img_real_det0_2D}.
However after inspection of the corresponding sign condition only
$\sqrt{s_2}=263.88\;$GeV (with  $x_1 \approx 0.533186, x_2 \approx
0.748618, x_3 \approx 0.774941$)  qualifies as a Landau
singularity. $\sqrt{s_2}=279.18\;$GeV has $x_1\approx -0.742921,
x_2 \approx -0.748618, x_3 \approx 1.06537$. The nature of the
leading Landau singularity in Fig.~\ref{gbbH_img_real_det0_2D} can
be extracted  by using the general formula (\ref{eq_T04}). With
the input parameters given above, the Landau matrix has only one
positive eigenvalue at the leading singular point, {\it i.e.}
$K=1$. The leading singularity behaves as
\bea D_0^{div}=-\fr{1}{16M_W^2m_t^2\sqrt{\det(S_4)-i\eps}}.
\eea
When approaching the singularity from the left, $\det(S_4)>0$,
the real part turns singular. When we cross the leading
singularity from the right, $\det(S_4)<0$, the imaginary part of
the singularity switches on, while the real part vanishes. In this
example, both the real and imaginary parts are singular because
$\det(S_4)$ changes  sign when  the leading singular point is
crossed.

\begin{figure}[httb!]
\centering \mbox{
\includegraphics[width=0.45\textwidth]{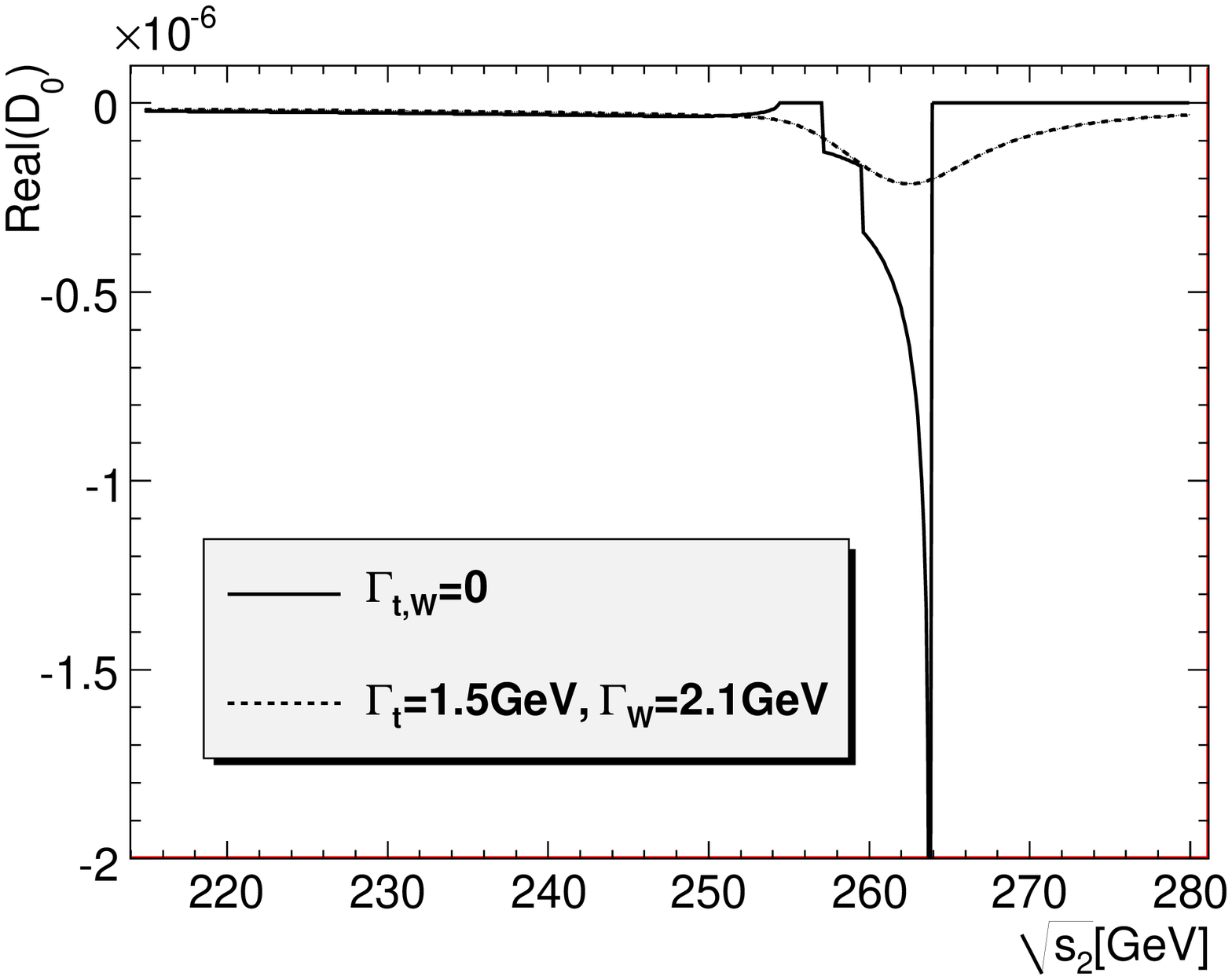}
\includegraphics[width=0.45\textwidth]{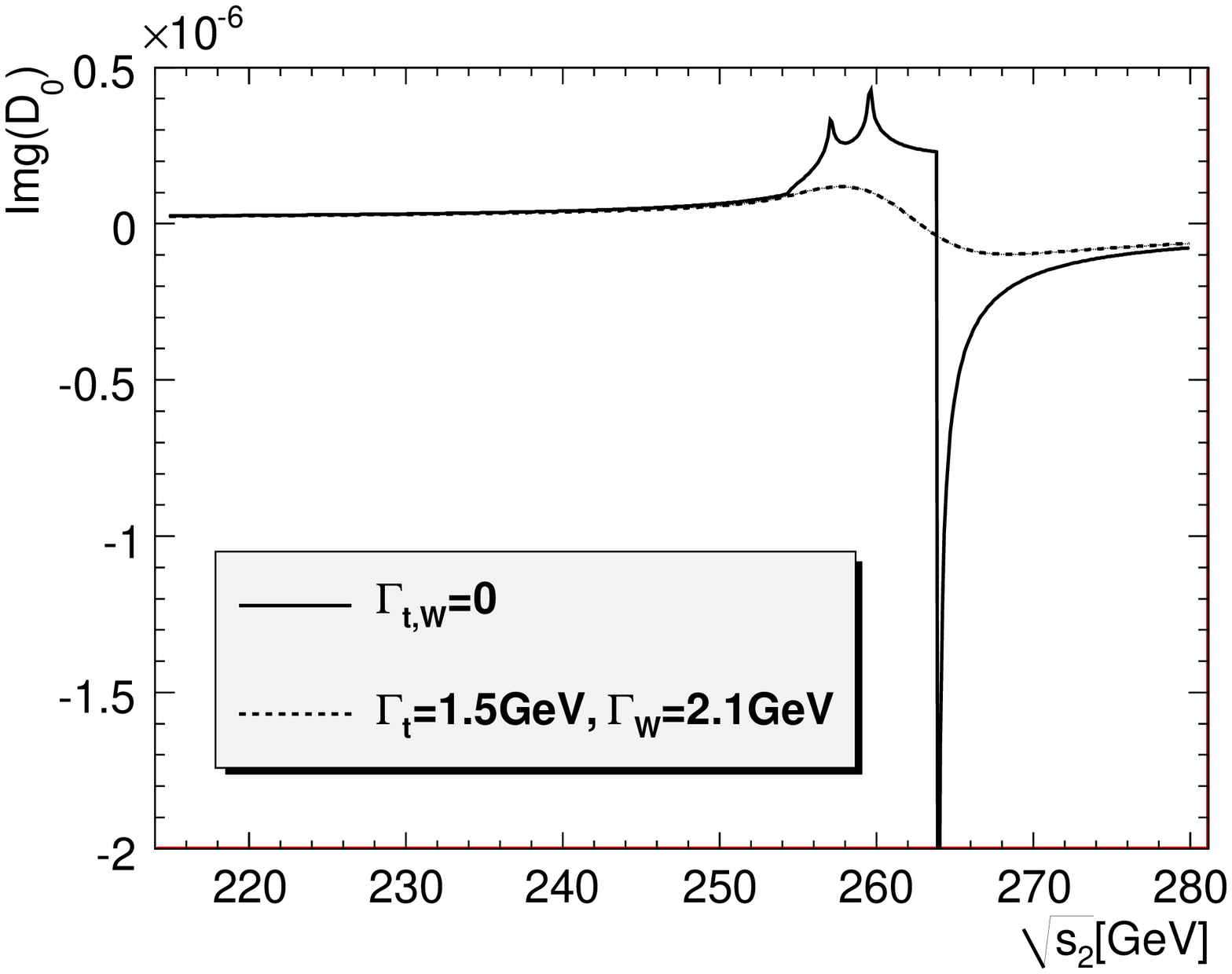}}
\caption{\scriptsize \textit{Effect of the combined width of the
$W$, $\Gamma_W$ and the top, $\Gamma-t$, to the real and imaginary
part of the scalar function.}} \label{dessin-width-smooth}
\end{figure}
The instabilities of the integral and the singularities are due to
the unstable internal particles. The problem can be remedied by
introducing the finite width of the $W$ and top. As seen from
Fig.~\ref{dessin-width-smooth}, introducing the finite width
effect in the scalar box gives a smooth behaviour.

\subsection{The six photon amplitude}
\label{sec:6photons}
 The second example concerns a case with
massless internal particles involving massless external particles:
the $6$-photon amplitude\cite{Bernicot:2007hs}, see
also\cite{Nagy:2006xy,Binoth:2007ca,Ossola:2007bb}. Although the
scalar integrals for the $6$-photon amplitude have a potential
Landau singularity that leads to some characteristic patterns of
the amplitude, direct calculations of the helicity amplitudes show
that the singularity is tamed by the dynamics of the gauge
interaction in a somehow unexpected way.  This is welcome since we
would not able, in this case, to revert to the trick of
introducing a width for the particles. This said, introducing
non-zero (internal) masses, as would be fit for the couplings of
the massless photons, would regulate a vanishing Landau
determinant, but would of course still pose a considerable
numerical problem if the singularity from the vanishing Landau
determinant is not counterbalanced by the spin and gauge algebra.

To be able to see the cancellation at the level of the amplitude
is only possible if one has very compact analytical expressions
for these amplitudes. In our investigation the expressions for the
amplitudes\cite{Bernicot:2007hs} are based on the unitarity-cut
methods and are made particularly simple thanks to the fact that
the six-photon amplitude has no IR/UV divergences and no rational
terms. The six-photon amplitude was calculated in three models: i)
scalar $QED$, $A_{6}^{scalar}$, ii) spinor $QED$:
$A_{6}^{fermion}$ and iii) supersymmetric $QED\;{\caln =1}$:
$A_{6}^{\caln=1}$. The three amplitudes
$A_{6}^{scalar},A_{6}^{fermion}$ and $A_{6}^{\caln =1}$ are in
fact related through:
\begin{align}
    A_{6}^{fermion} = -2 A_{6}^{scalar} + A_{6}^{\caln=1} \label{supersymetricdecomposition}
\end{align}

Full compact expressions for the amplitudes can be found in
\cite{Bernicot:2007hs}. The potential Landau singularity in the
$6$-photon amplitude reveals itself in the so-called double parton
scattering configuration\cite{Nagy:2006xy}, see
Fig.~\ref{dessin_doubleparton}.
\begin{figure}[httb!]
            \centering
\includegraphics[width=6cm]{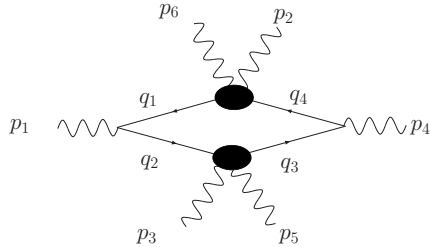}
\caption{\scriptsize \textit{Double parton scattering
configuration: $p_{1},p_{4}$ are incoming photons with $\vec{p}_1
+ \vec{p}_4 = \vec{0}$, each splits into a fermion pair which
rescatters to give photon pairs $(p_{2},p_6) ; (p_{3},p_{5})$ at
very small, vanishing, transverse momentum.}}
\label{dessin_doubleparton}
\end{figure}

The Landau conditions read
\begin{equation}
    \det(Q)= (s_{135}s_{435} -s_{35}s_{26})^{2}  \rightarrow 0  \;\;,
    \;\; s_{35},s_{26} >0 \;\;,\;\;  s_{135},s_{435}<0 
\end{equation}
where $s_{ijk} = (p_{i}+p_{j}+p_{k})^{2}$, all the $p_{i}$'s are
taken as incoming. Note the specific nature of $\det(Q)$ which has
a \underline{quadratic} form. This will lead to a double root
(eigenvalue) at the singularity, or in other words the derivative
of $\det(Q)$ at the singularity is also vanishing. In fact
$\det(Q)$ is proportional to the square of the Gram determinant,
$\det(G)$. To wit
\begin{equation}\label{gram}
\det(G) = - 2 s_{14} (s_{135}s_{435} -s_{35}s_{26}) \propto
\sqrt{\det(Q)}
\end{equation}
This property is due to the presence of many zeros, both from the
kinematics of the external photons and the masslessness of the
internal lines.

\noindent How does the singularity of the scalar integral
transpire at the level of the amplitude? Let us turn  to the NMHV
$({-}{-}{-}{+}{+}{+})$ six-photon helicity amplitude and specialise to the
kinematics\footnote{The correspondance between the kinematical
conventions of Nagy and Soper and the one used here are the
following: Nagy and Soper consider the reaction
$\gamma^{+}(p_4)+\gamma^{-}(p_1) \rightarrow
\gamma^{-}(-p_6)+\gamma^{+}(-p_2)+\gamma^{+}(-p_3)+\gamma^{-}(-p_5)$
i.e. their $k_{i}$'s and our $p_{j}$ are such that: $k_{1} =
p_{4}$, $k_{2} = p_{1}$, $k_{3} = - p_{2}$, $k_{4} = - p_{5}$,
$k_{5} = - p_{6}$ and $k_{6} = - p_{3}$ so that $k_1 + k_2 = k_3 +
k_4 + k_5 + k_6$. See \cite{Bernicot:2007hs} for more details.} of
the Nagy and Soper configuration \cite{Nagy:2006xy}. We start from
a fixed point in phase space in the centre of mass frame
$\vec{p}_{1}+\vec{p}_{4} =\vec{0}$ with $\vec{p}_{4}$ along the
$z$-axis:
\begin{equation}
\left\{
 \begin{array}{ll}
    \overrightarrow{p_{2}} = (-33.5,-15.9,-25.0)  & \overrightarrow{p_{3}} = ( 11.0, 13.2, 22.0) \\
    \overrightarrow{p_{5}} = ( 12.5,-15.3,- 0.3) &
    \overrightarrow{p_{6}} = ( 10.0, 18.0,  3.3)
 \end{array}
\right. \label{kinem-in}
\end{equation}
One can generate new  configurations  by rotating the final state
about the $y$-axis by an arbitrary angle $\theta$. We can then
study the behaviour of the amplitude in this parameter. It is
illuminating to rewrite $\det(Q)$ in terms of this parameter for
this particular configuration:
\begin{equation}\label{e-detS}
\det(Q) = \left( s_{14} \, k_{t}^{2} \right) ^{2} \; {\rm with} \;
k_{t}^{2} = p_{35 \, y}^{2} + \left( p_{35 \, x} \cos \theta +
p_{35 \, z} \sin \theta \right) ^{2}
\end{equation}
where $p_{35 \, i} = p_{3 \, i} + p_{5 \, i}$, $i=x,y,z$. The
minimum value of $k_{t}$ is given by $k_{t \, {\rm min}}^{2} =
p_{35 \, y}^{2}$.

The behaviour of the amplitude as a function of $\theta$ for this
particular configuration is shown in Fig.~\ref{dessin_Nagy}.
\begin{figure}[httb!]
\centering
\mbox{\includegraphics[width=0.48\textwidth]{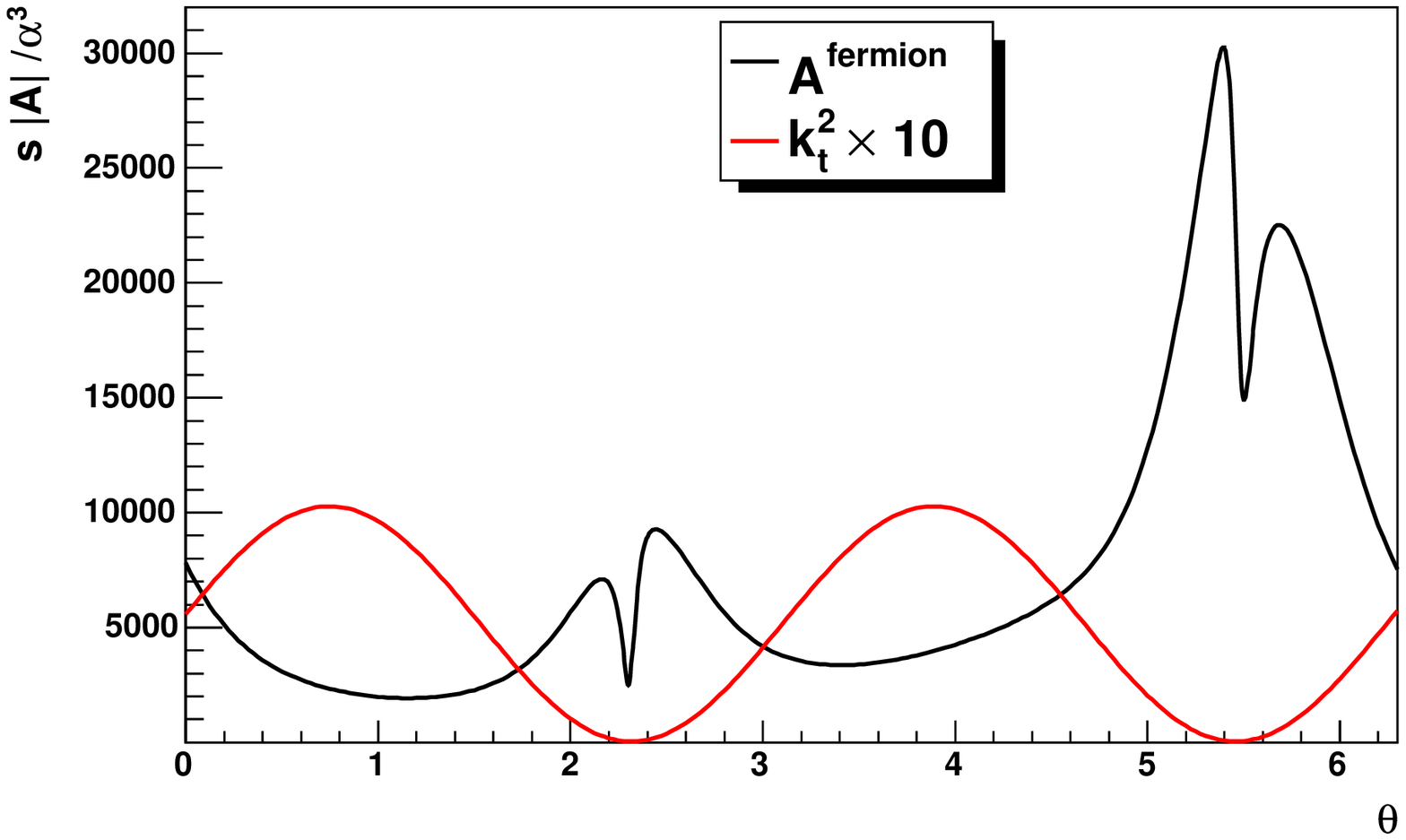}
 \includegraphics[width=0.48\textwidth]{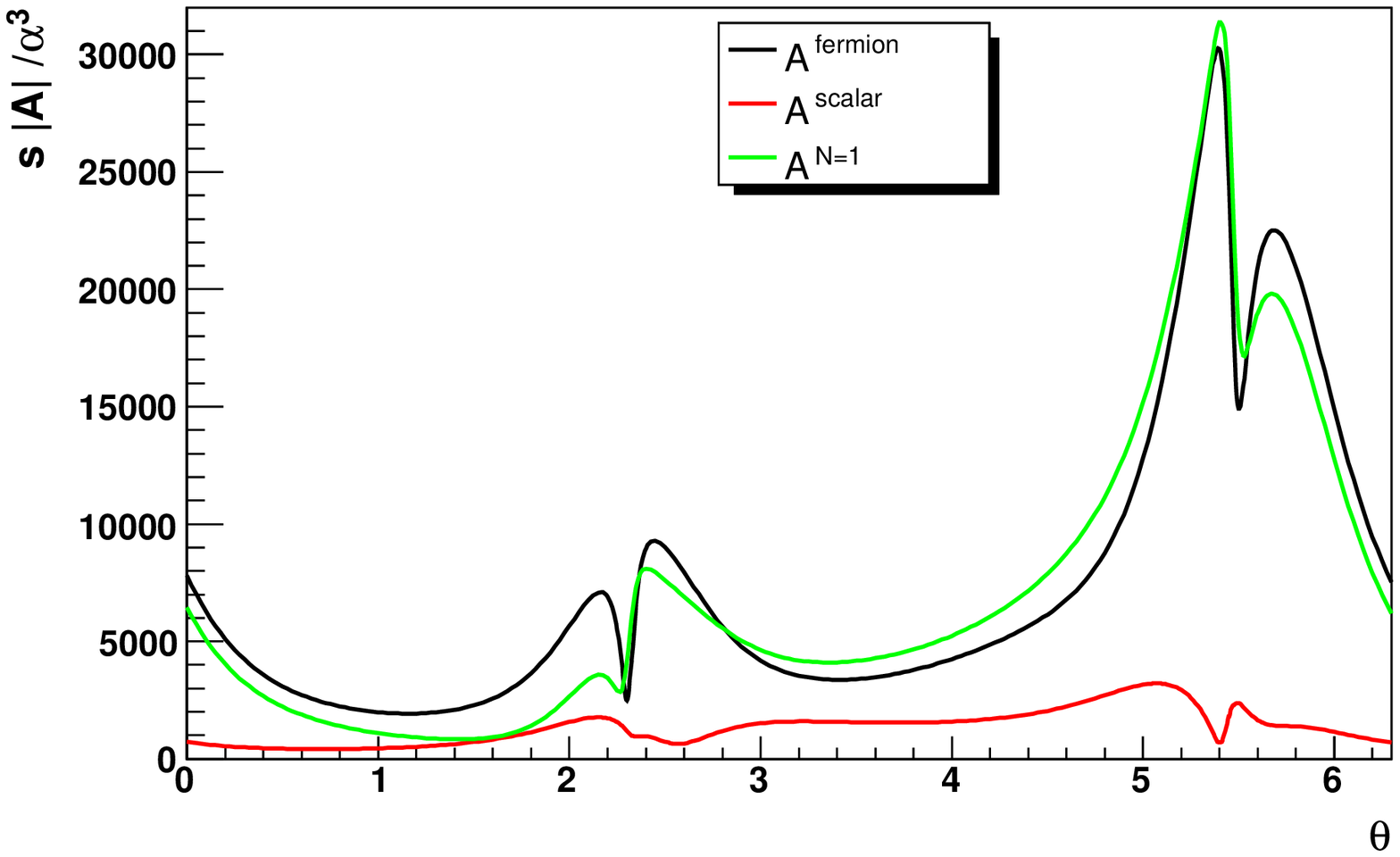}}
\caption{\scriptsize \textit{The NMHV amplitude as a function of
$\theta$ in the Nagy-Soper configuration in the case of QED (left)
as well as the scalar and ${\caln}=1$ SUSY(right). In the first
panel we also show the dependence of $k_t^2$ which is a good
measure of $\det(Q)$. }} \label{dessin_Nagy}
\end{figure}
The important conclusion to draw from Fig.~\ref{dessin_Nagy} is
that the structure of the amplitude, in particular the peculiar
dips, is well tracked by  $\det(Q)$. Indeed the dips that show in
the amplitude occur exactly at the points where $\det(Q)$ is
smallest. The dips occur at  $\theta \simeq 2.32$ and $\theta
\simeq 2.32 + \pi \simeq 5.46$. These values can be derived from
Eq.~(\ref{e-detS}) where $k_t=k_{t \, {\rm min}}$. \\
\noindent One can ask what would happen in a configuration where
$k_{t \, {\rm min}}$ and consequently $\det(Q) \ra 0$?  One can
arrive at this $\det(Q) \ra 0$ configuration by perturbing the
original kinematics in  Eq.~\ref{kinem-in}
\begin{equation}
\left\{
 \begin{array}{ll}
\overrightarrow{p_{2}}  \rightarrow  \overrightarrow{p_{2}^\prime}
= (-33.5,-15.9 - \Delta_{y},-25.0) &
\overrightarrow{p_{3}}  \rightarrow    \overrightarrow{p_{3}^\prime} = ( 11.0, 13.2 + \Delta_{y}, 22.0) \\
\overrightarrow{p_{5}}  \rightarrow \overrightarrow{p_{5}^\prime}
= ( 12.5,-15.3 + \Delta_{y},- 0.3) & \overrightarrow{p_{6}}
\rightarrow \overrightarrow{p_{6}^\prime} = ( 10.0, 18.0 -
\Delta_{y},  3.3)
 \end{array}
\right.
\end{equation}
The $\theta$ modulation is unchanged, such that the dips occur at
the same location in $\theta$. However now $\Delta_y$ can be
chosen such that $k_{t \, {\rm min}}=0$. This occurs for
$\Delta_y=1.05$.

\begin{figure}[httb!]
\centering
\mbox{\includegraphics[width=0.48\textwidth]{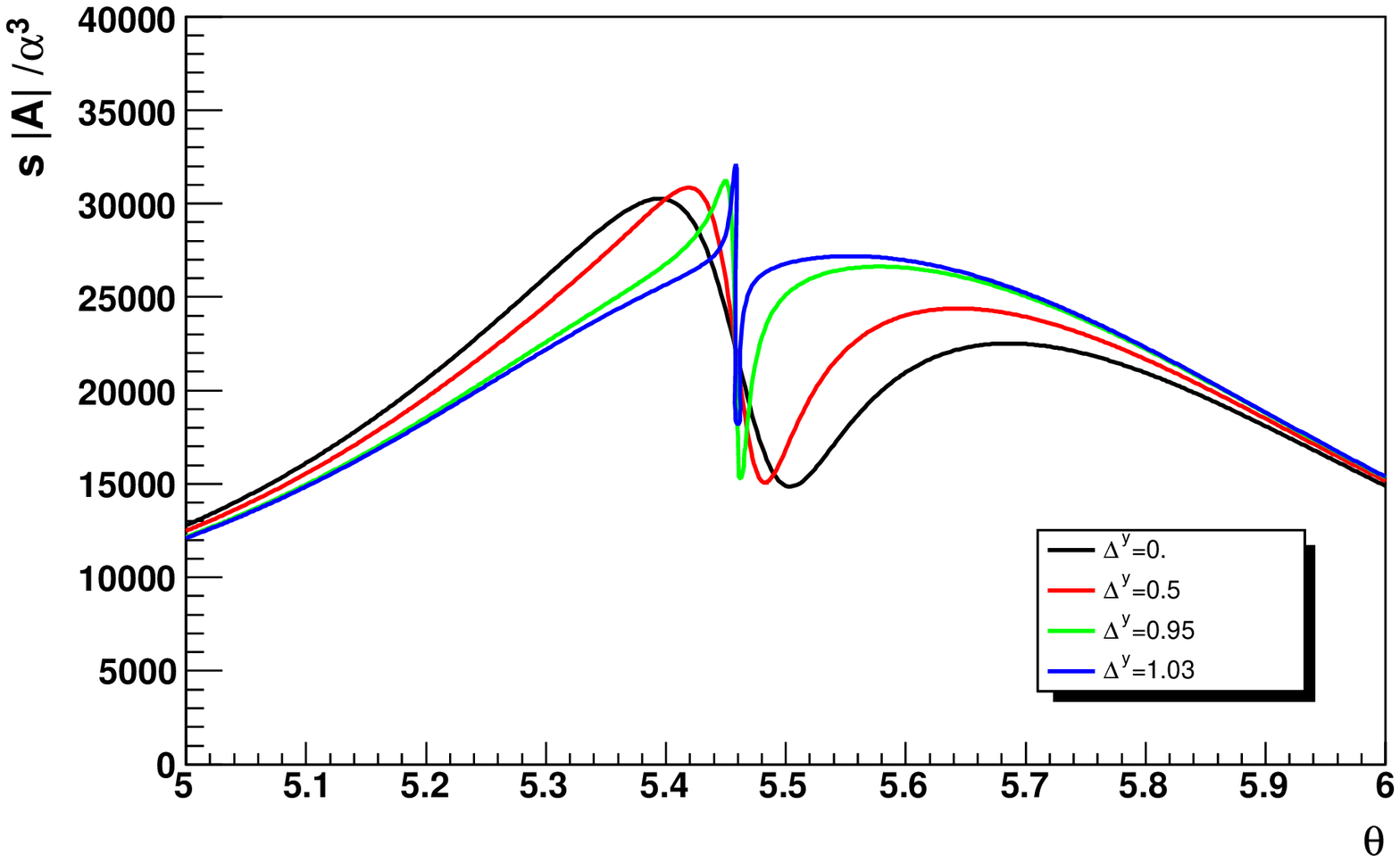}
\includegraphics[width=0.48\textwidth]{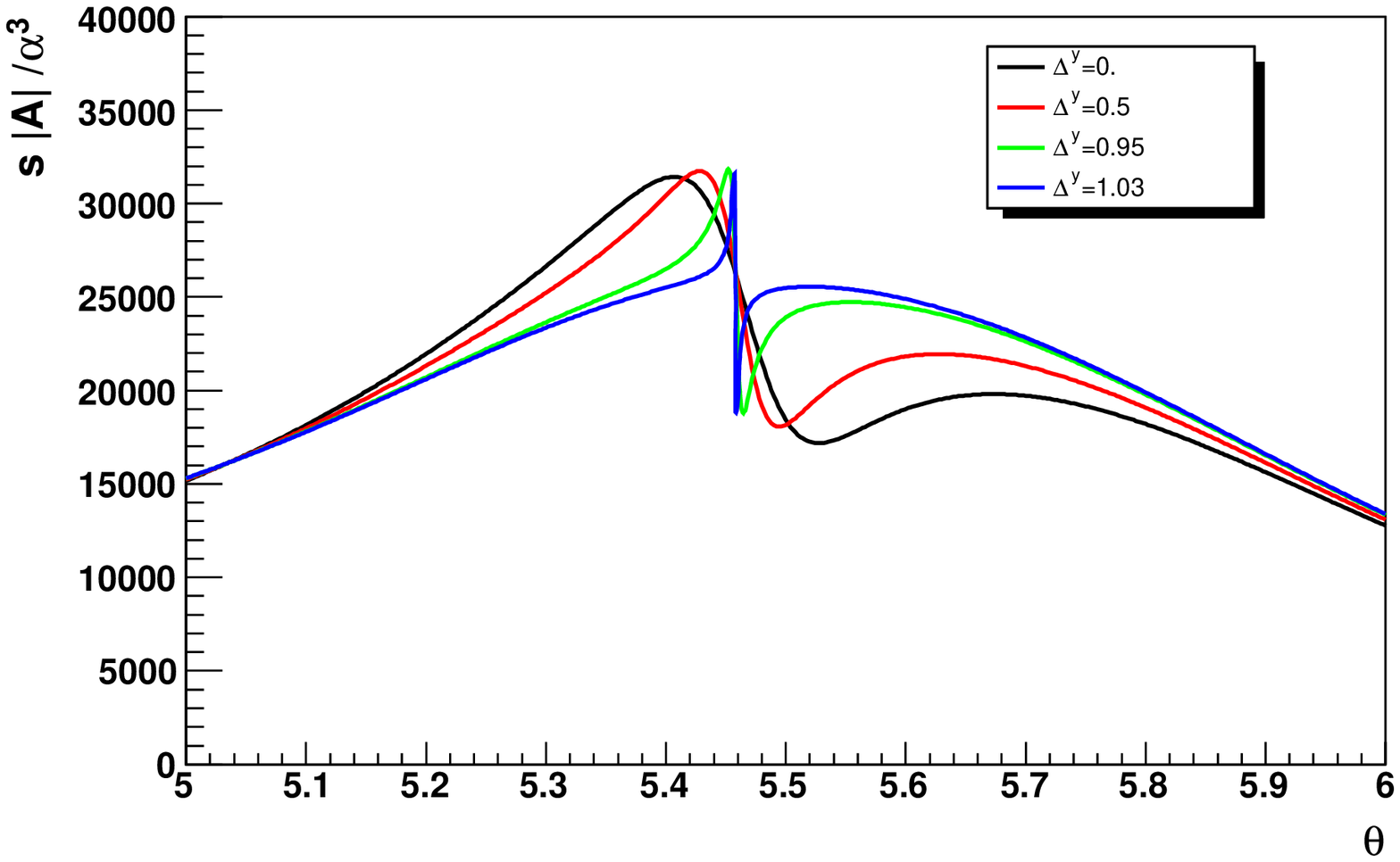}}
\caption{\scriptsize \textit{The six-photon amplitude around the
Landau singularity characterised by $\theta$ around $\theta=5.46$
and for different values of the parameter $\Delta_y$ that gives a
measure of $k_{t,{\rm min}}$ in spinor QED (left) and in
${\caln=1}$ susy QED (right). }} \label{Landausca}
\end{figure}
Figs.~\ref{Landausca}  show how the pattern of the amplitude, as
far as the dip around the singularity at $\theta=5.46$ is
concerned, evolves as $\Delta_y$ is varied from zero to $1.05$
where $\det(Q)$ and $k_{t \, {\rm min}}$ vanish.   It can be seen
that as $\det(Q) \rightarrow 0$ with increasing $\Delta_y$, the
width of the dip decreases more and more so as to behave as a
sudden jump, with the oscillation pattern disappearing completely
for $\Delta_y= 1.05$. The numerators of the six-photon amplitudes, 
reflecting the dynamics of the gauge interaction, vanish fast enough as
the Landau singularity is approached. Therefore the singularity
seems to be  {\em dynamically regulated} for the three cases of
the scalar, the fermion and the SUSY-amplitude.

It is also revealing to investigate how the apparent Landau
singularity is approached from different directions by considering
a two-dimensional parameterisation of $\det(Q)$ and the
kinematics.

\begin{figure}[httb!]
\centering
\includegraphics[width=0.80\textwidth,height=0.40\textwidth]{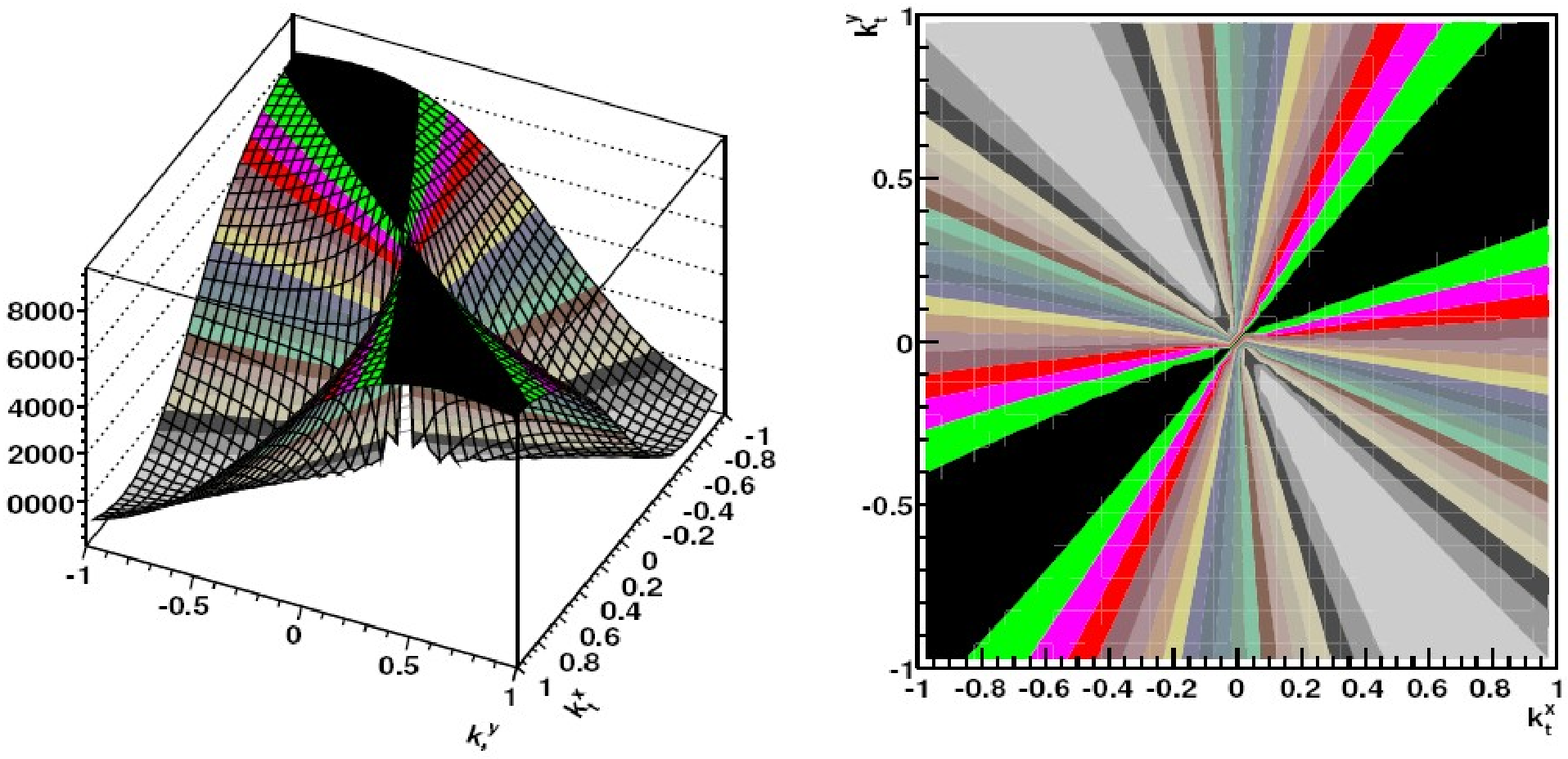}
\includegraphics[width=0.80\textwidth,height=0.40\textwidth]{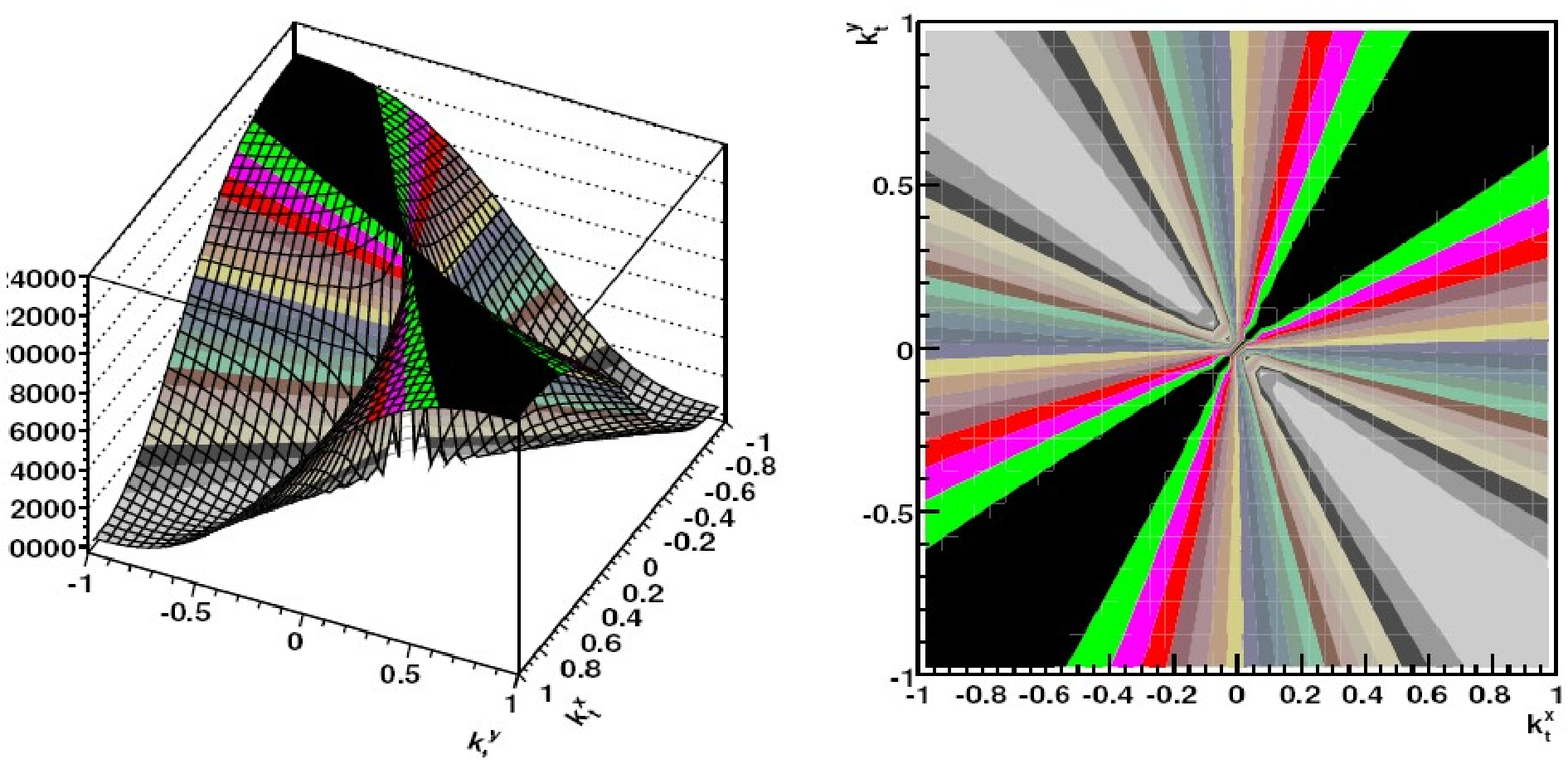}
\caption{\scriptsize \textit{The six-photon amplitude in spinor
QED (top) and in ${\caln=1}$ QED (bottom) around the Landau
singularity}} \label{Landausca2d}
\end{figure}

We therefore modify the original Nagy-Soper parameterisation such
as to generate a Landau singularity and add a $k_t$ variable both
along the $x$ and $y$ direction to follow the approach to the
singularity:
\begin{equation}
\left\{ \begin{array}{ll}
    \overrightarrow{p_{2}} = (-33.5 - k_{t \, x},-15.9 - k_{t \, y},-25.0)
    &
    \overrightarrow{p_{3}} = (-12.5 + k_{t \, x}, 15.3 + k_{t \, y}, 22.0) \\
    \overrightarrow{p_{5}} = ( 12.5 + k_{t \, x},-15.3 + k_{t \, y},- 0.3)
    &
    \overrightarrow{p_{6}} = ( 33.5 - k_{t \, x}, 15.9 - k_{t \, y},  3.3)
\end{array} \right.
\end{equation}
Figs.~\ref{Landausca2d} show the six-photon amplitudes
$A_{6}^{fermion/{\caln=1}}$ as functions of the two variables
$k_{t \, x}$ and $k_{t \, y}$. Up to an overall rotation, the
analytic structure of these amplitudes near the Landau singularity
at $k_{t \, x} = k_{t \, y} = 0$ can be  modelled as
\begin{equation}\label{model}
A_{6} \sim \frac{k_{t \, x}\,  k_{t \, y}}{k_{t \, x}^{2} + k_{t
\, y}^{2}}
 = \frac{1}{2} \sin (2 \alpha)
\end{equation}
where $k_{t \, x} = k_{t} \cos \alpha$, $k_{t \, y} = k_{t} \sin
\alpha$, $k_{t} = (k_{t \, x}^{2} + k_{t \, y}^{2})^{1/2}$.

The amplitudes
exhibit a valley and a ridge along mutually perpendicular axes
crossing each other at $k_{t \, x} = k_{t \, y} = 0$. The various
profiles shown in Fig.~\ref{Landausca} are nothing but cross
sections at fixed $k_{t \, y}$ of Fig~\ref{Landausca2d}. In
particular, the profiles for $\Delta_{y} = 1.05$ correspond to
$k_{t \, y} = 0$. More generally, when both $k_{t \, x}$ and $k_{t
\, y}$ approach $0$ simultaneously, $A_{6}$ remains finite: the
Landau singularity of the double parton scattering type does not
lead to a divergence $\sim 1/k_{t}^{2}$ as would have been naively
expected from a general power counting
argument\cite{Smatrix,Itzykson:1980rh}. Yet the limiting value
of $A_{6}$ depends on the direction $\alpha$ along which the
origin $k_{t \, x} = k_{t \, y} = 0$ is approached.

\subsection{Conclusions}
We foresee that in the calculations of multi-leg one-loop
processes the study of the Landau conditions will bring very
useful, if not crucial, information. More investigations of the
properties of these singularities need to be performed.

%
}


\section[Tensor one-loop integrals in exceptional phase-space regions]
{TENSOR ONE-LOOP INTEGRALS IN EXCEPTIONAL PHASE-SPACE REGIONS%
\protect\footnote{Contributed by: A.~Denner, S.~Dittmaier}}
{\graphicspath{{dittmaier/}}

\def\mathswitchr#1{\relax\ifmmode{\mathrm{#1}}\else$\mathrm{#1}$\fi}
\def\mathswitch#1{\relax\ifmmode#1\else$#1$\fi}
\newcommand{\Pe}{\mathswitchr e}
\newcommand{\Pep}{\mathswitchr {e^+}}
\newcommand{\Pem}{\mathswitchr {e^-}}
\newcommand{\Pd}{\mathswitchr d}
\newcommand{\Pu}{\mathswitchr u}
\newcommand{\PW}{\mathswitchr W}
\newcommand{\PZ}{\mathswitchr Z}
\newcommand{\PH}{\mathswitchr H}
\newcommand{\rd}{\mathrm{d}}
\newcommand{\ri}{\mathrm{i}}
\newcommand{\ina}{i_1}
\newcommand{\inb}{i_2}
\newcommand{\inc}{i_3}
\newcommand{\ind}{i_4}
\newcommand{\ine}{i_5}
\newcommand{\ing}{i_6}
\newcommand{\Zadj}{\tilde Z}
\newcommand{\Zadjadj}{\,\smash{\tilde{\!\tilde Z}}\vphantom{\tilde Z}}
\newcommand{\Ymod}{X}
\newcommand{\Ymodadj}{\tilde{\Ymod}}
\newcommand{\Ymodadjadj}{\,\smash{\tilde{\!\Ymodadj}}\vphantom{\Ymodadj}}
\newcommand{\Gramdet}{|Z|}
\def\nl{\nonumber\\*}
\newcommand{\MZ}{\mathswitch {M_\PZ}}
\newcommand{\Mu}{\mathswitch {m_\Pu}}
\newcommand{\GeV}{\unskip\,\mathrm{GeV}}

%
%
%
%

\subsection{Introduction}

At the LHC and ILC, many interesting processes involve more than four
external particles. 
A thorough description of such processes requires the evaluation
of strong and electroweak radiative corrections at least in
next-to-leading order (NLO). The most complicated part in such calculations
concerns the numerically stable evaluation of the one-loop tensor
integrals of the virtual corrections. 

For processes with up to four external particles the classical 
Passarino--Veltman (PV) reduction \cite{Passarino:1978jh}, which
recursively reduces tensor to scalar integrals,
is sufficient in practically all cases. 
This scheme, however, involves Gram determinants in the denominator,
which spoil the numerical stability if they become small. 
With up to four external particles this happens only near the
edge of phase space (forward scattering, thresholds). 
With more than four external particles, Gram determinants
also vanish within phase space, and methods 
are needed where Gram determinants can be small but still non-zero.
In this context it should be noticed that the described problem of
inverse Gram (and related) determinants occurs in {\it all} methods
that reduce loop diagrams or amplitudes to the basis set of standard
scalar integrals. This, in particular, also applies to unitarity-based
or bootstrap approaches that work at the 
analytical (see e.g.\ Ref.~\cite{Bern:2007dw} and references therein) or 
numerical 
\cite{Ossola:2006us,Forde:2007mi,Ellis:2007br,Kilgore:2007qr,Giele:2008ve}
level.
These methods certainly mitigate the problem of cancellations, but 
cannot avoid it completely. 

In this article we inspect two benchmark phase-space points that are
inspired from our calculation of electroweak (EW) ${\cal O}(\alpha)$ 
corrections to $\Pep\Pem\to4\,$fermions \cite{Denner:2005es,Denner:2005fg}.%
\footnote{Meanwhile the same methods have been successfully applied
to NLO EW and QCD corrections to the Higgs decay
$\PH\to\PW\PW/\PZ\PZ\to4f$ \cite{Bredenstein:2006rh,Bredenstein:2006ha}
and to Higgs production via vector-boson fusion at the LHC
\cite{Ciccolini:2007jr,Ciccolini:2007ec}.}
One of the two points involves a small Gram determinant, the other
involves both a small Gram and a small ``modified Cayley
determinant'' at the same time.
Although of course the real 
performance of proposed solutions can be only be found out in full
applications, i.e.\ when integrating loop corrections to
complicated processes over the whole phase space,
a selection of such benchmark points is certainly a useful testground
in the development of loop techniques.

Several solutions to the problem of numerical instabilities 
due to inverse Gram determinants have been proposed in recent years,
but not many of them have proven their performance in complicated
applications yet. For references and descriptions of some methods 
alternative to ours, we refer to Refs.~\cite{Denner:2005nn,Buttar:2006zd}.

\subsection{Tensor coefficients and their reduction}

We consistently follow the notations and conventions for scalar
and tensor one-loop integrals introduced in 
Refs.~\cite{Denner:2005nn,Denner:2002ii}.
Here we briefly repeat the conventions for 4-point integrals as
required in the considered examples. Tensor 4-point integrals of
rank~$P$ are defined as
\begin{equation}
D^{\mu_{1}\ldots\mu_{P}}
=\frac{(2\pi\mu)^{4-D}}{\ri\pi^{2}}\int \rd^{D}q\,
\frac{q^{\mu_{1}}\cdots q^{\mu_{P}}}
{N_0N_1N_2N_3},
\qquad
N_k= (q+p_k)^2-m_k^2+\ri0, \quad p_0=0,
\end{equation}
where $D$ is the number of space--time dimensions and $\mu$ the
reference scale of dimensionional regularization.
The tensor integrals are decomposed into covariants as follows,
\begin{eqnarray}
D^{\mu}&=&\sum_{\ina=1}^{3} p_{\ina}^{\mu}D_{\ina},
\qquad
D^{\mu\nu}=\sum_{\ina,\inb=1}^{3} p_{\ina}^{\mu}p_{\inb}^{\nu}D_{\ina\inb}
+g^{\mu\nu}D_{00},
\nonumber\\ 
D^{\mu\nu\rho}&=&\sum_{\ina,\inb,\inc=1}^{3} 
p_{\ina}^{\mu}p_{\inb}^{\nu}p_{\inc}^{\rho}D_{\ina\inb\inc}
+\sum_{\ina=1}^{3}
(g^{\mu\nu}p_{\ina}^{\rho}+g^{\nu\rho}p_{\ina}^{\mu}+g^{\rho\mu}p_{\ina}^{\nu})
D_{00\ina},
\end{eqnarray}
and so on for higher rank.
Up to rank~3, and only those are considered below, 4-point tensor integrals
are UV finite.
The kinematical arguments of the coefficients $D_{\dots}$,
which comprise all scalar products $p_i p_j$ and internal masses $m_k$,
are written as
\begin{equation}
D_{\dots} \equiv 
D_{\dots}(p_1^2,(p_2-p_1)^2,(p_3-p_2)^2,p_3^2,p_2^2,(p_3-p_1)^2,
m_0^2,m_1^2,m_2^2,m_3^2).
\end{equation}
Conventional PV reduction \cite{Passarino:1978jh}
expresses the rank-$P$ 4-point
coefficients in terms of lower-rank 4- and 3-point coefficients.
In each step $P\to(P-1)$
the inverse of the Gram matrix
\begin{equation}
Z = \left(\begin{array}{ccc}
2 p_1 p_1 & 2 p_1 p_2 & 2 p_1 p_3 \\
2 p_2 p_1 & 2 p_2 p_2 & 2 p_2 p_3 \\
2 p_3 p_1 & 2 p_3 p_2 & 2 p_3 p_3
\end{array} \right)
\end{equation}
occurs, which causes the above-mentioned numerical problems if the 
determinant $|Z|$ becomes small.
The highest negative power of $|Z|$ occurs in the calculation of
tensor coefficients $D_{\ina\inb\dots}$ without ``0'' indices, rendering
them numerically the most delicate.
In the following we also need the matrix
\begin{eqnarray}
X = \left(
\begin{array}{c|ccc}
\! 2m_0^2 & \! f_1 & \!\! f_2 & \!\! f_3 \! \\
\hline
\begin{array}{c} f_1 \\ f_2 \\ f_3 \end{array} 
& & Z & \\
\end{array}\right), \qquad
f_k=p_k^2-m_k^2+m_0^2.
\end{eqnarray}
The vanishing of the modified Cayley determinant $|X|$ corresponds
to necessary conditions
for true (Landau) singularities in a Feynman diagram.
The minors (i.e.\ determinants of submatrices where row $i$ and column
$j$ are discarded) of the matrices $Z$ and $X$, respectively, are called
$\Zadj_{ij}$ and $\Ymodadj_{ij}$ in the following.

\subsection{The ``DD'' approach}

One-loop tensor integrals can be naturally grouped into three categories,
which we have treated in completely different ways:

(i)
For {\it 1- and 2-point integrals} of arbitrary tensor rank, 
numerically stable analytical expressions are presented in 
Ref.~\cite{Denner:2005nn} (see also Ref.~\cite{Passarino:1978jh}).

(ii)
For {\it 3- and 4-point tensor integrals},
PV reduction \cite{Passarino:1978jh}
is applied for ``regular'' phase-space points where Gram determinants
are not too small. For the remaining problematic cases special reduction
techniques have been developed~\cite{Denner:2005nn}.

One of the techniques replaces the
standard scalar integral by a specific tensor coefficient that can be
safely evaluated numerically and reduces the remaining tensor
coefficients as well as the standard scalar integral to the new 
basis integrals. In this scheme no dangerous inverse Gram determinants
occur, but inverse modified Cayley determinants instead. 
We note that
the procedure is related to the fully numerical method described in
Ref.~\cite{Ferroglia:2002mz}. 

In a second class of techniques, 
the tensor coefficients are iteratively deduced up to terms that are
systematically suppressed by small Gram or other
kinematical determinants in specific kinematical configurations. The
numerical accuracy can be systematically improved upon
including higher tensor ranks.
In our previous applications the highest relevant tensor rank was improved 
only by one additional iteration; in the results shown below we employ
an new implementation of the methods where more than ten additional
iterations are included if relevant.
A similar idea, where tensor coefficients are iteratively determined
from higher-rank tensors has been described in Ref.~\cite{Giele:2004ub}
for the massless case.

(iii)
For {\it 5- and 6-point integrals}, direct
reductions to 5- and 4-point integrals, respectively, are possible
owing to the four-dimensionality of space-time. 
For scalar integrals such a reduction was already derived in the 
1960s \cite{Melrose:1965kb}.
In Refs.~\cite{Denner:2005nn,Denner:2002ii} we follow basically the same
strategy to reduce tensor integrals, which has the advantage that 
no inverse Gram determinants appear in the reduction.
Instead 
modified Cayley determinants occur in the denominator,
but we did not find numerical problems with these factors.
A reduction similar to ours
has been proposed in Ref.~\cite{Binoth:2005ff}.
\looseness-1

\vspace{1em}
We would like to stress two important features of our approach.

(i)
The methods are valid for massive and massless cases. 
The formulas given in Refs.~\cite{Denner:2005nn,Denner:2002ii} are valid
without modifications if IR divergences are regularized with mass
parameters or dimensionally.%
\footnote{For the method of Ref.~\cite{Denner:2002ii}, this has been
shown in Ref.~\cite{Dittmaier:2003bc}.}
Finite masses can be either real or complex.

(ii)
The in/out structure of the methods is the same as for
conventional PV reduction, i.e.\ 
no specific algebraic manipulations are needed in applications.
Therefore, the whole method can be (and in fact is) organized as a numerical
library for scalar integrals and tensor coefficients.

\vspace{1em}
We conclude this overview with some comments resulting from our
experience collected in the treatment of a full $2\to4$ scattering
reaction. 

(i) For a specific point in a multi-particle (multi-parameter) phase
space it is highly non-trivial to figure out which of the various
methods is the most precise.  It seems hopeless to split the phase
space into regions that are dedicated to a given method.  Therefore,
we estimate the accuracy for the different methods at each
phase-space point and take the variant promising the highest
precision. 
The accuracy of the PV method is valued
by checking symmetries and PV relations, and by estimating cancellations.
In the expansion approach, we estimate
the number of valid digits based on the expected accuracy of the
expansions and possible numerical cancellations before the evaluation
of the coefficients. 
In the seminumerical approach, the integration error is propagated to
the tensor coefficients, together with an estimate of possible cancellations.

(ii)       
In a complicated phase space it may happen that none of the
various methods is perfect or good in some exceptional situations.
Usually the corresponding events 
do not significantly contribute to cross sections. 
This issue can only be fathomed in actual applications. 
To be on the safe side, we employ the two independent ``rescue systems''
with different advantages and limitations.

(iii)
In view of this, figures as shown below are nice
illustrations, but should always be taken with a grain of salt.
No matter how many of such figures are shown, they will never be
exhaustive, so that no quantitative conclusions on the overall
precision of methods can be drawn.

\subsection{Two benchmark phase-space points}

In the following two examples of exceptional phase-space configurations
are considered:%
\footnote{We have to restrict the set of numerical results to a few
selected tensor coefficients; more results can be found under
{\tt http://wwwth.mppmu.mpg.de/members/dittmair/tensints/benchmarks.html} .}
one with small Gram determinant $|Z|$, another with 
both $|Z|$ and $|X|$ small. These two cases were already qualitatively
illustrated in Ref.~\cite{Denner:2006fy}, but without providing explicit
numbers. 
We also note that a complex Z-boson mass was used there.
Here we switch to a real-valued Z~mass to make it easier for other
groups to compare with our numbers.
For the sake of brevity, no results of the seminumerical method are 
included below; such results are illustrated in Ref.~\cite{Denner:2006fy}.

\subsubsection{A case with a small Gram determinant}

Figure~\ref{fig:DsmallGram} defines the first benchmark point for
a 4-point function in which the Gram determinant $|Z|$ becomes small.
\begin{figure}[p]
\hspace*{6em}
{\scriptsize\unitlength0.7pt 
\begin{picture}(150,95)(0,-50)
\SetScale{.7}
\put(-40,53){\small Full diagram:}
\ArrowLine(40,20)(10,20)
\ArrowLine(10,-20)(40,-20)
\ArrowLine(130,15)(100,15)
\ArrowLine(80,40)(130,40)
\ArrowLine(100,-15)(130,-15)
\ArrowLine(130,-40)(80,-40)
\SetColor{PineGreen}
\ArrowLine(100,15)(80,40)
\ArrowLine(40,-20)(40,20)
\ArrowLine(80,-40)(100,-15)
\Photon(40,20)(80,40){2}{4.5}
\SetColor{Black}
\Photon(100,-15)(100,15){2}{3.5}
\Photon(40,-20)(80,-40){-2}{4.5}
\Vertex(100,-15){2}\Vertex(100,15){2}
\Vertex(40,20){2}\Vertex(80,40){2}
\Vertex(40,-20){2}\Vertex(80,-40){2}
\Text(6,20)[r]{$\Pe^+$}
\Text(6,-20)[r]{$\Pe^-$}
\Text(35,0)[r]{$\Pe$}
\Text(135,40)[l]{$\mu^-$}
\Text(135,15)[l]{$\bar\nu_\mu$}
\Text(135,-15)[l]{$\Pu$}
\Text(135,-40)[l]{$\bar\Pd$}
\Text(60,40)[b]{$\PZ$}
\Text(60,-40)[t]{$\PZ$}
\Text(105,0)[l]{$\PW$}
\Text(98,27)[l]{$\mu$}
\Text(98,-27)[l]{$\Pd$}
\end{picture}}
\hspace*{8em}
{\scriptsize \unitlength .6pt
\begin{picture}(210,120)(10,-15)
\put(-35,105){\small Subdiagram:}
\SetScale{.6}
\Line(20, 85)( 60, 85)
\Line(20,  5)( 60,  5)
\Line(140, 85)(180, 85)
\Line(180,  5)(140,  5)
\SetColor{PineGreen}
\Line(60, 85)(140, 85)
\Line(60,  5)(140,  5)
\Line(60, 85)( 60,  5)
\Line(140,  5)(140, 85)
\SetColor{Black}
\Vertex( 60, 85){3}
\Vertex( 60,  5){3}
\Vertex(140,  5){3}
\Vertex(140, 85){3}
\put(64,40){$0$}
\put(95,11){$0$}
\put(92,67){$\MZ$}
\put(126,40){$0$}
\put( 6,80){$0$}
\put( 0,-5){$t_{\Pe\bar\Pd}$}
\put(185,80){$0$}
\put(185,-5){$s_{\bar\nu\Pu}$}
\CArc(100,50)(50,60,120)
\put(90,110){$t_{\bar\Pe\mu}$}
\CArc(110,45)(50,-30,30)
\put(165,40){$s_{\mu\bar\nu\Pu}$}
\end{picture} }
\\[1.5em]
\includegraphics[scale=0.7]{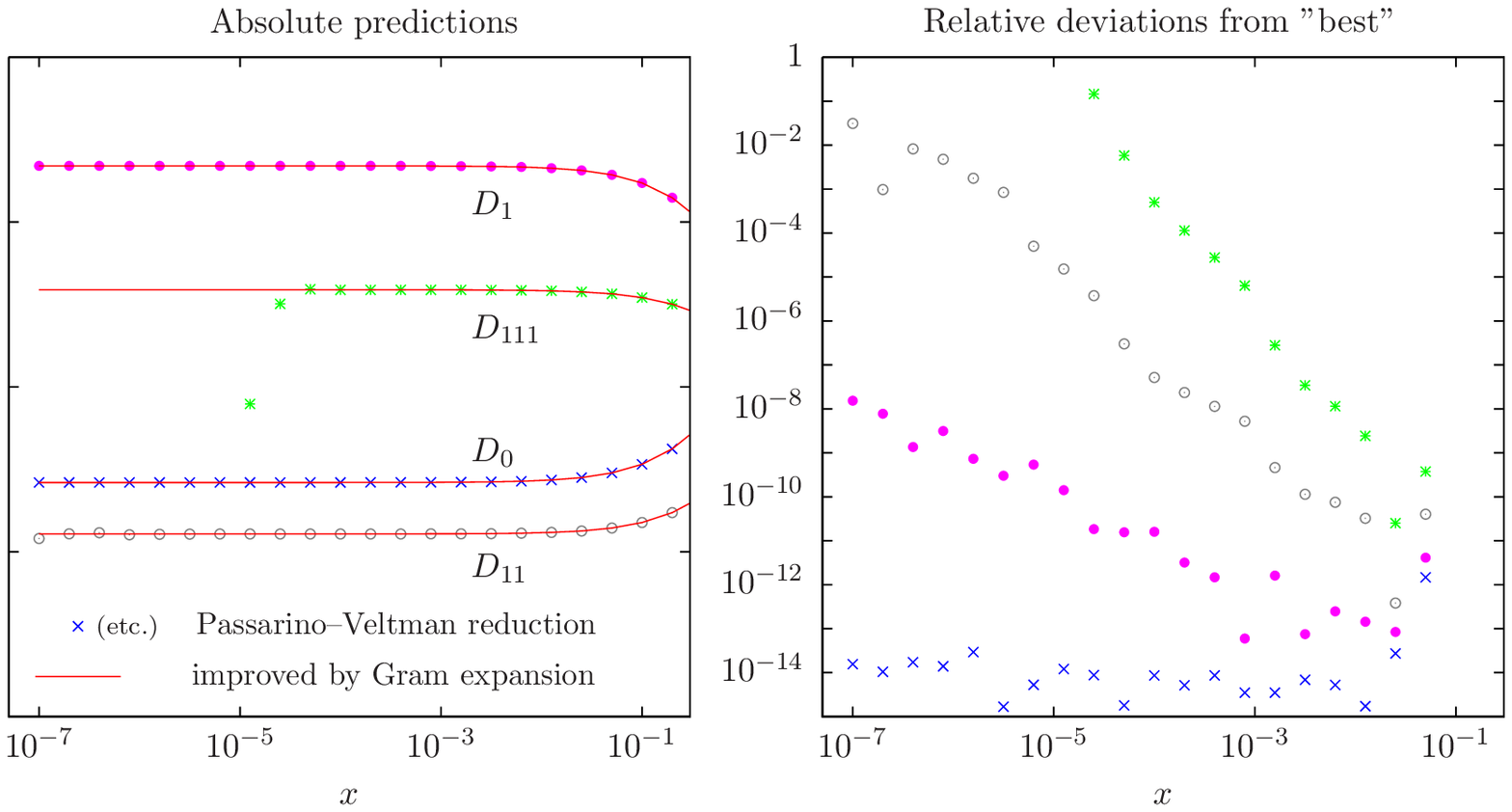}
\hspace*{0em}
\raisebox{3em}{\unitlength 1cm
\begin{picture}(3.7,4.2)
\put(0,4.4){\small Kinematics:}
\put(0.1,3.8){\scriptsize \parbox{.7cm}{$\MZ$}$=91.1876\GeV$}
\put(0.1,3.4){\scriptsize \parbox{.7cm}{$s_{\mu\bar\nu\Pu}$}$=+2{\times}10^4\GeV^2$}
\put(0.1,3.0){\scriptsize \parbox{.7cm}{$s_{\bar\nu\Pu}$}$=+1{\times}10^4\GeV^2$}
\put(0.1,2.6){\scriptsize \parbox{.7cm}{$t_{\bar\Pe\mu}$}$=-4{\times}10^4\GeV^2$}
\put(0.1,2.0){\scriptsize \parbox{.7cm}{$t_{\mathrm{crit}}$}$= \displaystyle
\frac{s_{\mu\bar\nu\Pu}(s_{\mu\bar\nu\Pu}-s_{\bar\nu\Pu}+t_{\bar\Pe\mu})}
{s_{\mu\bar\nu\Pu}-s_{\bar\nu\Pu}}$}
\put(0.1,1.4){\scriptsize \parbox{.7cm}{\mbox{}}$=-6{\times}10^4\GeV^2$}
\put(0.1,0.6){$|Z|\to0 \;\Leftrightarrow$}
\put(0.1,0.1){{$x\equiv {t_{\Pe\bar\Pd}}/{t_{\mathrm{crit}}}-1\to0$}}
\end{picture} }
\vspace{-1.0em}
\caption{A typical example for 4-point integrals with small
$|Z|$ ($x\to0$). 
The full diagram and the relevant subdiagram
are given above; absolute predictions (in arbitrary units) for some tensor 
coefficients, relative deviations from PV reduction, and
the kinematic specifications are shown below.
The precise kinematical assignment is
$D_{\dots}(t_{\Pe\bar\Pd},s_{\bar\nu\Pu},0,0,t_{\bar\Pe\mu},s_{\mu\bar\nu\Pu},
0,0,0,\MZ^2)$.
}
\label{fig:DsmallGram}
\end{figure}
\begin{table}[p]
{\small
\begin{tabular}{lccc}
& $x$ & $D_0[10^{-9}\GeV^{-4}]$& $D_1[10^{-9}\GeV^{-4}]$
\\
\hline
PV & $10^{-1}$ & $-0.67882897158103+\ri\, 6.0180488033754$ &
                 $1.7886414145138 -\ri\, 1.2549864424823$
\\
GE &           & $-0.67882877418780+\ri\, 6.0180477715020$ &
                 $1.7886420559893 -\ri\, 1.2549896774206$ 
\\
\hline
PV & $10^{-3}$ & $-0.83672359694266+\ri\, 6.2756930854749$ &
                 $1.9379452063976 -\ri\, 1.3078118992970$
\\
GE &           & $-0.83672359694268+\ri\, 6.2756930854749$ &
                 $1.9379452063946 -\ri\, 1.3078118992992$
\\
\hline
PV & $10^{-5}$ & $-0.83844622485772+\ri\, 6.2784151968393$ &
                 $1.9395624008169 -\ri\, 1.3083604510334$
\\
GE &           & $-0.83844622485773+\ri\, 6.2784151968392$ &
                 $1.9395624003839 -\ri\, 1.3083604516556$
\\
\hline
PV & $10^{-7}$ & $-0.83846346674121+\ri\, 6.2784424334401$ &
                 $1.9395786154611 -\ri\, 1.3083659591802$
\\
GE &           & $-0.83846346674123+\ri\, 6.2784424334401$ & 
                 $1.9395785857818 -\ri\, 1.3083659392409$
\\[1em]
& $x$ & $D_{11}[10^{-9}\GeV^{-4}]$& $D_{111}[10^{-9}\GeV^{-4}]$
\\
\hline
PV & $10^{-1}$ & $-1.1897035560343+\ri\, 0.24556726948834$ &
                 $0.78386334534494 +\ri\, 0.015037069443873$
\\
GE &           & $-1.1897015303789+\ri\, 0.24555744219672$ &
                 $0.78386954016210 +\ri\, 0.015008250147071$
\\
\hline
PV & $10^{-3}$ & $-1.2896489514112+\ri\, 0.24411794128315$ &
                 $0.85127803054027+\ri\, 0.030174795680439$
\\
GE &           & $-1.2896489629378+\ri\, 0.24411794473416$ & 
                 $0.85127066041158+\ri\, 0.030177001227644$
\\
\hline
PV & $10^{-5}$ & $-1.2906894073746+\ri\, 0.24417445247670$ &
                 $3.6185733047156+\ri\, 5.5143276069563$
\\
GE &           & $-1.2907326083248+\ri\, 0.24408881850424$ & 
                 $0.85200224111245+\ri\, 0.030350914400978$
\\
\hline
PV & $10^{-7}$ & $-1.3307540613183-\ri\, 0.18321620694255$ &
                 $-256227.63578209-\ri\, 2736466.9255631$
\\
GE &           & $-1.2907434539101+\ri\, 0.24408852556218$ & 
                 $0.85200956315901+\ri\, 0.030352656048116$
\end{tabular}
}
\caption{Numerical results corresponding to Fig.~\ref{fig:DsmallGram}.}
\label{tab:DsmallGram}
\end{table}
\begin{figure}[p]
\vspace*{1em}
\hspace*{6em}
\raisebox{-1em}{\scriptsize
\unitlength=1.1bp%
\begin{feynartspicture}(100,80)(1,1)
\FADiagram{}
\FAProp(6.,15.)(6.,5.)(0.,){/Straight}{-1}
\FAProp(0.,15.)(6.,15.)(0.,){/Straight}{-1}
\FALabel(3.,16.27)[b]{$\Pep$}
\FAProp(0.,5.)(6.,5.)(0.,){/Straight}{1}
\FAProp(20.,15.)(14.,15.)(0.,){/Straight}{-1}
\FALabel(3.,3.93)[t]{$\Pem$}
\FALabel(17.,16.07)[b]{$\Pu$}
\FAProp(20.,5.)(14.,5.)(0.,){/Straight}{1}
\FALabel(17.,3.73)[t]{$\bar\Pd$}
\FAProp(20.,10.)(14.,10.)(0.,){/Sine}{-1}
\FALabel(17.,11.07)[b]{$\PW$}
\FALabel(4.93,10.)[r]{$\Pe$}
\FALabel(10.,16.27)[b]{$\gamma$}
\Green{
\FAProp(6.,15.)(14.,15.)(0.,){/Sine}{0}
\FAProp(6.,5.)(14.,5.)(0.,){/Sine}{0}
\FAProp(14.,15.)(14.,10.)(0.,){/Straight}{-1}
\FAProp(14.,5.)(14.,10.)(0.,){/Straight}{1}
}
\FALabel(10.,3.93)[t]{$\PZ$}
\FALabel(12.73,12.5)[r]{$\Pu$}
\FALabel(12.73,7.5)[r]{$\Pd$}
\FAVert(6.,15.){0}
\FAVert(6.,5.){0}
\FAVert(14.,15.){0}
\FAVert(14.,5.){0}
\FAVert(14.,10.){0}
\FAVert(20.,10.){0}
\FAProp(20.,10.)(24.,12.)(0.,){/Straight}{1}
\FAProp(24.,8.)(20.,10.)(0.,){/Straight}{1}
\FALabel(26,13.5)[t]{$\mu^-$}
\FALabel(26,8)[t]{$\bar\nu_\mu$}
\end{feynartspicture}
}
\hspace*{8em}
{\scriptsize \unitlength .6pt
\begin{picture}(210,125)(10,-15)
\put(-320,100){\small Full diagram:}
\put(-30,100){\small Subdiagram:}
\SetScale{.6}
\Line(20, 85)( 60, 85)
\Line(20,  5)( 60,  5)
\Line(140, 85)(180, 85)
\Line(180,  5)(140,  5)
\SetColor{PineGreen}
\Line(60, 85)(140, 85)
\Line(60,  5)(140,  5)
\Line(60, 85)( 60,  5)
\Line(140,  5)(140, 85)
\SetColor{Black}
\Vertex( 60, 85){3}
\Vertex( 60,  5){3}
\Vertex(140,  5){3}
\Vertex(140, 85){3}
\put(64,40){$0$}
\put(92,11){$\MZ$}
\put(95,67){$\Mu$}
\put(126,40){$0$}
\put(-5,80){$\Mu^2$}
\put( 6,0){$s$}
\put(185,80){$s_{\mu\bar\nu}$}
\put(185,0){$0$}
\CArc(100,50)(50,60,120)
\put(90,110){$s_{\mu\bar\nu\Pu}$}
\CArc(110,45)(50,-30,30)
\put(165,40){$s_{\mu\bar\nu\Pd}$}
\end{picture} }
\\[.5em]
\includegraphics[scale=0.7]{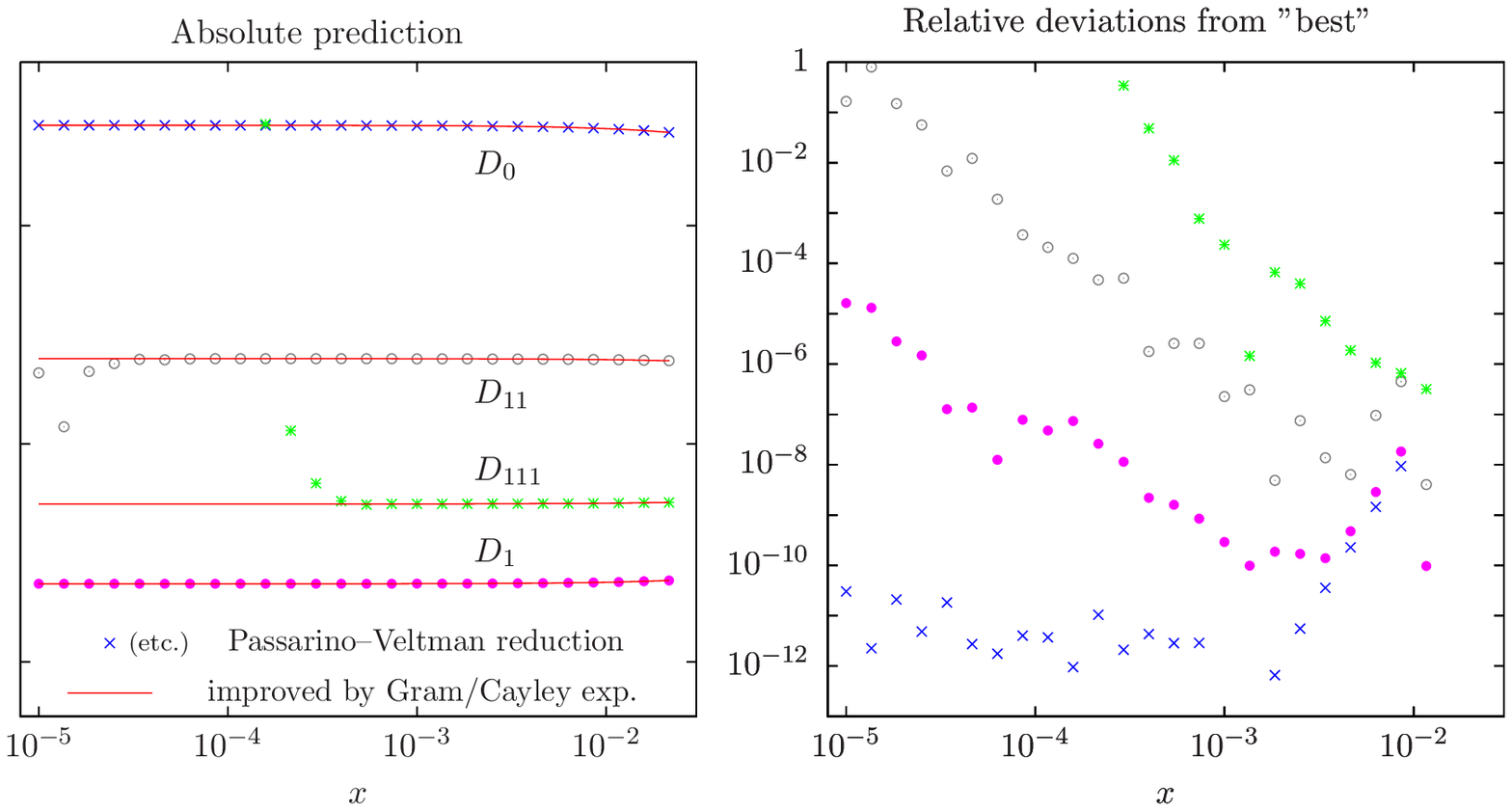}
\hspace*{1em}
\raisebox{3em}{\unitlength 1cm
\begin{picture}(3.7,4.4)
\put(0,4.6){\small Kinematics:}
\put(0.1,4.0){\scriptsize \parbox{.5cm}{$\Mu$}$=0.066\GeV$}
\put(0.1,3.6){\scriptsize \parbox{.5cm}{$\MZ$}$=91.1876\GeV$}
\put(0.1,3.2){\scriptsize \parbox{.5cm}{$s$}$=4{\times}10^4\GeV^2$}
\put(0.1,2.8){\scriptsize \parbox{.5cm}{$s_{\mu\bar\nu}$}$=64{\times}10^2\GeV^2$}
\put(0.1,2.0){$|Z|,|X|\to0 \;\Leftrightarrow$}
\put(0.1,1.6){$s_{\mu\bar\nu\Pd}\to s\;$ and $\;s_{\mu\bar\nu\Pu}\to s_{\mu\bar\nu}$}
\put(0.1,0.8){\small Considered limit:}
\put(0.1,0.3){{$x\equiv {s_{\mu\bar\nu\Pd}}/{s}-1$}}
\put(0.1,-.1){{$\phantom{x} \equiv {s_{\mu\bar\nu\Pu}}/{s_{\mu\bar\nu}}-1\to0$}}
\end{picture} }
\vspace{-1.0em}
\caption{An example for 4-point integrals with both 
$|Z|$ and $|X|$ small ($x\to0$). Details as in Fig. \ref{fig:DsmallGram}.
The precise kinematical assignment is
$D_{\dots}(\Mu^2,s_{\mu\bar\nu},0,s,s_{\mu\bar\nu\Pu},s_{\mu\bar\nu\Pd}
,0,\Mu^2,0,\MZ^2)$.}
\label{fig:DsmallGramCayley}
\end{figure}
\begin{table}[p]
{\small
\begin{tabular}{lccc}
& $x$ & $D_0[10^{-8}\GeV^{-4}]$& $D_1[10^{-8}\GeV^{-4}]$
\\
\hline
PV & $10^{-2}$ & $8.3606217876308 -\ri\,  3.0637590178519$ &
                 $-3.6746526331008 +\ri\, 0.92370985809148$
\\
GCE &          & $8.3605751148559 -\ri\, 3.0637472109275$ &  
                 $-3.6746146470383 +\ri\, 0.92369999581248$    
\\
\hline
PV & $10^{-3}$ & $8.4400974376543 -\ri\, 3.0949777817064$ &  
                 $-3.7124176130452 +\ri\, 0.93444204630892$    
\\
GCE &          & $8.4400974331251 -\ri\, 3.0949777805604$ &  
                 $-3.7124176082911 +\ri\, 0.93444204697694$    
\\
\hline
PV & $10^{-4}$ & $8.4481162422241 -\ri\, 3.0981290348801$ & 
                 $-3.7162301181594 +\ri\, 0.93552679201780$    
\\
GCE &          & $8.4481162422054 -\ri\, 3.0981290348524$ &  
                 $-3.7162304755308 +\ri\, 0.93552678170043$    
\\
\hline
PV & $10^{-5}$ & $8.4489188416187 -\ri\, 3.0984444568680$ &  
                 $-3.7165517842462 +\ri\, 0.93563927582254$    
\\
GCE &          & $8.4489188413614 -\ri\, 3.0984444566400$ &  
                 $-3.7166121290025 +\ri\, 0.93563537143079$    
\\[1em]
& $x$ & $D_{11}[10^{-8}\GeV^{-4}]$& $D_{111}[10^{-8}\GeV^{-4}]$
\\
\hline
PV & $10^{-2}$ & $2.2302468112479 -\ri\, 0.53202142768691$ &
                 $-1.5782872266397 +\ri\, 0.38602980478054$
\\
GCE &          & $2.2297642816234 -\ri\, 0.53189620367287$ &  
                 $-1.5778873843217 +\ri\, 0.38592377802513$
\\
\hline
PV & $10^{-3}$ & $2.2539023067993 -\ri\, 0.53805321575089$ &  
                 $-1.5955732338585+\ri\, 0.38916806038788$
\\
GCE &          & $2.2539023467387 -\ri\, 0.53805185525506$ &  
                 $-1.5951976445129 +\ri\, 0.39030849156415$
\\
\hline
PV & $10^{-4}$ & $2.2578016118662 -\ri\, 0.53856637974433$ &
                 $19.161260651686 +\ri\, 1.6687070921546$
\\
GCE &          & $2.2562925399069 -\ri\, 0.53866164959083$ &  
                 $-1.5969069247380 +\ri\, 0.39074113712771$
\\
\hline
PV & $10^{-5}$ & $1.8810483898149 -\ri\, 0.93548431089474$ &   
                 $492069.51092499 +\ri\, 67693.244541619$
\\
GCE &          & $2.2565317562670 -\ri\, 0.53872268382964$ &  
                 $-1.5970779937221 +\ri\, 0.39078443860164$ 
\end{tabular}
}
\caption{Numerical results corresponding to Fig.~\ref{fig:DsmallGramCayley}.}
\label{tab:DsmallGramCayley}
\end{table}
We compare results of PV reduction with results of the expansion in
the small Gram determinant as described in Section~5.4 of
Ref.~\cite{Denner:2005nn}.
In the upper half of the figure a hexagon diagram is shown that
contains a box subdiagram with the considered kinematical
configuration. The structural diagram illustrates the kinematical
assignment with internal masses and squared external momenta given
at the respective lines. The invariants near the arcs are the squares
of the sum of momenta flowing into the two neighbouring external lines.
The explicit values of the masses and invariants are given in the
figure. As indicated there, the Gram determinant vanishes if
the invariant $t_{\Pe\bar\Pd}$ approaches the critical value
$t_{\mathrm{crit}}$, corresponding to an inner phase-space point.
In the plots of Fig.~\ref{fig:DsmallGram} we show results on a few
tensor coefficients when $t_{\Pe\bar\Pd}$ is varied while keeping
all other invariants fixed. The variation in $t_{\Pe\bar\Pd}$ is
translated into a variation of the dimensionless variable
$x=t_{\Pe\bar\Pd}/t_{\mathrm{crit}}-1$ where the exceptional point 
with $|Z|=0$ corresponds to $x=0$. 

It is clearly seen in the plot on the l.h.s.\ that the tensor
coefficients calculated with PV reduction show numerical instabilities
for small $x$, while the results of the expansion method behave smoothly.
The PV instabilities increase with increasing tensor rank.
The plot on the r.h.s.\ shows the relative difference between the PV
results and the corresponding ``best'' predictions, which are either
obtained with the PV or the expansion method. With decreasing $x$
this difference rises because of the PV instabilities, and for
a sufficiently high $x$ the difference becomes zero (and falls out of
the plot range), because PV reduction promises better accuracy there.
It is essential to see a broad region in $x$ where the 
difference is small for each tensor coefficient. This region
corresponds to the overlap in which both PV reduction and the
expansion method are trustworthy, the difference reflecting the
uncertainty of the less precise result. The plot suggests that
both methods should be precise within a relative accuracy of about
$10^{-9}$ for the considered coefficients which go up to rank~3.
As already mentioned for the $x$ values of the shown points,
the error estimate of the expansion promises better precision,
otherwise (for larger $x$) PV reduction seems to be better.
Table~\ref{tab:DsmallGram} provides explicit numbers for the
considered tensor coefficients at some $x$ values. These 
numbers could serve as a benchmark also for other methods.

We recall that
the expansion for small $|Z|$ is limited to the case where
$\Ymodadj_{0j}$ and $\Zadj_{kl}$ are not too small for at
least one set of indices $j$, $k$, $l$. If all $\Ymodadj_{0j}$ are small,
then $|X|$ is small, too. Such a case is considered in the next subsection.
The case in which all $\Zadj_{kl}$ are small is elaborated in
Section~5.6 of Ref.~\cite{Denner:2005nn}.
\looseness-1

\subsubsection{A case with small Gram and modified Cayley determinants}

Figure~\ref{fig:DsmallGramCayley} defines the second benchmark point for
a 4-point function in which both determinants $|Z|$ and $|X|$
become small.
Here we compare results of PV reduction with results of a
simultaneous expansion in $|Z|$ and $|X|$ as described in Section~5.5 of
Ref.~\cite{Denner:2005nn}.
In the upper half of the figure a pentagon diagram is shown that
contains a box subdiagram with the considered kinematical
configuration. The structural diagram again illustrates the kinematical
situation as in the previous case and the explicit values of the masses 
and invariants are given in the figure. The u-quark mass $\Mu$ is kept
only as regulator of the mass singularity, i.e.\ it is only kept
non-zero in the logarithm $\ln\Mu$, but set to zero otherwise.
The two determinants $|Z|$ and $|X|$ vanish if the two conditions
$s_{\mu\bar\nu\Pd}=s$ and $s_{\mu\bar\nu\Pu}=s_{\mu\bar\nu}$ are
fulfilled. We explore the neighbourhood of this exceptional
configuration on the specific line parametrized by the dimensionless
variable $x={s_{\mu\bar\nu\Pd}}/{s}-1={s_{\mu\bar\nu\Pu}}/{s_{\mu\bar\nu}}-1$,
while keeping the internal masses and the squares of the external momenta
fixed.

The plot on the l.h.s.\ again illustrates the instabilities for small $x$
in the PV reduction that become more serious for higher tensor ranks, 
while the results of the expansion method behave smoothly.
The relative difference between the PV
and the corresponding ``best'' prediction is shown on the r.h.s.,
revealing the expected increase for $x\to0$.
For a sufficiently high $x$ the difference becomes zero,
because PV reduction is more accurate than the expansion.
In the overlap region 
both methods should be precise within a relative accuracy of about
$10^{-6}$ for the considered coefficients.
Table~\ref{tab:DsmallGramCayley} provides explicit numbers for the
considered tensor coefficients at some $x$ values. 

The expansion method fails if either all $\Zadj_{k i}$ or all 
$\Ymodadj_{ij}$ are small. Possible treatments of these exceptional cases
are also described in Ref.~\cite{Denner:2005nn}.

\subsection*{Acknowledgements}
This work is supported in part by the European
Community's Marie-Curie Research Training Network HEPTOOLS under
contract MRTN-CT-2006-035505.

%
%
}


\section[Singularities in one-loop amplitudes from the point of view of reduction methods]
{SINGULARITIES IN ONE-LOOP AMPLITUDES FROM THE POINT OF VIEW OF REDUCTION 
METHODS%
\protect\footnote{Contributed by: G.~Duplan\v ci\' c}}
{%
%
\newcommand{\ri}{{\rm i}}
\newcommand{\e}{{\rm i}\epsilon}
\newcommand{\nn}{\nonumber}
\newcommand{\eir}{\varepsilon}

%
%
%
%
%

\subsection{Introduction}

Obtaining radiative corrections requires the evaluation of loop
Feynman integrals. The simplest, but also the most important, loop
integrals are one-loop Feynman integrals. Considerable progress
has recently been made in developing various approaches for
calculating one-loop integrals. Today, at least in principle, it
is possible to calculate any of them to arbitrary precision no
matter how many external legs the corresponding Feynman diagram
has. Unfortunately, despite huge development, for a practitioner,
the calculation of amplitudes up to one-loop contributions is still a
difficult task. With the increasing complexity of the process
under consideration, the number of Feynman diagrams whose
contributions have to be obtained rises very quickly, as does the
complexity of the corresponding one-loop Feynman integrals which have
to be calculated. Therefore, we are forced to automatize our
calculations. Use of available automatized algorithms helps
tremendously, but the moment when calculations of physically
relevant processes will demand for practical use unacceptable
amounts of computer time and memory is not far away. To surpass
this problem it is necessary to look for new approaches for
calculating one-loop amplitudes, but also to implement algorithms
in a more "computer friendly" way, which means less computer algebra
and more numerics. Unfortunately, numerically oriented codes
increase our chances to face numerical instabilities. This problem
is usually connected with the presence of singularities in the functions
under consideration. It is known that loop Feynamn
integrals have rich singularity structures. For that reason, it
is important to summarize all that is known about the problem as well as to
share experience from previously performed calculations of
amplitudes. Since reduction to the set of basic scalar Feynman
integrals is at the heart of most methods for calculating Feynman
integrals, here we discuss singularities from that point of view.
Despite different approaches which can be taken, the final
decomposition of the given Feynman integral, in terms of
predefined set of basic integrals, should be unique. Therefore,
any approach taken to discuss the singularity structure of the final
decomposition is equally valid. Here the reduction method based on
Refs.
\cite{Davydychev:1991va,Tarasov:1996br,Fleischer:1999hq,Duplancic:2003tv}
is used.

\subsection{Definitions and reduction method}

In order to obtain one-loop amplitudes, integrals of the
following type are required,
\begin{eqnarray}
I^N_{\mu _1\cdots \mu_P}(D;\{\nu_i\}) &\equiv&
(\mu ^2)^{2-D/2}\int
\frac{{\rm d}^D l}{(2\pi)^D}
\frac{l_{\mu_1}\cdots l_{\mu_P}}{\prod_{i=1}^{N}\left[(l+r_i)^2-m_i^2+\e \right]^{\nu_i}}~,
\label{f1}
\\
I^N_{0}(D;\{\nu_i\}) &\equiv&
(\mu ^2)^{2-D/2}\int
\frac{{\rm d}^D l}{(2\pi)^D}
\frac{1}{\prod_{i=1}^{N}\left[(l+r_i)^2-m_i^2+\e \right]^{\nu_i}}~.
\label{f2}
\end{eqnarray}
The integral $I^N_{\mu _1\cdots \mu_P}$~ ($I^N_{0}$) is a rank $P$
tensor (scalar) one-loop $N$-point Feynman integral in
$D$-di\-men\-si\-o\-nal space-time, where $\nu_i$ are powers of
propagators and $l+r_i$~ ($m_i$) is the momentum (mass) of particle
propagating along the corresponding internal line. The momentum $l$ is
the loop momentum and the $r_i$ are linear combinations of external
momenta. The scale $\mu$ is the usual dimensional regularization
scale and the quantity $\e$ ($\epsilon >0$) represents an infinitesimal
imaginary part which ensures causality and, after the integration,
determines the correct sign of the imaginary part of the
logarithms and dilogarithms. It is customary to choose the loop
momentum in such a way that one of the momenta $r_i$ vanishes.
However, for general considerations, it is convenient to keep the
symmetry of the integral with respect to the indices $1,\cdots
,N$.

It can be shown that every tensor one-loop integral can be
expressed as a linear combination of scalar one-loop integrals by the
following equation,
\begin{eqnarray}
I^N_{\mu _1\cdots \mu_P}(D;\{ \nu_i\})& = &
\sum_{k, j_1,\cdots,j_{N}\geq 0\atop 2 k+{\scriptstyle
 \Sigma} j_i=P} \left \{ [g]^k [r_1]^{j_1}\cdots
 [r_{N}]^{j_{N}}\right\}_{\mu _1\cdots \mu_P}
 \frac{( 4 \pi \mu^2)^{P-k}}{(-2)^{k}}
\left[\prod_{i=1}^N \frac{\Gamma (\nu_i+j_i)}{
\Gamma(\nu_i)}\right]
\nn \\
& & {}\times I^N_0(D+2(P-k);\{
\nu_i+j_i\}), \label{f15}
\end{eqnarray}
where $ \{ [g]^k [r_1]^{j_1}\cdots [r_{N}]^{j_{N}}\}_{\mu _1\cdots
\mu_P}$ represents a symmetric (with respect to $\mu _1\cdots
\mu_P$) combination of tensors, each term of which is composed of
$k$ metric tensors and $j_i$ momenta $r_i$. Therefore, the problem
of calculating tensor integrals has been reduced to the
calculation of the general scalar integral, which is the most
convenient to evaluate from the following representation,
\begin{eqnarray}
\lefteqn{I^N_0(D;\{ \nu_i\})=\frac{
\ri }{(4\pi)^2}
(4\pi \mu ^2)^{2-D/2}\frac{\Gamma\left(
\sum\nolimits_{i=1}^{N} \nu_i-D/2\right)}{
\prod\nolimits_{i=1}^{N} \Gamma
 (\nu_i)}(-1)^{\Sigma_{i=1}^N \nu_i}}\nn \\
& &\times\,\int_0^1\! \left( \prod_{i=1}^{N}
 {\rm d}y_i
y_i^{\nu_i-1}\right)
\delta\left(
\sum_{i=1}^{N}y_i-1\right)
\left[ \,-\sum_{i,j=1\atop i<j}^N y_i y_j
\left( r_i-r_j\right)^2 + \sum_{i=1}^N y_i m_i^2
-\e  \, \right]^{D/2-\Sigma_{i=1}^{N}
 \nu_i}\hspace{-1.5cm}.\label{f13}
\end{eqnarray}
Direct evaluation of the general scalar integral represents a
non-trivial problem. However, with the help of the recursion
relations, the problem can be simplified in the sense that the
calculation of the original scalar integral can be reduced to the
calculation of a certain number of simpler basic integrals. All
relevant recursion relations for scalar integrals can be written
in matrix notation as
\begin{eqnarray}
\left( \begin{array}{ccccc}
0 & 1 & 1 & \cdots & 1 \\
1 & R_{11}+ 2\e & R_{12}+ 2\e & \cdots & R_{1N}+ 2\e \\
1 & R_{12}+ 2\e & R_{22}+ 2\e & \cdots & R_{2N}+ 2\e \\
\vdots & \vdots & \vdots & \ddots & \vdots \\
1 & R_{1N}+ 2\e & R_{2N}+ 2\e & \cdots & R_{NN}+ 2\e
\end{array} \right)  \cdot
\left( \begin{array}{c}
(D-1-\sum_{j=1}^{N}\nu_j ) I^N_{0}(D;\{\nu_i\})\\
\nu_1 ~I^N_{0}(D;\{\nu_i+\delta_{i1}\})\\
\nu_2 ~I^N_{0}(D;\{\nu_i+\delta_{i2}\})\\ \vdots \\
\nu_N ~I^N_{0}(D;\{\nu_i+\delta_{iN}\})
\end{array} \right) = &  & \nn \\
 =
\left( \begin{array}{c}
-(4\pi{\mu}^2)^{-1} I^N_{0}(D-2;\{\nu_i\}) \\ -(4\pi{\mu}^2)^{-1}
~I^N_{0}(D-2;\{\nu_i-\delta_{i1}\})
\\ -(4\pi{\mu}^2)^{-1} ~I^N_{0}(D-2;\{\nu_i-\delta_{i2}\})
\\ \vdots \\ -(4\pi{\mu}^2)^{-1}
~I^N_{0}(D-2;\{\nu_i-\delta_{iN}\})
\end{array} \right),  \label{f32}
\end{eqnarray}
 where $R_{ij}=(r_i-r_j)^2-m_i^2-m_j^2$.
In the following we introduce the notation $S_N$ for the
$(N+1)\times(N+1)$ matrix in Eq. (\ref{f32}). Making use of
relations which follow from Eq.(\ref{f32}), each scalar integral
$I_0^N(D;\{\nu_i\})$ can be represented as a linear combination of
integrals $I_0^N(D';\{ 1\})$ and integrals with the number of
propagators which is less than $N$ (it has be understood that $
I_0^N(D;\{\cdots \nu_{l-1}, 0, \nu_{l+1}\cdots \})$ $\equiv
I_0^{N-1}(D;\{\cdots \nu_{l-1}, \nu_{l+1}\cdots \})$~). For the
dimension $D'$, one usually chooses $4+2 \eir$, where $\eir$ is
the infinitesimal parameter regulating the divergences. By
successively applying the above mentioned procedure to the remaining
less-than-$N$-point integrals it is at the end possible to express
the integral $I_0^N(D;\{\nu_i\})$ as a linear combination of integrals
$I_0^k(D';\{ 1\})$, $k=1,\ldots,N$. It is convenient to write
these basic integrals as
\begin{eqnarray}
\lefteqn{I^N_0(D';\{ 1\})=\frac{
\ri }{(4\pi)^2}
(4\pi \mu ^2)^{2-D'/2} \Gamma\left( N-D'/2\right)
(-1)^N} \nn \\
& & \,\times \int_0^1\!  \prod_{i=1}^{N}
 {\rm d}y_i
\delta\left( \sum_{i=1}^{N}y_i-1\right) \left[ \, -\frac{1}{2}
\sum_{i,j=1}^N y_i \left( R_{ij} +2\e \right) y_j \,
\right]^{D'/2-N},\label{f16}
\end{eqnarray}
where the properties of the $\delta$ function were used.

\subsection{Singularities}

The necessary conditions for Feynman integrals to have
singularities are given by the Landau equations. In the integral
representations given by Eqs. (\ref{f13}) and (\ref{f16}), the
singularity conditions \cite{Smatrix} are given by
\begin{equation}
\sum_{i,j=1}^N y_i \left(R_{ij}+2\e\right) y_j =0 \label{f20}
\end{equation}
and
\begin{equation}
{\bf either}
~~y_i=0 ~~{\bf or} ~~ \sum_{j=1}^N \left(R_{ij}+2\e\right) y_j =0
~~\mbox{for each $i$.}\label{f21}
\end{equation}
Notice that condition (\ref{f20}) is automatically satisfied
when conditions (\ref{f21}) are.
The singularity of the given Feynman integral corresponding to
all $y_i\neq 0$ is called the {\it leading singularity} of the integral,
 while those corresponding to some $y_i=0$ are called
{\it lower-order singularities} of the integral. Lower-order
singularities are leading singularities of integrals where all
propagators associated with vanishing $y_i$s have been omitted. In
the language of Feynman diagrams this translates as contraction to
a point of all lines associated with $y_i=0$.

Finding the general solution of the Landau equations is non-trivial
task. Here we consider only real singularities. Real singularities
are those occurring for real values of the invariants $R_{ij}$ on
the physical sheet. Notice that these real values of the
invariants do not necessarily correspond to a physically possible
kinematical configuration.

Due to presence of the $\e$, no singularity appears along the real
contour of integration in the parametric space in Eqs. (\ref{f13}) and
(\ref{f16}). It should be understood that singularities appear
only in the limit $\e\to 0$.

In the previous section, it was described how to express an arbitrary
Feynman integral as a linear combination of the basic scalar
integrals. The question arises if all singularities of the
starting integral correspond to singularities of the basic scalar
integrals or some of them correspond to singularities of
coefficients of the decomposition. To answer that question, it is
enough to check if $\e$ appears in denominators of the
coefficients. That is, singularities appear only in the limit
$\e\to 0$ and if some of the coefficients diverge independently of
that limit, then the corresponding singularity is artificial in the sense that it
is not a singularity of the starting Feynman integral. Consequently,
such a divergence should cancel in sum of all terms in the
decomposition.

The simplest way to see when $\e$ appears in denominators is to
invert Eq. (\ref{f32}) by multiplying it by inverse of $S_N$. The
resulting equation is
\begin{eqnarray}
\lefteqn{\left( \begin{array}{c}
(D-1-\sum_{j=1}^{N}\nu_j ) I^N_{0}(D;\{\nu_i\})\\
\nu_1 ~I^N_{0}(D;\{\nu_i+\delta_{i1}\})\\
\vdots \\
\nu_N ~I^N_{0}(D;\{\nu_i+\delta_{iN}\})
\end{array} \right) =-
\frac{(4\pi{\mu}^2)^{-1}}{\mbox{Det}[S_N]}\,\times} \label{inv} \\
& &{}\hspace{-7mm} \left( \hspace{-1mm}\begin{array}{cccc}
\mbox{Det}[R_N]-2\e \,\mbox{Det}[S_N] & -S_{N}^{2\,1}
 & \cdots & (-1)^{N+2}S_{N}^{N+1\,1} \\
-S_{N}^{1\,2} & S_{N}^{2\,2} &  \cdots &
(-1)^{N+3}S_{N}^{N+1\,2}\\
\vdots & \vdots & \ddots & \vdots \\
(-1)^{N+2}S_{N}^{1\,N+1} & (-1)^{N+3}S_{N}^{2\,N+1} & \cdots &
(-1)^{2N+2}S_{N}^{N+1\,N+1}
\end{array} \hspace{-2mm}\right)\hspace{-2mm}
\left( \hspace{-2mm}\begin{array}{c}  I^N_{0}(D-2;\{\nu_i\})
\\ I^N_{0}(D-2;\{\nu_i-\delta_{i1}\})
\\ \vdots \\
I^N_{0}(D-2;\{\nu_i-\delta_{iN}\})
\end{array} \hspace{-2mm}\right)\!, \nn
\end{eqnarray}
where $S_N^{i\,j}$ is minor of $S_N$ obtained by removing $i$th
row and $j$th column, and $R_N$ is an $N\times N$ matrix with
elements equal to $R_{ij}$. The matrix $R_N$ is sometimes called
modified Cayley matrix and its determinant the modified Cayley
determinant. All minors appearing in Eq. (\ref{inv}) as well as 
$\mbox{Det}[S_N]$ are $\e$ independent. In all that determinants
the first row or column can be simply used to remove completely the $\e$
dependance. Therefore, only the recursion relation following from the
first row of Eq. (\ref{inv}) will have an $\e$ dependent
coefficient. From the form of that relation it follows that real
singularities can appear in the coefficients of decomposition only
if relations of that type are used during reduction to increase
dimension of integrals. In that case, a singularity can appear when
$\mbox{Det}[R_N]=0$. As expected, the singularity is related to the
same matrix which appears in Landau equations (\ref{f20}) and
(\ref{f21}).

What happens if we calculate the integral exactly for kinematical
variables and masses for which $\mbox{Det}[R_N]$ vanishes? In that
case the limit $\e\to 0$ should produce a divergence. But, from the
beginning, dimensional regularization was introduced exactly to
avoid explicit appearance of divergences. Hence, the limit $\e\to
0$ can be applied and divergences appear in the form of powers of
$1/\eir$. It follows that the term $\mbox{Det}[R_N]-2\e
\,\mbox{Det}[S_N]$ vanishes and the first row from Eq. (\ref{inv})
can be used to reduce the $N$-point integral to a linear combination
of $(N-1)$-point integrals.

To complete the discussion, it is necessary to comment on how
reduction works for vanishing $\mbox{Det}[S_N]$. Let us first
express $\mbox{Det}[S_N]$ in a better known form. By subtracting the
last column from the second, third, $\ldots$, and $N$th column,
and then the last row from the second, third, $\ldots$, and $N$th
row, $\mbox{Det}[S_N]$ is given by
 \begin{equation}
\mbox{Det} [S_N]=-\mbox{Det} \left[ -2 (r_i-r_N)\cdot (r_j-r_N)
\right],\qquad i,j=1,\ldots, N-1 .  \label{f42}
\end{equation}
The determinant on the right hand side of Eq. (\ref{f42}) is known
as the Gram determinant. If $\mbox{Det}[S_N]$, i.e. the Gram determinant,
vanishes, then the rows (columns) of the matrix in  Eq. (\ref{f32})
should be linearly dependent. That is, there are real constants $-C$,
$z_1$, \ldots, $z_N$, not all of them equal zero, which satisfy the
equation
\begin{equation}
\left( \begin{array}{ccccc}
0 & 1 & 1 & \cdots & 1 \\
1 & R_{11}+ 2\e & R_{12}+ 2\e & \cdots & R_{1N}+ 2\e \\
1 & R_{12}+ 2\e & R_{22}+ 2\e & \cdots & R_{2N}+ 2\e \\
\vdots & \vdots & \vdots & \ddots & \vdots \\
1 & R_{1N}+ 2\e & R_{2N}+ 2\e & \cdots & R_{NN}+ 2\e
\end{array} \right)\cdot
\left(\begin{array}{c} -C \\ z_1 \\ z_2 \\ \vdots \\ z_N
\end{array}
\right) =\left(
\begin{array}{c}
0 \\ 0 \\ 0 \\ \vdots \\ 0
\end{array}
\right). \label{gg1}
\end{equation}
To see that the constants $-C$, $z_1$, \ldots, $z_N$ should be real,
just remove the complete $\e$ dependance from the system in Eq.
(\ref{gg1}) by subtracting the equation from the first row
multiplied by $2\e$ from equations in all other rows. After
multiplying Eq. (\ref{f32}) by row $\left(
\begin{array}{ccccc}
-C & z_1 & z_2 & \cdots & z_N
\end{array}
\right)$, the following relation emerges,
\begin{equation}
C ~ I^N_{0}(D-2;\{\nu_i\}) = \sum_{j=1}^N z_j
~I^N_{0}(D-2;\{\nu_i-\delta_{ij}\}).\label{gr0}
\end{equation}
It is easy to see that by using above relation it is always
possible to reduce relevant $N$-point scalar integral to a linear
combination of $N-1$-point scalar integrals. For details see
\cite{Duplancic:2003tv}.

From the considerations above, we can conclude that vanishing of the Gram
determinant is not related to the singularities of Feynman
integrals. It is important to point out that the situation is not so
simple in the case of diagrams with more than one loop. There, Gram
determinants are related to so-called {\it second-type
singularities}.

\subsection{Practice and problems}

In practice we deal with 4-dimensional Minkowski space. An
immediate consequence of this is that, for all integrals with $N>5$,
$\mbox{Det}[S_N]$ vanishes due to the linear dependence of the
vectors $r_i$ and all integrals with $N>5$ can be reduced to the
integrals with $N\leq 5$. In view of what has been said above, all
one-loop integrals are expressible in terms of the integrals
$I_0^k(4+2\eir;\{ 1\})$ with nonvanishing $\mbox{Det}[S_k]$ and
$\mbox{Det}[R_k]$, where $k=1,\ldots,5$. In fact, for practical
calculations, also the 5-point basic scalar integral is reducible.
That is because we are interested in calculations up to
$\cal{O}(\eir)$. Details can be found in the literature
\cite{Fleischer:1999hq,Bern:1993kr,Binoth:1999sp}.

For most practical calculations the starting Feynman integrals
obtained from Feynman diagrams by using Feynman rules are in
$4+2\eir$ dimensions and with $\nu_i=1$. In the next step, tensor
decomposition, Eq. (\ref{f15}), will produce scalar integrals with
higher dimensions and powers of propagators. By successively using
all recursion relations following from Eq. (\ref{inv}), except the one
coming from the first row, in the cases of nonvanishing Gram
determinants and recursion relations following from Eqs.
(\ref{gg1}) and (\ref{gr0}) in the cases of vanishing Gram
determinants, it is possible to express an arbitrary Feynman integral
as a linear combination of integrals $I_0^k(2 n+2\eir;\{ 1\})$
with nonvanishing $\mbox{Det}[S_k]$ and $\mbox{Det}[R_k]$, where
$k=1,\ldots,5$. The possible values for parameter $n$ depend on
kinematics involved. If the kinematics is such that during reduction
no case appears where the constant $C$ in Eqs. (\ref{gg1}) and
(\ref{gr0}) vanishes, the parameter $n$ is an integer greater than 1.
Now, the recursion from the first row of Eq. (\ref{inv}) can be
successively used to lower all dimensions down to $4+2\eir$. Since
in the above procedure that relation was never used to increase
dimension, from what has been said in the previous section, it
follows that all singularities are in basic scalar integrals and
divergences appearing in coefficients should cancel in the sum. The
cases with vanishing $C$ appear regularly when dealing with
diagrams containing collinear external lines, i.e. for exceptional
kinematics.

Assuming the situation described in the previous paragraph, many Gram
determinants to different powers will appear in denominators of
the coefficients when an arbitrary Feynman integral is decomposed into
the basic integrals. The real problem in practice is when one has
to calculate in a kinematical region where some of those
determinants are small. Since vanishing of the Gram determinant does
not correspond to a singularity, one faces cancellation of big
numbers and consequently numerical instabilities. In principle, if
one is using methods where all Feynman integrals are expressed as linear
combinations of basic integrals, this problem is unavoidable no
matter in which framework coefficients are calculated. That is
because the decomposition into the basic integrals is unique. However,
there are some hints from experience as to where one should look
to soften this problem. The main guideline is to try to avoid
separate calculation of diagrams contributing to the process under
consideration. Namely, powers of determinants in denominators tend
to be smaller if a group of diagrams (for example, a gauge
invariant group) is calculated together. Additionally, one has to
use all symmetries of the basic integrals to reduce the basic set as
much as possible. Of course, at the end, to get more precision, it
is always necessary to make an expansion around a point where the Gram
determinant vanishes. However, if calculating in the neighborhood
of the point where both Gram and Cayley determinants vanish
simultaneously, the expansion is problematic because the decomposition
is not analytic at that point. One can hope that such regions will
not give sizable contribution to calculated physical quantities.

\subsection{Conclusion}
Vanishing of various Gram and modified Cayley determinants will
always produce numerical instabilities if reduction methods are
used to perform the calculation. The instabilities can be softened by
using various clever approaches but the question remains, will
that work for all practical cases? One can also doubt if reduction
to basic integrals is the optimal approach to perform calculations
which, due to their complexity, become more and more numerically
oriented. Maybe some kind of direct numerical integration of the
Feynman integrals is more efficient. Surely this is a more natural
approach for numerical calculations.

\subsection*{Acknowledgements}
Author is thankful to P. Mastrolia for useful discussions and
organizers of the workshop for support.

%
}

\part[CROSS SECTIONS]{CROSS SECTIONS}

\section[Tuned comparison of QCD corrections to 
\boldmath{$pp\to WW{+}jet{+}X$} at the LHC]
{TUNED COMPARISON OF QCD CORRECTIONS TO 
\boldmath{$pp\to WW{+}jet{+}X$} AT THE LHC%
\protect\footnote{Contributed by: T.~Binoth, J.~Campbell, S.~Dittmaier,
R.K.~Ellis, J.-P.~Guillet, S.~Kallweit, S.~Karg, N.~Kauer, G.~Sanguinetti,
P.~Uwer, G.~Zanderighi}}
{
\def\slsh{\rlap{$\;\!\!\not$}}     

\def\as{\alpha_s}
\def\eps{\epsilon}

\hfuzz 0.5pt

%
%
%
%
%

\subsection{Introduction}

The complicated hadron collider environment of the LHC requires not only 
sufficiently precise predictions for the expected signals, but also
reliable rates for complicated background reactions, especially
for those that cannot be entirely measured from data.
Among such background processes, several involve
three, four, or even more particles in the final state, rendering
the necessary next-to-leading-order (NLO) calculations in QCD 
technically challenging. At the previous Les Houches workshop
this problem lead to the creation of a list of calculations that are 
a priority for LHC analyses, the so called 
''experimenters' wishlist for NLO calculations''
\cite{Buttar:2006zd,Campbell:2006wx}.
The process $pp\to W^+W^-{+}{jet}{+}X$ made it to the top of this list.

The process of WW+jet production
is an important source for background to the
production of a Higgs boson that subsequently decays into a W-boson
pair, where additional jet activity might arise from the
production~\cite{Mellado:2007fb}.
WW+jet production
delivers also potential background to new-physics searches, such as
the search for supersymmetric particles, because of leptons and missing transverse
momentum from the W~decays. Last, but not least, the process is
interesting in its own right, since W-pair production processes
enable a direct precise analysis of the non-abelian
gauge-boson self-interactions, and a large fraction of W~pairs
will show up with additional jet activity at the LHC.

First results on the calculation of NLO QCD corrections to
WW+jet production have been presented by two groups in
Refs.~\cite{Dittmaier:2007th,Campbell:2007ev}.
A third calculation is in progress \cite{ppwwj_binoth}.
In the following the key features of these three independent
calculations are described and results of an ongoing tuned comparison
are presented.

\subsection{Descriptions of the various calculations}

At leading order (LO), hadronic WW+jet production receives contributions
from the partonic processes $q\bar q\to W^+W^- g$, $qg\to W^+W^- q$,
and $\bar qg\to W^+W^- \bar q$, where $q$ stands for up- or down-type
quarks. All three channels are related by crossing symmetry.

The virtual corrections modify the partonic processes that are
already present at LO. At NLO these corrections
are induced by self-energy, vertex,
box (4-point), and pentagon (5-point) corrections, the latter being
the most complicated loop diagrams.
Apart from an efficient handling of the huge amount of algebra,
the most subtle point certainly is the numerically stable
evaluation of the numerous tensor loop integrals, in particular
in the vicinity of exceptional phase-space points.
The three calculations described below employ completely different
loop methods. Some of them are already briefly reviewed in
Ref.~\cite{Buttar:2006zd}, where
more details on problems in multi-leg loop calculations
and brief descriptions of proposed solutions can be found.

The real corrections are induced by the large variety of processes
that result from crossing any pair of QCD partons in
$0 \to W^+W^-   q \bar q g g$ and
$0 \to W^+W^-   q \bar q q' \bar q'$ into the initial state.
Here the main complication in the evaluation is connected to an
efficient phase-space integration with a proper separation
of soft and collinear singularities. For the separation of
singularities the three calculations
all employ the subtraction method\cite{Ellis:1980wv}
using the dipole subtraction formalism of Catani and Seymour\cite{Catani:1996vz}.

\subsection*{The calculation of DKU \cite{Dittmaier:2007th}}

This calculation is actually based on two completely independent evaluations of the
virtual and real corrections. The W~bosons are taken to be on~shell,
but the results on cross sections presented in Ref.~\cite{Dittmaier:2007th}
do not depend on the details of the W~decays.

Both evaluations of loop diagrams start with an amplitude generation
by {\sl Feyn\-Arts}, using the two independent versions~1.0
\cite{Kublbeck:1990xc} and 3.2 \cite{Hahn:2000kx}.
One of the calculations essentially follows the same strategy already
applied to the related processes of $\mathrm{t\bar tH}$
\cite{Beenakker:2002nc} and $\mathrm{t\bar t{+}jet}$ \cite{Dittmaier:2007wz}
production. Here the amplitudes are
further processed with in-house {\sl Mathematica} routines,
which automatically create an output in {\sl Fortran}.
The IR (soft and collinear) singularities are treated in dimensional
regularization and analytically separated
from the finite remainder as described in
Refs.~\cite{Beenakker:2002nc,Dittmaier:2003bc}.
The pentagon tensor integrals are directly reduced to box
integrals following Ref.~\cite{Denner:2002ii}.
Box and lower-point integrals are reduced
\`a la Passarino--Veltman \cite{Passarino:1978jh} to scalar integrals,
which are either calculated analytically or using the results of
Refs.~\cite{'tHooft:1978xw,Beenakker:1988jr,Denner:1991qq}.
The second loop calculation is based on
{\sl FormCalc}~5.2 \cite{Hahn:1998yk}, which
automatically produces {\sl Fortran} code.
The reduction of tensor to scalar integrals is done with the
help of the {\sl LoopTools} library \cite{Hahn:1998yk},
which also employs the method of Ref.~\cite{Denner:2002ii} for the
5-point tensor integrals, Passarino--Veltman \cite{Passarino:1978jh}
reduction for the lower-point tensors, and the {\sl FF} package
\cite{vanOldenborgh:1990wn,vanOldenborgh:1991yc} for the evaluation
of regular scalar integrals.
The dimensionally regularized soft or collinear singular 3- and 4-point
integrals had to be added to this library.

One calculation of the real corrections employs analytical results
for helicity amplitudes obtained in a spinor formalism.
The phase-space integration is performed by a
multi-channel Monte Carlo integrator~\cite{Berends:1994pv}
with weight optimization~\cite{Kleiss:1994qy}
written in {\sl C++}.
The results for cross sections with two resolved hard jets
have been checked against results obtained with
{\sl Whizard}~1.50~\cite{Kilian:2007gr}
and {\sl Sherpa}~1.0.8~\cite{Gleisberg:2003xi}.
Details on this part of the calculation can be found in
Ref.~\cite{SK-diplomathesis}.
The second evaluation of the real corrections
is based on scattering amplitudes calculated
with {\sl Madgraph} \cite{Stelzer:1994ta} generated code.
The code has been modified to allow for a non-diagonal
quark mixing matrix and the extraction of the required colour and
spin structure. The latter enter the evaluation of the dipoles in the
Catani--Seymour subtraction method. The evaluation of the individual dipoles
was performed using a {\sl C++} library developed during the calculation of
the NLO corrections for $\mathrm{t\bar t{+}jet}$ \cite{Dittmaier:2007wz}.
For the phase-space integration a
simple mapping has been used where the phase space is generated from
a sequential splitting.

\subsection*{The calculation of CEZ \cite{Campbell:2007ev}}

The method of choice for calculation of the virtual corrections of
Ref.~\cite{Campbell:2007ev} is similar to the techniques adopted by
the other groups and is based on the semi-numerical method of
Ref.~\cite{Giele:2004iy} augmented with a mechanism to handle
exceptional configurations~\cite{Ellis:2005zh}. This method has
already been used for the NLO calculation of Higgs
plus dijet production via gluon-gluon fusion~\cite{Campbell:2006xx}.
Tree-level matrix elements for real radiation have been checked
against the results of {\sl Madgraph}~\cite{Maltoni:2002qb}. Soft and
collinear singularities are handled using the dipole subtraction
scheme~\cite{Catani:1996vz}.
As for the other authors, CEZ have performed several checks to test the
reliability of their code. These include checks of Ward identities of
the amplitudes containing external gluons.

The calculation of Ref.\cite{Campbell:2007ev} is however
different from the other two in that the decay of the W~bosons is
included from the outset. Rather than summing over the polarizations
of a W~boson of momentum $k$ with
\begin{equation} \label{Wpolsum}
\sum \varepsilon^\mu \varepsilon^\nu  = \Bigg[-g^{\mu \nu}
+ \frac{k^\mu k^\nu}{M_W^2} \Bigg]\; ,
\end{equation}
the authors of this paper project out the combination of polarizations
which occurs in the physical decay of the W~boson, $W^-(k) \to
e^-(l_1)+\bar{\nu}(l_2)$,
\begin{equation} \label{CEZpolsum}
\sum \varepsilon^\mu \varepsilon^\nu  \sim
\frac{1}{2 l_1.l_2} {\mbox{Tr}} [\slsh{l_1} \gamma^\mu \slsh{l_2}
\gamma^\nu  \gamma_L],\;\;\; \gamma_L=(1-\gamma_5)/2.
\end{equation}
The inclusion of the decay is well-motivated from a physical point of
view, because it allows phenomenological analyses which include cuts
on the decay leptons.

For the purposes of the comparison of virtual matrix elements for a
fixed phase-space point, the results including the decays can be used
to extract the result
for the amplitude squared summed over the polarization of the vector
boson, as would be obtained using Eq.~(\ref{Wpolsum}). This is
achieved by performing 6$\times$6=36 evaluations of the amplitude
squared\cite{Dixon:1998py} in which each lepton is emitted along three
orthogonal axes (in both positive and negative directions) in the
corresponding vector-boson center-of-mass frame. The results of this
comparison, with input parameters tuned for the comparison, will be
given below.

\subsection*{The calculation of BGKKS \cite{ppwwj_binoth}}

This calculation is also done in two independent ways. The graph
generation is based on {\sl QGRAF} \cite{Nogueira:1991ex} and was cross checked
by having two independent codes.
All diagrams neglect the quarks of the 3rd generation.

Up to now the LO part and the virtual corrections are evaluated.
By using the spinor helicity formalism, projectors on the different helicity
amplitudes are defined. In this way all Lorentz indices can be saturated such
that the complexity of the one-loop 5-point tensor reduction is such that
at most rank-1 5-point integrals appear. For each helicity amplitude an algebraic
representation in terms of certain basis functions is obtained by using
the reduction methods developed in Refs.~\cite{Binoth:1999sp,Binoth:2005ff}.
The whole algebra is done in an automated way
by using {\sl FORM} \cite{Vermaseren:2000nd} and {\sl MAPLE}. In both approaches
the IR divergent integrals are isolated  by using 6-dimensional IR finite box functions
such that IR poles are
in 3-point functions only. One implementation uses the function set defined in
Appendix~C of Ref.~\cite{Binoth:2005ff}, and uses the implementation of the {\sl Fortran 90} code \texttt{golem90}.
The other computation uses standard scalar 2- and 3-point functions as a basis.
The complete algebraic reduction to d=6 scalar box and d=n scalar 2- and 3-point
functions is largely equivalent to a standard Passarino--Veltman reduction. Only
the 5-point functions are treated differently \cite{Binoth:2005ff}. Tractable
analytical expressions of the coefficients to the two sets of basis functions are
obtained for each independent helicity amplitude.

Discrete symmetries (Bose,C,P) are used to check and relate
helicity amplitudes with each other. The coefficients are exported to a {\sl Fortran} code and used
to evaluate the loop correction of the process.

For the treatment of $\gamma_5$ the 't~Hooft--Veltman scheme is applied.
The $\gamma$-algebra and the loop momenta are split into $4$- and $(D-4)$-dimensional
parts. Whereas the $\gamma_5$ anti-commutes with the $D=4$ matrices, it commutes
with the gamma matrices defined in $d=D-4$. As is well known the QCD corrections 
of an axial vector current
are different from the vector part and a finite renormalisation has to be performed.
The following gauge boson vertex which includes a finite counterterm for the axial part
(see e.g. Refs.~\cite{Larin:1993tq,Trueman:1995ca,Harris:2002md}) is used,
\begin{eqnarray}
V^\mu_{Vq\bar{q}} \sim g_v \, \gamma^\mu + Z_5 \, g_a \, \gamma^\mu \gamma_5 &\mbox{with}&
 Z_5 = 1 - C_F \,\frac{\alpha_s}{\pi},
\end{eqnarray}
to reinforce the correct chiral structure of the amplitudes. Note that
the 't~Hooft--Veltman scheme treats the observed particles in 4 dimensions but the
soft/collinear gluons in $D$ dimensions. This guarantees that for the IR subtractions
the same Catani--Seymour dipole terms as for conventional dimensional regularisation
can be used \cite{Catani:1996pk}.

\subsection{Tuned comparison of results}

The following results essentially employ the setup of
Ref.~\cite{Dittmaier:2007th}.
The CTEQ6~\cite{Pumplin:2002vw,Stump:2003yu}
set of parton distribution functions (PDFs) is used throughout, i.e.\
CTEQ6L1 PDFs with a 1-loop running $\alpha_{\mathrm{s}}$ are taken in
LO and CTEQ6M PDFs with a 2-loop running $\alpha_{\mathrm{s}}$ in NLO.
Bottom quarks in the initial or final
states are not included, because the bottom PDF is suppressed
w.r.t.\ to the others.
Quark mixing between the first two generations is introduced via
a Cabibbo angle $\theta_{\mathrm{C}}=0.227$.
In the strong coupling constant
the number of active flavours is $N_{F}=5$, and the
respective QCD parameters are $\Lambda_5^{\mathrm{LO}}=165$~MeV
and $\Lambda_5^{\overline{\mathrm{MS}}}=226$~MeV,
leading to
$\alpha_{\mathrm{s}}^{\mathrm{LO}}(M_W)=0.13241687663294$ and
$\alpha_{\mathrm{s}}^{\mathrm{NLO}}(M_W)=0.12026290039064$.
The top-quark loop in the gluon self-energy is
subtracted at zero momentum. The running of
$\alpha_{\mathrm{s}}$ is, thus, generated solely by the contributions of the
light quark and gluon loops. 
In all results shown in the following, the
renormalization and factorization scales are set to $M_W$.
The top-quark mass is
$m_t=174.3$~GeV, the masses of all other quarks are neglected.
The weak boson masses
are $M_W=80.425$~GeV, $M_Z=91.1876$~GeV, and $M_H=150$~GeV.
The weak mixing angle is set to its on-shell value, i.e.\
fixed by $c_w^2=1-s_w^2=M_W^2/M_Z^2$, and the electromagnetic
coupling constant $\alpha$ is derived from Fermi's constant
$G_\mu=1.16637\cdot10^{-5}$~GeV$^{-2}$ according to
$\alpha=\sqrt{2}G_\mu\/M_W^2s_w^2/\pi$.

We apply the jet algorithm of Ref.~\cite{Ellis:1993tq}
with $R=1$ for the definition of the tagged hard jet and
restrict the transverse momentum of the hardest jet by
$p_{T,jet}>100$~GeV. 

\subsection{Results for a single phase-space point}

For the comparison the following set of four-momenta is chosen,
\begin{eqnarray}
\label{eq:ppwwj_momenta}
p_1^\mu &=& (7000,0,0,7000), \qquad p_2^\mu = (7000,0,0,-7000),
\\
p_3^\mu &=& (6921.316234371218,3840.577592920205,0,5757.439881432096),
\nonumber\\
p_4^\mu &=& (772.3825553565997,-67.12960601170266,-279.4421082776151,-712.3990141151700),
\nonumber\\
p_5^\mu &=& (6306.301210272182,-3773.447986908503,279.4421082776151,-5045.040867316925),
\nonumber
\end{eqnarray}
where the momentum assignment is for $a(p_1)b(p_2)\to W^+(p_3)W^-(p_4)c(p_3)$.

Table~\ref{tab:ppwwj_single_lo} shows some results for the
(spin- and colour-summed) squared LO matrix elements, as obtained
with {\sl Madgraph} \cite{Stelzer:1994ta}. The results of all three groups
agree with these numbers within about 13 digits.
\begin{table}
\centerline{
\begin{tabular}{cc}
&  $|{\cal M}_{\mathrm{LO}}|^2/e^4/g_{\mathrm{s}}^2[\mathrm{GeV}^{-2}]$
\\ \hline
$u\bar u\to W^+W^- g$      & $0.9963809154477200\cdot10^{-3}$
\\ \hline
$d\bar d\to W^+W^- g$      & $0.3676289952184384\cdot10^{-5}$
\\ \hline
$ug\to W^+W^- u$           & $0.1544340549124799\cdot10^{-3}$
\\ \hline
$dg\to W^+W^- d$           & $0.1537758419168101\cdot10^{-5}$
\\ \hline
$g\bar u\to W^+W^- \bar u$ & $0.7491333451663728\cdot10^{-4}$
\\ \hline
$g\bar d\to W^+W^- \bar d$ & $0.2776156068243590\cdot10^{-4}$
\end{tabular} }
\caption{Results for squared LO matrix elements at the phase-space point
(\ref{eq:ppwwj_momenta}).}
\label{tab:ppwwj_single_lo}
\end{table}

Because of the different treatment of the number of active flavours in
the calculations of DKU and CEZ and in order to be independent of the
subtraction scheme to cancel IR divergences,
we found it useful to compare virtual results prior
to any subtraction. The ${\cal O}(\alpha_s)$ contribution to the
virtual, renormalized squared amplitude is given by the interference
between tree-level and one-loop virtual amplitude, which we denote
schematically as
\begin{equation}
\label{eq:virtonlydef}
2 \mathrm{Re}\{M_{V}^*\cdot M_{\mathrm{LO}}\} = e^4 g_s^2 
f(\mu_{\mathrm{ren}})
\left(c_{-2} \frac{1}{\eps^2} +c_{-1} \frac{1}{\eps} + c_0
\right),
\end{equation}
with%
\footnote{Note that this factor differs from the overall factor
$c_\Gamma$ extracted when quoting results for one phase-space point in
the CEZ paper.} 
$f(\mu_{\mathrm{ren}}) = \Gamma(1+\eps) (4\pi\mu_{\mathrm{ren}}^2/M_W^2)^\eps$
and the number of space--time dimensions $D=4-2\eps$.
In the following we split the coefficients of
the double and single pole and for the constant part, $c_{-2}, c_{-1}$, and
$c_0$, into bosonic contributions (``bos'') without closed fermion loops
and the remaining fermionic parts. The fermionic corrections are further
split into contributions from the first two generations (``ferm1+2'')
and from the third generation.

Table~\ref{tab:virtbosonly} shows the results for the bosonic parts of
the coefficients $c_{-2}$, $c_{-1}$, and $c_0$ ($c_{-2}$ does not
receive fermion-loop corrections). 
\begin{table}
\centerline{
\begin{tabular}{cccc}
& $c_{-2}[\mathrm{GeV}^{-2}]$ & $c^{\mathrm{bos}}_{-1}[\mathrm{GeV}^{-2}]$ & $c^{\mathrm{bos}}_{0}[\mathrm{GeV}^{-2}]$
\\ \hline
\multicolumn{4}{l}{$u\bar u\to W^+W^- g$}
\\
DKU & $ -1.080699305508758\cdot 10^{-4} $ & $ 7.842861905263072\cdot 10^{-4} $ &
 $ -3.382910915425372\cdot 10^{-3} $
 \\
CEZ & $-1.080699305505865\cdot 10^{-4} $ &  $ 7.842861905276719\cdot 10^{-4} $ &
 $-3.382910915464027\cdot 10^{-3} $
 \\
BGKKS & $ -1.080699305508814\cdot 10^{-4} $ & $ 7.842861905263293\cdot 10^{-4} $ 
& $-3.382910915616242\cdot10^{-3}$
\\ \hline
\multicolumn{4}{l}{$d\bar d\to W^+W^- g$}
\\
DKU & $-3.987394716797222\cdot 10^{-7}$  &  $2.893736116870099\cdot 10^{-6}$ & $
-1.252531649334637\cdot 10^{-5}$
\\
CEZ & $-3.987394716665197\cdot 10^{-7} $ &  $ 2.893736115389983\cdot 10^{-6} $ &
 $-1.252531614999332\cdot 10^{-5} $
\\
BGKKS & $-3.987394716798342\cdot 10^{-7}$  &  $2.893736117550454\cdot 10^{-6}$ & 
$-1.252531647620369\cdot 10^{-5}$
\\ \hline
\multicolumn{4}{l}{$ug\to W^+W^- u$}
\\
DKU & $ -1.675029833503229\cdot 10^{-5}$ & $ 1.236268430131559\cdot 10^{-4}$ & $
 -5.417120947927877\cdot 10^{-4}$
\\
CEZ & $-1.675029833501256\cdot 10^{-5} $ &  $1.236268430124113\cdot 10^{-4 } $ &  $-5.417120948004078\cdot 10^{-4} $
\\
BGKKS & $ -1.675029833503285\cdot 10^{-5}$ & $ 1.236268430131930\cdot 10^{-4}$ & $ -5.417120948184518\cdot 10^{-4}$
\\ \hline
\multicolumn{4}{l}{$dg\to W^+W^- d$}
\\
DKU & $ -1.667890693078443\cdot 10^{-7}$ &$ 1.231000679615805\cdot 10^{-6}$ & $-5.402644808236175\cdot 10^{-6}$
\\
CEZ & $-1.667890693268847\cdot 10^{-7} $ & $1.230999331981130\cdot 10^{-6 } $ &  $-5.402644353170802\cdot 10^{-6} $
\\
BGKKS & $ -1.667890693077475\cdot 10^{-7}$ &$ 1.230999333576065\cdot 10^{-6}$ & $-5.402644211736123\cdot 10^{-6}$
\\ \hline
\multicolumn{4}{l}{$g\bar u\to W^+W^- \bar u$}
\\
DKU & $ -8.125284951799448\cdot 10^{-6}$ &  $7.047108864062224\cdot 10^{-5}$&$-3.525581727244482\cdot 10^{-4}$
\\
CEZ & $-8.125284951286924\cdot 10^{-6} $ & $ 7.047108863931619\cdot 10^{-5} $ & $-3.525581728065669\cdot 10^{-4} $
\\
BGKKS & $ -8.125284951799859\cdot 10^{-6}$ &  $7.047108864102780\cdot 10^{-5}$&$-3.525581727287365\cdot 10^{-4}$
\\ \hline
\multicolumn{4}{l}{$g\bar d\to W^+W^- \bar d$}
\\
DKU & $-3.011087314520321\cdot 10^{-6} $ &$2.611534269956032\cdot 10^{-5} $&$ -1.326197552139531\cdot 10^{-4} $
\\
CEZ & $-3.011087314528406\cdot 10^{-6} $ &   $ 2.611534269870494\cdot 10^{-5} $ &    $-1.326197549152728\cdot 10^{-4} $
\\
BGKKS & $-3.011087314520429\cdot 10^{-6} $ &$2.611534269951226\cdot 10^{-5} $&$ -1.326197552106838\cdot 10^{-4} $
\end{tabular} }
\caption{Results for the bosonic virtual corrections at the phase-space point 
(\ref{eq:ppwwj_momenta}) with $c_{-2}$, $c_{-1}$ and $c_0$ are 
defined in Eq.~(\ref{eq:virtonlydef}).} 
\label{tab:virtbosonly}
\end{table}
\begin{table}
\centerline{
\begin{tabular}{ccc}
& $c^{\mathrm{ferm1+2}}_{-1}[\mathrm{GeV}^{-2}]$ & $c^{\mathrm{ferm1+2}}_{0}[\mathrm{GeV}^{-2}]$
\\ \hline
\multicolumn{3}{l}{$u\bar u\to W^+W^- g$}
\\
DKU & $2.542821895320379\cdot10^{-5}$ & $4.372323372044527\cdot10^{-7}$
\\
CEZ & $2.542821895311753\cdot10^{-5}$ & $4.372790378087550\cdot10^{-7}$
\\
BGKKS & $2.542821895314862 \cdot10^{-5}$ & $4.372324288356448 \cdot10^{-7}$
\\ \hline
\multicolumn{3}{l}{$d\bar d\to W^+W^- g$}
\\
DKU & $9.382105211529244\cdot10^{-8}$ & $2.383985481697933\cdot10^{-8}$
\\
CEZ & $9.382105220158816\cdot10^{-8}$ & $2.381655056763332\cdot10^{-8}$
\\
BGKKS & $9.382105215996126\cdot10^{-8}$ & $2.383986138730693\cdot10^{-8}$
\\ \hline
\multicolumn{3}{l}{$ug\to W^+W^- u$}
\\
DKU & $3.941246664484964\cdot10^{-6}$ & $2.261655163318730\cdot10^{-7}$
\\
CEZ & $3.941246667066658\cdot10^{-6}$ & $2.261900862449825\cdot10^{-7}$
\\
BGKKS & $3.941246667066566 \cdot10^{-6}$ & $2.261651778836927\cdot10^{-7}$
\\ \hline
\multicolumn{3}{l}{$dg\to W^+W^- d$}
\\
DKU & $3.924449049876280\cdot10^{-8}$ & $-3.340508442179341\cdot10^{-8}$
\\ 
CEZ & $3.924448807787651\cdot10^{-8}$ & $-3.341842650545260\cdot10^{-8}$
\\ 
BGKKS & $3.924448689594072\cdot10^{-8}$ & $-3.340505335889721\cdot10^{-8}$
\\ \hline
\multicolumn{3}{l}{$g\bar u\to W^+W^- \bar u$}
\\
DKU & $1.911831753319591\cdot10^{-6}$ & $-3.332688444715011\cdot10^{-7}$
\\
CEZ & $1.911831753400357\cdot10^{-6}$ & $-3.332770821153847\cdot10^{-7}$
\\
BGKKS & $1.911831753364673 \cdot10^{-6}$ & $-3.332688443882355\cdot10^{-7}$
\\ \hline
\multicolumn{3}{l}{$g\bar d\to W^+W^- \bar d$}
\\
DKU & $7.084911328500216\cdot10^{-7}$ & $-3.420298601940541\cdot10^{-7}$
\\
CEZ & $7.084911328417316\cdot10^{-7}$ & $-3.419939732016338\cdot10^{-7}$
\\
BGKKS & $7.084911328283340 \cdot10^{-7}$ & $-3.420298578631734\cdot10^{-7}$
\end{tabular} }
\caption{Results for the fermionic contributions of the first two
quark generations to $c_{-1}$ and $c_0$ at the phase-space point
(\ref{eq:ppwwj_momenta}).}
\label{tab:virtfermonly}
\end{table}
\begin{table}
\centerline{
\begin{tabular}{cccc}
& $c_{-2}[\mathrm{GeV}^{-2}]$ & $c_{-1}[\mathrm{GeV}^{-2}]$ & $c_{0}[\mathrm{GeV}^{-2}]$
\\ \hline
\multicolumn{4}{l}{$u\bar u\to W^+W^- g$}
\\
DKU & $-1.080699305508778\cdot10^{-4}$ & $8.160714642177893\cdot10^{-4}$ 
& $-3.382201173786996\cdot10^{-3}$
\\ \hline
\multicolumn{4}{l}{$d\bar d\to W^+W^- g$}
\\
DKU & $-3.987394716797186\cdot10^{-7}$ & $3.011012432041691\cdot10^{-6}$ 
& $-1.248828433702770\cdot10^{-5}$
\\ \hline
\multicolumn{4}{l}{$ug\to W^+W^- u$}
\\
DKU & $-1.675029833503229\cdot10^{-5}$ & $1.285534013444099\cdot10^{-4}$ 
& $-5.413834847221341\cdot10^{-4}$
\\ \hline
\multicolumn{4}{l}{$dg\to W^+W^- d$}
\\
DKU & $-1.667890693078551\cdot10^{-7}$ & $1.280056291844283\cdot10^{-6}$
& $-5.452219162448072\cdot10^{-6}$
\\ \hline
\multicolumn{4}{l}{$g\bar u\to W^+W^- \bar u$}
\\
DKU & $-8.125284951799523\cdot10^{-6}$ & $7.286087833227389\cdot10^{-5}$
& $-3.528788476602400\cdot10^{-4}$
\\ \hline
\multicolumn{4}{l}{$g\bar d\to W^+W^- \bar d$}
\\
DKU & $-3.011087314520238\cdot10^{-6}$ & $2.700095661561590\cdot10^{-5}$
& $-1.331943241722592\cdot10^{-4}$
\end{tabular} }
\caption{Results for the full bosonic+fermionic contributions 
to $c_{-2}$, $c_{-1}$ and $c_0$ at the phase-space point
(\ref{eq:ppwwj_momenta}).}
\label{tab:virtonly}
\end{table}
The results on $c_0$ obtained by the different groups typically agree 
within $7{-}11$ digits; the ones on $c_{-2}$ and $c_{-1}$ agree much
better, because they are much easier to calculate.
The results for the fermionic contributions of the first two generations 
are given in Table~\ref{tab:virtfermonly}. 
Compared to the bosonic corrections these contributions are suppressed
by three orders of magnitude. Counting this suppression factor, which results
from cancellations, as significant digits,
the finite parts agree within $6{-}9$ digits. The agreement is somewhat better
in the coefficients of the single pole, which entirely stems from
the counterterm of the fermion-loop part of the gluon self-energy.
The remaining contributions from closed loops of the third quark generation
are not compared yet. For future reference we show the full corrections 
including all bosonic and fermionic contributions in 
Table~\ref{tab:virtonly}.

\subsection{Results for integrated cross sections}

A tuned comparison of integrated cross sections is still in progress.
Table~\ref{tab:ppwwj_cs} illustrates the agreement in the LO cross
section obtained by the different groups
and provides the DKU result in NLO for future comparisons.
The subcontribution $\sigma_{\mathrm{virt+I}}$ corresponds to the
IR-finite sum of the virtual corrections and the contribution
of the $I$ operator that is extracted from the real corrections with
the dipole subtraction formalism \cite{Catani:1996vz}.
\begin{table}
\centerline{
\begin{tabular}{rlll}
$pp\to W^+W^-{+}jet{+}X$
& $\sigma_{\mathrm{LO}}$[fb]
& $\sigma_{\mathrm{NLO}}$[fb]
& $\sigma_{\mathrm{virt+I}}$[fb]
\\ \hline
  DKU & $10371.7(12)$~~ & $14677.6(98)$ & $-881.5(42)$
\\
 CEZ & $10372.26(97)$
\\
 BGKKS & $10371.7(11)$~~
\end{tabular} }
\caption{Results for contributions to the integrated pp cross sections
at the LHC in LO and NLO.}
\label{tab:ppwwj_cs}
\end{table}

\subsection{Conclusions}

We have reported on an ongoing tuned comparison of NLO QCD calculations
to WW+jet production at the LHC. For a fixed phase-space point,
the virtual corrections obtained by three different groups 
using different calculational techniques agree within 6--9 digits.
The comparison of full NLO cross sections, which involve the non-trivial
integration of the virtual corrections over the phase space, is still
in progress.

The agreement found so far gives us confidence in
the conclusions drawn for physical quantities,
which were reported in 
Refs.~\cite{Dittmaier:2007th,Campbell:2007ev}.

\subsection*{Acknowledgements}
P.U.\ is supported as Heisenberg Fellow of the Deutsche
Forschungsgemeinschaft DFG.
This work is supported in part by the European
Community's Marie-Curie Research Training Network HEPTOOLS under
contract MRTN-CT-2006-035505 and
by the DFG Sonderforschungsbereich/Transregio 9
``Computergest\"utzte Theoretische Teilchenphysik'' SFB/TR9.
T.B.\ is supported by the British Science and Technology 
Facilities Council (STFC),
the Scottish Universities Physics Alliance (SUPA). 
T.B., S.~Karg and N.K.\ are supported
by the Deutsche Forschungsgemeinschaft (DFG, BI-1050/1).
N.K.\ is supported by the Bundesministerium f\"{u}r
Bildung und Forschung (BMBF) under contract 05HT1WWA2.
G.Z.\ is supported by the British Science and Technology Facilities Council 
(STFC).

%
}


\section[From the high energy limit of massive QCD amplitudes to the full mass
dependence]
{FROM THE HIGH ENERGY LIMIT OF MASSIVE QCD AMPLITUDES TO THE FULL MASS
DEPENDENCE%
\protect\footnote{Contributed by: M.~Czakon}}
{\graphicspath{{czakon/}}
%
%
%
%
%

\subsection{Introduction}

It is clear that the physics program of the LHC poses new challenges to the
theory. In fact, the description of hadronic collisions involves several
quantities, both non-perturbative and perturbative, the determination of which
is a highly non-trivial task. As far as the perturbative part is concerned, we
are  still a long way of having the partonic cross sections predicted at a
suitable level of accuracy. Whereas most processes will have to be known to
next-to-leading order, there are some for which the experimental precision
grants a study going one order higher in the strong coupling
constant. Particularly interesting here is the top quark pair production cross
section. With statistics going into millions of events, a systematics
dominated error of under 10\% is expected already in the first phase of the
LHC. Despite years of efforts, the appropriate complete NNLO prediction is not
yet available. The bottleneck, as in most such cases is the evaluation of the
two-loop virtual corrections.

Recently, the high energy limit of the amplitudes in the quark annihilation
and gluon fusion channels has been derived \cite{Czakon:2007ej,Czakon:2007wk} by a mixture of direct
evaluation of Feynman graphs and an approach based on factorization properties
of QCD (see A. Mitov's and S. Moch's contribution). The knowledge gained can
already be used for the description of high $p_T$ events and as a test of a
future complete prediction. Clearly due to the behavior of the particle
fluxes, what is needed is a calculation covering the whole range of variation
of the  kinematical parameters. It is interesting that one can actually use
the high energy limit to deal with this problem. Unfortunately, it is not
enough to have the whole amplitude, but it is rather necessary to know all of
the master integrals. In the following, I describe the steps that lead to the
complete result.

\subsection{The high energy limit}

By the high energy limit, I understand the limit where all the invariants are
much larger than the mass. A direct approach to the evaluation of the
amplitude under this assumption has been devised in \cite{Czakon:2004wm,Czakon:2006pa}. As a first step,
one uses the Laporta algorithm to reduce all of the integrals occurring to a
small set of masters. In the case at hand, the number of integrals is 145 and
422 for the quark annihilation and gluon fusion channels respectively.

Subsequently, Mellin-Barnes representations are constructed for all the
integrals \cite{Smirnov:1999gc,Tausk:1999vh}. This can be done by an automatic package, here by one written
by the Author and G. Chachamis\footnote{There is a public package available
\cite{Gluza:2007rt} that constructs representations for planar graphs. In the present
case, also non-planar graphs occur, however.}. After analytic continuation in
the dimension of space-time performed with the MB package \cite{Czakon:2005rk}, the integrals
have the following general form
\begin{equation}
I = (m^2)^{n-2\epsilon}\int_{-i \infty}^{i \infty} d z \;
\left(-\frac{m^2}{s}\right)^z f\left(\frac{t}{s},z\right),
\end{equation}
where
the $f$ function contains, amongst others, a product of $\Gamma$, or
possibly $\psi$ functions, which have poles in $z$. The desired expansion is
obtained by closing the contour and taking residues. As a result, one obtains
integrals which have lower dimension and a simpler structure. These still
require evaluation. Due to the fact, that there is  a relation between the
massive and the massless cases, the result must have a similar structure. In
particular, it has to be given by harmonic polylogarithms, and therefore it
should be possible to resum the integrals by further closing contours and
evaluating the resulting series. This can again be achieved automatically with
the help of the XSummer package \cite{Moch:2005uc}.

What remains at the end are integrals, which are pure numbers, but do not have
a structure suggesting a solution in terms of harmonic series. The same
argument as before shows, however, that this must be the case. Instead of
working out specific methods for particular integrals, it turned out to be
possible to evaluate them to very high precision and subsequently use the PSLQ
algorithm to reconstruct the solution in terms of Riemann zeta values.

It has to be noted, that the procedure sketched above works for the majority
of cases, but some remain at the end. For these, it is usually necessary to
change the basis of integrals, in order to obtain expressions of suitable
structure and/or size for evaluation. At present this program has been
completed for the quark annihilation channel, and thus all the color
structures given in \cite{Czakon:2007ej} have been computed directly with agreement with
the factorization approach. The gluon fusion channel is still under way.

\subsection{Power corrections}

As explained in the introduction, the high energy limit by itself is not
enough for practical applications. To go one step further, it is possible to
compute power corrections in the mass. These will then cover most of the
range, apart from the threshold region and the small angle region, where the
series is not convergent any more.

The main idea is as follows. The derivative of any Feynman integral with
respect to any kinematical variable is again a Feynman integral with possibly
higher powers of denominators or numerators. These can, however, be
reduced to the same master integrals. This means that one can construct a
partially triangular system of differential equations in the mass \cite{Kotikov:1990kg,Remiddi:1997ny},
which can subsequently be solved in the form of a power series.

\begin{figure}
\begin{center}
\includegraphics[width=7cm]{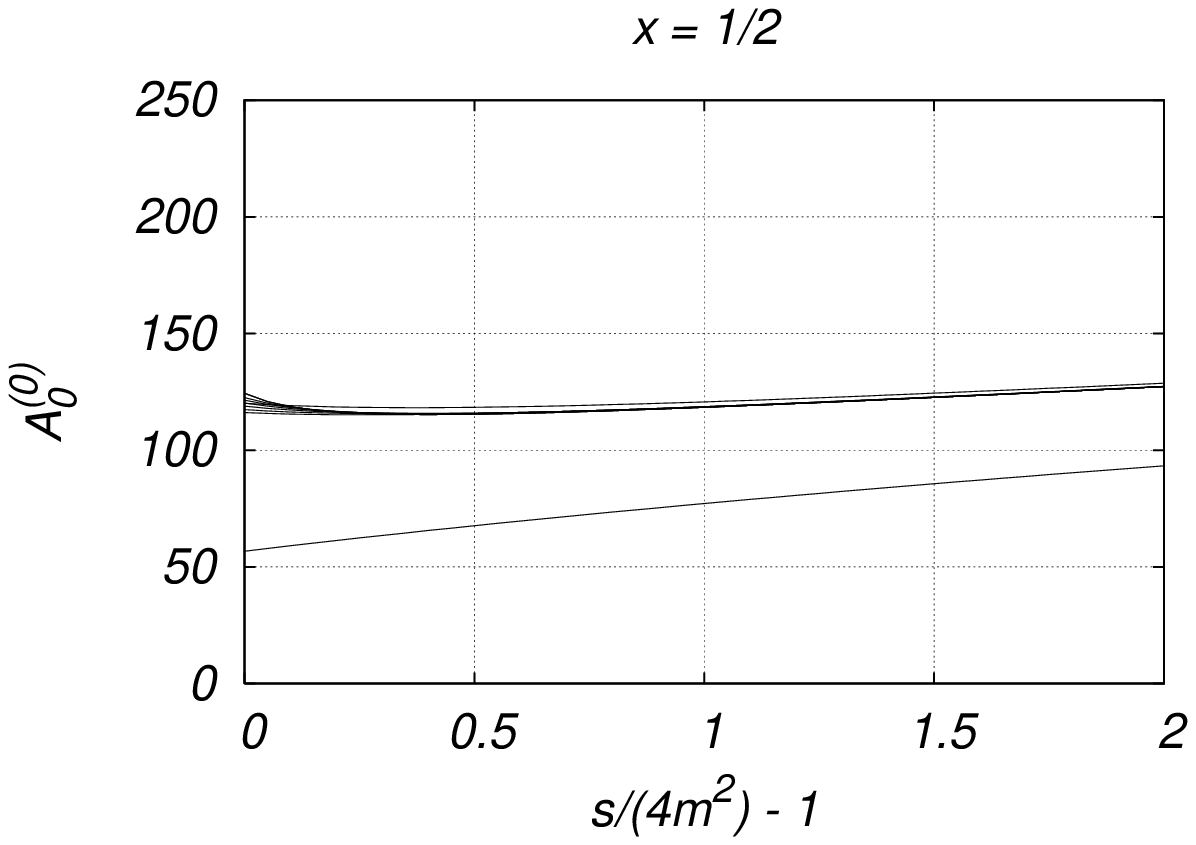}
\includegraphics[width=7cm]{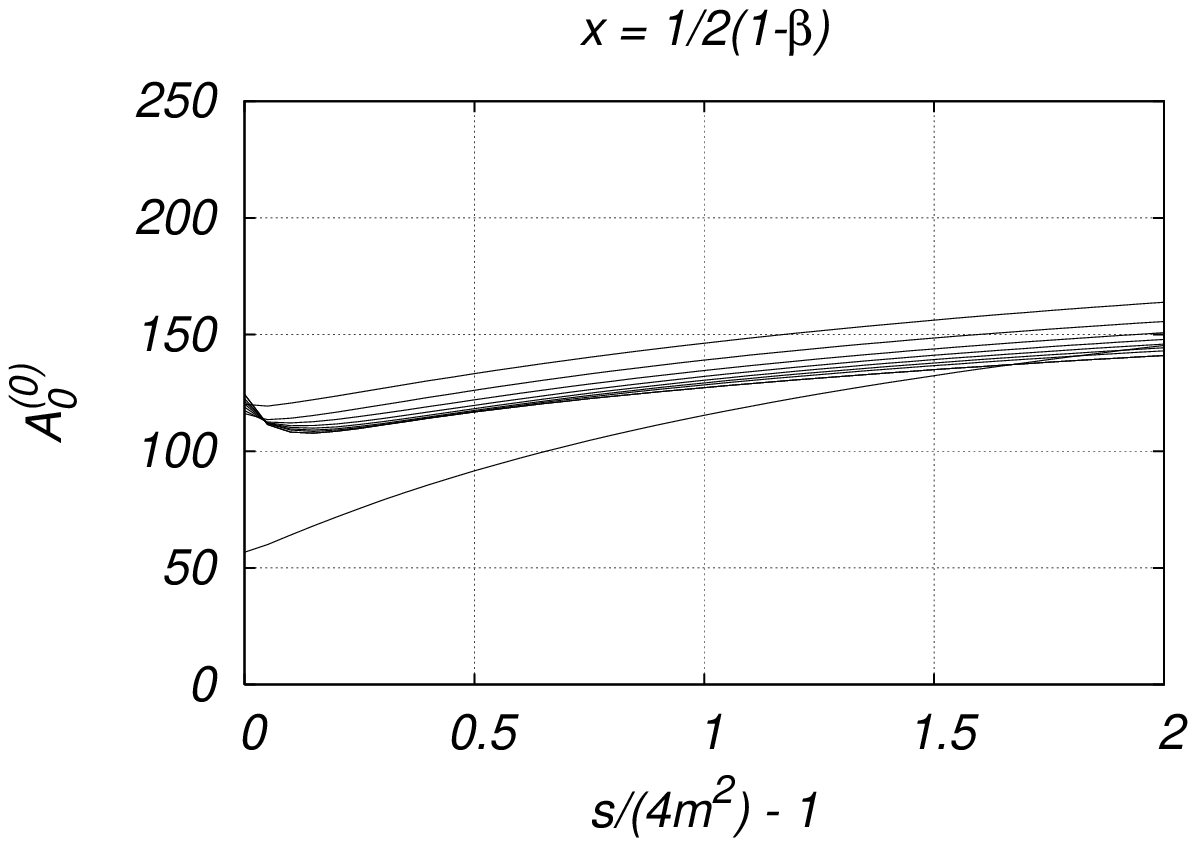}
\caption{Bare leading color amplitude for top quark pair production in the
  quark annihilation channel expanded in the mass. The more divergent terms at
  threshold (left of the plots) correspond to higher orders of expansion. The
  left panel corresponds to 90 degree scattering, whereas the right to forward
  scattering. The variables are defined in the text.}
\label{czakon_expansion}
\end{center}
\end{figure}

In Fig.~\ref{czakon_expansion}, I show the result of expansion for the
leading color term. The kinematic variable $x$ is
\begin{equation}
  x = -\frac{t}{s}, \;\; t = (p_3-p_1)^2-m_t^2,
\end{equation}
and its variation within the range $ [ 1/2(1-\beta), 1/2(1+\beta) ] $, where
$\beta=\sqrt{1-4m_t^2/s}$ is the velocity, corresponds to angular variation
between the forward and backward scattering.

The series appears to be asymptotic at the boundaries. Unfortunately, the
behavior is worse for the subleading color terms, as a consequence of the
Coulomb singularity among others.

\subsection{Numerical evaluation}

Using the same system of differential equations one can obtain a full
numerical solution to the problem. The only requirement is to have the
boundary conditions to suitable accuracy. These are provided by the series
expansions of the previous section. It is crucial to perform the numerical
integration along a contour in the complex plane, since there are spurious
singularities along the real axis. Here, I chose an ellipse, because of the
improved control on the integration error that one gets from the software used
(ODEPACK).

Fig.~\ref{czakon_full} shows the solution in the range, where the
expansion of the previous section starts to diverge. The achievable precision,
if double precision arithmetic is used, is about 10 digits for most points,
with evaluation times of the order of a second. This is going to be
substantially slower, when subleading color terms will be added. However, the
method is fast and precise enough to be sufficient for practical
applications. In particular it is possible to construct grids of solutions, which
will be subsequently interpolated when implemented as part of a Monte Carlo
program.

\begin{figure}
\begin{center}
\includegraphics[width=12cm]{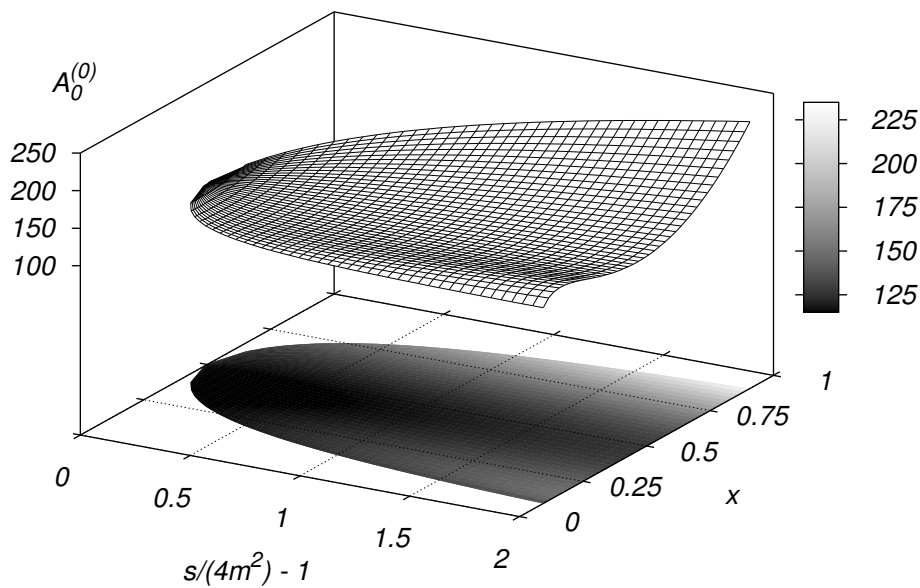}
\caption{Full mass dependence of the bare leading color amplitude in the quark
annihilation channel.}
\label{czakon_full}
\end{center}
\end{figure}

It is clear that the method is suitable for problems, which have a relatively
small number of scales, and seems to be perfect for $2 \rightarrow 2$ QCD
processes at the two-loop level. The main drawback is the size of the
expressions, and the difficulties connected to the derivation of the boundary
conditions.

\subsection{Conclusions}

I have described an approach for the evaluation of massive QCD amplitudes
starting from the high energy limit and its application to the NNLO
corrections to the top quark pair production cross section. Needless to say,
the same procedure can be applied to other problems of interest. At present
the Author, together with G. Chachamis and D. Eiras, is working on the
corrections to gauge boson pair production.

\subsection*{Acknowledgements}

This work was supported by the Sofja Kovalevskaja Award of the Alexander von
Humboldt Foundation sponsored by the German Federal Ministry of Education and
Research.

%
}


\section[Much can be said about massive amplitudes just from knowing their massless limit]
{MUCH CAN BE SAID ABOUT MASSIVE AMPLITUDES JUST FROM KNOWING THEIR MASSLESS LIMIT%
\protect\footnote{Contributed by: A.~Mitov, S.~Moch}}
{
%
%
%
%
%
%

\subsection{The high-energy limit}

For the precise evaluation of collider observables the knowledge of
the pure virtual correction to the corresponding Born process is
required. This is true at any order (NLO, NNLO, etc). In presence of
heavy flavors, and especially at higher orders, the problem of their
evaluation becomes acute.

There are very important applications awaiting such results. An
example of central importance is top production at LHC which is
one of the few eagerly awaited {\it precision} observables at this
collider. One of the peculiar features of top production at LHC,
and in contrast to the situation at the Tevatron, is that no
specific kinematical region dominates the cross-section. This is due
to the shape of the luminosity function for LHC kinematics.

Direct calculation of the amplitudes is certainly a very demanding
task and it seems that one can hope that numerical results in some,
hopefully easy to handle form, will become available soon (see the
contribution by M.~Czakon for progress in this direction). Here we consider
an alternative approach which explores the special properties of the
gauge theory amplitudes in the high-energy limit and easily provides
(partial) results for the heavy flavor amplitudes even at higher
perturbative orders.

In the following we start by introducing the concept of {\it
high-energy limit} with the help of simple and physically motivated
arguments.
By high-energy limit one means a kinematical situation where the corresponding 
invariants are much larger than the masses of the heavy particles of
interest. In the following we will consider the case of a single
massive fermion with mass $m$ in presence of a typical large
kinematical invariant $Q$. Specific examples are detailed in
section~\ref{sec:mmitovmochapplications}. 
If the quantity of interest
(like total or differential cross-section, amplitude, etc.) is
regular in the limit $m/Q\to 0$ then the high-energy limit is quite
trivial: it is an $m$-independent function of the kinematical
invariants which coincides with the one evaluated in the massless
limit. Therefore it can be computed by setting the mass $m$ to zero
from the very beginning.

Such a situation is, however, relatively rare. In most quantities of
interest, like the differential ones, the limit $m/Q\to 0$ is
singular. The obvious manifestation of that singularity in the results 
is the presence of terms of the type $\sim\ln^n(m/Q)$. When such
contributions appear (and in fact this is the typical
situation) the high-energy limit is defined as the full result with
all power corrections in the mass neglected, i.e. it contains all
logarithms (not multiplied by powers of the mass) as well as the
so-called ``constant" or mass-independent terms. Clearly, in such
cases the high-energy limit is different from the massless limit.

Before we detail the relation between these two limits, we would
first like to clarify the origin and meaning of the logarithmic
terms mentioned above. These terms are known as (quasi-) collinear
logs since they originate from emissions of collinear radiation. To
be precise, the role of the mass is to regulate small angle
emissions that would otherwise diverge in a massless theory; see
Ref.~\cite{Catani:2000ef} for a detailed exposition. In this regard, a
parton's mass gets dual significance, since one can take one of the
following two viewpoints:

$\bullet$ small or large, the mass is nevertheless non-zero,
therefore the result is always (collinearly) finite;

$\bullet$ the small mass is just a formal regulator for collinear
singularities much like dimensional regularization in the purely
massless case.

In this write-up we take the unifying viewpoint that both approaches
are useful and do not have to be considered as alternatives to each
other. One can think of the mass as a regulator which is helpful in
deriving certain properties of the theory but it can also be thought
of as an approximation to the full massive result which is surprisingly
good in many physical applications.

The prominent role these logarithmic terms play in physical
applications has been acknowledged long ago, and has been formalized
in the so-called Perturbative Fragmentation Function approach
\cite{Mele:1990cw} now known through two-loops
\cite{Melnikov:2004bm,Mitov:2004du}; for a recent review see
\cite{Mitov:2005vk}. The idea behind this formalism is the fact that
up to power corrections in the mass, a differential with respect to
some kinematical parameter $z$ cross-section for the production of a
massive parton $h$, can be written as:
\begin{equation}
{d\sigma_h\over dz}(z,Q,m) = \sum_a {d\hat\sigma_a\over dz}(z,Q)
\otimes D_{a\to h}(z,m) + {\cal O}(m)\; . \label{PFF}
\end{equation}

The function $D_{a\to h}(z,m)$ does not depend on the hard scale $Q$
and is thus a process independent object that can be computed to any
fixed order. It has the important property that it contains all the
mass dependence within the approximation indicated in
Eq.~(\ref{PFF}). On the other side the partonic cross-section
$d\hat\sigma_a$ for the production of any parton $a$ is intrinsically
massless, i.e. it is obtained from a calculation where $m=0$ is set
from the very beginning. Of course, collinear singularities are
still present in a massless calculation but they are regulated
dimensionally, i.e. they appear as poles in $\epsilon$, where
$d=4-2\epsilon$:
\begin{equation}
{d\sigma_a\over dz}(z,Q,\epsilon) = \sum_b {d\hat\sigma_b\over
dz}(z,Q) \otimes \Gamma_{ba}(z,\epsilon)\; .
\label{collinear-factorization}
\end{equation}
The explicit expression for the collinear counterterm $\Gamma$
contains arbitrariness; the only condition on it is that it contains
all poles in $\epsilon$. It has become a standard practice in 
recent years to work in the $\overline{\mathrm{MS}}$ scheme where
$\Gamma$ contains only poles. The choice of a subtraction scheme is of
course also implicit in the definition of the function $D$ in
Eq.~(\ref{PFF}).

From Eqs.~(\ref{PFF}), (\ref{collinear-factorization}) it is quite
clear that one can obtain a massive cross-section in the small-mass
(or high-energy) limit by performing a purely massless calculation.
The calculational simplifications following from this can be
enormous, especially at higher orders. The usefulness of such an 
approach has been appreciated in the past in many applications
related to heavy quark production (typically $b$ and $c$) at special
kinematics like large $P_T$ hadroproduction and $e^+e^-$
annihilation at the $Z$-pole (see~\cite{Kluth:2006bw} for a review). In such
kinematical configurations the neglected power corrections can be as
low as a few percent effect and are often totally negligible.

A second virtue of Eq.~(\ref{PFF}), and one that cannot be matched in
conventional perturbation theory, is that it allows resummation of
large collinear logs $\ln(Q/m)$ to all perturbative orders. This
feature is due to the fact that the function $D$ satisfies the DGLAP
evolution equation, or in other words one achieves exponentiation of
the (remnants of) soft and collinear singularities.

As we will demonstrate in the next section, all these features of
massive cross-sections in the small-mass limit can be translated to
massive amplitudes in gauge theories where similar properties can be uncovered.
Moreover, one can exploit these properties in much the same way;
this is illustrated by the physical applications we consider in
section~\ref{sec:mmitovmochapplications}.

\subsection{Factorization in massive amplitudes}

As was indicated above, in the following we will be concerned with
the factorization properties of massive QCD amplitudes in the
high-energy limit. Since one of our main objectives is to relate the
small-mass limit of an amplitude with its massless limit, we start
our discussion with a brief review of the well-known factorization
properties of massless amplitudes
~\cite{Catani:1998bh,Sterman:2002qn}.

The scattering amplitude ${\cal M}_{\rm p}$
\begin{eqnarray}
\label{eq:QCDamplitude} | {\cal M}_{\rm p} \rangle &\equiv& {\cal
M}_{\rm p}\left(\{ k_i \},\{ m_i \},{\{c_i\}},{Q^2 \over
\mu^2},\alpha_{\rm s}(\mu^2),\epsilon \right) \, ,
\end{eqnarray}
for a general $2 \to n$ scattering processes of on-shell partons
$p_i$
\begin{equation}
\label{eq:QCDscattering} {\rm p}: \qquad p_1 + p_2
\:\:\rightarrow\:\: p_3 + \dots + p_{n+2} \, .
\end{equation}
with a set of fixed external momenta $\{ k_i \}$, masses $\{ m_i \}$
and color quantum numbers $\{ c_i \}$, can be written in the
massless case $m_i=0$ as a product of three functions ${\cal J}_{\rm
p}^{(m=0)}$, ${\cal S}_{\rm p}^{(m=0)}$ and ${\cal H}^{\rm[p]}$,
\begin{eqnarray}
\label{eq:QCDfacamplitude-zero} | {\cal M}_{\rm p}\rangle^{(m=0)} \,
=  {\cal J}_{\rm p}^{(m=0)}\left({Q^2 \over \mu^2},\alpha_{\rm
s}(\mu^2),\epsilon \right) {\cal S}_{\rm p}^{(m=0)}\left(\{ k_i
\},{Q^2 \over \mu^2},\alpha_{\rm s}(\mu^2),\epsilon \right) | {\cal
H}_{\rm p} \rangle \, ,
\end{eqnarray}

The decomposition Eq.~(\ref{eq:QCDfacamplitude-zero}) can be
understood with simple physical arguments. The jet function ${\cal
J}_{\rm p}^{(m=0)}$ contains all collinearly sensitive
contributions, is color-diagonal and depends only on the external
partons. On the other side the soft function ${\cal S}_{\rm
p}^{(m=0)}$ contains all soft radiation interferences and is
therefore process specific. Finally, the short-distance dynamics of
the hard scattering is described by the (infrared finite) hard
function ${\cal H}_{\rm p}$. To leading order this function is just
proportional to the Born amplitude. More details about the above
expressions can be found in the review \cite{Moch:2007pj}.

As was explained in \cite{Sterman:2002qn}, the decomposition
Eq.~(\ref{eq:QCDfacamplitude-zero}) contains arbitrariness related
to subleading soft as well as finite contributions, which can be
removed by fixing a prescription. A convenient and natural choice is
to identify the jet function with the massless form factor for the
flavor corresponding to any particular leg, i.e.:
\begin{equation}
  \label{eq:jetfactor}
  {\cal J}_{\rm p}^{(m=0)}
  \, = \,
  \prod_{i\in\ \{{\rm all}\ {\rm legs}\}}\,
  {\cal J}_{[i]}^{(m=0)}
  \, = \,
  \prod_{i\in\ \{{\rm all}\ {\rm legs}\}}\,
  \left(
    {\cal F}_{[i]}^{(m=0)}
  \right)^{1 \over 2}
  \, ,
\end{equation}
where $i=q,g$ for quarks and gluons. ${\cal J}_{[i]}^{(m=0)}$ is the
individual jet function of each external parton. The needed massless
jet factors are known through three-loops and the soft functions
through two-loops for any $2\to n$ scattering process
~\cite{MertAybat:2006mz}.

We are now ready to consider the massive case. Based on our
discussion in the previous sections in the small-mass limit one
should expect a decomposition of massive amplitudes similar to the one
in Eq.~(\ref{eq:QCDfacamplitude-zero}). Let us be more specific: we
know that in the massive case collinear logs do appear but we also
know that they should be absorbed in a corresponding jet function.
On the other side, up to power corrections, the soft and hard
functions in the massive case should be the same as in the massless
case since, by construction, they are not sensitive to collinear
emissions. With the exception of contributions related to heavy
quark loops (to be discussed below) in the presence of a hard scale
$Q$ we write for the massive amplitudes~(\ref{eq:QCDscattering}):
\begin{eqnarray}
\label{eq:QCDfacamplitude} | {\cal M}_{\rm p}\rangle^{(m)} \, =
{\cal J}_{\rm p}^{(m)}\left({Q^2 \over \mu^2},\{ m_i \},\alpha_{\rm
s}(\mu^2),\epsilon \right) {\cal S}_{\rm p}^{(m=0)}\left(\{ k_i
\},{Q^2 \over \mu^2},\alpha_{\rm s}(\mu^2),\epsilon \right) | {\cal
H}_{\rm p} \rangle + {\cal O}(m)\, .
\end{eqnarray}

It is very easy to find out what the jet function in the massive
case should be. Working in the prescription chosen for the massless
case, one can apply the amplitude decomposition to the form factor
itself; the latter has no nontrivial soft or hard functions.
Therefore, in the massive case the jet function must be nothing but
the massive form factor evaluated in the small-mass limit.

Combining Eqs.~(\ref{eq:QCDfacamplitude-zero}), 
(\ref{eq:QCDfacamplitude}) one gets the following very suggestive
relation \cite{Mitov:2006xs}:
\begin{eqnarray}
\label{eq:Mm-M0} {\cal M}_{\rm p}^{(m)} &=& \prod_{i\in\ \{{\rm
all}\ {\rm legs}\}}\,
  \left(
    Z^{(m\vert0)}_{[i]}
  \right)^{1 \over 2}\,
  \times\
{\cal M}_{\rm p}^{(m=0)} + {\cal O}(m,\dots) \, ,
\end{eqnarray}
where,
\begin{equation}
\label{eq:Z} Z^{(m\vert0)}_{[i]}\left({m^2 \over \mu^2},\alpha_{\rm
s},\epsilon \right) \, = \, {\cal F}_{[i]}^{(m)}\left({Q^2\over
\mu^2},{m^2\over\mu^2},\alpha_{\rm s},\epsilon \right) \left({\cal
F}_{[i]}^{(m=0)}\left({Q^2\over \mu^2},\alpha_{\rm s},\epsilon
\right)\right)^{-1} + \dots\, ,
\end{equation}
is a universal, process independent factor. It is sensitive to the
definition of the mass $m$ as well as the coupling constant (see
\cite{Mitov:2006xs} for details). The process-independence in
Eq.~(\ref{eq:Z}) is manifest because $Z^{(m\vert0)}_{[i]}$ is only a
function of the process-independent ratio of scales $\mu^2/m^2$. The
process-dependent scale $Q$ cancels completely between the massive
and the massless form factors.

The last statement, however, requires one important clarification.
From the explicit results for the massive and massless form factors
one can easily see that starting from two loops the ratio indicated
above contains also $Q$-dependent logarithmic terms originating from
diagrams with the heavy parton in loops. It is these terms that we have indicated
with dots in Eq.~(\ref{eq:Z}). Luckily, these terms are easy to
recognize and to separate since in the color decomposition of the
amplitudes they are proportional to the number of heavy flavors
${n^{}_{\! h}}$. For that reason in the definition of $Z$-factor
given originally in Ref.~\cite{Mitov:2006xs} contributions
proportional to the number of heavy flavors have been excluded, as
indicated by the dots in Eq.~(\ref{eq:Mm-M0}). A first step in the
understanding of the loop contributions and their incorporation into
the factorization approach was made in Ref.~\cite{Becher:2007cu} in
the context of Bhabha scattering. We discuss this process as well as
other applications in the next section.

Comparing the results of this section with the ones in the previous
section, we can clearly see the similarities offered by QCD
factorization between small-mass limits of amplitudes and
cross-sections. In both cases the small-mass results are
proportional to the corresponding massless results. The
proportionality factors are process independent and universal. The
proportionality is in the sense of usual multiplication for
amplitudes and convolution for cross-sections, as usual. Moreover, 
it was explained in Ref.~\cite{Mitov:2006xs} the so-called
$Z$-factor in Eq.~(\ref{eq:Z}) seems in fact to equal the pure
virtual contributions to the perturbative fragmentation function $D$
in Eq.~(\ref{PFF}).

\subsection{Applications}\label{sec:mmitovmochapplications}

The results in the previous section have been cross-checked at the
amplitude level with the general-mass predictions for the structure
of the $\epsilon$-poles and $\ln(m)$ terms of any one-loop amplitude
\cite{Catani:2000ef}. Complete agreement was found. We have also
checked that for the process $q\bar{q}\to h\bar{h}$ the prediction
based on Eq.~(\ref{eq:Mm-M0}) completely agrees with the results from
the one-loop calculation of Ref.~\cite{Korner:2002hy}. We want to
stress that we have compared not only the singular terms but also
all terms that are finite in the limits $\epsilon \to 0$ and $m\to
0$. The agreement applies to all color structures of the amplitude
as well as for both its real and imaginary parts.

In subsequent work~\cite{Czakon:2007wk,Czakon:2007ej} a prediction for
the small-mass limit of all two-loop heavy quark production squared
amplitudes at hadron colliders has been made, while the terms
proportional to ${n^{}_{\! h}}$ were obtained from a direct
calculation. We will not go into details here (they can be found for
example in the recent review \cite{Moch:2007pj}) but will only
summarize the main features of the result: several of the color
structures were calculated both directly as well as predicted and we
observed full agreement between the two approaches. Therefore, this is a
first two-loop check for the factorization approach and represents a
direct confirmation of its validity.

Another obvious application where the small-mass limit plays
important role is Bhabha scattering. The knowledge of the two-loop
QED massive amplitudes in the small-mass limit there is needed for
achieving the intended precision of the luminosity measurement; see for example
\cite{Balossini:2006sd}. Complete results for the photonic
corrections to large-angle Bhabha scattering were first obtained by
Penin \cite{Penin:2005kf} and later confirmed in
Ref.~\cite{Becher:2007cu} in the approach discussed in the previous
section. Therefore, this is yet another example of its usefulness and
power.

\subsection{Conclusions}

We have presented a newly developed relation between massive and
massless QCD amplitudes. We have emphasized its relevance for
physical applications and its ability to seamlessly produce results
for processes that cannot be calculated currently by direct means.

The relation was introduced based on the idea for massless limit of
a massive amplitude and was given in parallel to the much better
known relation between massive and massless differential
cross-sections.

The new relation between massive and massless amplitudes represents
the proper generalization of the naive textbook replacement relation
$1/\epsilon \to \ln(m) + \dots$ to all perturbative orders and for
any process. Moreover, with the obvious identification of the color
factors, the relation is applicable to any $SU(N)$ gauge theory, QCD
being a prominent example. QCD and QED applications like heavy quark
production at hadron colliders at two loops and two-loop corrections
to Bhabha scattering were briefly discussed.

%
}


\section[NNLO predictions for hadronic event shapes in 
\boldmath{$e^+e^-$} annihilations]
{NNLO PREDICTIONS FOR HADRONIC EVENT SHAPES IN \boldmath{$e^+e^-$} 
ANNIHILATIONS%
\protect\footnote{Contributed by: G.~Dissertori, A.~Gehrmann--De Ridder,
T.~Gehrmann, E.W.N.~Glover, G.~Heinrich, H.~Stenzel}}
{\graphicspath{{gehrmann/}}


\newcommand\new{\newcommand}         
\newcommand\ren{\renewcommand}       

\def\ph#1{{\rm{#1}}}

\def\beq{\begin{equation}}   
\def\eeq{\end{equation}}
\def\bea{\begin{eqnarray}}  
\def\eea{\end{eqnarray}} 
\def\nn{\nonumber}
\def\eps{\epsilon}

\def\MA#1#2{{\cal M}^{#1}_{A,#2}}
\def\MB#1#2{{\cal M}^{#1}_{B,#2}}
\def\MC#1#2{{\cal M}^{#1}_{C,#2}}
\def\O{y}
\def\A{{\cal A}}
\def\B{{\cal B}}
\def\L{{\cal L}}
\def\CA{C_A}
\def\CF{C_F}
\def\TF{T_F}
\def\NF{N_F}
\def\Flavour{F}
\def\Poles{{\cal P}oles}
\def\Finite{{\cal F}inite}
\def\Re{\mbox{Re}}
\def\bom#1{{\mbox{\boldmath $#1$}}}
\def\MSbar{$\overline{{\rm MS}}$}
\def\ket#1{|{#1}\rangle}
\def\bra#1{\langle{#1}|}
\def\braket#1#2{\langle #1 |#2 \rangle}
\def\cm{{\cal M}}
\def\cmb{\overline{{\cal M}}}
\def\bt{B_T}
\def\bw{B_W}

\def\JET{J}
\def\sgn{{\rm sgn}}
\def\sab{s_{12}}
\def\sac{s_{13}}
\def\sbc{s_{23}}
\def\yab{y_{12}}
\def\yac{y_{13}}
\def\ybc{y_{23}}
\def\sij{s_{ij}}\def\sabc{s_{123}}
\def\Sab{s_{12}}
\def\Sac{s_{13}}
\def\Sad{s_{14}}
\def\Sbc{s_{23}}
\def\Sbd{s_{24}}
\def\Scd{s_{34}}
\def\Sae{s_{15}}
\def\Sbe{s_{25}}
\def\Sce{s_{35}}
\def\Sde{s_{45}}
\def\Sabc{s_{123}}
\def\Sabd{s_{124}}
\def\Sacd{s_{134}}
\def\Sbcd{s_{234}}

\def\Sabe{s_{125}}
\def\Sace{s_{135}}
\def\Sade{s_{145}}
\def\Sbce{s_{235}}
\def\Sbde{s_{245}}
\def\Scde{s_{345}}

\def\H{\hbox{H}} \def\Hp{\hbox{H}_+}
\def\Hm{\hbox{H}_-}
\def\e{\epsilon}
\def\d{\hbox{d}}
\def\dd{\partial}
\def\f{\hbox{f}}
\def\FORM{{\tt F}}
\def\g{\hbox{g}}
\def\S{\hbox{S}}
\def\Li{\hbox{Li}}
\def\ln{\hbox{ln}}
\def\hpl{HPL}
\def\tdhpl{2dhpl}
\def\G{\hbox{G}}
\def\Gp{\hbox{G}_+}
\def\Gm{\hbox{G}_-}
\def\ie{i.e.}
\def\iep{i\epsilon}
\def\nn{\nonumber}
\renewcommand{\topfraction}{1.0}
\renewcommand{\bottomfraction}{1.0}
\renewcommand{\textfraction}{0.0}

\new{\emem}{{\ifmmode\mathrm{e}^-\else e$^-$\fi}}
\new{\epem}{{\ifmmode\mathrm{e}^+\else e$^+$\fi}}
\new{\zo}  {{\ifmmode\mathrm{Z}\else Z\fi}}
\new{\epm} {{\ifmmode\mathrm{e^+e^-}\else $\mathrm{e^+e^-}$\fi}}
\new{\qq}  {{\ifmmode\mathrm{q}\else q\fi}}
\new{\qqb} {{\ifmmode\bar{\mathrm{q}}\else $\bar{\mathrm{q}}$\fi}}
\new{\bq}  {{\ifmmode\mathrm{b}\else b\fi}}
\new{\bqb} {{\ifmmode\bar{\mathrm{b}}\else $\bar{\mathrm{b}}$\fi}}
\new{\qqbar}{\qq\qqb}

\new{\LEP}        {\mbox{\small\textsc{LEP}}}
\new{\LEPONE}     {\mbox{\small\textsc{LEP1}}}
\new{\LEPTWO}     {\mbox{\small\textsc{LEP2}}}
\new{\CERN}       {\mbox{\small\textsc{CERN}}}
\new{\ALEPH}      {\mbox{\small\textsc{ALEPH}}}
\new{\DELPHI}     {\mbox{\small\textsc{DELPHI}}}
\new{\LD}         {\mbox{\small\textsc{L3}}}
\new{\OPAL}       {\mbox{\small\textsc{OPAL}}}


\new{\eV}         {{\ifmmode {\mathrm{ eV}}\else ${\mathrm{ eV}}$\fi}}
\new{\MeV}        {{\ifmmode {\mathrm{ MeV}}\else ${\mathrm{ MeV}}$\fi}}
\new{\MeVc}       {{\ifmmode {\mathrm{ MeV}}/c\else ${\mathrm{ MeV}}/c$\fi}}
\new{\MeVcc}      {{\ifmmode {\mathrm{ MeV}}/c^2\else ${\mathrm{ MeV}}/c^2$\fi}}
\new{\GeV}        {{\ifmmode {\mathrm{ GeV}}\else ${\mathrm{ GeV}}$\fi}}
\new{\GeVc}       {{\ifmmode {\mathrm{ GeV}}/c\else ${\mathrm{GeV}}/c$\fi}}
\new{\GeVcc}      {{\ifmmode {\mathrm{ GeV}}/c^2\else ${\mathrm{GeV}}/c^2$\fi}}
\new{\TeV}        {{\ifmmode {\mathrm{ TeV}}\else ${\mathrm{ TeV}}$\fi}}


\new{\JS}         {\mbox{\small\textsc{JETSET}}}
\new{\HW}         {\mbox{\small\textsc{HERWIG}}}
\new{\AR}         {\mbox{\small\textsc{ARIADNE}}}
\new{\PY}         {\mbox{\small\textsc{PYTHIA}}}
\new{\JSv}        {\mbox{\small\textsc{JETSET\ 7.405}}}
\new{\HWo}        {\mbox{\small\textsc{HERWIG\ 5.8}}}
\new{\HWn}        {\mbox{\small\textsc{HERWIG\ 5.9}}}
\new{\ARv}        {\mbox{\small\textsc{ARIADNE\ 4.05}}}
\new{\PYv}        {\mbox{\small\textsc{PYTHIA\ 5.7}}}

\new{\Mh}         {{\ifmmode M_{\mathrm{ H}}
                    \else $M_{\mathrm{H}}$\fi}}

\new{\Mz}         {{\ifmmode M_{\mathrm{Z}}
                    \else $M_{\mathrm{Z}}$\fi}}
\new{\Mzsq}       {{\ifmmode M^2_{\mathrm{ Z}}
                    \else $M^2_{\mathrm{Z}}$\fi}}

\new{\as}[1]      {{\ifmmode\alpha^{#1}_s
                    \else$\alpha^{#1}_s$\fi}}
\new{\asx}[1]      {{\ifmmode a^{#1}_s
                    \else $a^{#1}_s$\fi}}
\new{\asb}[1]     {{\ifmmode\overline{\alpha}^{#1}_s
                    \else $\overline{\alpha}^{#1}_s$\fi}}
\new{\asmz}       {{\ifmmode\alpha_s(\Mzsq)
                    \else $\alpha_s(\Mzsq)$\fi}}
\new{\lqcd}       {{\ifmmode\Lambda_{\mathrm{ QCD}}
                    \else $\Lambda_{\mathrm{ QCD}}$\fi}}
\new{\lqcdsq}     {{\ifmmode\Lambda^2_{\mathrm{ QCD}}
                    \else $\Lambda^2_{\mathrm{ QCD}}$\fi}}
\new{\llla}       {{\ifmmode\Lambda_{\mathrm{ LLA}}
                    \else $\Lambda_{\mathrm{ LLA}}$\fi}} 
\new{\lmsbar}[1]  {{\ifmmode \Lambda^{(#1)}_{\overline{\mathrm{MS}}}
                    \else $\Lambda^{(#1)}_{\overline{\mathrm{MS}}}$\fi}}
\new{\lmsb}       {{\ifmmode \Lambda_{\overline{\mathrm{MS}}}
                    \else $\Lambda_{\overline{\mathrm{MS}}}$\fi}}
\new{\lmsbsq}     {{\ifmmode \Lambda^{2}_{\overline{\mathrm{MS}}}
                    \else $\Lambda^{2}_{\overline{\mathrm{MS}}}$\fi}}

%
%
%
%
%
%
\subsection{Introduction}

For more than a decade experiments at LEP (CERN) and SLC (SLAC) 
gathered  a wealth of high precision high energy hadronic data
from electron-positron annihilation at a range of centre-of-mass energies~\cite{Buskulic:1996tt,Heister:2003aj,Acton:1993zh,Alexander:1996kh,Ackerstaff:1997kk,Abbiendi:1999sx,Abbiendi:2004qz,Acciarri:1995ia,Acciarri:1997xr,Acciarri:1998gz,Acciarri:2000hm,Achard:2002kv,Achard:2004sv,Abreu:1999rc,Abdallah:2003xz,Abdallah:2004xe,Abe:1994mf}. 
This data provides one of the    
 cleanest
ways of probing our quantitative understanding of QCD. 
This is particularly so because the strong interactions occur only in 
the final state and are not entangled with the parton density functions associated 
with beams of hadrons.
As the understanding of the strong interaction, and the capability of 
making more precise theoretical predictions, develops, 
more and more stringent comparisons of theory and experiment are possible,
leading to improved measurements
of fundamental quantities such as the strong 
coupling constant~\cite{Biebel:2001dm,Kluth:2006bw}.

In addition
to measuring multi-jet production rates, more specific information  about the
topology of the events can be extracted. To this end, many variables  have been
introduced which characterise the hadronic structure of an event. 
With the precision data from LEP and SLC, experimental
distributions for such event shape variables have been extensively  studied and
have been compared with theoretical calculations based on next-to-leading order
(NLO)  parton-level event generator  programs~\cite{Ellis:1980wv,Kunszt:1980vt,Vermaseren:1980qz,Fabricius:1981sx,eventKN,Giele:1991vf,Catani:1996jh}, 
  improved by
resumming kinematically-dominant leading and next-to-leading logarithms
(NLO+NLL)~\cite{Catani:1992ua,Catani:1991kz,Catani:1992jc,Dokshitzer:1998kz,Banfi:2001bz,Dissertori:2003pj}  and by the inclusion of  
non-perturbative models of power-suppressed hadronisation
effects~\cite{Korchemsky:1994is,Dokshitzer:1997ew,Dokshitzer:1995zt,Dokshitzer:1998pt}. 

Comparing the different sources of error in the extraction of $\alpha_s$
from hadronic data,
one finds that the purely experimental error is negligible compared to
the theoretical uncertainty. There are two sources of theoretical
uncertainty: the theoretical description of the parton-to-hadron
transition (hadronisation uncertainty) and the uncertainty stemming from the 
truncation of the perturbative series at a certain order, as estimated by scale
variations (perturbative or scale uncertainty).  Although the precise
size of the hadronisation uncertainty is debatable and perhaps often
underestimated, it is conventional to consider the scale
uncertainty as the dominant source of theoretical error on the precise
determination of  $\alpha_s$ from three-jet observables. This 
scale uncertainty can be lowered only by including perturbative QCD 
corrections beyond NLO. 

We report here on the  computation of NNLO corrections to 
event shape distributions, and discuss the impact of these corrections on 
the extraction of   $\alpha_s$ from LEP data.

\subsection{Event shape variables}
\label{sec:shapes}

In order to characterise hadronic final states in electron-positron
annihilation, a variety of event shape variables have been proposed in 
the literature, for a review see e.g.~\cite{Ellis:1991qj,Dissertori:2003pj}. These variables can be categorised 
into different classes, 
according to the minimal number of final-state particles required for them 
to be non-vanishing: In the following we shall only consider three particle final states which are thus closely related to three-jet final states.

Among those shape variables,
six variables were studied in great detail: the thrust $T$~\cite{Brandt:1964sa,Farhi:1977sg}, the
normalised heavy jet mass $\rho$~\cite{Clavelli:1981yh}, 
the wide and total jet
broadenings $B_W$ and $B_T$~\cite{Rakow:1981qn},  
the $C$-parameter~\cite{Parisi:1978eg,Donoghue:1979vi} and the transition from three-jet to 
two-jet final states in the Durham jet algorithm $Y_3$~\cite{Catani:1991hj,Brown:1990nm,Stirling:1991ds,Bethke:1991wk,Bethke:1998ue}.

The perturbative expansion for the distribution of a 
generic observable $\O$ up to NNLO at \epm\ centre-of-mass energy $\sqrt{s}$, 
for a renormalisation scale $\mu^2$,  is given by
\begin{eqnarray}
\frac{1}{\sigma_{{\rm had}}}\, \frac{\d\sigma}{\d y} (s,\mu^2,y) &=& 
\left(\frac{\as{}(\mu^2)}{2\pi}\right) \frac{\d \bar A}{\d y} +
\left(\frac{\as{}(\mu^2)}{2\pi}\right)^2 \left( 
\frac{\d \bar B}{\d y} + \frac{\d \bar A}{\d y} \beta_0 
\log\frac{\mu^2}{s} \right)
\nonumber \\ &&
+ \left(\frac{\as{}(\mu^2)}{2\pi}\right)^3 
\Bigg(\frac{\d \bar C}{\d y} + 2 \frac{\d \bar B}{\d y}
 \beta_0\log\frac{\mu^2}{s}
\nonumber \\ &&
\hspace{24mm} + \frac{\d \bar A}{\d y} \left( \beta_0^2\,\log^2\frac{\mu^2}{s}
+ \beta_1\, \log\frac{\mu^2}{s}   \right)\Bigg)+ {\cal O}(\as{4})  \;.
\label{eq:NNLOmu} 
\end{eqnarray}
The dimensionless 
perturbative coefficients $\bar A$, $\bar B$ and $\bar C$ depend only 
on the event shape variable $y$. They are computed by a fixed-order 
parton-level calculation, which includes final states with three partons 
at LO, up to four partons at NLO and up to five partons at NNLO. 
LO and NLO corrections to event shapes have been available already for 
a long time~\cite{Ellis:1980wv,Kunszt:1980vt,Vermaseren:1980qz,Fabricius:1981sx,eventKN,Giele:1991vf,Catani:1996jh}. 

 The calculation of the NNLO corrections is carried out using 
a newly developed
parton-level event generator programme {\tt EERAD3} which contains 
the relevant 
matrix elements with up to five external partons~\cite{Garland:2002ak,Garland:2001tf,Moch:2002hm,Glover:1996eh,Bern:1996ka,Campbell:1997tv,Bern:1997sc,Hagiwara:1988pp,Berends:1988yn,Falck:1989uz}. 
Besides explicit infrared divergences from the loop integrals, the 
four-parton and five-parton contributions yield infrared divergent 
contributions if one or two of the final state partons become collinear or 
soft. In order to extract these infrared divergences and combine them with 
the virtual corrections, the antenna subtraction method~\cite{Kosower:1997zr,Kosower:2003bh,Campbell:1998nn} 
was extended to NNLO level~\cite{GehrmannDeRidder:2005cm,GehrmannDeRidder:2004tv,GehrmannDeRidder:2005hi,GehrmannDeRidder:2005aw} and implemented
for $\epm \to 3\,\mathrm{jets}$ and related event-shape variables~\cite{GehrmannDeRidder:2007jk}. The analytical cancellation of all 
infrared divergences serves as a very strong check on the implementation. 
{\tt EERAD3} yields the perturbative  $A$, $B$ and $C$ 
coefficients\footnote{$A$, $B$ and $C$ differ from $\bar A$, $\bar B$ and $\bar C$ in their normalisation to $\sigma_0$ instead of $\sigma_{{\rm had}}$~\cite{GehrmannDeRidder:2007hr}.} as 
histograms for all infrared-safe event-shape variables related to three-particle 
final states at leading order. 
As a cross check, the $A$ and $B$  coefficients have also been obtained from an independent integration~\cite{eventKN,Giele:1991vf,Catani:1996jh}
of the NLO matrix elements~\cite{Ellis:1980wv}, showing excellent agreement. 

For small values of the event shape variable $y$, the fixed-order expansion, 
eq.\ (\ref{eq:NNLOmu}), fails to converge, 
because the fixed-order coefficients are enhanced by powers of $\ln(1/y)$.
In order to obtain reliable predictions
in the region of $y \ll 1$ it is necessary to resum entire sets of logarithmic terms at all orders in \as{}. 
A detailed description of the predictions at next-to-leading-logarithmic approximation (NLLA) can
be found in Ref.\ \cite{Jones:2003yv}.

\subsection{Generic features of the NNLO corrections}

The precise size and shape of the NNLO corrections depend on the observable 
in question. Common to all observables is the divergent behaviour of 
the fixed-order prediction in the two-jet limit, where soft-gluon effects 
at all orders become important, and where resummation is needed. For several 
event shape variables 
 (especially $T$ and $C$) the full kinematical range is not yet covered 
for three partons, but attained only in the multi-jet limit. In this case,
the fixed-order description is also not applicable since it is  limited 
to a fixed multiplicity (five partons at NNLO). Consequently, the 
fixed-order description is expected to be reliable in a restricted 
interval bounded by the two-jet limit on one side and the multi-jet 
limit on the other side. 

In this intermediate region, we observe that 
inclusion of  NNLO corrections (evaluated at the $Z$-boson mass, and 
for a fixed value of the strong coupling constant) typically increases 
the previously available NLO prediction. 
The magnitude of this increase differs considerably between 
different observables\cite{GehrmannDe Ridder:2007bj,GehrmannDeRidder:2007hr}, 
it is substantial for $T$ (18\%), $B_T$ (17\%)  and 
$C$ (15\%), moderate for $\rho$ and $B_W$ (both 10\%) and small for 
$Y_3$ (6\%). For all shape variables, we observe that the renormalisation
scale uncertainty of the NNLO prediction is reduced by a factor of two 
or more compared to the NLO prediction. 
Inclusion of the NNLO corrections also modifies the shape of the event shape 
distributions. We observe that 
the NNLO prediction describes the shape of the measured event shape 
distributions over a wider kinematical range than the NLO prediction, both 
towards the two-jet and the multi-jet limit. To illustrate the 
impact of the NNLO corrections, we compare the fixed-order predictions 
for $Y_3$ to LEP2-data obtained by the ALPEH experiment in 
Figure~\ref{fig:y23}, which illustrates especially the improvement
when approaching  the two-jet region, corresponding to large $-\ln(Y_3)$.   
\begin{figure}[t]
\begin{center}
\includegraphics[angle=-90,width=10cm]{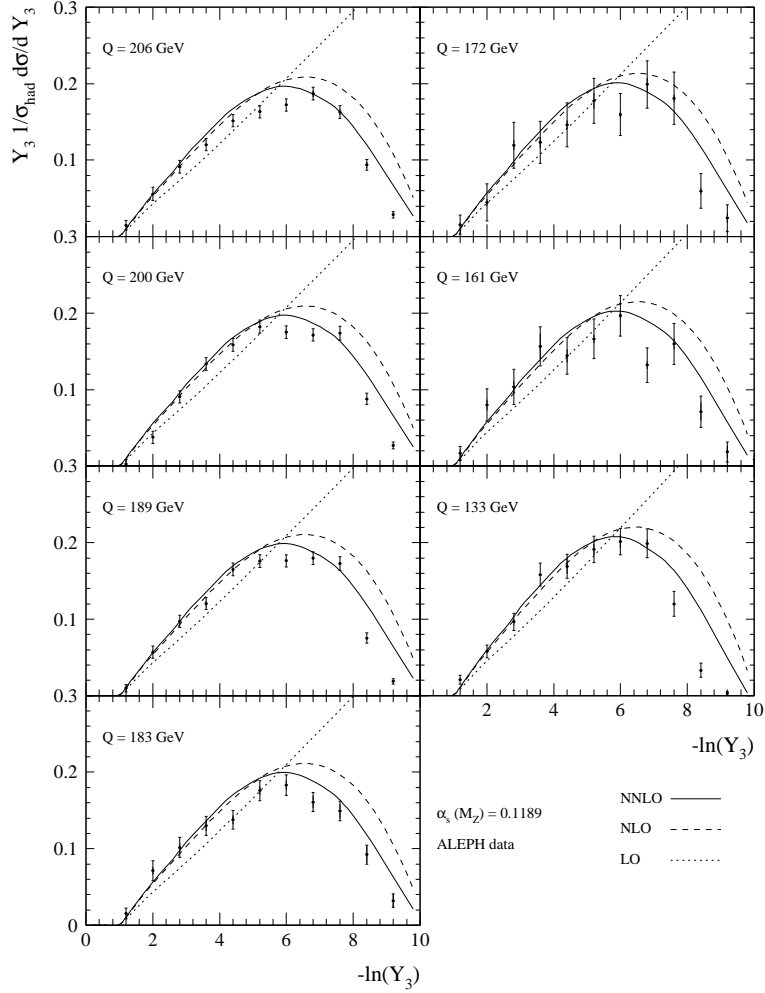}
\end{center}
\caption{\small Perturbative fixed-order predictions 
for the $Y_3$-distribution.} \protect\label{fig:y23}
\end{figure}

\subsection{Determination of the strong coupling constant}
Event shape data from  LEP and LEP2 were used in the past for a
precise determination of the strong coupling constant $\alpha_s$. These
studies were based on the previously available NLO  results, improved by 
NLLA resummation; the resulting error on $\alpha_s$ was completely 
dominated by the renormalisation scale uncertainty inherent to the 
NLO calculation. 
Using the newly computed NNLO corrections to event shape variables, we
performed a new extraction of $\alpha_s$ from data on the standard set of 
six event shape variables, measured 
 by the \ALEPH\ collaboration \cite{Heister:2003aj}
at centre-of-mass energies of 91.2, 133, 161, 172, 183, 189, 200 and 206 \GeV.
The event-shape distri\-butions were obtained using the 
reconstructed momenta and energies of charged and
neutral particles. The measurements have been corrected for detector effects, 
ie., the final distributions correspond to a so-called particle 
(or hadron) level (stable hadrons and leptons after hadronisation).

The coupling constant \as{}\ is
determined from a fit of the perturbative QCD predictions to measured
event-shape distributions. The procedure adopted here 
follows closely the one described in Ref.~\cite{Heister:2003aj}. 
Event-shape distributions are fitted in a central region of
the three-jet production, where a good perturbative description is
available. The fit range is
placed inside the region where hadronisation and detector
corrections are below 25$\%$ and the signal-to-background ratio at LEP2 is
above one.  At the higher LEP2 energies the 
good perturbative description extends further into the two-jet region, 
while in the four-jet region the background becomes large. 
Thus the fit range is selected as a result 
of an iterative procedure balancing theoretical, experimental and 
statistical uncertainties.  
\begin{figure}[t]
\begin{center}
\includegraphics[width=14cm]{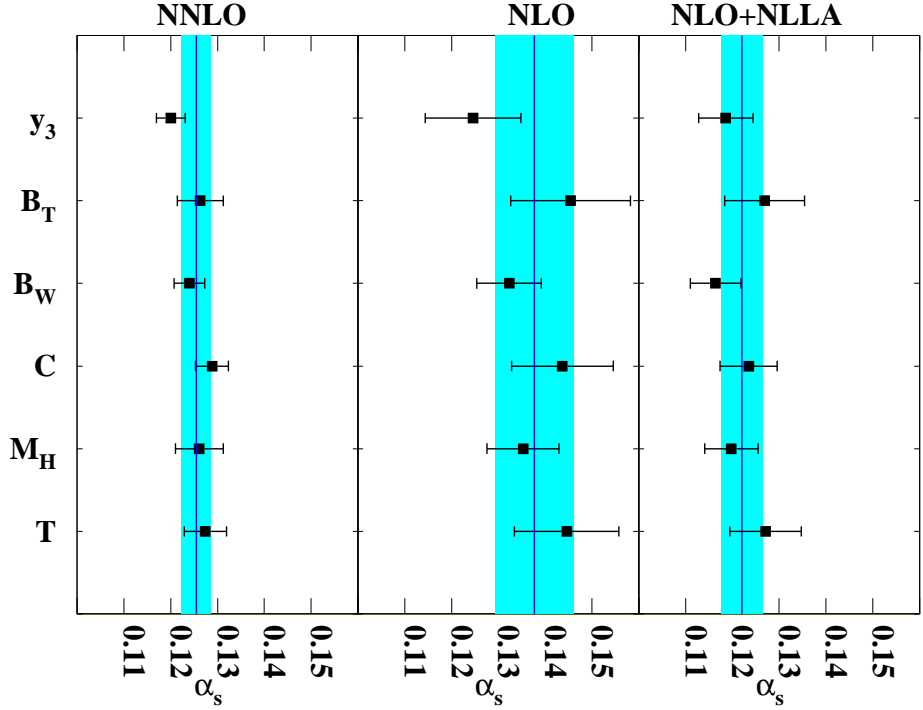}
\end{center}
\caption{\small The measurements of the strong coupling constant $\alpha_s$
for the six event shapes, at $\sqrt{s}=\Mz$, when using QCD predictions at different
approximations in perturbation theory.} \protect\label{fig:scatter}
\end{figure}

Here we concentrate on fits of NNLO 
predictions~\cite{Dissertori:2007xa}
 and compare them to pure NLO and matched NLO+NLLA predictions 
as used in the analysis of Ref.\ \cite{Heister:2003aj}. Results from 
individual event shapes are displayed in Figure~\ref{fig:scatter}.
The combination of 
all NNLO determinations from all shape variables  yields 
\begin{equation}
    \asmz = 0.1240 \;\pm\; 0.0008\,\mathrm{(stat)}
     					 \;\pm\; 0.0010\,\mathrm{(exp)}
                                   \;\pm\; 0.0011\,\mathrm{(had)}
                                   \;\pm\; 0.0029\,\mathrm{(theo)} ,
 \end{equation}
which is indicated by the error band on Figure~\ref{fig:scatter}.
We observe a clear improvement in the fit quality when going to
NNLO accuracy. Compared to NLO the value of \as{} is lowered 
by about 10\%, but still higher than for NLO+NLLA~\cite{Heister:2003aj},
 which 
shows the obvious need for a matching of NNLO+NLLA for a fully reliable 
result.  
 The scatter among the
 $\alpha_s$-values extracted from different shape variables is 
lowered considerably, and the theoretical uncertainty is decreased by 
a factor 2 (1.3) compared to NLO (NLO+NLLA). 

These observations visibly illustrate the improvements gained from 
the inclusion of the NNLO corrections, and highlight the need for 
further studies on the matching of NNLO+NLLA, and on the 
derivation of NNLLA resummation terms.


\subsection*{Acknowledgements}

This research was supported in part by the Swiss National Science Foundation
(SNF) under contract 200020-117602,  
 by the UK Science and Technology Facilities Council and  by the European Commision's Marie-Curie Research Training Network under contract
MRTN-CT-2006-035505 ``Tools and Precision Calculations for Physics Discoveries
at Colliders''.


%
%

}

\part[PARTON SHOWERS]{PARTON SHOWERS}

\section[Developments in leading order parton showers]
{DEVELOPMENTS IN LEADING ORDER PARTON SHOWERS%
\protect\footnote{Contributed by: D.E.~Soper, P.Z.~Skands}}
{

\definecolor{red}{rgb}{1,0,0}
\newcommand{\NOTE}[1]{\textcolor{red}{ \bf[NOTE: #1]}}

\newcommand{\La}{\mathrm{a}}
\newcommand{\Lb}{\mathrm{b}}
\def\ket#1{\big|{#1}\big\rangle}
\def\bra#1{\big\langle{#1}\big|}
\def\brax#1{\big\langle{#1}}   

\def\sket#1{\big|{#1}\big)}
\def\sbra#1{\big({#1}\big|}
\def\sbrax#1{\big({#1}}        

%
%
%

At the Les Houches workshop, there was lively discussion of parton
showers as represented in Monte Carlo event generators. One of the
main current issues in this field is the problem of matrix-element /
parton-shower matching, and the workshop saw several
reports on ways of (re-)formulating parton showers that could make this
problem easier to deal with, a trend one might denote with a fancy
word as ``designer showers''. In this section, we review the part of 
that discussion that relates to how a leading order parton shower can
be organized. Despite the apparent differences, all the new approaches
can be discussed at a common footing if we adopt a little bit of notation
(adapted from \cite{Nagy:2007ty}). 

A typical parton shower algorithm for hadron-hadron collisions works
with states with two initial state partons, $\La $ and $\Lb$, and some
number $m$ of final state partons that we can label with integers
$1,2,\dots,m$. The momenta of these partons can then be specified by
giving $\{p\}_m = \{p_\La,p_\Lb,p_1,\dots,p_m$\}. Each parton also
carries a flavor $f \in \{{\rm g},{\rm u},\bar{\rm u},{\rm d},\bar{\rm
  d},\dots\}$, so that the momenta and flavors can be specified with
$\{p,f\}_m$. Typically, we also keep track of color connections (the
labels of one or two partons to which parton $i$ is connected in the
leading-color limit). We may therefore denote the complete set of $m+2$
partons by ${\{p,f,c\}_{m}}$, where $c$ denotes the color
connections.  

We can now consider the states $\sket{\{p,f,c\}_{m}}$
to form a basis for a vector space in the sense of
statistical mechanics. After some amount of shower evolution starting
from a basis state $\sket{\{p,f,c\}_{2}}$ with two final state
partons, one reaches a state $\sket{\rho}$ that is a linear
combination of basis states, so that
$\sbrax{\{p,f,c\}_{m}}\sket{\rho}$ represents the probability, in the
shower model, for the state $\sket{\rho}$ to consist of $m+2$ partons
with momenta, flavors, and colors $\{p,f,c\}_{m}$. 

As the state develops, partons split. 
The evolution of the state is tracked with a shower ``time'' $t$ which can be
interpreted as (the logarithm of) a typical time for a quantum process such as a parton splitting. In most of the current algorithms, the shower time is the logarithm of the virtuality or transverse momentum in 
a splitting. (In \textsc{Herwig}, the shower time represents
  the energy of the mother parton times the square of the splitting
  angle. In order to cast \textsc{Herwig} into the form presented
  here, one also needs a cut on virtuality such that splittings with
  too small virtuality are not allowed. In other parton shower
  algorithms, there is also a smallest virtuality allowed, but that
  can be obtained by simply stopping the shower evolution at some
  point.) 

The
evolution starts with the hard process and works forward in physical
time for final state evolution and backwards in physical time for the
evolution of the initial state. Thus we take the shower time for a
splitting $\{p,f,c\}_{m} \to \{\hat p,\hat f,\hat c\}_{m+1}$ to be $t
= t(\{\hat p,\hat f,\hat c\}_{m+1})$ where, for instance if $i$ and
$j$ are the daughter partons and $Q$ represents the virtuality scale
for the hard process that starts the shower, 
\begin{equation}
t(\{\hat p,\hat f,\hat c\}_{m+1}) = \log\left(\frac{Q^2}{(\hat p_i+\hat p_j)^2}\right)
\;\;.
\end{equation}
It can, and should, be debated whether there is a preferred choice for
the shower evolution variable and, if so, what it is. 

Using this notation, we can represent what a typical parton shower
Monte Carlo does. This representation is an approximation to what real
computer codes do. 
We assume that each stage of evolution is
independent of what happened at previous stages, depending instead
only on the shower time $t$ and the partonic state at that stage. This is not the case if, for
instance, we do not exactly conserve four-momentum at each stage and
then adjust the parton four-momenta at the end. 

If we start with a particular basis state $\sket{\{p,f,c\}_{m}}$ at shower time $t_0$, then at a later time $t'$ we get a state related to $\sket{\{p,f,c\}_{m}}$ by an evolution operator ${\cal U}(t',t_0)$. In the notation of conventional parton showers, based on collinear DGLAP splitting kernels, the form of the evolution operator would be 
\begin{equation}
\begin{split}
{\cal U}&(t',t_0)\sket{\{p,f,c\}_{m}}  =  
\Delta(t',t_0;\{p,f,c\}_{m})\sket{\{p,f,c\}_{m}}
\\& 
+ 
\sum_{i,j,k}\ \int_{t_0}^{t'} \!\!
dt_1 \ \Delta(t_1,t_0;\{p,f,c\}_{m}) \
\int_{z_\mathrm{min}(t_1)}^{z_\mathrm{max}(t_1)} \hspace*{-0.5cm}dz 
\int\frac{d\phi}{2\pi}\
  \frac{\alpha_s}{2\pi} \ P_{i\to
  jk}(z) \ {\cal U}(t',t_1)
 \sket{\{\hat p,\hat f,\hat c\}_{m+1}}
\;\;,
\label{eq:traditional}
\end{split}
\end{equation}
where $dt_1=dQ^2/Q^2$ is the differential of the 
evolution variable, $z$ is an energy-momentum
sharing fraction,  $P(z)$ are the DGLAP splitting
kernels, and we include an integral over angle that is usually 
uniformly distributed. Once the algorithm picks which
parton splits, the flavors, and the splitting variables $t,z,\phi$, the new state $\{\hat p,\hat f,\hat c\}_{m+1}$ is known. Reformulating Eq.~(\ref{eq:traditional}) in the notation outlined above, the second term changes appearance slightly,
\begin{equation}
\begin{split}
\label{eq:evolutiondetail}
{\cal U}(t',t_0)\sket{\{p,f,c\}_{m}} ={}& 
\Delta(t',t_0;\{p,f,c\}_{m})\sket{\{p,f,c\}_{m}} 
\\
& \hspace*{-2cm}+ 
\int_{t_0}^{t'}\!\! dt_1\  \Delta(t_1,t_0;\{p,f,c\}_{m})
\\&\hspace*{-2cm}
\quad \times
\int \frac{\big[d\{\hat p,\hat f,\hat c\}_{m+1}\big]}{(m+1)!}\
\sbra{\{\hat p,\hat f,\hat c\}_{m+1}}
{\cal H}_{\rm I}(t_1)\sket{\{p,f,c\}_{m}}
\ {\cal U}(t',t_1)\,
\sket{\{\hat p,\hat f,\hat c\}_{m+1}}
\,
\;\;.
\end{split}
\end{equation}
In either notation, the second term represents that at a shower time
$t_1 > t_0$, the first splitting occurs. This splitting time is 
determined on a probabilistic basis, so $t_1$ is integrated over. The
probability to get a particular state $\{\hat p,\hat f,\hat c\}_{m+1}$
is given by  
\begin{equation}
\sbra{\{\hat p,\hat f,\hat c\}_{m+1}}
{\cal H}_{\rm I}(t_1)\sket{\{p,f,c\}_{m}}
\;\;,
\end{equation}
where ${\cal H}_{\rm I}$ is the splitting operator, analogous to the
interaction hamiltonian in quantum mechanics. 
There is an integration over the possible
outcomes $\{\hat p,\hat f,\hat c\}_{m+1}$. The requirement that the 
splitting at shower time $t_1$ be the first after $t_0$ 
means that we must include the probability
that there is no earlier splitting. This ``no-branching'' probability
is given by a function (the Sudakov form factor) 
\begin{equation}
\Delta(t_1,t_0;\{p,f,c\}_{m})
\;\;.
\end{equation}
In a lowest order shower, this function is fixed so that the probability not to split in shower time interval $dt_1$ is 1 minus the probability to split,
\begin{equation}
\begin{split}
\Delta(t_1,t_0;& \{p,f,c\}_{m})
=
\\&
\exp\left(-\int_{t_0}^{t_1}\! d \tau\
\frac{1}{(m+1)!}
\int \big[d\{\hat p,\hat f,\hat c\}_{m+1}\big]\
\sbra{\{\hat p,\hat f,\hat c\}_{m+1}}
{\cal H}_{\rm I}({\tau})\sket{\{p,f,c\}_{m}}
\right)
\;\;.
\end{split}
\end{equation}
The last ingredient in line two of Eq.~(\ref{eq:evolutiondetail}) is
the evolution operator ${\cal U}(t',t_1)$. This says that further
splittings can happen, in the same way, once the first splitting has
occurred. It can also happen that there is {\em no} splitting
generated between shower times $t_0$ and $t_1$. This is represented in
the first term of Eq.~(\ref{eq:evolutiondetail}). 

Evidently, the main content of a parton shower resides in the
generator ${\cal H}_{\rm I}(t)$ of the evolution. This has two main
parts: a splitting function and a momentum mapping.  

Consider first the splitting functions, functions of the daughter
parton momenta that give the probability to split. If a parton 
splits into two nearly collinear partons, 
then the splitting function must match the probability given by
Feynman graphs in the collinear limit. For the moment, we discuss a
spin averaged, leading color shower. Then the splitting function
matches the result from Feynman graphs averaged over the mother parton
spin and summed over the daughter spins in the approximation of
neglecting contributions that are suppressed by $1/N_{\rm c}^2$, where
$N_{\rm c}$ is the number of colors. When the emitted parton is a soft
gluon, the splitting function should match the probability given by
Feynman graphs in the limit $p_{m+1} \to 0$. Away from the soft and
collinear limits, however, one can choose what functional form to use and one
can debate the merits of different choices. 

This can be illustrated by the case of \textsc{Vincia}
\cite{Giele:2007di}, which represents a new development and is
discussed in more detail later in this section. 
In a leading order shower in the leading
color limit, the fundamental object that emits gluons is a color
dipole, that is, two partons, say $l$ and $k$, that are
color-connected (i.e., adjoining on a
color string). The basic idea here goes back to the Lund dipole
\cite{Gustafson:1986db}, 
implemented in \textsc{Ariadne} \cite{Lonnblad:1992tz}. (We shall
henceforth refer to such showers as \emph{dipole-antenna} showers, in order
to disambiguate them from what we shall call \emph{partitioned-dipole} 
showers below.)
The relevant Feynman graphs
in the amplitude are those in which the gluon $m+1$ is emitted from
parton $l$ and those in which it is emitted from parton $k$. In the
squared amplitude, one has a contribution $l$-$l$, corresponding to
the square of the graph for emission from $l$, a similar contribution
$k$-$k$, and two interference contributions $l$-$k$ and $k$-$l$. The
approximation of keeping only the leading color contribution restricts
us to the case that $l$ and $k$ are color-connected. In dipole-antenna
showers, each dipole is treated as a unit, an antenna that radiates
gluons, and the splitting functions can be chosen such as to 
match the perturbative result in all the relevant limits, i.e.\ 
gluon $m+1$ collinear to $l$, collinear to $k$, or soft. There are two 
main differences between \textsc{Vincia} and \textsc{Ariadne} (and also a
recent \textsc{Sherpa} implementation \cite{Winter:2007ye}). The first
is that an explicit possibility to vary the shower ambiguities away
from the singular regions is retained in \textsc{Vincia}, and the
second is that it combines the original dipole shower with the 
antenna subtraction formalism of
Refs.~\cite{Kosower:1997zr,Campbell:1998nn,GehrmannDeRidder:2005cm} to
match to fixed-order matrix elements.

Another new development is what we can call the {\it partitioned-dipole} shower \cite{Schumann:2007mg, Dinsdale:2007mf}, which is discussed in more detail later in this section. Here one partitions the splitting function
into two parts. One part contains the singularity corresponding to
parton $m+1$ being collinear with parton $l$ and part of the soft
singularity. The other part contains the singularity corresponding to
parton $m+1$ being collinear with parton $k$, along with the remaining
part of the soft singularity. Away from these singularities, one has a
choice. A sensible choice (as suggested in
Ref.~\cite{Nagy:2006kb}) is to take the splitting functions to
be precisely those defined by the Catani-Seymour dipole subtraction
scheme \cite{Catani:1996vz} that is widely used for next-to-leading
order perturbative calculations. This has the advantage that it should
be fairly straightforward to match these NLO calculations to a
Catani-Seymour dipole shower. It has the disadvantage that the
splitting of the emission probability from a dipole antenna into two
parts is perhaps a bit artificial. There is more than one way to
accomplish this splitting. 

The second part of the generator ${\cal H}_{\rm I}(t)$ of shower
evolution is the specification of the momentum mapping. In
Eq.~(\ref{eq:evolutiondetail}), there is a nominal integration over
the momenta of all the partons after the splitting. However the matrix
element of ${\cal H}_{\rm I}(t)$ contains delta functions that, for
given starting momenta $\{p\}_m$, restrict the new momenta $\{\hat
p\}_{m+1}$ to lie on a three dimensional surface. This surface could
be parametrized by splitting variables $t,z,\phi$, as in
Eq.~(\ref{eq:traditional}). In the case of the timelike dipole-antenna
showers in \textsc{Ariadne}, \textsc{Vincia}, and
\textsc{Sherpa}, the momenta of all of the
partons not part of the dipole remain the same before and after the
parton emission. For the partons $l$ and $k$ that form the dipole, the
momenta $p_l$ and $p_k$ plus three splitting variables are mapped
reversibly to the momenta of three daughter partons, $\hat p_l$, $\hat
p_k$, and $\hat p_{m+1}$ after the splitting, with all of the parton
momenta being on-shell. This mapping is symmetric under label
interchange $l \leftrightarrow k$. In the special case that $\hat
p_{m+1}$ is collinear with $\hat p_l$, we have $p_l = \hat p_l + \hat
p_{m+1}$ and $p_k = \hat p_k$. Similarly, if $\hat p_{m+1}$ is
collinear with $\hat p_k$, we have $p_k = \hat p_k + \hat p_{m+1}$ and
$p_l = \hat p_l$. In the soft limit, $\hat p_{m+1} = 0$, we have $p_l
= \hat p_l$ and $p_k = \hat p_k$. Away from these limits, the mapping
is necessarily not so trivial, leading to a further non-singular
ambiguity which \textsc{Vincia} attempts to explore. 
For the partioned-dipole shower, there is a
similar but simpler mapping, this time not symmetric under $l
\leftrightarrow k$. The splitting function that includes the
singularity for $\hat p_{m+1}$ collinear with $p_l$, comes with a
momentum mapping for which $p_l = \hat p_l + \hat p_{m+1}$ and $p_k =
\hat p_k$ when $\hat p_{m+1}$ is collinear with $\hat p_l$ or
soft. Away from these limits, the mapping takes some momentum from
parton $k$ in order to keep momentum conserved and all partons on
shell. The choice here is to use the same momentum mapping as was defined
by Catani and Seymour for the subtractions in next-to-leading order
calculations. In the case of splittings involving an initial state
splitting (spacelike showers), 
the momentum mappings are a little more complicated than
sketched here. We should mention that it is also possible to take the
momentum needed to keep all partons on-shell from {\em all} of the
final state partons in what might be called a democratic way
\cite{Nagy:2007ty}. 

We hope that this comparative discussion may be useful as a guide to the more detailed presentations later in this section. We may also mention the published work \cite{Nagy:2007ty} that was discussed at Les Houches but is not separately presented in this document. The idea here is to extend the idea of a leading order parton shower so that one does {\em not} average over spins or take just the leading color limit. In this case, there is an evolution equation similar to Eq.~(\ref{eq:evolutiondetail}) but with spin indices and a more detailed specification of the color state.  The solution of the evolution equation yields integrals that could, in principle, be computed numerically. However, an algorithm that is likely to be usefully convergent with finite computer resources is still under development \cite{Nagy:2008ns}.

%
}


\section[Time-like showers based on dipole-antenna radiation functions]
{TIME-LIKE SHOWERS BASED ON DIPOLE-ANTENNA RADIATION FUNCTIONS
\protect\footnote{Contributed by: R.~Frederix, W.T.~Giele, D.A.~Kosower,
P.Z.~Skands}}
{\graphicspath{{skands/}}
\renewcommand{\d}{\ensuremath{\mathrm{d}}}
\newcommand{\GeV}{\,\mbox{Ge\kern-0.2exV}}
\def\lsim{\mathrel{\raise.3ex\hbox{$<$\kern-.75em\lower1ex\hbox{$\sim$}}}}
\def\gsim{\mathrel{\raise.3ex\hbox{$>$\kern-.75em\lower1ex\hbox{$\sim$}}}}
%
%
%
%
%
%
%

\subsection{Introduction}
In this report we take the next step in the development of the
\textsc{Vincia}\ shower towards a full-fledged parton shower, embedded
into the \textsc{Pythia} 8 generator
\cite{Giele:2007di,Sjostrand:2007gs}.  Previously, we included only 
the gluonic time-like shower \cite{Giele:2007di}.  By including
massless quarks we can start making comparisons at \textsc{LEP}\
energies and make quantitative studies for future linear colliders.
As the \textsc{Vincia}\ shower is a dipole-antenna shower, we can make
direct comparisons with the dipole-antenna functions used in
\textsc{Ariadne}\ \cite{Lonnblad:1992tz}.

We also make a phenomenological comparison with the \textsc{Pythia} 8
shower. For this purpose, we choose the evolution variable, the
hadronization boundary and other parameters in \textsc{Vincia}\ as
close as possible to the default \textsc{Pythia} 8 settings. In this
emulation mode we compare a few representative distributions, both
infrared safe and infrared regulated observables, such as jet rates,
thrust, and parton multiplicities for hadronic $Z$ decays at
$\sqrt{s}=m_Z$.
 
\subsection{Dipole-antenna functions}
The most general form for a leading-log antenna
function for massless parton splitting, $\hat{a}\hat{b}\to arb$,
can be represented by a double Laurent series in the two branching invariants, 
\begin{equation}
a(y_{ar},y_{rb} ; s) = \ \ \ \frac{1}{s}
\!\!\sum_{\alpha,\beta=-1}^\infty\!\! C_{\alpha,\beta} \ 
y_{ar}^\alpha\ y_{rb}^{\beta} ~,\label{eq:a}
\end{equation}
where 
\begin{equation}
s = s_{\hat{a}\hat{b}}=s_{arb}~~~\mbox{and}~~~
y_{ij} = \frac{s_{ij}}{s} \le 1 
\end{equation}
are the invariant mass squared of the antenna and the scaled branching
invariants, respectively. In principle, eq.~(\ref{eq:a}) could also be
multiplied by an overall phase space veto function, restricting the
radiation to specific ``sectors'' of phase space, but we shall here
use so-called ``global'' antenna functions which are summed together
without such cuts.  Note that we have here written the antenna
function stripped of color factors, to emphasize that this part of the
discussion is not limited to the leading-color limit.

The coefficient of the most singular term, $C_{-1,-1}$, controls the
strength of the double (soft) singularity (the ``double log'' term)
and the coefficients $C_{-1,j\ge0}$ and $C_{i\ge0,-1}$ govern the single
(collinear) singularities (``single log'' terms). These, in parton
shower terminology collectively labeled ``leading log'' terms, are
universal, whereas the polynomial coefficients $C_{i\ge0,j\ge0}$ are
arbitrary, corresponding to beyond-leading-log ambiguities in the
shower or, equivalently, different NLO subtraction terms in the
fixed-order expansion.

We take the Gehrmann-de-Ridder-Glover (``GGG'') antenna functions
\cite{GehrmannDeRidder:2005cm} as our starting point. The
corresponding coefficients $C_{\alpha,\beta}$ for the the five antennae
that occur in massless QCD at LL are collected in
tab.~\ref{tab:coefficients}.
\begin{table}[t]
\begin{center}
\begin{tabular}{lr|rrrrrr|rrr}
\hline 
& $C_{-1,-1}$ & $C_{-1,0}$ & $C_{0,-1}$ & $C_{-1,1}$ & $C_{1,-1}$ &
$C_{-1,2}$ & $C_{2,-1}$ &  $C_{0,0}$ & $C_{1,0}$ & $C_{0,1}$ \\
\hline
GGG\\
$q\bar{q}\to qg\bar{q}$ & 2 & -2 & -2 & 1 & 1 & 0 & 0 & 0 & 0 & 0\\
$qg\to qgg$& 2 & -2 & -2 & 1 & 1 & 0 & -1 & $\frac52$ & -1 & $\frac32$\\
$gg\to ggg$& 2 & -2 & -2 & 1 & 1 &-1 & -1 & $\frac83$ & -1 & -1 \\
$qg\to q\bar{q}'q'$& 0 & 0 & $\frac12$ & 0 & -1 & 0 & 1 & -$\frac12$ & 1& 0 \\
$gg\to g\bar{q} q$& 0 & 0 & $\frac12$ & 0 & -1 & 0 & 1 & -1 & 1 & $\frac12$\\
\hline
\textsc{Ariadne} \\
$q\bar{q}\to qg\bar{q}$ & 2 & -2 & -2 & 1 & 1 & 0 & 0 & 0 & 0 & 0\\
$qg\to qgg$& 2 & -2 & -3 & 1 & 3 & 0 & -1 & 0 & 0 & 0\\
$gg\to ggg$& 2 & -3 & -3 & 3 & 3 &-1 & -1 & 0 & 0 & 0\\
$qg\to q\bar{q}'q'$& 0 & 0 & $\frac12$ & 0 & -1 & 0 & 1 & -1 & 1  & $\frac12$\\
$gg\to g\bar{q} q$& 0 & 0 & $\frac12$ & 0 & -1 & 0 & 1 & -1 & 1  & $\frac12$\\
\hline
\multicolumn{11}{l}{\textsc{Ariadne}2 (re-parameterization of \textsc{Ariadne}\ functions \`a la
  GGG, for comparison)}\\
$q\bar{q}\to qg\bar{q}$ & 2 & -2 & -2 & 1 & 1 & 0 & 0 & 0 & 0 & 0\\
$qg\to qgg$& 2 & -2 & -2 & 1 & 1 & 0 & -1 & -1 & 0 &0\\
$gg\to ggg$& 2 & -2 & -2 & 1 & 1 &-1 & -1 & -$\frac43$ & -1 &-1\\
\hline
\end{tabular}
\caption{Laurent coefficients for
  massless LL QCD antennae (${\hat{a}\hat{b}\to arb}$). The
  coefficients with at least one negative index are universal (apart
  from a re-parameterization ambiguity for gluons). 
  For ``GGG'' (the defaults in \textsc{Vincia}), 
  the finite terms correspond to 
  the specific matrix elements considered in
  \cite{GehrmannDeRidder:2005cm}. In particular, the $q\bar{q}$ antenna
  absorbs the tree-level $Z\to qg\bar{q}$ matrix element
  \cite{GehrmannDeRidder:2004tv} and the $gg$
  antennae absorb the tree-level $h^0\to gg\to ggg$ and $h^0 \to gg
  \to g\bar{q} q$ matrix elements \cite{GehrmannDeRidder:2005aw}. 
  The $qg$ antennae are derived from a neutralino decay process 
  \cite{GehrmannDeRidder:2005hi}. 
\label{tab:coefficients}}
\end{center}
\end{table}
For reference, we also compare to the radiation functions
\cite{Gustafson:1986db,Gustafson:1987rq,Andersson:1989ki} used in the
\textsc{Ariadne}\ dipole shower \cite{Lonnblad:1992tz}, which are also
the ones used in a recent study by the \textsc{Sherpa} group
\cite{Winter:2007ye}.  Note that the single log terms have a slight
ambiguity when gluons are involved, arising from the arbitrary choice
of how to decompose the radiation off the gluon into the two antennae
it participates in. Nominally, the \textsc{Ariadne}\ single log
coefficients therefore look different from the GGG ones. However, a
re-parameterization of the total gluon radiation, which we label
\textsc{Ariadne}2, reveals that the only real difference lies in the
choice of finite terms. Interestingly, while all the \textsc{Ariadne}\
radiation functions are positive definite, the equivalent
\textsc{Ariadne}2 one for $gg\to ggg$ is not and hence could not be
used as a basis for a shower Monte Carlo.

In modern versions of \textsc{Ariadne}, gluon splitting to quarks has
an additional pre-factor
$2/(1+s_{\hat{a}\hat{b}}/s_{\hat{b}\hat{c}})$, where $\hat{c}$ is the
neighbor on the other side of the splitting gluon. This is based on
comparisons to $e^+e^-\to q\bar{q}' q'\bar{q}$ matrix elements and implies
that the smaller dipole takes the larger part of the $g\to q\bar{q}$
branching. Such effects are not included in \textsc{Vincia}\ at this
point.

Our convention for color factors is that they count color degrees of
freedom. Their normalization should therefore be such that, in the
large-$N_C$ limit, they tend to $N_C$ raised to the power of the
number of new color lines created in the splitting. In particular,
\begin{equation}
\begin{array}{rclclcl}
\hat{C}_F & = &\frac{N_C^2-1}{N_C} &=& \frac83~, \\ 
C_A & = &N_C &=& 3~. \\
\end{array}
\end{equation}
For gluon splitting to quarks, the antenna shower explicitly sums over
each flavor separately, hence the relevant antenna functions should be
normalized to one flavor, $\hat{T}_R=1$. (We use the hatted symbols
$\hat{C}_F$ and $\hat{T}_R$ to distinguish this normalization from the
conventional parton-shower one in which $C_F=4/3$ and $T_R=1/2$.)

The complete antenna functions, in the notation of \cite[eqs.~(2) and
(11)]{Giele:2007di}, are then
\begin{equation}
\begin{array}{rcl}
A({q\bar{q}\to qg\bar{q}}) & = & 4\pi\alpha_s \ 
  \hat{C}_F \ a({q\bar{q}\to qg\bar{q}})~,\\
A({qg\to qgg}) & = & 4\pi\alpha_s \ 
  \hat{C}_F \ a({qg\to qgg})~,\\
A({gg\to ggg}) & = & 4\pi\alpha_s \ N_C \ a({gg\to ggg})~,\\
A({qg\to q\bar{q}'q'})& = & 4\pi\alpha_s \ a({qg\to q\bar{q}'q'})~,\\
A({gg\to g\bar{q} q}) & = & 4\pi\alpha_s \ a({gg\to g\bar{q} q})~,
\end{array} \label{eq:A}
\end{equation}
where $\alpha_s=\alpha_s(\mu_R)$ may depend on the branching
kinematics. If so, we use a nominal $\hat{\alpha}_s=1$ for generating trial
branchings, which are then accepted with probability $\alpha_s(\mu_R)$
at the point when the full kinematics have been constructed (see below). The
possibilities for $\mu_R$ currently implemented in \textsc{Vincia} are
\begin{equation}
\mu_R = \left\{
\begin{array}{lcl}
\mbox{type 0} & : & K_R\ 2p_{\perp} \\
\mbox{type 1} & : & K_R\ Q_E \\
\mbox{type 2} & : & K_R\ \sqrt{s_{\hat{a}\hat{b}}}~, 
\end{array}\right.
\end{equation}
where $K_R$ is an arbitrary constant, $p_{\perp}$ is defined as in
\textsc{Ariadne} with $p_{\perp}^2=s_{ar}s_{rb}/s_{\hat{a}\hat{b}}$
\cite{Lonnblad:1992tz}, $Q_E$ is the evolution variable, and
$\sqrt{s_{\hat{a}\hat{b}}}$ is the invariant mass of the mother
dipole-antenna. The default is a 1-loop running five-flavor $\alpha_s$
with $\mu_R=p_{\perp}$ (i.e., Type 0 above, with $K_R=\frac12$) and
$\alpha_s(m_Z)=0.137$ (the default in \textsc{Pythia} 8, making
comparisons simpler).  Alternatively, both fixed and 2-loop running
options are available as well \cite{Sjostrand:2007gs}. For the pure
shower, the dependence on the renormalization scheme of $\alpha_s$ is
beyond the required precision and hence we do not insist on an
$\overline{\mathrm{MS}}$ definition here. Indeed, the default value of
$\alpha_s(m_Z)$ in \textsc{Pythia} 8 is determined from tuning to LEP
event shapes. Though beyond the scope of the present paper, we note
that in the context of higher-order matching, one should settle on a
specific scheme, and should then see the dependence on both the scheme
and scale choices start to cancel as successive orders are included.

\subsection{Shower implementation}
Brief descriptions of the \textsc{Vincia} switches and parameters are
contained in the program's XML ``manual'', by default called
\texttt{Vincia.xml}, which is included together with the code.  This
file also contains the default values and ranges for all adjustable
parameters, which may subsequently be changed by the user in exactly
the same way as for a standard \textsc{Pythia} 8 run
\cite{Sjostrand:2007gs}.

The default antenna functions are contained in a separate XML file,
\texttt{Antennae-GGG.xml}.  Antennae that are related by charge
conjugation to the 
ones listed tab.~\ref{tab:coefficients} are obtained by simple
swapping of invariants (e.g., $g\bar{q}$ antennae are obtained from the
$qg$ ones). Similarly, antenna functions that are permutations of the
ones in tab.~\ref{tab:coefficients}, such as $gg\to \bar{q} q g$, are
obtained by swapping. In view of the probabilistic nature of the
shower, all antenna functions are checked for positivity during
initialization. If negative regions are found, the constant term
$C_{0,0}$ is increased to offset the difference and a warning is given,
stating the new value of $C_{0,0}$.

We use the \textsc{Pythia} 8 event record \cite{Sjostrand:2007gs},
which includes Les Houches color tags \cite{Boos:2001cv,Alwall:2006yp}
for representing color connections. At every point during the event
evolution, leading-color antennae are spanned between all pairs of
(non-decayed) partons for which the color tag of one matches the
anti-color tag of the other.

Shower generation proceeds largely as for the pure-gluon case
described in \cite{Giele:2007di}, including the choice between two
evolution variables
\begin{equation}
y_E = \left\{\begin{array}{lclclcl} 
\mbox{type I ($p_{\perp}$-ordering)}&: & y^2_\mathrm{I} 
 & = &\displaystyle \frac{Q_\mathrm{I}^2}{s} 
 & = & \displaystyle 4\frac{s_{ar}s_{rb}}{s^2} = 4 y_{ar} y_{rb} \\[3mm]
\mbox{type II (dipole-mass-ordering)} & :& y^2_\mathrm{II} 
 & = & \displaystyle \frac{Q_\mathrm{II}^2}{s} 
 & = & \displaystyle 2\mathrm{min}(y_{ar},y_{rb}) 
\end{array}\right.~~~.
\label{eq:y-ordering}
\end{equation}
Note that we do not include an ``angular-ordering'' option. In
  conventional parton showers, which use collinear splitting
  functions, angular ordering gives a good
  approximation of the coherent dipole radiation patterns we here
  describe by the antenna functions $A$. Since dipole-antenna showers
  use $A$ directly, coherence is thus independent of the choice of
  evolution variable to first order in this formulation (see, e.g.,
  \cite{Gustafson:1986db}).

For the phase space map an optimal choice for the functional form of
the ``recoil angle'' $\psi_{\hat{a}a}$ (see
\cite{Giele:2007di,Lonnblad:1992tz}) away from the soft and collinear
limit exists for $q\bar{q}$ antennae \cite{Kleiss:1986re}.  However, we
have not yet implemented this particular subtlety in the
\textsc{Vincia} code.  The default choice for all antennae is thus
currently the same as for the $gg\to ggg$ splitting in
\textsc{Ariadne} \cite{Lonnblad:1992tz}
\begin{eqnarray}
\psi_{\mbox{\textsc{Ariadne}}} & = &
\frac{E^2_b}{E_a^2+E_b^2}(\pi-\theta_{ab})~, 
\end{eqnarray}
with alternative choices listed in \cite{Giele:2007di}.

Trial branchings are generated by numerically solving for
$y_{\mathrm{trial}}$ in the equation $R=\widehat{\Delta}(y_\mathrm{trial})$,
where $R$ is a random number uniformly distributed between zero and
one, and the trial Sudakov is \cite[eq.~(51)]{Giele:2007di}
\begin{eqnarray}
\hspace*{-5mm}\widehat{\Delta}(y_\mathrm{trial}) \!\! & = & \!\!
\exp\Bigg[-\int_{y_\mathrm{trial}}^{1}\hspace*{-3mm}\! \d{y_E} \int_0^1 
\!\!\!\d y_{ar}\int_0^{1-y_{ar}}\hspace*{-5mm} \d y_{rb} \
\delta(y_E-y_E(y_{ar},y_{rb}))
\frac{\hat{\mathcal{A}}(y_{ar},y_{rb})}{16\pi^2}
\Bigg],\label{eq:trialSudakov}
\end{eqnarray}
with $\mathcal{A}$ an overestimate of the ``true'' antenna function
such that 
\begin{equation}
\hat{\mathcal{A}}(y_{ar},y_{rb}) \equiv 
s_{arb}\hat{A}(y_{ar},y_{rb};s_{arb},1) > s_{arb} A(y_{ar},y_{rb};s_{arb},1)
\end{equation}
only depends on the rescaled invariants 
(for instance by using a fixed overestimate of
$\hat{\alpha}_s=1$ here). Once the full kinematics are known (see below) 
the trial branching can be vetoed with probability $1-A/\hat{A}$, which by
the veto algorithm changes the resulting distribution back to that of
$A$, as desired.

During program execution, cubic splines of $\widehat{\Delta}$ and
$\widehat{\Delta}^{-1}$ are used for the actual trial
generation. These splines are constructed on the fly, with the
2-dimensional phase space integrals in eq.~(\ref{eq:trialSudakov})
carried out either by 2-dimensional adaptive Gaussian quadrature (AGQ)
on $\hat{\mathcal{A}}$ directly or (substantially faster) by
1-dimensional AGQ on the primitive function along a contour of fixed
$y_{ar}$, defined by
\begin{eqnarray}
I_a(y_{ar},y_1,y_2) & = & 
  \int_{y_1}^{y_2} \!\!\! \mathrm{d}y_{rb} \ 
  \frac{\hat{\mathcal{A}}(y_{ar},y_{rb})}{16\pi^2 } \nonumber\\
& = & \frac{\hat{\alpha}_s\mathcal{C}_i}{4\pi} 
  \sum_{\alpha=-1}^\infty y_{ar}^\alpha\left[C_{\alpha,-1} 
  \ln\left(\frac{y_{2}}{y_{1}}\right) ~ + ~ \sum_{\beta=0}^\infty
  C_{\alpha,\beta} \frac{y_{2}^{\beta+1}-y_{1}^{\beta+1}}{\beta+1}
  \right]~,\label{eq:primitive}
\end{eqnarray}
where $\hat{\alpha}_s$ is the overestimate of $\alpha_s$ discussed
earlier, $\mathcal{C}_i$ represents the color factors appearing in
eq.~(\ref{eq:A}), and the phase space limits $y_{1,2}$ depend on the choice of
evolution variable, see below. During initialization, the program
checks for consistency between the analytic and numeric integrals and
a warning is issued if the numerical precision test fails. 

The antenna with the largest trial scale is then selected for further
inspection. A $\phi$ angle distributed uniformly in $[0,2\pi]$ 
is generated, and a complementary phase space invariant, $z$, is chosen
according to the probability distribution
\begin{eqnarray}
I_z(y_E,z) 
& = & \int_{z_\mathrm{min}(y_E)}^z \hspace*{-9mm}\d z' |J(y_E,z')|
\frac{\hat{\mathcal{A}}(y_{ar},y_{rb})}{16\pi^2 } ~~~, 
\end{eqnarray}
where $|J(y_E,z)|$ is the Jacobian arising from translating
$\{y_{ar},y_{rb}\}$ to $\{y_E,z\}$ and $z_\mathrm{min}(y_E)$ is the
smallest value $z$ attains inside the physical phase space 
for a given $y_E$. Depending on the type of evolution variable, as
defined in eq.~(\ref{eq:y-ordering}), we choose $\{y_E,z\}(y_{ar},y_{rb})$ as 
\begin{eqnarray}
\mbox{type I} & : & y_E = 4y_{ar}y_{rb}~,~z = y_{rb}\nonumber\\
& \Rightarrow & |J_\mathrm{I}| = 1/(4z)~,~z_\mathrm{max,min}(y_E) =
\frac12(1\pm\sqrt{1-y_E})~,\\
\mbox{type II} & : & \begin{array}[t]{lcl}
    y_E=2y_{ar}~,~z=y_{rb}~\mbox{for}~z\le1-\frac12y_E\\[2mm]
    y_E=2y_{rb}~,~z=y_{ar} + (1-2y_{rb})~\mbox{for}~z>1-\frac12y_E
                        \end{array}\nonumber\\
&\Rightarrow&|J_\mathrm{II}| = 1/2~,~z_\mathrm{min}(y_E)
 = \frac12 y_E~,~z_\mathrm{max}(y_E) =  2-\frac32y_E
\label{eq:zdef}
\end{eqnarray}
where, for type II, we have arranged the two separate branches
$y_{ar}<y_{rb}$ and $y_{rb}<y_{ar}$ one after the other by a trivial
parallel displacement in the $z$ coordinate. Using the Laurent
representation of the antenna functions, the analytical forms of $I_z$
become
\begin{eqnarray}
\mbox{type I} & : & \frac{\hat{\alpha}_s\mathcal{C}_i}{16\pi} 
\sum_{\alpha=-1}^\infty
\left(\frac{y_R}{4}\right)^\alpha  \left[ C_{\alpha,\alpha}
\ln\frac{z}{z_{\mathrm{min}}(y_E)} + \sum_{\beta\ne\alpha} \frac{z^{\beta-\alpha}-z_\mathrm{min}(y_E)^{\beta-\alpha}}{\beta-\alpha}\right]\\
\mbox{type II} & : & \frac{\hat{\alpha}_s\mathcal{C}_i}{8\pi}  \left[
I_a\left(\frac12y_E,z_\mathrm{min}(y_E),\min(z,1-z_\mathrm{min}(y_E))\right)\right.
\nonumber \\ & & \hspace*{1cm}+
\left.I_a^T\left(\frac12y_E,1-z_\mathrm{min}(y_E),\max(z,1-z_\mathrm{min}(y_E)\right)
\right]~,
\end{eqnarray}
where the $I_a$ is defined in eq.~(\ref{eq:primitive}) and $I_a^T$ is
the primitive along a direction of fixed $y_{rb}$
\begin{equation}
I^T_a(y_{rb},y_1,y_2) 
= \sum_{\beta=-1}^\infty y_{rb}^\beta\left[C_{-1,\beta} 
\ln\left(\frac{y_{2}}{y_{1}}\right) ~ + ~ \sum_{\alpha=0}^\infty
C_{\alpha,\beta} \frac{y_{2}^{\alpha+1}-y_{1}^{\alpha+1}}{\alpha+1}
\right]~.
\end{equation}

\subsection{Numerical results}
We now turn to a quantitative comparison between \textsc{Pythia} 8 and
\textsc{Vincia}\ for $e^+e^-\to Z \to q\bar{q}$ at $\sqrt{s}= m_Z$.  We
use a 1-loop running $\alpha_s$ with $\alpha_s(m_Z)=0.137$ (the
default in \textsc{Pythia} 8), with a 5-flavor running matched to 4
and 3 flavors at the $b$ and $c$ thresholds, but to eliminate the
question of explicit quark mass effects we only allow $d$ and $u$
quarks in the $Z$ decay and subsequent shower evolution.  The
evolution is terminated at $p_{\perp\mathrm{had}} = 0.5 \GeV$, and we
have switched off hadronization so as not to unintentionally obscure
the differences between the partonic evolutions. Likewise, photon
radiation is switched off in all cases, and in \textsc{Pythia} 8 we
further switch off gluon polarization effects.  For \textsc{Vincia},
we use three different settings: transverse-momentum ordering with
``GGG'' antenna functions, dipole-mass ordering with ``GGG'' antenna
functions, and transverse-momentum ordering with the
``\textsc{Ariadne}'' antenna functions.

Fig.~\ref{fig:jets} shows the 3-, 4-, and 5-jet inclusive fractions as
functions of the logarithm of Durham $k_T$, using the default
\textsc{Pythia} 8 Durham clustering algorithm
\cite{Sjostrand:2007gs}. In \textsc{Pythia} 8, the 3-jet rate (the set
of curves furthest to the right) is matched to the tree-level 3-parton
matrix element, whereas the GGG and \textsc{Ariadne}\ antenna
functions in \textsc{Vincia}\ reproduce it by construction.  The
general agreement on the 3-jet rate is therefore a basic validation of
the $q\bar{q}\to qg\bar{q}$ antenna implementation.  Higher-order effects
appear to make the mass-ordered \textsc{Vincia}\ slightly softer,
which we tentatively conclude is due to this variable favoring soft
wide-angle radiation over high-$p_\perp$ collinear radiation (as
illustrated by fig.~2 in \cite{Giele:2007di}).

Similarly, the 4-jet fractions (the middle set of curves in
 fig.~\ref{fig:jets}) test the $qg$ antennae in \textsc{Vincia}, with
 the GGG showers here slightly higher and the \textsc{Ariadne}\ one slightly
 lower, in agreement with the differences in $qg$ antenna finite
 terms, cf.~tab.~\ref{tab:coefficients}. This trend becomes more
 pronounced in the 5-jet fraction, since also the $gg\to ggg$ function
 in \textsc{Ariadne}\ is softer than GGG.
\begin{figure}[t]\begin{center}\vspace*{-8mm}
\includegraphics*[scale=0.5]{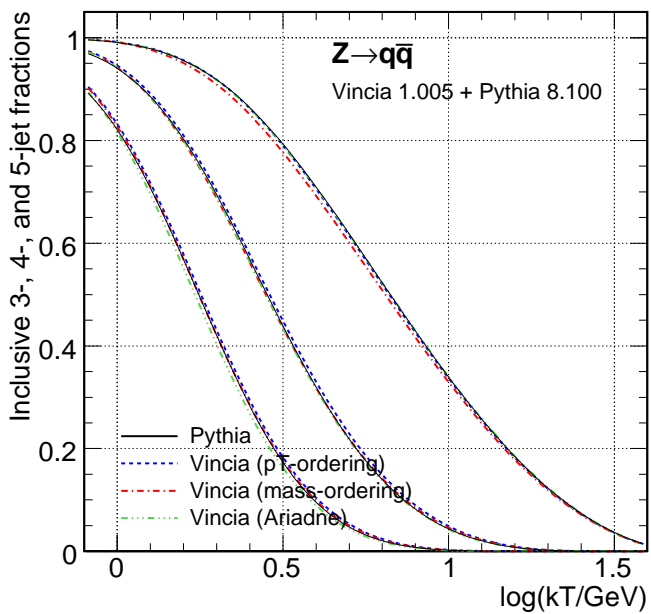}\vspace*{-1cm}
\caption{Inclusive 3-, 4-, and 5-jet fractions. \label{fig:jets}}
\end{center}
\end{figure}

We may now study further distributions, as a representative example of
which we take thrust, illustrated as $1-T$ in the top row of
fig.~\ref{fig:thrustnpartons}. The full distribution is shown to the
left with a closeup of the region $1-T<0.1$ to the right. The region
$0.1<1-T<\frac13$ is dominated by well-separated three-jet
configurations. In the tail, $1-T>\frac13$, a matching to $e^+e^-\to4$
jets would be required to improve the accuracy. In the region below
$1-T=0.1$, however, this would not help. These are three-jet
configurations which are ``nearly two-jet''.  Here, the type and size
of the Sudakov suppression is essential, the first fixed order of
which could be accessed by 1-loop matching, but since the fixed-order
expansion is poorly convergent in this region anyway, the disagreement
is more likely to be cured by a systematic inclusion of
higher-logarithmic effects in the showers (either implicitly, by
``clever choices'' of evolution, renormalization, and kinematic
variables in the LL shower, or explicitly, by a systematic inclusion
of NLL splittings).  It should be noted, however, that hadronization
and hadron decay effects are important in the region below
\begin{equation}
1-T \sim 1-\max(x_{k}) = \min(y_{ij}) \lsim
\frac{(\mathrm{A~few~\GeV})^2}{m_Z^2} \lsim 0.01~, 
\end{equation}
where the $x$ and $y$ fractions pertain to 3-jet configurations. This
complicates the separation of genuine non-trivial higher-log
effects from non-perturbative effects when comparing to experimental
data at currently accessible collider energies.
\begin{figure}[t]\vspace*{-0.7cm}
\hspace*{-3mm}\includegraphics*[scale=0.44]{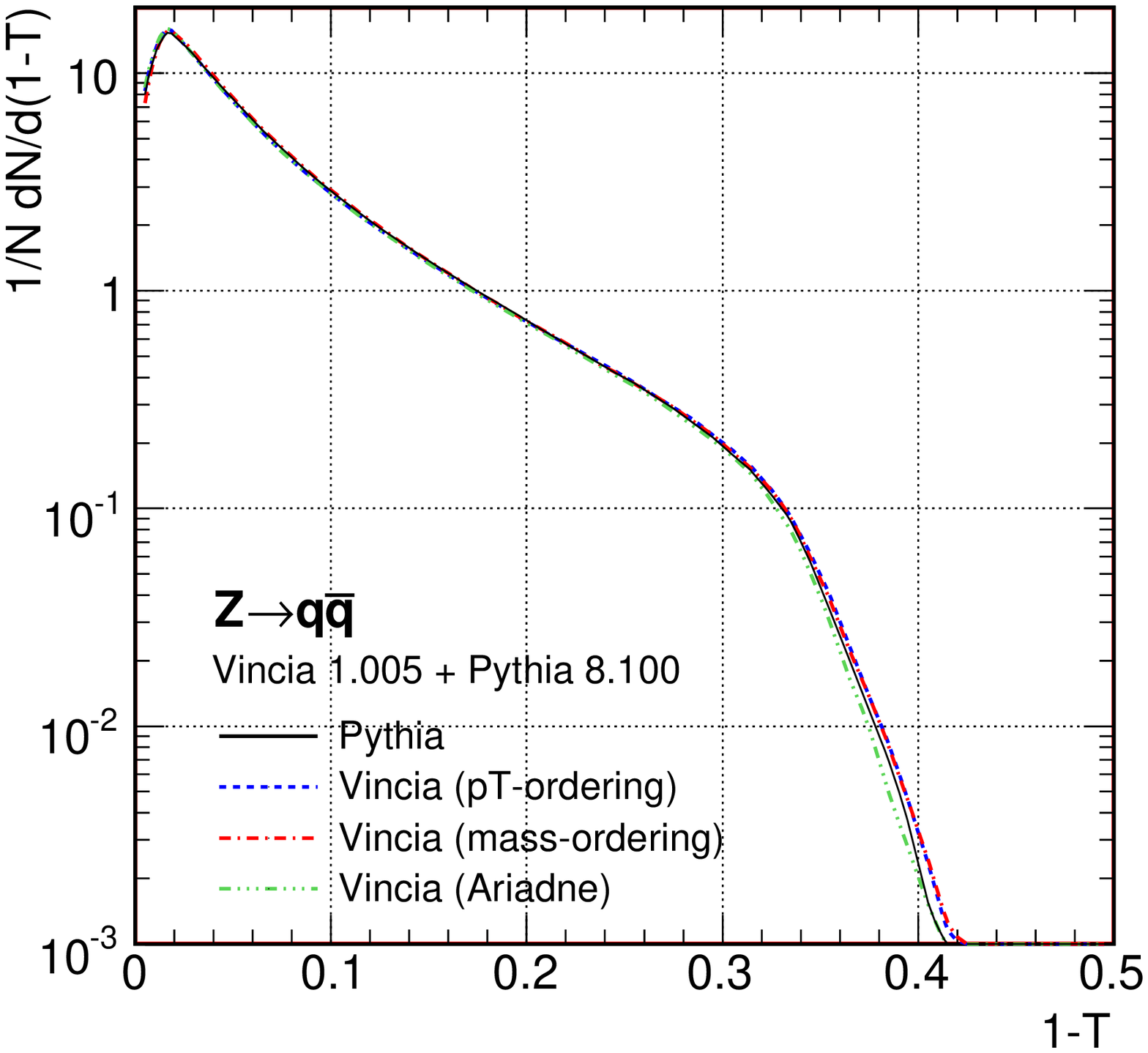}\hspace*{-0.9cm}
\raisebox{0mm}{\includegraphics*[scale=0.44]{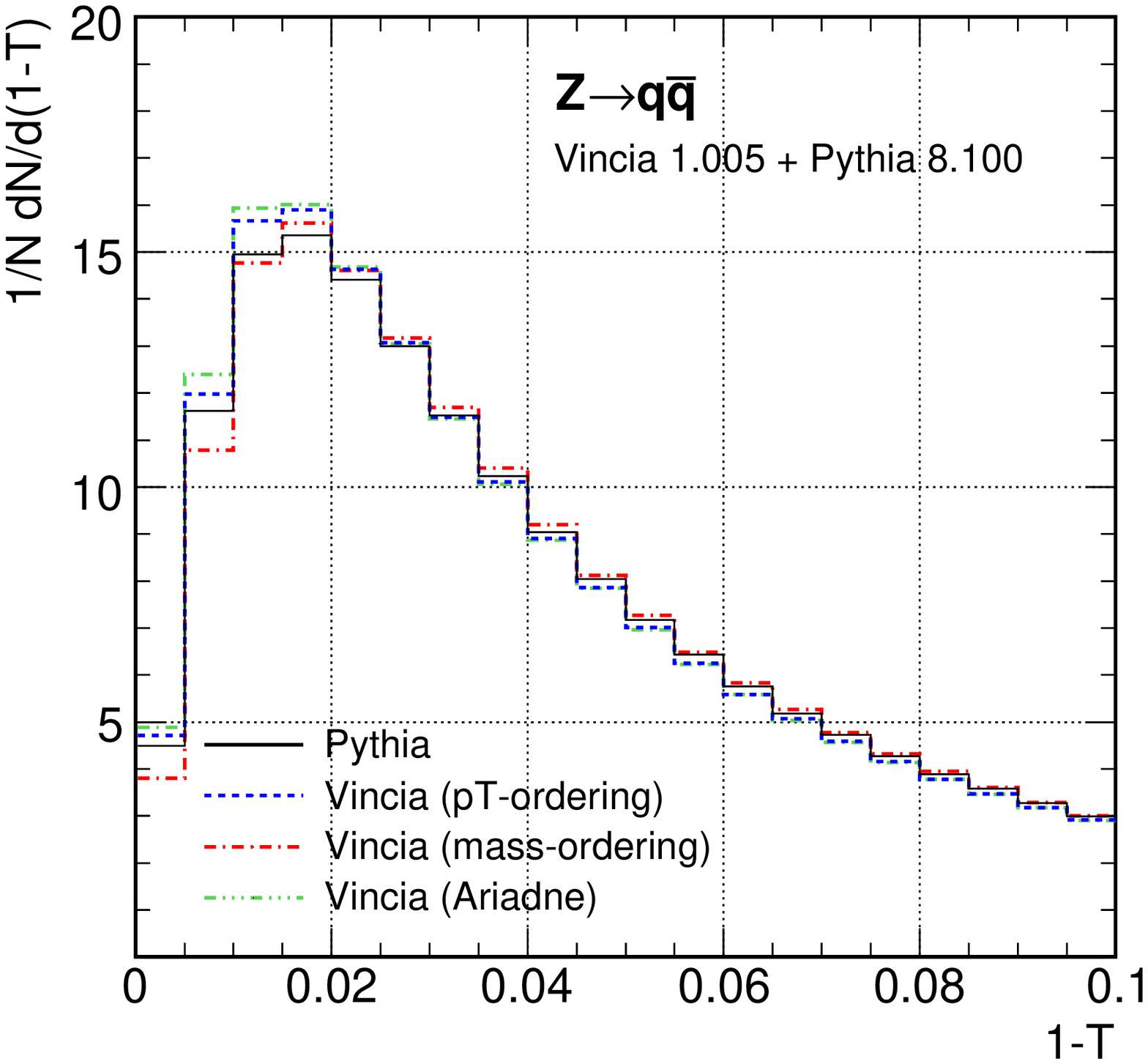}}\vspace*{-0.7cm}\\
\hspace*{-3mm}\includegraphics*[scale=0.44]{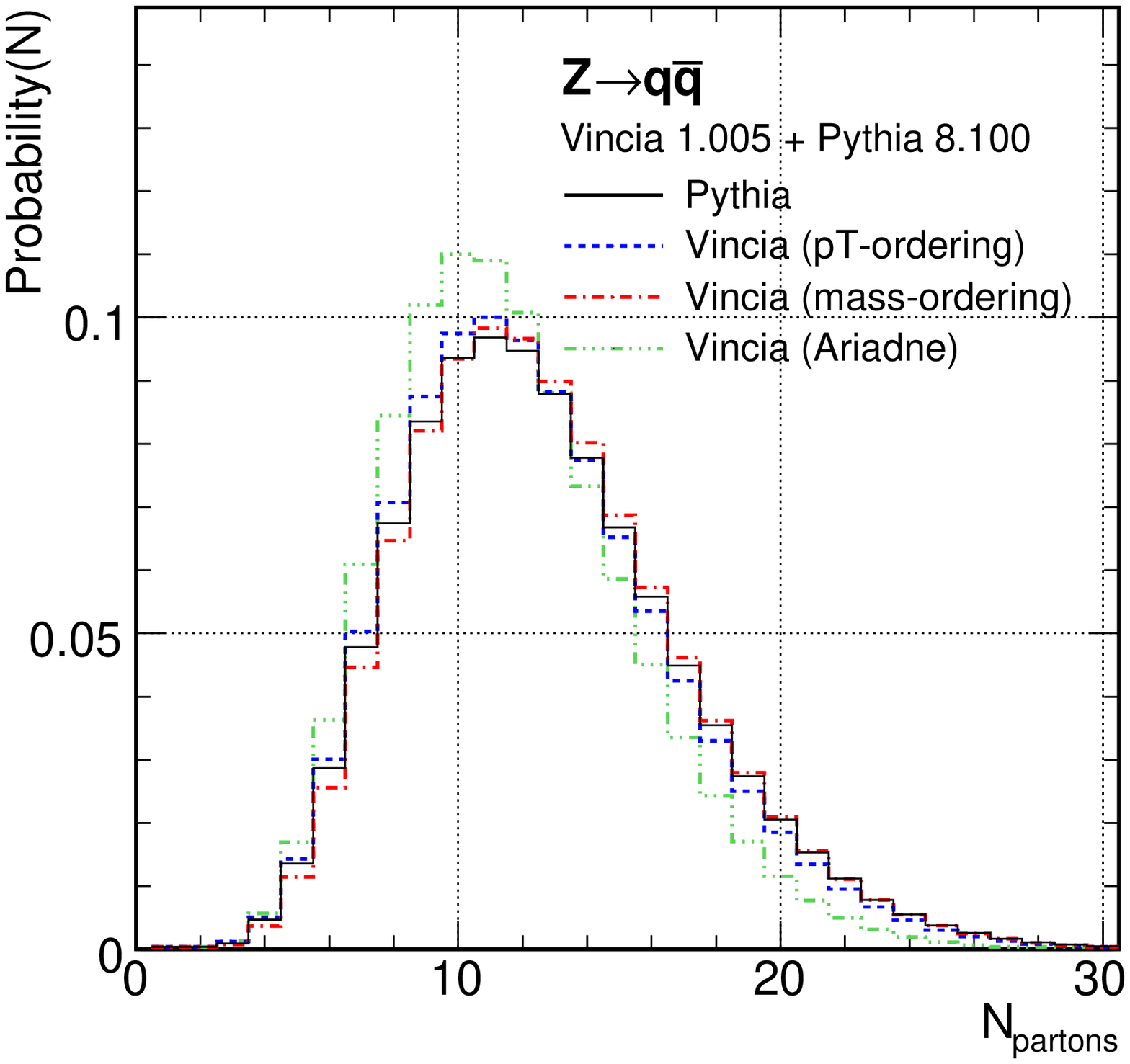}\hspace*{-0.9cm}
\includegraphics*[scale=0.44]{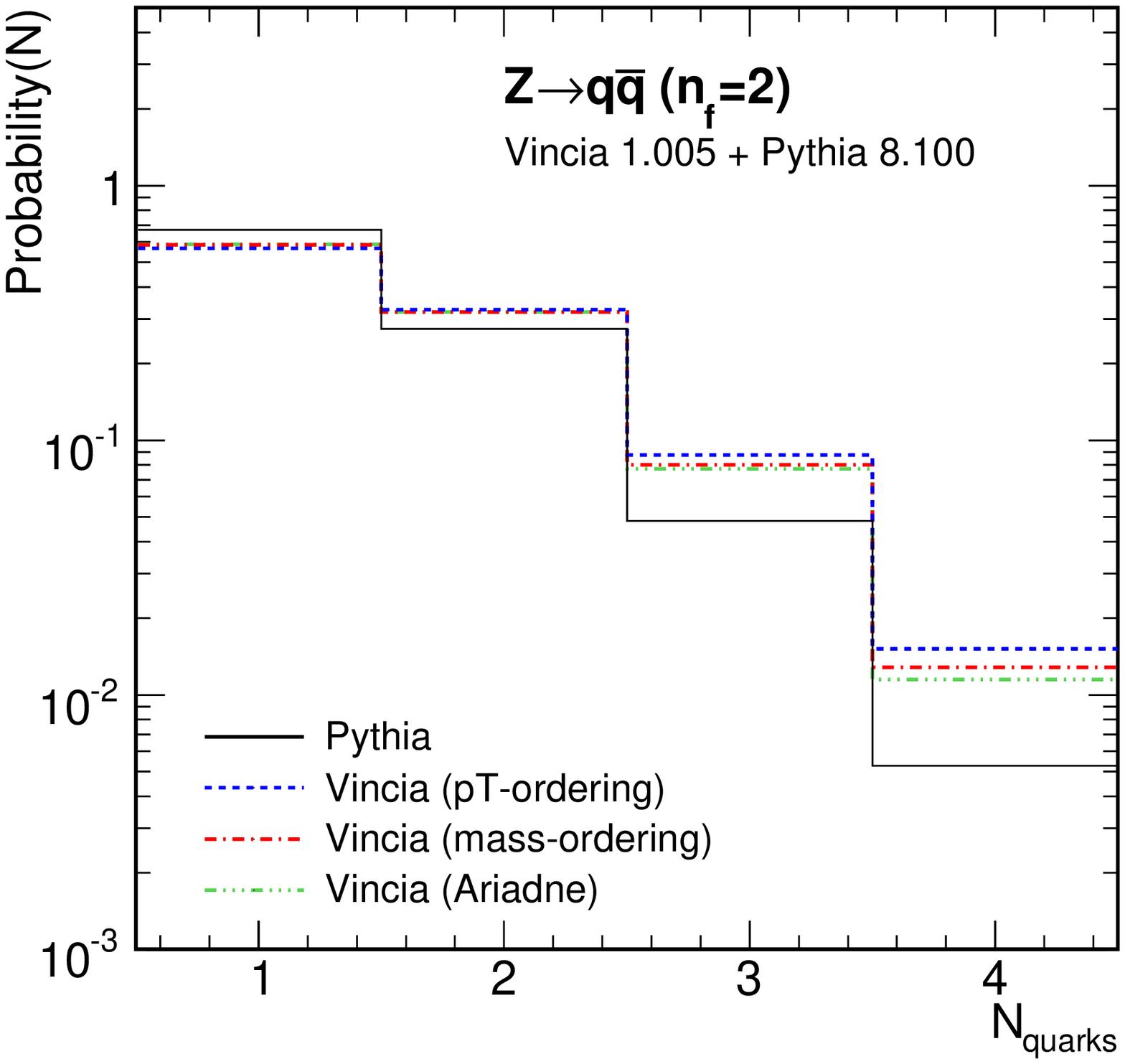}\vspace*{-1cm}
\caption{Top row: Thrust, $1-T$. Bottom row: Number of partons (left)
  and number of quarks (right) at shower termination, with 2 massless
  quark flavors.\label{fig:thrustnpartons}} 
\end{figure}

Finally, as illustration of an infrared sensitive quantity, in the
bottom row of fig.~\ref{fig:thrustnpartons} we plot the probability
distribution of the number of partons produced at the shower
termination for each of the four models.  The total number of partons
is shown to the left and the number of quarks (not counting
anti-quarks) to the right.  The definitions of $p_{\perp}$ in
\textsc{Pythia}\ and in \textsc{Vincia}/\textsc{Ariadne},
respectively, are not exactly identical, but they have the same
infrared limiting behavior \cite{Sjostrand:2004ef}, and hence a
comparison of the number of resolved partons with a cutoff at
$p_{\perp\mathrm{had}}=0.5\GeV$ should be meaningful.  Since we have
also chosen the 
same $\alpha_s$ values etc., the basic agreement between the models in
the lower left-hand plot in fig.~\ref{fig:thrustnpartons} reconfirms
that there are no large differences between the showers, even at the
infrared sensitive level. \textsc{Ariadne}\ produces somewhat fewer
partons, consistent with the \textsc{Ariadne}\ radiation functions
being slightly softer. On the right-hand plot, however, it is
interesting to note the first substantial difference between
\textsc{Pythia} 8 and the \textsc{Vincia}\ showers. The
\textsc{Pythia}\ shower produces significantly fewer quarks than any
of the \textsc{Vincia}\ showers, despite its being higher or
comparable on the total number of partons (cf.~the left-hand plot). A
similar difference between parton and dipole-antenna showers was
observed in an earlier \textsc{Ariadne} study \cite{Andersson:1989ki},
in which a comparison was made to the virtuality-ordering of
traditional parton showers. It is interesting that we here observe the
same trend when comparing to the \textsc{Pythia} 8 shower which is
ordered in $p_\perp$.  Finally, we note that this difference will also
have practical consequences; in the context of tuning of hadronization
models, the \textsc{Vincia}\ showers will presumably need a stronger
suppression of non-perturbative strangeness production to make up for
the larger perturbative production rate, as compared to
\textsc{Pythia} 8.

\subsection{Conclusions}
We have presented the inclusion of massless quarks into the
\textsc{Vincia}\ shower algorithm, implemented as a plug-in to the
\textsc{Pythia} 8 event generator. The dipole-antenna radiation
functions are expressed as double Laurent series in the branching
invariants, with user-specifiable coefficients. At the analytical
level, we compare the coefficients of the ``GGG'' antenna functions
\cite{GehrmannDeRidder:2005cm} used by default in \textsc{Vincia} to
the \textsc{Ariadne} ones \cite{Lonnblad:1992tz}. Modulo a
re-parameterization of emissions from gluons, we find the double and
single log coefficients to be identical, as expected. The finite
terms, however, are generally somewhat smaller for the
\textsc{Ariadne} functions. This represents a genuine shower ambiguity
which can only be systematically addressed by matching to fixed-order
matrix elements.

At the phenomenological level, we have also compared to the hybrid
parton-dipole shower in \textsc{Pythia} 8 \cite{Sjostrand:2007gs} for
$e^+e^-\to Z \to q\bar{q}$ at $\sqrt{s}=m_Z$. We find a good overall
agreement, even at the level of an infrared sensitive quantity such as
the final number of partons. For the number of quarks produced,
however, \textsc{Pythia} 8 is markedly lower than any of the
\textsc{Vincia}\ showers we have compared to here.

\subsection*{Acknowledgements}
We thank J.~Andersen, Y.~Dokshitzer, G.~Marchesini, Z.~Nagy,
T.~Sj\"ostrand, D.~Soper, and G.~Zanderighi for enlightening
discussions. DAK is supported in part by the Agence Nationale de la
Recherche of France under grant ANR-05-BLAN-0073-01. This work has
been partially supported by Fermi Research Alliance, LLC, under
Contract No.\ DE-AC02-07CH11359 with the United States Department of
Energy.  


}


\section[LLL subtraction and PS kinematics]
{LLL SUBTRACTION AND PS KINEMATICS%
\protect\footnote{Contributed by: S. Odaka}}
{\graphicspath{{odaka/}}
%
%
%
%
%

\subsection{Introduction}
We are developing NLO event generators for hadron collision interactions 
based on GRACE \cite{Ishikawa:1993qr},
using the Limited Leading-Log (LLL) subtraction technique \cite{Odaka:2007gu}
for the parton radiation matching.
The matching technique is crucial 
since the contributions of an additional QCD parton radiation in NLO 
are also involved in the evolution of Parton Distribution Functions (PDFs) 
in a collinear approximation.
A naive application of a PDF to NLO calculations results in an apparent 
double-counting.
We avoid the double-counting by subtracting Leading-Log (LL) collinear 
contributions from the matrix element (ME) of radiative processes.
The subtraction is stopped ("limited") at the factorization scale ($\mu_{F}$)
since PDFs do not involve any radiation harder than this energy scale.
The LL contribution of the radiation is easy to calculate \cite{Kurihara:2006kt}, 
though an appropriate care is necessary in the kinematical mapping 
to non-radiative processes \cite{Odaka:2007gu}.
The subtracted LL terms are formally moved to non-radiative processes 
and to be cancelled with divergences in virtual corrections.

\begin{figure}
\begin{center}
\includegraphics[width=0.5\textwidth]{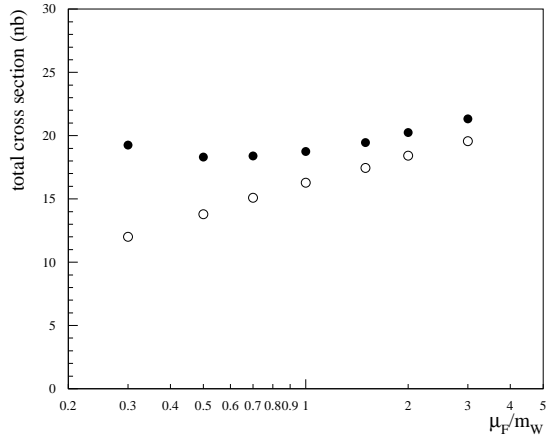}
 \caption{Factorization-scale ($\mu_{F}$) dependence of the total cross section 
 for the $W$-boson production at LHC.
 An apparent $\mu_{F}$ dependence of the inclusive $W$ ($W$ + 0 jet) production 
 cross section (open circles) is greatly reduced when we add the LLL-subtracted 
 $W$ + 1 jet production cross section. The summed cross sections are shown with 
 filled circles.
}
\label{fig:total-xsec}
\end{center}
\end{figure}

Figure \ref{fig:total-xsec} shows the sum of the total cross sections for inclusive 
$W$-boson production and LLL-subtracted $W$ + 1 jet production evaluated 
for the LHC condition (proton-proton collisions at $\sqrt{s} = 14$ TeV). 
Here, "jet" denotes a gluon or a light quark in the final state.
The cross sections are calculated using the tree-level MEs for $W$ production 
and $W$ + 1 jet production, respectively, convoluted with the CTEQ5L PDF 
\cite{Lai:1999wy}.
Results are shown as a function of the factorization scale ($\mu_{F}$).
We can see a strong $\mu_{F}$ dependence of the inclusive $W$ production 
cross section (open circles) is greatly reduced by adding the LLL-subtracted 
radiative cross section.
This shows a good matching between the ME and PDF; 
{\it i.e.}, the LL contents in ME and PDF are nearly the same.

The virtual corrections are yet to be included in the results shown 
in Fig. \ref{fig:total-xsec}.
They can also be evaluated automatically in the framework of GRACE \cite{Kurihara:2006kt}.
Divergent terms in these corrections are to be cancelled with those moved from 
radiative processes.
Remaining finite terms will alter the normalization of non-radiative processes,
and will result in a substantial mismatch 
since there is no such correction in radiative processes.
However, this mismatch is at the level of NNLO.
It will be possible to restore the matching within the accuracy of NLO.
The simplest way would be to change the normalization of LL components 
of the hard radiation remaining in radiative processes 
by the same amount as applied to non-radiative processes.
This is actually a modification at the NNLO ($\alpha_{s}^{2}$) level.

So far we have discussed the matching in the integrated cross section.
We have to achieve a good matching in differential cross sections, as well, 
in order to construct practical event generators.
The QCD evolution evaluated in PDFs is simulated by means of a parton shower (PS) 
in event generators for hadron collisions.
PS and PDF are based on the same factorization theory.
However, since theoretical arguments are given only at the collinear limit, 
the theory gives us predictions only at the first-order approximation 
for the transverse behavior.
It is necessary to introduce a model of 3-dimensional kinematics 
in order to construct a practical PS conserving the energy and momenta.
The introduction of a suitable model is crucial for achieving 
a good matching in differential cross sections.
We discuss about such models in the following sections.

\subsection{Initial-state PS kinematics}
We have constructed an initial-state Leading-Log (LL) PS program for the use 
in NLO event generation.
The program is based on the simplest expression of the LL Sudakov form factor 
employing $Q^{2}$ as the evolving parameter,
\begin{equation}\label{sudakov1}
	S(Q_{1}^{2}, Q_{2}^{2}) = \exp\left[ - \int_{Q_{1}^{2}}^{Q_{2}^{2}}
	{dQ^{2} \over Q^{2}} \int_{\epsilon}^{1-\epsilon} dz \  
	{\alpha_{s}(Q^{2}) \over 2\pi}\ P(z) \right] .
\end{equation}
The details are described in our paper \cite{Odaka:2007gu}.
We stay in a naive LL implementation without introducing corrections partially 
incorporating higher order effects, such as the angular ordering, 
because we plan to extend our PS to a true Next-to-Leading-Log (NLL) 
approximation \cite{Tanaka:2003gk}.

We first tested the kinematics model employed in the "old" PYHTIA-PS 
\cite{Sjostrand:1985xi,Bengtsson:1986gz}, 
since the theoretical bases is nearly the same. 
We found this model gives a very soft transverse activity. 
It results in an apparent mismatch in the transverse momentum ($p_{T}$) 
distribution of $W$ bosons, 
when we tried to merge the inclusive $W$ production with the LLL-subtracted 
$W$ + 1 jet production by applying this PS to both processes.
The starting assumptions of the "old" PYHTIA-PS kinematics are 
that the $z$ parameter of a branch is the ratio of squared cm energies after and 
before each branch instead of the fraction of light-cone momenta, 
and that the $Q^{2}$ is identical to the virtuality of the evolving partons.
The first assumption requires the definition of a "target" parton; 
thus, it is model dependent.
However, this $z$ definition ensures a simple relation 
between squared cm energies of a hard interaction and the beam collision; 
$s_{\rm hard} = x_{1} x_{2} s_{\rm beam}$ 
where $x_{1}$ and $x_{2}$ are given by the product of all $z$ values in each beam.
From a simple kinematical argument we found this model gives a relation, 
\begin{equation}
p_{T}^{2} = (1-z)^{2}Q^{2},
\label{eq:pt-oldpythia}
\end{equation}
for each branch at the soft limit \cite{Odaka:2007gu}.

On the other hand, ordinary arguments based on the massless approximation 
give a slightly different relation, 
\begin{equation}
p_{T}^{2} = (1-z)Q^{2}
\label{eq:pt-prefix}
\end{equation}
at the soft limit.
Apparently Eq. (\ref{eq:pt-oldpythia}) gives a smaller $p_{T}$ value than 
Eq. (\ref{eq:pt-prefix}) for a given set of $Q^{2}$ and $z$.
The relation (\ref{eq:pt-prefix}) must be better for the matching 
since external partons are nearly massless in ME calculations.
We have introduced a new kinematics model where $p_{T}$ of each branch is 
given ("prefixed") by Eq. (\ref{eq:pt-prefix}). 
We keep the definition of the $z$ parameter.
The momenta of evolving partons are calculated from this $p_{T}$ value 
and the $z$ value.
Thus, the virtuality is not necessarily equal to the $Q^{2}$ of a branch.
This new PS gives a harder $W$-boson $p_{T}$ spectrum than the 
"old" PYTHIA-PS in the inclusive $W$ production simulation, 
showing a better matching to the LLL-subtracted $W$ + 1 jet simulation.
The sum of the two simulations gives a smooth $p_{T}$ spectrum stable against 
a variation of the factorization scale ($\mu_{F}$) \cite{Odaka:2007gu}.

After the submission of the paper \cite{Odaka:2007gu} 
we tried another definition of the "prefixed" $p_{T}$, 
\begin{equation}
p_{T}^{2} = (1-z-Q^{2}/\hat{s})Q^{2}.
\label{eq:pt-prefix-new}
\end{equation}
The parameter $\hat{s}$ is the squared cm energy before the branch.
This is the result of the massless approximation of branching kinematics 
before taking the soft limit ($Q^{2}/\hat{s} \rightarrow 0$).
This definition is ugly in some sense since $\hat{s}$ is model dependent, 
but gives us a better matching than Eq. (\ref{eq:pt-prefix}).
We plot the summed $p_{T}$ spectra of $W$-bosons for three different 
$\mu_{F}$ values ($\mu_{F}/m_{W}$ = 0.5, 1.0 and 1.5) 
in Fig. \ref{fig:pt-spectrum}.
We can see almost no variation of the spectrum except for a small difference 
around $p_{T} = m_{W}$ in this $\mu_{F}$ range.

\begin{figure}
\begin{center}
\includegraphics[width=0.5\textwidth]{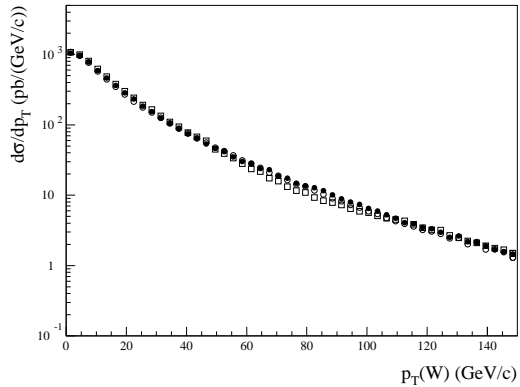}
 \caption{Sum of the simulated $p_{T}$ spectra of $W$ bosons 
 for the inclusive $W$ production and the LLL-subtracted $W$ + 1 jet 
 production at LHC.
 The new $p_{T}$-prefixed PS described in the text is applied to
 both processes.
 Results are plotted for three different choices of $\mu_{F}$: 
 $\mu_{F}/m_{W}$ = 0.5 (open circles), 1.0 (filled circles) 
 and 1.5 (open squares).
}
\label{fig:pt-spectrum}
\end{center}
\end{figure}

\subsection{Prospects for the final-state PS matching}
It is enough to consider the initial-state matching 
if we consentrate ourselves to NLO corrections for color singlet or heavy particle 
productions.
However, once we go to NLO for those processes having 
a gluon or a light quark ("jet") in the final state, 
we also need to consider the matching in the final state.

We plan to use a simple LL parton shower employing $Q^{2}$ as the evolving parameter 
also for the final state.
We need to introduce an appropriate kinematics model to this PS, too.
In the initial-state PS, 
models in which the definition of $p_{T}$ precedes that of $Q^{2}$ give us 
a better matching as we have discussed in the previous section.
This is because $p_{T}$ is in principle an observable quantity 
while $Q^{2}$ is not physically well-defined for initial-state partons.
Similar arguments should be done also for the final state.

In the final state, 
the virtuality is in principle an observable as an invariant mass of particles 
even after the hadronization and decays.
Therefore, it must be natural to identify $Q^{2}$ as the virtuality 
of the evolving partons.
The transverse momentum ($p_{T}$) is also an observable in principle.
Thus, $z$ should be treated as an unphysical parameter.
It should be used as a hidden parameter only to give $p_{T}$ values 
according to the relation, 
\begin{equation}
p_{T}^{2} = z(1-z)Q^{2}.
\label{eq:pt-final}
\end{equation}
In this kinematics model, 
PS is a process to give final-state partons additional masses 
equal to the $Q^{2}$ of their first branches.
The invariant mass of the hard interaction system should be unchanged 
even after the application of PS, 
since it is a very fundamental parameter for the evaluation of matrix elements.
We also want to keep the production angles in the cm frame unchanged.
These requirements can be fulfilled by introducing a common multiplication factor 
to the momenta of all final-state particles.

We need to apply a proper mapping of non-radiative subsystem in a radiative 
event to an on-shell non-radiative event in the LL subtraction.
A mapping using momenta of the branched parton and the target parton works well 
for the initial-state radiation \cite{Odaka:2007gu}.
The subsystem is boosted and rotated to its cm frame where the momenta of 
two incoming partons are aligned along the $z$ axis.
This is the process exactly reversing the kinematical rearrangement 
in our initial-state PS.

The mapping should be done in the same concept also in the final state, 
exactly reversing the rearrangement in PS.
It can be done as follows: pick up an arbitrary pair of final-state partons.
If they can be considered as products of a PS branch, 
replace them with the parent parton having the invariant mass 
of the pair as its virtuality ($Q^{2}$).
If not, skip this pair.
Rescale the momenta of all particles in the cm frame with a common factor 
to make the replaced parent parton become on-shell.
Evaluate the matrix element of the non-radiative process based on these 
rearranged momenta, multiply it with the LL radiation factor proportional to 
$1/Q^{2}$, then we get an LL approximation of a final-state radiation.
This procedure should be applied to all possible combinations 
if we have more than two partons in the final state.

We expect that the LL contribution can be evaluated 
in such a systematic way, including the initial-state contributions, as well.
All contributions should be summed to evaluate the total LL contribution.
A program is under development based on these concepts.

\subsection{Conclusions}
We have achieved a good matching between PDF and matrix-element (ME) evaluations 
for the parton radiation in NLO QCD corrections, 
by using the Limited Leading-Log (LLL) subtraction technique.
It has been demonstrated as a good stability of the $W$ production cross section 
against a variation of the factorization scale ($\mu_{F}$), 
where the total cross section is evaluated by the sum of the cross sections 
for inclusive $W$ production and the LLL-subtracted $W$ + 1 jet production.

We have to achieve a good matching between the parton shower (PS) and ME, as well,
in order to construct practical NLO event generators. 
The transverse activity of PS depends on the applied kinematics model of parton branches.
We have successfully built a suitable model for our Leading-Log (LL) 
initial-state PS, 
where $p_{T}$ is prefixed according to the relation in the massless approximation
of branching kinematics.
The simulation employing this PS shows a good matching 
between the inclusive $W$ production and the LLL-subtracted $W$ + 1 jet production 
in the $p_{T}$ spectrum of $W$ bosons.
The spectrum is stable against the variation of $\mu_{F}$ in a wide range.

It is necessary to achieve a good PS-ME matching for the final-state radiation, as well, 
when we construct NLO event generators for those processes including "jet(s)" 
in the final state. 
A study is in progress for the final state based on the experience 
on the initial-state radiation.

\subsection*{Acknowledgements}
This work has been carried out as an activity of the NLO Working Group, 
a collaboration between the Japanese ATLAS group and the numerical analysis 
group (Minami-Tateya group) at KEK.
The author wishes to acknowledge useful discussions with the members: 
Y. Kurihara, J. Kodaira, J. Fujimoto, T. Kaneko and T. Ishikawa of KEK, 
and K. Kato of Kogakuin University.

%
}


\section[A parton-shower model based on Catani--Seymour dipole factorisation]
{A PARTON-SHOWER MODEL BASED ON CATANI--SEYMOUR DIPOLE FACTORISATION%
\protect\footnote{Contributed by: S.~Schumann, F.~Krauss}}
{\graphicspath{{schumann/}}
%
%
%
%
%

\subsection{Introduction}

Parton-shower models form an indispensable building block of Monte Carlo event 
generators, such as Herwig \cite{Corcella:2000bw}, Pythia \cite{Sjostrand:2000wi} 
and Sherpa \cite{Gleisberg:2003xi}, that aim at the realistic description 
of multi-particle final states as they are observed in high-energy collider 
experiments. By accounting for QCD bremsstrahlung processes, parton showers relate a 
small number of partons emerging from a hard interaction, defined at scale 
$Q_{\rm hard}$ and theoretically described through a fixed order calculation, to a 
larger set of partons at scales $Q_o\ll Q_{\rm hard}$. The parton-shower approach 
relies on the universal pattern of QCD emission processes once soft or collinear 
parton kinematics are considered. The soft and collinear phase-space regions are 
singular and obtain large corrections order by order in perturbation theory what 
makes an all-orders resummation of the associated kinematical logarithms essential. 
Most shower algorithms rely on collinear factorisation of QCD matrix elements and 
are accurate to the leading-logarithmic level. The Ariadne approach, however, 
is based around the soft limits \cite{Lonnblad:1992tz}.

\noindent
The parton-shower approach being perturbative it cannot be extended to arbitrary
small scales but has to be stopped at some infrared cut-off scale 
$Q_o \geq \Lambda_{\rm QCD}$. Below that scale event generators model the transition
of QCD partons into the experimentally observed hadrons through non-perturbative
hadronisation models. In fact, only through the incorporation of parton showers 
these hadronisation models can be made universal or independent of the 
underlying hard process. This, however, assumes that perturbative QCD between
scales $Q_{\rm hard}$ and $Q_o$ is appropriately described by the parton-shower 
model used.

\noindent
In the past few years there have been lots of major improvements related to 
parton-shower Monte Carlos. This includes the incorporation of exact multi-leg 
tree-level matrix elements for the description of the first few hardest emissions 
from a given hard process, know as ``matrix element parton shower merging'', 
see e.g. \cite{Catani:2001cc,Alwall:2007fs}, or the consistent matching of 
next-to-leading order calculations with parton showers, know as 
``Monte Carlo at NLO'', see for instance \cite{Frixione:2002ik,Frixione:2007vw}. 
In addition the available shower algorithms of Herwig and Pythia have been revised 
and improved \cite{Sjostrand:2004ef,Gieseke:2003rz}.

\noindent 
Only very recently new shower algorithms emerged that are based on formalisms used to
construct subtraction terms that allow for a numerical cancellation of infrared 
singularities in NLO QCD calculations 
\cite{Nagy:2005aa,Nagy:2006kb,Giele:2007di,Winter:2007ye,Dinsdale:2007mf,Schumann:2007mg}. 
There exist now implementations of such shower algorithms for two commonly used 
subtraction schemes, the antenna subtraction method \cite{Kosower:1997zr} 
and the Catani--Seymour dipole formalism \cite{Catani:1996vz,Catani:2002hc}. Besides 
incorporating the last knowledge on the infrared behaviour of QCD matrix elements, these 
models should largely facilitate the matching with NLO calculations carried out in the 
respective scheme. In this note we briefly report on the construction of a parton-shower 
algorithm relying on Catani--Seymour subtraction that has more extensively been presented in 
\cite{Schumann:2007mg}.

\subsection{The shower model}

The Catani--Seymour formalism provides all the ingredients to construct a local 
approximation to the real-correction matrix element in {\it any} QCD NLO calculation. 
These subtraction terms, that can be constructed in a process-independent way, possess 
exactly the same infrared divergences as the real-emission correction, such that the 
difference of the two is infrared finite and can safely be (numerically) integrated 
in four dimensions. In addition, the subtraction terms are chosen such, that they 
can be analytically integrated in $d=4-2\epsilon$ dimensions over the phase space of
the produced soft or collinear parton that causes the divergences. The occurring 
$1/\epsilon^2$ and $1/\epsilon$ poles exactly cancel the ones from the loop integration 
in the virtual part when adding the two pieces. Such, the Catani--Seymour method provides 
a way to construct a parton-level Monte Carlo program for a NLO calculations once the 
one-loop and real-emission corrections to the Born process are known. 

\noindent 
In the Catani--Seymour approach the additional soft or collinear parton is emitted 
from an emitter-spectator pair (called dipole). Considering both the emitter and
the spectator to be either in the final or initial state, four configurations have 
to be considered, representing the singularities associated to emissions from the final
or initial state. Labelling final-state particles by $i,j$ and $k$ and initial-state 
partons by $a$ and $b$ the real-emission matrix element can always be approximated
by the sum over all the possible dipoles, 
\begin{eqnarray}\label{eq:CSMaster1}
|{\cal M}_{m+1}|^2=
 \sum_{i,j}\sum_{k\neq i,j}{\cal D}_{ij,k}
+\left[\sum_{i,j}{\cal D}_{ij}^a
+\sum_{i}\sum_{k\neq i}{\cal D}_k^{ai}
+\sum_{i}{\cal D}^{ai,b} +(a\leftrightarrow b)\right]
\,.
\end{eqnarray}
Hereby, ${\cal D}_{ij,k}$ describe splittings of a final-state parton $\widetilde{ij}$ into the
pair $i,j$ accompanied by a spectator $k$. Due to the presence of the spectator, 
four-momentum conservation and on-shell momenta can be accomplished locally for each 
individual splitting. The terms ${\cal D}_{ij}^{a}$ represent final-state splittings
with an initial-state spectator, while ${\cal D}_k^{ai}$ and ${\cal D}^{ai,b}$ 
correspond to a splitting initial-state line accompanied by a final- and initial-state
parton, respectively. The individual dipole terms are constructed from the Born 
matrix element by inserting colour- and spin-dependent operators that describe the 
actual splitting. For massless final-state emitters and final-state spectators, 
for instance, the dipole contributions read
\begin{eqnarray}\label{eq:FFdipolecontrib}
{\cal D}_{ij,k} = -\frac{1}{2p_ip_j}\hspace*{4mm}
\langle_{\!\!\!\!\!\!\!\!\!\!\;m}\,\, 1,\,\dots,\,\widetilde{ij}\,\dots,\,\tilde k,\,\dots|
             \frac{{\bf T}_k\cdot{\bf T}_{ij}}{{\bf T}^2_{ij}}{\bf V}_{ij,k}
             |1,\,\dots,\,\widetilde{ij}\,\dots,\,\tilde k,\,\dots\rangle_m\,.
\end{eqnarray}
The ${\bf T}_{ij}$ and ${\bf T}_k$ thereby denote the colour charge operators of the 
emitter and spectator, respectively, they lead to colour correlations in the full
amplitude. The ${\bf V}_{ij,k}$ are $d$-dimensional matrices in the emitter's spin 
space that induce spin correlations. 

\noindent
For the construction of a parton-shower algorithm from the dipole formula 
Eq.~(\ref{eq:CSMaster1}) certain approximations are needed that finally 
allow for an exponentiation of the splitting operators to derive the Sudakov
form factors central for a shower implementation. In addition, the splitting
kinematics, choices on scale settings and the actual shower-ordering parameter 
have to be fixed.

\subsubsection{Shower construction criteria}

The full colour correlations present in the $|{\cal M}_{m+1}|^2$ matrix element
have to be discarded in the shower picture, instead the leading terms in $1/N_c$ are 
considered only \footnote{Although formally subleading, we consider splittings of 
the type $g\to q\bar q$ as well}. In this approximation a colour flow can be assigned 
to each parton configuration. Motivated by considerations on the colour dynamics for
soft emissions, we choose the emitter and spectator to be colour connected in the shower 
formalism. The colour-charge operators simplify to
\begin{eqnarray}
-\frac{\bf{T}_{k}\cdot\bf{T}_{ij}}{\bf{T}^2_{ij}} \to 
\frac{1}{{\cal N}^{spec}_{ij}}\,,
\end{eqnarray}
with ${\cal N}^{spec}_{ij} =1,2$ in case the emitter has one ($SU(3)$ (anti-)triplet) 
or two ($SU(3)$ octet) possible spectators. The four-dimensional dipole functions 
${\bf V}$ are used as the shower splitting functions. Furthermore, we neglect spin 
correlations by using spin-averaged splitting functions $\langle{\bf V}\rangle$ 
\footnote{Some of the dipole functions can become negative in non-singular phase-space
region, prohibiting a simple probabilistic interpretation. We choose to set them to zero 
in these cases.}. 

\noindent
As shower evolution variable we choose the transverse momentum between the splitting
products for branching final-state partons and the transverse momentum with respect to
the beam for emissions from the initial state, collectively denoted by $\bf{k}_\perp$.
This scale is also employed as the scale of the running coupling and the parton 
distributions, once initial-state partons are present.

\noindent
Based on the above approximations and choices Sudakov form factors corresponding
to the different types of Catani--Seymour dipoles can be derived, that determine
the probability for a certain branching not to occur for a given range of the 
evolution variable. The four generic cases are briefly reviewed in the following. For
simplicity, here we consider massless partons only, the massive case is discussed
in \cite{Schumann:2007mg}.
\subsubsection{Final-state emitter -- final-state spectator}
Consider the final-state splitting $\{\widetilde{ij},\tilde{k}\} \to \{i,j,k\}$ with 
the four-momentum constraint $\tilde p_{ij}+\tilde p_k = p_i+p_j+p_k \equiv Q\,$ and
all momenta being on their mass-shell. The branching can be characterised by the 
Lorentz invariant variables 
\begin{eqnarray}
        y_{ij,k}=\frac{p_ip_j}{p_ip_j+p_ip_k+p_jp_k}\,,\quad 
        \tilde z_i=1-\tilde z_j= \frac{p_ip_k}{p_ip_k+p_jp_k}\,.
\end{eqnarray}
The factorised form of the fully differential $(m+1)$-parton cross section that 
exactly reproduces the corresponding soft and collinear divergences of the 
real-emission process reads
\begin{eqnarray}
\mbox{d}\hat\sigma_{m+1} 
= 
\mbox{d}\hat\sigma_m\,\sum\limits_{ij}\sum\limits_{k\neq ij}\,
\frac{\mbox{d} y_{ij,k}}{y_{ij,k}}\,d\tilde z_i\,\frac{\mbox{d}\phi}{2\pi}\,
\frac{\alpha_{\rm s}}{2\pi}\,\frac{1}{{\cal N}^{spec}_{ij}}\, (1-y_{ij,k})
\langle {\bf{V}}_{ij,k}(\tilde z_i,y_{ij,k}) \rangle\,. 
\end{eqnarray}
The spin-averaged splitting kernels $\langle{\bf{V}}_{ij,k}\rangle$ for the branchings 
$q\to qg$, $g\to gg$ and $g\to q\bar q$ read
\begin{eqnarray}
\langle {\bf{V}}_{q_ig_j,k}(\tilde z_i,y_{ij,k})\rangle 
&=&
C_{\rm F}\left\{\frac{2}{1-\tilde z_i+\tilde z_iy_{ij,k}}-(1+\tilde z_i)\right\}
\label{eq:FFPqq}\,,\\
\langle {\bf{V}}_{g_ig_j,k}(\tilde z_i,y_{ij,k})\rangle 
&=&
2C_{\rm A}\left\{\frac{1}{1-\tilde z_i+\tilde z_iy_{ij,k}}+
                \frac{1}{\tilde z_i+y_{ij,k}-\tilde z_iy_{ij,k}}-
                2+\tilde z_i\,(1-\tilde z_i)\right\}
\label{eq:FFPgg}\,,\\
\langle {\bf{V}}_{q_iq_j,k}(\tilde z_i)\rangle
&=&
T_{\rm R}\left\{1-2\tilde z_i\,(1-\tilde z_i)\right\}
\label{eq:FFPgq}\,.
\end{eqnarray}
In terms of the splitting variables the transverse momentum between the splitting 
products $i$ and $j$ (our shower evolution variable) can then be written as  
\begin{eqnarray}
{\bf k}_\perp^2 = 2\tilde p_{ij}\tilde p_{k}\,\,y_{ij,k}\,\tilde z_i\,(1-\tilde z_i)\,,
\end{eqnarray}
and accordingly 
\begin{eqnarray}
\frac{\mbox{d} y_{ij,k}}{y_{ij,k}}=\frac{\mbox{d}{\bf k}_\perp^2}{{\bf k}_\perp^2}\,.
\end{eqnarray}
Setting the infrared shower cut-off equal to ${\bf k}_{\perp,0}^2$ and the upper limit
to ${\bf k}_{\perp,{\rm max}}^2$ the $\tilde z_i$ integration is constrained to  
\begin{eqnarray}
z_\mp({\bf k}_{\perp,\rm max}^2,{\bf k}_{\perp,0}^2) &=& 
\frac12\left(1\mp\sqrt{1-\frac{{\bf k}_{\perp,0}^2}{{\bf k}_{\perp,{\rm max}}^2}}\right)\,.\label{eq:FFzboundaries}
\end{eqnarray}
The kinematics of the splitting are fixed through
\begin{eqnarray}
p_i 
&=& 
\hphantom{(1-\tilde)}\tilde z_i \,\tilde p_{ij}+
\frac{\hphantom{(1}{\bf k}_\perp^2\hphantom{(())}}
     {\hphantom{(1-}\tilde z_i\,2\tilde p_{ij}\tilde p_k\hphantom{))}}\,
\tilde p_k + k_\perp\,,
\label{eq:FFpiKINmassless}\\
p_j 
&=& 
(1-\tilde z_i) \,\tilde p_{ij}+
\frac{{\bf k}_\perp^2}{(1-\tilde z_i)\,2\tilde p_{ij}\tilde p_k}\,\tilde p_k - k_\perp\,,
\label{eq:FFpjKINmassless}\\
p_k 
&=& 
\left(1-y_{ij,k}\right)\tilde p_k
\label{eq:FFpkKINmassless}\,,
\end{eqnarray}
with $k_\perp$ the spacelike transverse-momentum vector perpendicular to $\tilde p_{ij}$
and $\tilde p_k$ and $k_\perp\cdot k_\perp = -{\bf k}_{\perp}^2$. The Sudakov form factor
for having no final-state splitting with a final-state spectator between 
${\bf k}_{\perp,{\rm max}}^2$ and ${\bf k}_{\perp,0}^2$ reads 
\begin{eqnarray}
 \lefteqn{\Delta_{\rm FF}({\bf k}_{\perp,{\rm max}}^2,{\bf k}_{\perp,0}^2)}\nonumber\\
        &=& 
        \exp\left(-\sum\limits_{ij}\sum\limits_{k\neq ij}\,
                \frac{1}{{\cal N}^{spec}_{ij}}\,
                \int\limits_{{\bf k}_{\perp,0}^2}^{{\bf k}_{\perp,{\rm max}}^2}
                \frac{\mbox{d} {\bf k}_{\perp}^2}{{\bf k}_\perp^2}\,
                \int\limits_{z_-}^{z_+}\mbox{d}\tilde z_i\,
                \frac{\alpha_{\rm s}({\bf k}_\perp^2)}{2\pi}\,(1-y_{ij,k})
                \langle {\bf{V}}_{ij,k}(\tilde z_i,y_{ij,k}) \rangle\right)\,.
\end{eqnarray}
\subsubsection{\label{sec:FI}Final-state emitter -- initial-state spectator}
In the presence of initial-state partons a final-state splitter may be colour
connected to one of the incoming lines. We consider the splitting 
$\{\widetilde{ij},\tilde a\}\to \{i,j,a\}$, with 
$\tilde p_{ij}-\tilde p_a = p_i+p_j-p_a \equiv Q\,$. This time the branching is 
parameterised by the quantities
\begin{eqnarray}
x_{ij,a} = 
        \frac{p_ip_a+p_jp_a-p_ip_j}{p_ip_a+p_jp_a}\,,\quad
        \tilde z_i 
        = 1-\tilde z_j =
        \frac{p_ip_a}{p_ip_a+p_jp_a}\,.
\end{eqnarray}
The relative transverse momentum of the new emerging final-state partons is given by
        \begin{eqnarray}\label{eq:FIKTmassive}
        {\bf k}_\perp^2 
        = 
        2\tilde p_a\tilde p_{ij}\,\frac{1-x_{ij,a}}{x_{ij,a}}\,
                \tilde z_i\,(1-\tilde z_i) \,.
\end{eqnarray}
The derived Sudakov form factor for this splitting type reads
\begin{eqnarray}
\label{eq:FIsudakov}
        \lefteqn{\Delta_{\rm FI}({\bf k}_{\perp,{\rm max}}^2,{\bf k}_{\perp,0}^2)}\nonumber\\
        &=&
        \exp\left(-\sum\limits_{ij}\sum\limits_{a}\,
                \frac{1}{{\cal N}^{spec}_{ij}}\,
                \int\limits_{{\bf k}_{\perp,0}^2}^{{\bf k}_{\perp,{\rm max}}^2}\frac{\mbox{d} {\bf k}_{\perp}^2}{{\bf k}_\perp^2}\,
                \int\limits_{z_-}^{z_+}\mbox{d}\tilde z_i\,\frac{\alpha_{\rm s}({\bf k}_\perp^2)}{2\pi}\,
                \frac{f_a(\eta_a/x_{ij,a},{\bf k}_\perp^2)}{f_a(\eta_a,{\bf k}_\perp^2)}\,
                \langle {\bf V}^a_{ij}(\tilde z_i,x_{ij,a})\rangle\right)\,.\nonumber\\
\end{eqnarray}
Here, $\eta_a$ is the momentum fraction of the spectator parton $a$ and 
$f_a(\eta_a,{\bf k}_\perp^2)$ the corresponding hadronic PDF evaluated at some 
scale $\mu_F^2= {\bf k}_\perp^2$. The parton-distribution function 
$f_a(\eta_a/x_{ij,a},{\bf k}_\perp^2)$ accounts for the new incoming momentum.
The $\tilde z_i$ integration boundaries are given by Eq.~(\ref{eq:FFzboundaries}) and
the concrete splitting functions, $\langle {\bf V}^a_{ij}(\tilde z_i,x_{ij,a})\rangle$,  
can be found in Ref.~\cite{Schumann:2007mg}. The branching kinematics are fixed to 
\begin{eqnarray}
p_i 
&=& 
\hphantom{(1-\tilde)}\tilde z_i\,\tilde p_{ij} + 
\frac{{\bf k}_\perp^2}
        {\hphantom{(1-}\tilde z_i\, 2\tilde p_{ij}\tilde p_a\hphantom{))}}\,\tilde p_a + 
k_\perp\,,\\
p_j 
&=& 
(1-\tilde z_i)\,\tilde p_{ij} + 
\frac{{\bf k}_\perp^2}{(1-\tilde z_i)\,2\tilde p_{ij}\tilde p_a}\,\tilde p_a - k_\perp\,,
\end{eqnarray}
with $k_\perp$ perpendicular to both the emitter and the spectator momentum. 
The new spectator momentum is given by
\begin{eqnarray}
p_a &=& \frac{1}{x_{ij,a}}\tilde p_a\,.
\end{eqnarray}
\subsubsection{Initial-state emitter -- final-state spectator}
Once a final-state line is colour connected to the initial state, besides the situation
discussed in Sec.~\ref{sec:FI}, the reversed case occurs as well. Namely, the initial-state
line can split and emit a new final-state parton while the spectator is in the final state. 
The momentum-conservation condition for such a branching $\{\widetilde{ai},\tilde k\} 
\to \{a,i,k\}$ reads $\tilde p_k - \tilde p_{ai} = p_i+p_k-p_a \equiv Q\,$. The splitting 
variables are defined as 
\begin{eqnarray}
x_{ik,a}=\frac{p_ip_a+p_kp_a-p_ip_k}{p_ip_a+p_kp_a}\,,\quad 
u_i = \frac{p_ip_a}{p_ip_a+p_kp_a}\,,
\end{eqnarray}
and the transverse-momentum squared of parton $i$ with respect to the beam becomes
\begin{eqnarray}
{\bf k}_\perp^2 = 2\tilde p_{ai}\tilde p_k\,\frac{1-x_{ik,a}}{x_{ik,a}}\,u_i(1-u_i)\,.
\end{eqnarray}
The Sudakov form factor associated with this splitting type reads
\begin{eqnarray}
\lefteqn{\Delta_{\rm IF}({\bf k}_{\perp,{\rm max}}^2,{\bf k}_{\perp,0}^2)}\nonumber\\
&=& 
\exp\left(-\sum\limits_{ai}\sum\limits_{k}\,\frac{1}{{\cal N}^{spec}_{ai}}\,
\int\limits_{{\bf k}_{\perp,0}^2}^{{\bf k}_{\perp,{\rm max}}^2}\frac{\mbox{d} {\bf k}_{\perp}^2}{{\bf k}_\perp^2}\,
\int\limits_{x_-}^{x_+}\mbox{d} x_{ik,a}\,\frac{\alpha_{\rm s}({\bf k}_\perp^2/4)}{2\pi}\,
\tilde J(x_{ik,a},u_i;{\bf k}_\perp^2)\,
\langle {\bf V}_k^{ai}(x_{ik,a},u_i)\rangle\right)\,,\nonumber\\
\end{eqnarray}
with $x_-=\eta_{ai}$ and $x_+=Q^2/(Q^2+4{\bf k}_{\perp,0}^2)$ and
\begin{eqnarray}
\tilde J(x_{ik,a},u_i;{\bf k}_\perp^2) = 
\frac{1-u_i}{1-2u_i}\,\frac{1}{x_{ik,a}}\,
\frac{f_a(\eta_{ai}/x_{ik,a},{\bf k}_\perp^2)}{f_{ai}(\eta_{ai},{\bf k}_\perp^2)}\,,
\end{eqnarray}
accounting for a possible flavour change of the incoming line through the 
backward-evolution step. The complete list of splitting kernels can again be 
found in Ref.~\cite{Schumann:2007mg}. The branching kinematics are given by 
\begin{eqnarray}
p_a &=& \frac{1}{x_{ik,a}}\,\tilde p_{ai}\,,\\
p_i &=& (1-u_i)\,\frac{1-x_{ik,a}}{x_{ik,a}}\,\tilde p_{ai}+
                \hphantom{(1-))}u_i\,\tilde p_k+k_\perp\,,\\
p_k &=& \hphantom{(1-))}u_i\,\frac{1-x_{ik,a}}{x_{ik,a}}\,\tilde p_{ai}+
                (1-u_i)\,\tilde p_k-k_\perp\,.
\end{eqnarray}
\subsubsection{Initial-state emitter -- initial-state spectator}
The last case to be considered is the splitting of an initial-state line that is 
colour connected to the second incoming parton. The branching is parametrised through
\begin{eqnarray}
x_{i,ab}=\frac{p_ap_b-p_ip_a-p_ip_b}{p_ap_b}\,,\quad \tilde v_i = \frac{p_ip_a}{p_ap_b}\,,
\end{eqnarray}
such that the transverse-momentum squared of the new final-state parton becomes
\begin{eqnarray}
{\bf k}_\perp^2 = 2\tilde p_{ai}p_b\,\tilde v_i\,\frac{1-x_{i,ab}-\tilde v_i}{x_{i,ab}}\,.
\end{eqnarray}
The Sudakov form factor of this configuration reads 
\begin{eqnarray}
\lefteqn{\Delta_{\rm II}({\bf k}_{\perp,{\rm max}}^2,{\bf k}_{\perp,0}^2)}\nonumber\\
&=&
\exp\left(-\sum\limits_{ai}\sum\limits_{b\neq ai}\,\frac{1}{{\cal N}^{spec}_{ai}}\,
\int\limits_{{\bf k}_{\perp,0}^2}^{{\bf k}_{\perp,{\rm max}}^2}\frac{\mbox{d}{\bf k}_\perp^2}{{\bf k}_\perp^2}\,
\int\limits_{x_-}^{x_+}\mbox{d} x_{i,ab}\,\frac{\alpha_{\rm s}({\bf k}_\perp^2/4)}{2\pi}\,
\tilde J(x_{i,ab},\tilde v_i;{\bf k}_\perp^2)\,\langle 
{\bf V}^{ai,b}(x_{i,ab})\rangle\right)\,,\nonumber\\
\end{eqnarray}
with 
\begin{eqnarray}
\tilde J(x_{i,ab},\tilde v_i;{\bf k}_\perp^2) = 
\frac{1-x_{i,ab}-\tilde v_i}{1-x_{i,ab}-2\tilde v_i}\,\frac{1}{x_{i,ab}}\,
\frac{f_a(\eta_{ai}/x_{i,ab},{\bf k}_\perp^2)}{f_{ai}(\eta_{ai},{\bf k}_\perp^2)}\,,
\end{eqnarray}
and $x_- = \eta_{ai}$ and $x_+= 2\tilde p_ap_b/(2\tilde p_ap_b+4{\bf k}_{\perp,0}^2)$.
For the kinematics of the emission process it is convenient to keep the spectator
momentum fixed and to align the new incoming parton $a$ with the old incoming momentum
according to $p_a=1/x_{i,ab}\cdot\tilde p_{ai}$. The momentum of 
the newly emerged final-state parton $i$, is given by
\begin{eqnarray}
 p_i &=& \frac{1-x_{i,ab}-\tilde v_i}{x_{i,ab}}\,\tilde p_{ai}+
                        \tilde v_i\,p_b+k_\perp\,.
\end{eqnarray}
Its transverse momentum has to be balanced by the entire set of final-state 
particles of the $m$-parton process (including all non-QCD particles).

\subsubsection{The algorithm}

Having at hand factorised expressions for all possible emission processes
and corresponding Sudakov form factors a probabilistic shower algorithm of 
independent emissions can be formulated. The start seed forms a $2\to 2$ 
core event with fixed colour flow and a process dependent shower start
scale ${\bf k}_{\perp,{\rm max}}^2$. 

\begin{enumerate}
\item The scale of the next emission is chosen according to the Sudakov form
factors of all contributing emitter--spectator pairs. The dipole that yields
the highest transverse momentum is picked to split. 
\item The value of the second splitting variable is chosen according to the 
splitting kernel.
\item The splitting kinematics are determined, the new particle is inserted 
and the colour flow gets adapted.
\item Start from step 1 as long as ${\bf k}_\perp^2 > {\bf k}_{\perp,0}^2$ and 
replace ${\bf k}_{\perp,{\rm max}}^2$ by the transverse momentum of the last 
splitting.
\end{enumerate}
This yields a chain of subsequent emissions strictly ordered in transverse 
momenta. There is no formal subdivision of initial and final state evolution, 
instead, all dipoles are treated on equal footing.

\subsection{Comparison with experimental data}

The ultimate test of a theoretical model is a direct comparison with experimental 
measurements. Here we compare the newly developed and implemented 
parton-shower algorithm (called CS shower in the following) with some experimental 
data on hadron production in $e^+e^-$ annihilation, and Drell-Yan and jet production 
in $p\bar p$ collisions. Therefore the shower simulation has been supplemented with 
the string fragmentation routines of Pythia-6.2 \cite{Sjostrand:2001yu} to account 
for hadronisation.

\noindent
We begin with some of the most precisely measured quantities, event-shape 
observables in $e^+e^-$ annihilation at the $Z^0$ pole. Fig.~\ref{eshapes} contains 
a comparison for the normalised 1-Thrust ($1-T$) and C-parameter ($C$) distributions 
with LEP1 Delphi data \cite{Abreu:1996na}. Both observables obtain large 
higher-order corrections for two-jet like events that appear as $1-T \approx 0$ and 
$C\approx 0$. In addition, there is a singularity in the C-parameter also in the region
$C\approx 0.75$ that requires a resummation of large kinematical 
logarithms \cite{Catani:1997xc,GehrmannDeRidder:2007hr}. The CS shower yields a good 
agreement with the experimental data. Only very pencil like events, that are 
sensitive to hadronisation corrections, are overestimated in the Monte Carlo. 
We believe that this can be improved through a more detailed tuning of the 
hadronisation model parameters.

\begin{figure}
\begin{center}
\begin{tabular}{cc}
\includegraphics[width=0.5\textwidth]{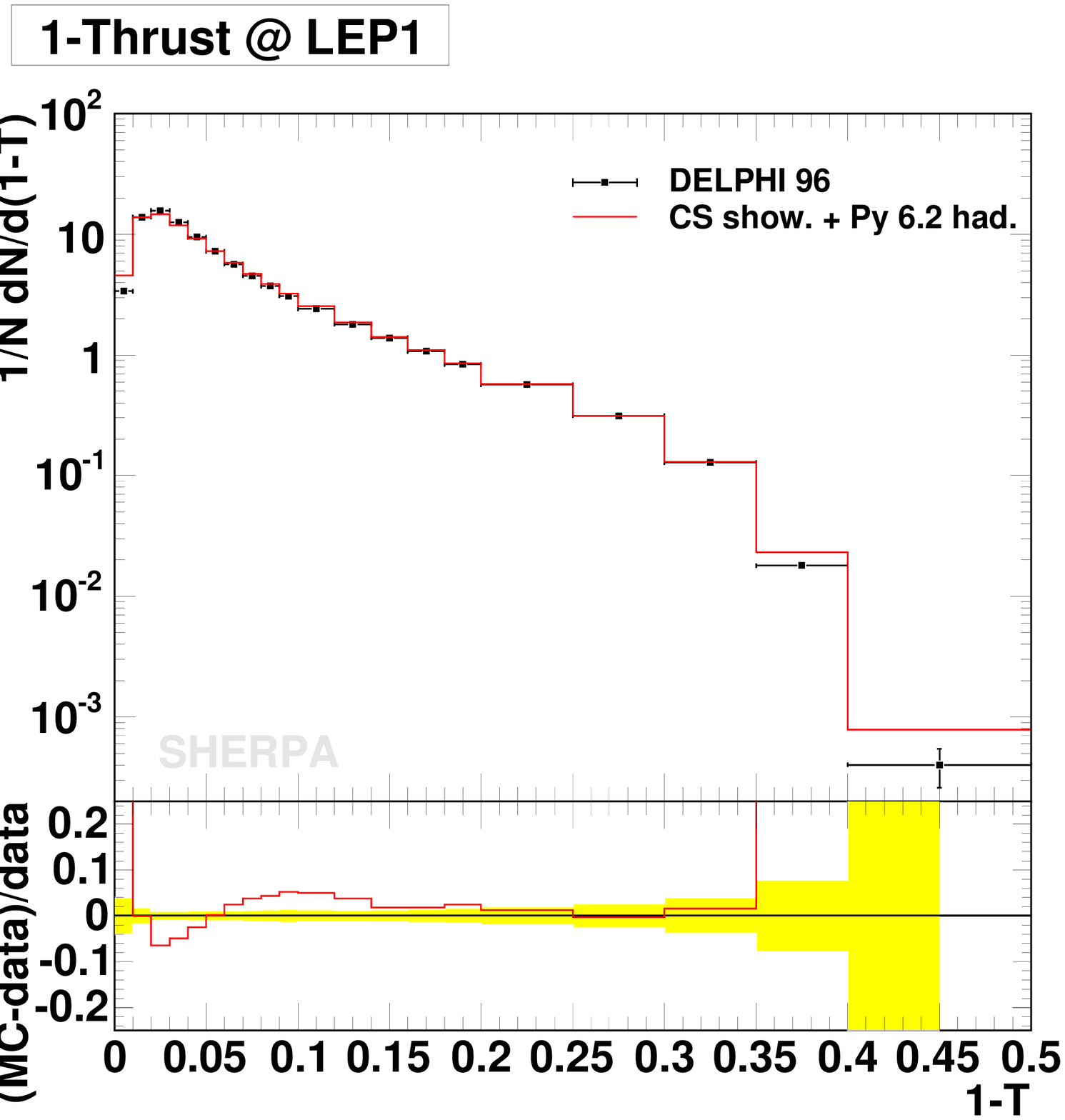}
&
\includegraphics[width=0.5\textwidth]{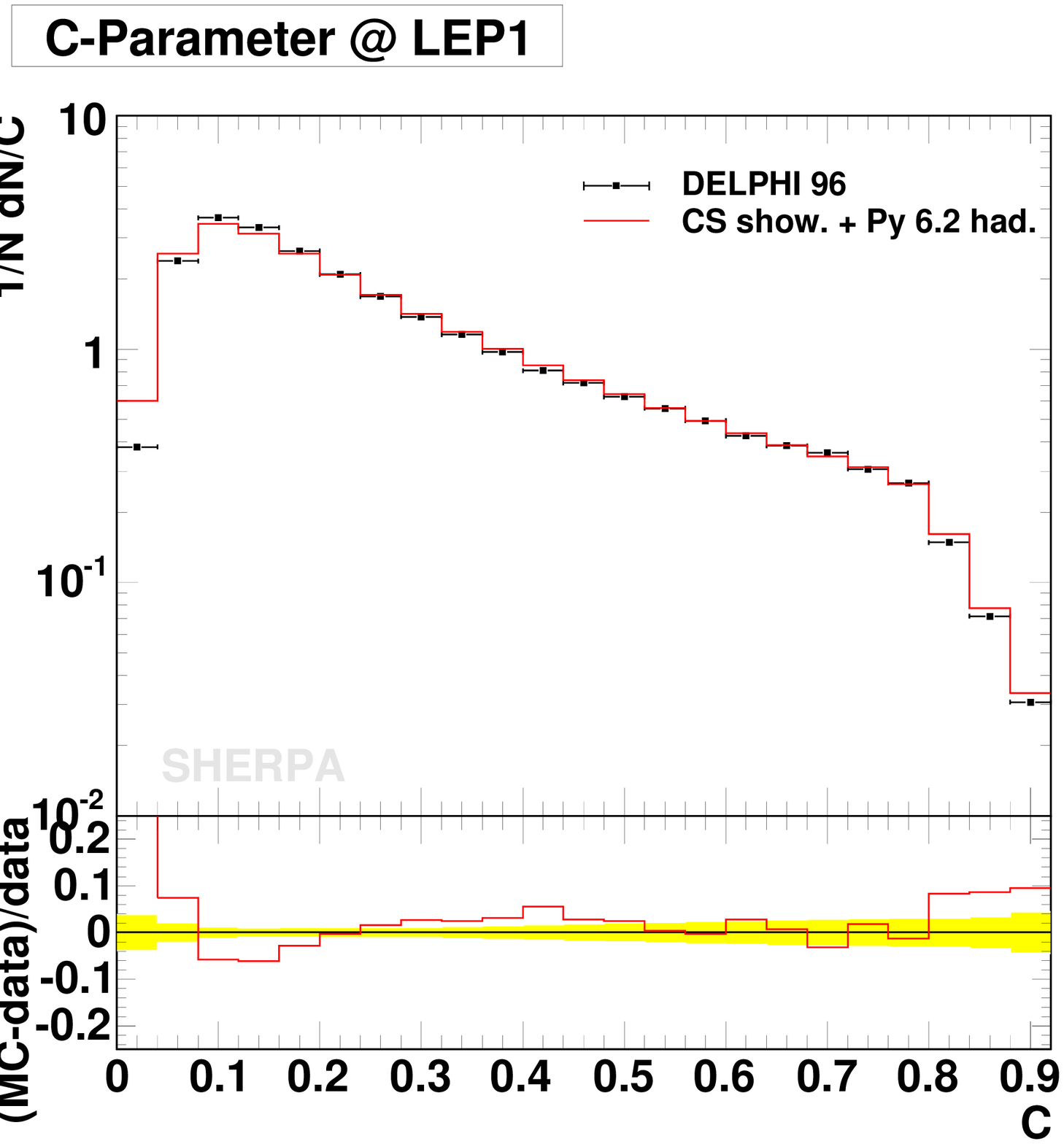}
\label{eshapes}
\end{tabular}
 \caption{The event-shape variables 1-Thrust $(1-T)$ and C-parameter $(C)$
in comparison with Delphi LEP1 data \cite{Abreu:1996na}.}
\end{center}
\end{figure}

\begin{figure}
\begin{center}
\begin{tabular}{cc}
\label{ptZ_CDF}
\includegraphics[width=0.5\textwidth]{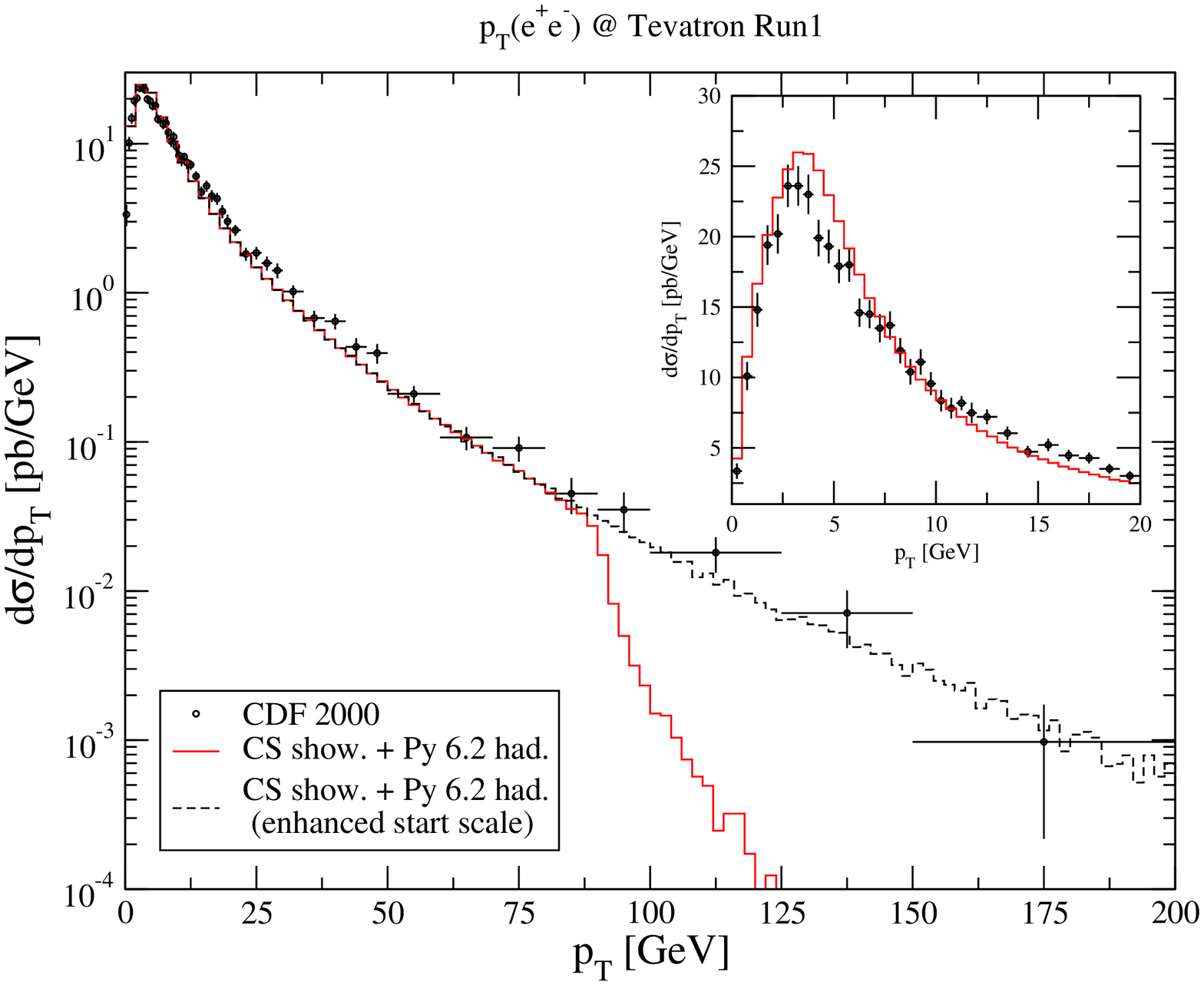}
&
\includegraphics[width=0.5\textwidth]{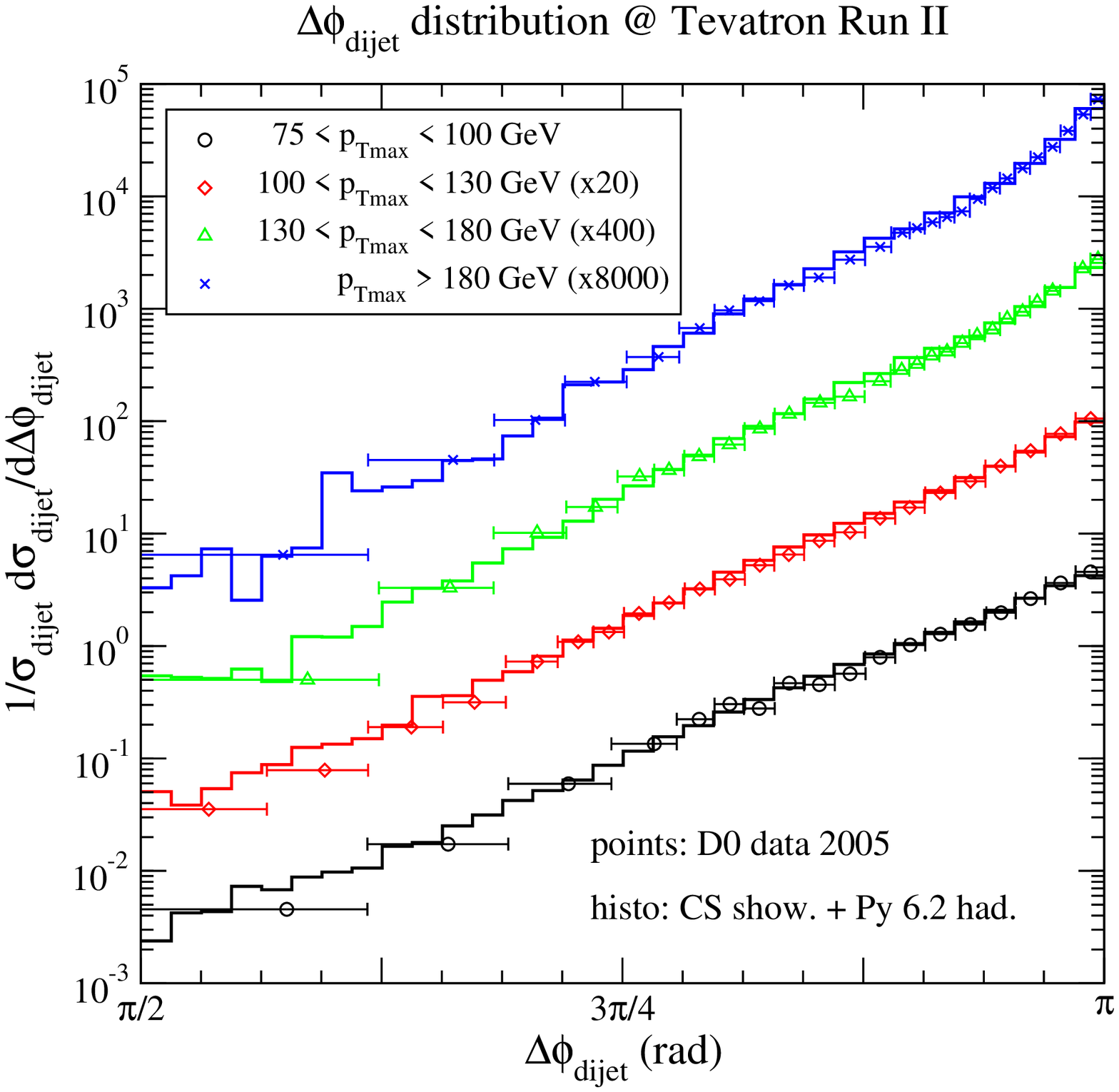}
\end{tabular}
\caption{The transverse-momentum distribution of $e^+e^-$ Drell-Yan pairs 
compared with CDF Run I data \cite{Affolder:1999jh} (left panel) and the
Dijet Azimuthal decorrelation measured by D\O\ at Tevatron Run II 
\cite{Abazov:2004hm} (right panel).}
\end{center}
\end{figure}

\noindent 
In Fig.~\ref{ptZ_CDF} we present the predictions of our model for the lepton-pair
transverse-momentum distribution in Drell-Yan production and for the azimuthal
decorrelation of inclusive dijet events in $p\bar p$ collisions. Both observables are 
nontrivial only if additional QCD radiation is produced and thereby test the emission 
pattern of the shower ansatz. We observe a good agreement with data for both observables 
in phase-space regions dominated by rather soft or collinear emissions but the agreement 
outside this range, i.e. large $p^Z_T$ or small $\Delta\phi_{\rm dijet}$, is also very 
satisfactory.


\subsection{Conclusions}

We have presented a new parton-shower algorithm that uses fully factorised
versions of the Catani--Seymour dipole functions to describe multi-parton
production processes in a probabilistic manner. The model encodes exact 
four-momentum conservation on the level of each individual splitting due to 
the notion of splitting emitter--spectator pairs. Subsequent emissions are
ordered in transverse momenta and the evolution of initial- and final-state 
partons is done in a unified way. Comparison with experimental data yields
very encouraging results. In a next step we will combine this new shower 
approach with exact multi-leg tree-level matrix elements. 
Moreover, this model should facilitate a matching with exact NLO QCD 
calculations.   
 
\subsection*{Acknowledgements}
We would like to thank Zoltan Nagy and Davison Soper for fruitful discussions.
S. Schumann would like to thank the organisers of the Les Houches workshop.

%
}


\clearpage

\bibliography{nlm07}

\end{document}